\documentclass[11pt]{article}

\pdfoutput=1 % if your are submitting a pdflatex (i.e. if you have images in pdf, png or jpg format)
\usepackage{./aux/jheppub}
\usepackage[utf8]{inputenc}
\usepackage{graphicx,tikz}
\usepackage{amsfonts,amsmath,amssymb}
\usepackage{esint,array,mathrsfs,yfonts,dsfont,bbm,colonequals,amscd,mathtools}
\usepackage{relsize,suffix,cancel,bbm,capt-of,upgreek,verbatim}

\usepackage{tikz}
\usetikzlibrary{trees}
\usetikzlibrary{decorations.pathmorphing}
\usetikzlibrary{decorations.markings}
\usetikzlibrary{shapes.misc}

\newcommand{\mS}{\mathrm{s}}
\newcommand{\mT}{\mathrm{t}}
\newcommand{\mU}{\mathrm{u}}

\newcommand{\be}{\begin{equation}}
\newcommand{\ee}{\end{equation}}
\newcommand{\ba}{\begin{equation}\begin{aligned}}
\newcommand{\ea}{\end{aligned}\end{equation}}

\newcommand{\zb}{\bar{z}}
\newcommand{\wb}{\bar{w}}
\newcommand{\Wb}{\bar{W}}
\newcommand{\hb}{\bar{h}}

\newcommand{\Df}{\Delta_{\phi}}

\newcommand{\cO}{\mathcal{O}}
\newcommand{\cP}{\mathcal{P}}
\newcommand{\cF}{\mathcal{F}}
\newcommand{\cG}{\mathcal{G}}
\newcommand{\Disc}{\mathrm{Disc}}
\newcommand{\dDisc}{\mathrm{dDisc}}
\renewcommand{\Re}{\mathrm{Re}}

\usepackage{mathtools}          %loads amsmath as well

\DeclareMathOperator*{\Res}{Res}

\colorlet{darkblue}{blue!70!black}
\colorlet{darkgreen}{green!70!black}
\colorlet{darkred}{red!70!black}

\title{Dispersive CFT Sum Rules}

\preprint{
YITP-SB-20-24\\
\vspace{-0.34in}
\begin{flushright}CALT-TH 2020-034\end{flushright}
}

\author[a]{Simon Caron-Huot,\!}
\author[b, c]{Dalimil Maz\'{a}\v{c},\!}
\author[c]{Leonardo Rastelli,\!}
\author[d]{and David Simmons-Duffin}

\affiliation[a]{Department of Physics, McGill University, 3600 Rue University, Montr\'eal, QC Canada}
\affiliation[b]{Simons Center for Geometry and Physics, Stony Brook University,
Stony Brook, NY 11794, USA}
\affiliation[c]{C. N. Yang Institute for Theoretical Physics, Stony Brook University,
Stony Brook, NY 11794, USA}
\affiliation[d]{Walter Burke Institute for Theoretical Physics, Caltech, Pasadena, CA 91125, USA }

\abstract{
We give a unified treatment of  dispersive sum rules for four-point correlators in conformal field theory. We call a sum  rule ``dispersive'' if it has double zeros at all double-twist operators above a fixed twist gap.
Dispersive sum rules have their conceptual origin in Lorentzian kinematics and absorptive physics (the notion of double discontinuity). 
They have been  discussed using three seemingly different methods: analytic functionals dual to double-twist operators,  dispersion relations in position space, and dispersion relations in Mellin space.
We show that these three approaches can be mapped into one another and lead to completely equivalent sum rules. A central idea of our discussion 
is a fully nonperturbative expansion of the correlator as a sum over  {\it Polyakov-Regge blocks}.
Unlike the usual OPE sum, the Polyakov-Regge expansion utilizes
the data of {\it two} separate channels, while having (term by term) good Regge behavior in the third channel.
We construct sum rules which are non-negative above the double-twist gap; they have the physical interpretation
of a subtracted version of  ``superconvergence'' sum rules.
We expect dispersive sum rules to be a very useful tool to study expansions
around mean-field theory, and to constrain the low-energy description of holographic CFTs with a large gap. We give examples of the first kind of applications, notably we
exhibit a candidate extremal functional for the spin-two gap problem. 
}


\usepackage[T1]{fontenc} 
\usepackage{dsdshorthand}

\begin{document} 

\maketitle
\flushbottom

%%%%%%%%%%%%%%%%%%%%%%%%%%%%%%%%

% !TEX root = ../main.tex

\section{Introduction}\label{sec:Introduction}

The modern conformal bootstrap program was kindled by the observation~\cite{Rattazzi:2008pe} that the constraints of unitarity and crossing, even when applied to a small number of correlators, are surprisingly effective in carving out the space of consistent CFTs.  Unitarity allows to replace the infinite set of bootstrap equalities 
with a finite set of rigorous inequalities, which can be studied numerically. The numerical bootstrap has  matured into a very  powerful and flexible toolkit, which can both put general constraints on theory space and determine  the low-lying data of specific models with unprecedented accuracy. See~\cite{Poland:2016chs, Poland:2018epd, Chester:2019wfx} for reviews and e.g.~\cite{Afkhami-Jeddi:2019zci,Homrich:2019cbt,Agmon:2019imm,Lin:2019vgi,Rong:2019qer,Reehorst:2019pzi,Chester:2019ifh,Henriksson:2020fqi,He:2020azu,Li:2020bnb,Afkhami-Jeddi:2020hde,Bonifacio:2020xoc} for a partial list of recent results.  

 \smallskip

In the past several years we have also gained a much better {\it analytic} understanding of the bootstrap constraints.  An important circle of ideas revolve around Lorentzian kinematics and in particular the study of commutators,
which are nonvanishing only at timelike separation.
Commutators suppress many contributions to the OPE and allow to focus on the irreducible physics.
A particularly fruitful construct has been the expectation value of a commutator squared.
This expectation value
enjoys especially useful properties: it is positive, and bounded, even in
extreme kinematics such as the Regge limit (large boost). Schematically:
\be
 0\leq -\frac{1}{4}\<[X,Y]^2\> \leq \langle XYYX\rangle\,.
\ee
These twin properties have been the key to a number of analytic results, such as
the bound on chaos \cite{Maldacena:2015waa} and a proof of the average null energy condition (ANEC) \cite{Hartman:2016lgu}.
The proof exploits that the ANEC operator is a universal contribution in the lightcone OPE of two scalars,
whose contribution satisfies a sum rule expressing it in terms of the commutators squared,
establishing its positivity.  The sum rule converges precisely thanks to the upper bound.
Physically, the commutator squared has been interpreted as an absorption probability and
the upper bound is simply the conservation of probability.

\smallskip
Another prominent example is large-spin perturbation theory \cite{Alday:2007mf, Fitzpatrick:2012yx, Komargodski:2012ek, Alday:2013cwa, Alday:2015eya, Alday:2015ewa, Alday:2016njk, Alday:2016jfr, Simmons-Duffin:2016wlq}.  Any CFT at large spin
contains multi-twist families which behave approximately like non-interacting products of simpler operators. 
Their properties can be predicted quantitatively by studying lightcone singularities \cite{Fitzpatrick:2012yx, Komargodski:2012ek}. It turns out that
carefully chosen double commutators precisely capture those singularities while discarding regular terms;
in turn, the vacuum expectation values of those double commutators
are easily computed by extracting ``double discontinuities'' (dDisc) of the four-point function. 
Remarkably,
no actual information is lost:
the position-space dispersion relation derived in~\cite{Carmi:2019cub}
 reconstructs the full correlator as an integral transform of two independent
 double discontinuities (around the cuts of two of its three OPE channels).\footnote{This is true as stated only for correlators that are ``Regge superbounded''. Dispersion relations for physical correlators, which in general are ``just bounded'' in the Regge limit, require subtractions. We will briefly come back to this important technical point below in the Introduction, and discuss it at full length in Section~\ref{sec:Subtractions} of the paper.} Schematically,
 \be
 \cG(z, \bar z) =
 \int_{w, \bar w}  K_s(z, \bar z ;w, \bar w )\, \dDisc_s[\cG(w,\wb)] +  \int_{w, \bar w} K_t(z, \bar z ;w, \bar w )\, \dDisc_t[\cG(w,\wb)]  \, , 
 \ee
 where $K_{s, t}$ are certain explicit kinematic kernels in cross-ratio space.
A closely related result is the  Lorentzian inversion formula (LIF)~\cite{Caron-Huot:2017vep, Simmons-Duffin:2017nub, Kravchuk:2018htv}, which reconstructs the coefficient function $c(\Delta, J)$ of the conformal partial wave expansion in a given OPE channel from the double discontinuities around  the other two channels.
What is more, the LIF reveals a powerful organizing principle for the spectrum a  unitary CFT: all operators lie in a set of Regge trajectories, analytic in spin (see e.g.~\cite{Simmons-Duffin:2016wlq, Albayrak:2019gnz, Liu:2020tpf, Caron-Huot:2020ouj} for several concrete illustrations of this idea). The LIF has been given a  physical interpretation and extended to spinning operators through the novel concept of light-ray operators \cite{Kravchuk:2018htv}, a vast generalization of the average null energy operator.\footnote{A variety of other results can be interpreted as sum rules arising from similar Lorentzian kinematics, e.g.~\cite{Cordova:2017zej, Gillioz:2018kwh}.}

\smallskip

A second, parallel line of work has been the development of exact ``double-twist functionals''. A complete basis of double-twist functionals was constructed recently  for CFTs in general spacetime dimension  \cite{Mazac:2019shk}, building on previous foundational work \cite{Mazac:2016qev, Mazac:2018mdx, Mazac:2018ycv, Mazac:2018qmi} in $d=1$ CFTs. In \cite{Hartman:2019pcd}, it was shown that these functionals can also be used for modular bootstrap, and to solve the sphere-packing problem in 8 and 24 dimensions.\footnote{See also \cite{Kaviraj:2018tfd, Mazac:2018biw} for the
boundary CFT case.} The defining property of these functionals is that they exhibit double zeros on double-twist operators of mean field theory. The action of each of these functionals can be represented in terms of double-contour integral (in the space of complexified cross ratios $z$, $\bar z$) with a suitable kernel, with a contour prescription that mimics the double discontinuity.\footnote{The idea to define $d>1$ functionals as double contour integrals appeared independently in \cite{Paulos:2019gtx} and examples of genuine extremal functionals were also constructed there. Another intersting example of an extremal functional in the two-variable situation appeared in the context of the modular bootstrap \cite{Afkhami-Jeddi:2020hde}.} 
 As already noted in \cite{Mazac:2019shk}, there appears to be a close relationship of this approach with the position space dispersion relation reviewed above.

  \smallskip

A third track of the analytic bootstrap  takes place in Mellin space \cite{Mack:2009mi, Penedones:2010ue, Fitzpatrick:2011ia}. 
   The Mellin amplitude   ${\cal M} (\mS, \mT)$  is a rather close CFT analog of the flat space S-matrix. This analogy is particularly sharp  for holographic CFTs but holds more generally.
The authors of~\cite{Penedones:2019tng} have studied the validity of the Mellin representation for general CFTs and derived 
non-perturbative dispersion relations in Mellin space, analogous to the familiar dispersion relations obeyed by the S-matrix.  They have obtained interesting  sum rules for CFT data by insisting  (as befits a generic interacting theory) that the spectrum does {\it not} contain operators with exact  double-twist quantum numbers. This spectral assumption translates  in what they dub the ``non-perturbative Polyakov conditions'' for the Mellin amplitude.

\smallskip

\begin{figure}
\centering{\def\svgwidth{8.2cm}%% Creator: Inkscape 1.0beta1 (32d4812, 2019-09-19), www.inkscape.org
%% Accompanies image file 'triangle.pdf' (pdf, eps, ps)
\begingroup%
  \makeatletter%
  \providecommand\color[2][]{%
    \errmessage{(Inkscape) Color is used for the text in Inkscape, but the package 'color.sty' is not loaded}%
    \renewcommand\color[2][]{}%
  }%
  \providecommand\transparent[1]{%
    \errmessage{(Inkscape) Transparency is used (non-zero) for the text in Inkscape, but the package 'transparent.sty' is not loaded}%
    \renewcommand\transparent[1]{}%
  }%
  \providecommand\rotatebox[2]{#2}%
  \newcommand*\fsize{\dimexpr\f@size pt\relax}%
  \newcommand*\lineheight[1]{\fontsize{\fsize}{#1\fsize}\selectfont}%
  \ifx\svgwidth\undefined%
    \setlength{\unitlength}{389.99998558bp}%
    \ifx\svgscale\undefined%
      \relax%
    \else%
      \setlength{\unitlength}{\unitlength * \real{\svgscale}}%
    \fi%
  \else%
    \setlength{\unitlength}{\svgwidth}%
  \fi%
  \global\let\svgwidth\undefined%
  \global\let\svgscale\undefined%
  \makeatother%
  \begin{picture}(1,0.51923079)%
    \lineheight{1}%
    \setlength\tabcolsep{0pt}%
    \put(0.34771306,0.42023627){\makebox(0,0)[lt]{\lineheight{1.25}\smash{\begin{tabular}[t]{l}Position space\end{tabular}}}}%
    \put(0,0){\includegraphics[width=\unitlength,page=1]{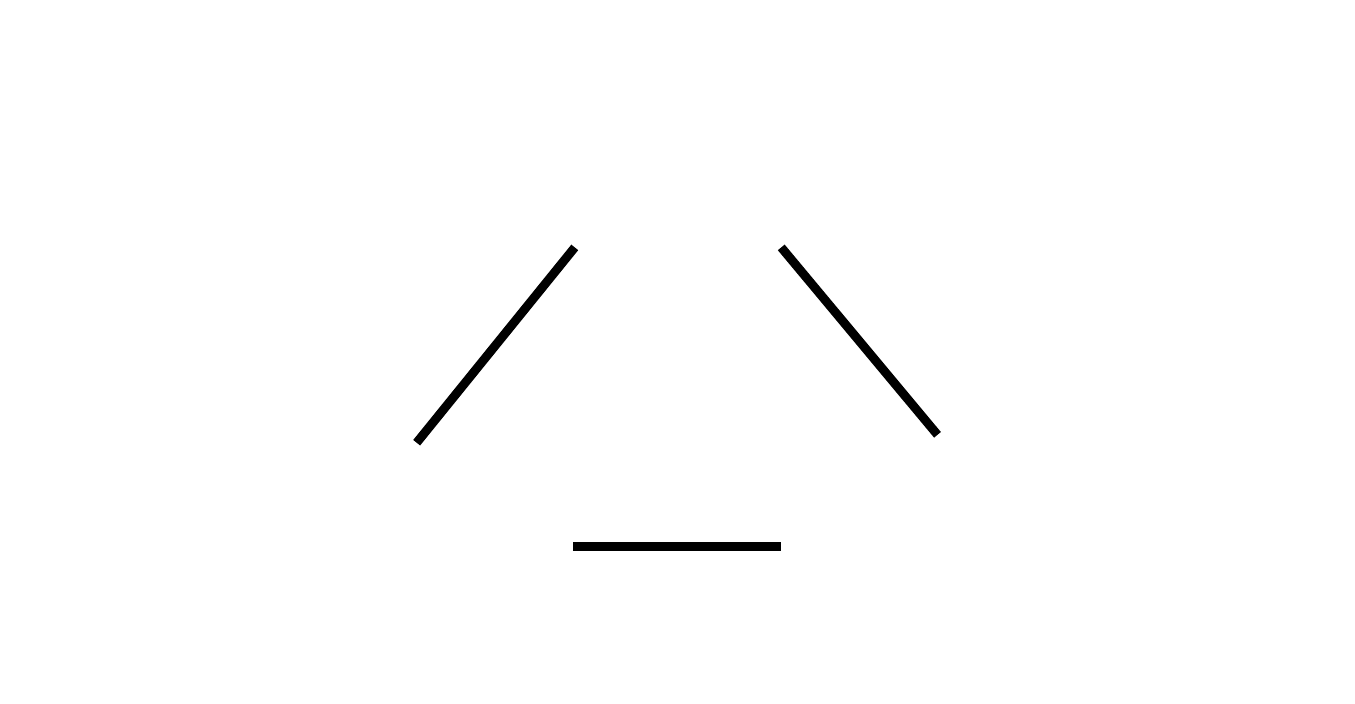}}%
    \put(0.07133096,0.11926982){\makebox(0,0)[lt]{\lineheight{1.25}\smash{\begin{tabular}[t]{l}Double twists\end{tabular}}}}%
    \put(0.65049454,0.11860316){\makebox(0,0)[lt]{\lineheight{1.25}\smash{\begin{tabular}[t]{l}Mellin space\end{tabular}}}}%
    \put(0,0){\includegraphics[width=\unitlength,page=2]{triangle.pdf}}%
    \put(0.44529061,0.36481565){\makebox(0,0)[lt]{\lineheight{1.25}\smash{\begin{tabular}[t]{l}$(z,\zb)$\end{tabular}}}}%
    \put(0,0){\includegraphics[width=\unitlength,page=3]{triangle.pdf}}%
    \put(0.15682906,0.06673873){\makebox(0,0)[lt]{\lineheight{1.25}\smash{\begin{tabular}[t]{l}$(n,\ell)$\end{tabular}}}}%
    \put(0.72480339,0.06834797){\makebox(0,0)[lt]{\lineheight{1.25}\smash{\begin{tabular}[t]{l}$(\mS,\mT)$\end{tabular}}}}%
    \put(0,0){\includegraphics[width=\unitlength,page=4]{triangle.pdf}}%
  \end{picture}%
\endgroup%
}
\caption{\label{fig:triangle}
Three natural spaces in which we consider conformal bootstrap constraints:
Position space $(z,\zb$), the set of double-twist labels $(n,\ell)$ and Mellin space $(\mS,\mT)$.
Dispersion relations in position space and Mellin space will turn out to be equivalent to each other,
and to provide generating functions for functionals dual to double-twist operators.
}
\end{figure}

In summary, 
three analytic approaches to non-perturbative bootstrap sum rules have been developed recently. They are formulated in three different spaces (Figure \ref{fig:triangle}):
\begin{enumerate}
\item[(i)]  Position space $(z, \bar z)$. Sum rules are obtained by  requiring compatibility of  the position space dispersion relation \cite{Carmi:2019cub} with crossing symmetry. 
\item[(ii)] Double-twist ``space'' $(n, \ell)$, where $n$ and $\ell$ are the non-negative integers that label
double-twist operators $[{\cal O}_1 \Box^{n} \partial^\ell {\cal O}_2]$. Sum rules are obtained by applying the dual double-twist functionals \cite{Mazac:2019shk} to the crossing equation.
\item[(iii)] Mellin space $(\mS, \mT)$.  Sum rules follow from the Mellin dispersion relations  \cite{Penedones:2019tng}  by imposing  crossing and the ``non-perturbative Polyakov conditions''.
\end{enumerate}
These three methods  have a clear family resemblance. They all rely (more or less directly) on Lorentzian kinematics and the notion of dDisc. Consequently, they all lead to sum rules with double-zeros on all mean field theory double-twist operators above a certain minimal twist,  $n \geq n_0$.  We will refer to sum rules of this kind (in the precise sense just stated)
as ``dispersive''.
The goal of this paper is to systematically classify and study dispersive sum rules. We will show that the three approaches  outlined above can be precisely mapped into one another and lead to completely equivalent sum rules. In particular, 
  the Mellin space sum rules of  \cite{Penedones:2019tng} can all be understood in the conventional language of analytic functionals, and follow from nothing more than the usual requirements of unitarity and crossing -- there is no need to make the additional spectral assumptions encoded in the non-perturbative Polyakov conditions. 

\smallskip

A common thread of our discussion will be an alternative, fully nonperturbative expansion of the correlator, as a sum over  {\it Polyakov-Regge blocks} \cite{Sleight:2019ive, Mazac:2019shk}.\footnote{The term ``Polyakov-Regge block'' was introduced in \cite{Mazac:2019shk}.  Polyakov-Regge blocks have good Regge behavior 
 in one channel (in the conventions of the present paper, the u-channel). Essentially the same notion appeared independently in \cite{Sleight:2019ive}, where it was called ``cyclic Polyakov block''. More precisely, the cyclic Polyakov block of \cite{Sleight:2019ive} is the sum of the s-channel and the t-channel Polyakov-Regge blocks of \cite{Mazac:2019shk}. This  is to be contrasted with the ``Polyakov blocks'' of \cite{Gopakumar:2016wkt, Dey:2016mcs, Dey:2017fab, Dey:2017oim, Gopakumar:2018xqi}, which are fully crossing symmetric but are not superbounded in Regge limit if the exchanged spin is greater than zero.} 
Unlike the usual OPE sum, the Polyakov-Regge expansion utilizes
the data of {\it two} separate channels, while having (term by term) good Regge behavior in the third channel.\footnote{Drawing on a suggestion in Polyakov's classic paper \cite{Polyakov:1974gs}, 
a fully crossing symmetric version of the Polyakov expansion has been proposed and developed in \cite{Gopakumar:2016wkt, Dey:2016mcs, Dey:2017fab, Dey:2017oim, Gopakumar:2018xqi}. It has been very inspirational for our work. We will comment on it in Section \ref{ssec:symmetricPolyakov}}
We will explain that the Polyakov-Regge expansion follows by expanding the dispersion relation using the OPE, and is thus rigorously established.

\smallskip

We expect dispersive sum rules  to be very useful for at least two broad classes of physical problems.
First, in mean field theory all the operators contributing to the OPE of a four-point correlator
are double-twists. Sum rules with zeros on all double-twists are then trivially satisfied,
making them a natural starting point to construct extremal functionals that implement expansions
around mean-field theory. A very natural conjecture is that mean field theory maximizes the twist gap for every even spin $\ell \geq 2$, and it is of great interest to construct the corresponding extremal functionals.
Second, in holographic theories, a large gap separates
the first higher-spin single-trace operator from the light spectrum.
Double-trace operators below the gap dominate the OPE, and projecting them out is a prerequisite
to access physics above the gap and study the implications of UV unitarity to low-energy physics.
This projection is automatically carried out by the commutator squared.
In this paper we develop the general formalism and give examples of  the first class of applications.
The application of  dispersive sum rules 
to AdS effective field theory and the holographic bootstrap is the subject 
of a forthcoming article \cite{CHMRSD}.

\smallskip

In order to discuss physical applications, we need to overcome an additional technical hurdle.   Physical correlators are bounded in the Regge limit \cite{Caron-Huot:2017vep}, but the simplest versions
of the dispersive sum rules discussed above only apply to ``superbounded''  functions with a slightly improved Regge behavior, see (\ref{superbounded}).  
Indeed,  the position space dispersion relation  uniquely reconstructs superbounded correlators from their double discontinuities;  equivalently, the LIF applies all the way down to $J = 0$. In Mellin space, superboundedness
allows to drop the arc at infinity in the standard contour manipulation of the Cauchy formula that  yields a dispersion relation. Finally, when acting on superbounded functions the functionals dual to double-twists are valid functionals which can be ``swapped'' with the conformal block expansion. To derive valid sum rules for a physical correlator, we need ``subtracted'' dispersion relations, i.e.~dispersion relations for 
a rescaled correlator with improved Regge behavior.  Subtractions can be implemented in each of the three spaces, see Figure~\ref{fig:triangle sub}. In practice, we have found that Mellin space is a good starting point to motivate subtraction schemes that can then be translated into explicit   functionals defined by position-space kernels. While the nonpertubative validity of the Mellin  representation is very subtle~\cite{Penedones:2019tng}, it is generally easy to demonstrate rigorously the ``swappability'' of position-space functionals and to analyze other properties such as positivity.

\smallskip
For many applications, it is essential consider functionals whose action is {\it non-negative} above a fixed twist gap. It turns out that the individual functionals dual to double-twists 
do not enjoy this property, but we have found infinite linear combinations that do. 
One of our main concrete results is the explicit construction of a one-parameter family of swappable functionals,\footnote{The subscript ``2'' indicates that these are  ``twice-subtracted'' functionals; $v$ is the parameter. Non-negativity above the double-twist gap holds for $v \geq 1$.} called $B_{2, v}$, that are non-negative above the double-twist gap, i.e. for $\Delta  \geq 2 \Delta_\phi +J$, where $\Delta_\phi$ is the external dimension.
The $B_{2, v}$ sum rule can be motivated by a simple subtraction  in Mellin space, and it has a  neat  physical interpretation as a subtracted version of the ``superconvergence'' sum rule of~\cite{Kologlu:2019bco}.

\smallskip

Using the $B_{2, v}$  family as a basic toolkit,  we can start looking for functionals that answer interesting physical questions. First, and most straightforwardly,  we exhibit a swappable functional that is (experimentally) non-negative on all scalar primaries and on all spinning primaries above the double-twist gap, with zeros at the double-twists. The corresponding sum rule is automatically satisfied by mean field theory, and shows that any other CFT 
 must contain a spinning primary {\it below} the double-twist gap --  an eminently plausible (if somewhat weak) result. 
 A more challenging task is to disentangle the spin dependence. We should construct, for every even spin $\ell \geq 2$, an extremal functional 
 for the spin-$\ell$ gap problem, which as we have mentioned
is expected to be maximized by mean-field theory.  
Such an extremal functional 
 is required to have zeros at all mean-field operators 
 and be non-negative everywhere above the unitarity bound, except in the spin $\ell$ sector below the lowest double twist. For  $\ell =2$, we
   exhibit a candidate functional $\Phi_2$ that appears to do the job in some range of  $\Delta_\phi$ and  $d$.
   We strongly believe (but have not rigorously shown) that  it is a swappable functional, and  have checked numerically that it has the requisite positivity properties. 
   We have also compared the analytic functional $\Phi_2$ with an approximate extremal functional for the spin-two gap problem obtained by  numerical bootstrap methods, and found spectacular agreement.\footnote{As we will explain in detail, agreement is expected (and found) only for the action on scalar blocks -- there are several distinct extremal functionals for the spin-two gap problem, which differ in their action on spinning operators but must agree on scalars.}   
   When applied to the   $\langle \sigma \sigma \sigma \sigma \rangle$ correlator of the 3D Ising model, whose low-lying data are very precisely known from the numerical bootstrap,
 the $\Phi_2$ sum rule  converges rapidly, being saturated at the 95\% level just by the stress tensor and $\epsilon$ exchanges. We will comment how $\Phi_{2}$ and other dispersive functionals can be used to set up a version of the lightcone bootstrap with rigorously controlled errors.

\bigskip
\noindent
The organization of the paper is most accurately gleaned from the Table of Contents.  In Section~\ref{sec:Vertices} we review  three analytic bootstrap methods (Figure \ref{fig:triangle}) and show their complete equivalence, emphasizing the unifying theme of the Polyakov-Regge bootstrap.  In Section~\ref{sec:Edges} we discuss sum rules  in the simplified setting of superbounded correlators. 
In Section~\ref{sec:Subtractions} we discuss subtracted dispersion relations and sum rules valid for physical correlators; in particular, we construct and study the $B_{2, v}$ family of sum rules.
In Section~\ref{sec:AppliedSubtractions} we tie a few conceptual loose ends, related to the tension between Regge boundedness and full crossing symmetry:  we show how to classify contact diagrams using functionals; we relate our formalism to the lightcone bootstrap; and we comment on the s-t-u  symmetric Polyakov bootstrap.
 In Section~\ref{sec:MFTbounds} we apply our new tools
to the construction of extremal functionals that show optimality of mean field theory for various maximization problems, notably we propose an extremal functional $\Phi_2$ for the spin-two gap problem.
We conclude in Section~\ref{sec:Conclusions} with a discussion and open questions.

Several technical appendices complement the main text. Appendix \ref{app:contours} contains computations demonstrating the validity of the position space dispersion relation. In Appendix \ref{app:decompositions}, we illustrate the position space dispersion relation on simple example correlators. In Appendix \ref{app:mellinFunctionals}, the position space kernel of the Mellin space dispersion relation is computed. In Appendix \ref{app:numericalbtwov}, we explain a method to efficiently evaluate the action of dispersive functionals on conformal blocks. Another general method, based on weight-shifting operators, is explained in Appendix \ref{app:wsrecursion}. Finally, Appendix \ref{app:extremalfunctionalnumerics} includes the details of our numerical implementation of the twist gap maximization problem.

% !TEX root = ../main.tex

\section{Three approaches to dispersive CFT sum rules}\label{sec:Vertices}

In this section we review three analytic methods to study CFT correlators: the position space dispersion relation  \cite{Carmi:2019cub}, the basis of analytic functionals \cite{Mazac:2019shk}, and the Mellin space dispersion relation \cite{Penedones:2019tng}.
We argue that these three approaches are just different ways to encode the constraints of unitarity and crossing
into dispersive functionals, and are completely equivalent to one other.
A unifying idea is that of the {\it Polyakov-Regge bootstrap}:
all three formalisms lead naturally to an expansion of the correlator as a sum over ``Polyakov-Regge blocks''. 
Unlike the familiar conformal block expansion, the Polyakov-Regge expansion utilizes
the data of {\it two} separate OPE channels and manifests good Regge behavior in a third channel.

\smallskip
 
The aim of this section is to present the three different formalisms and describe their relations.   In the next section we
use this machinery to derive non-perturbative sum rules for CFT data.
For simplicity, both in this section and the next we restrict attention to  correlators that are  ``superbounded'' in the u-channel Regge limit. The extension to the physically relevant case of correlators that are
``just'' Regge bounded involves a few additional technicalities, and is postponed to Section \ref{sec:Subtractions}.

%%%%%%%%
\subsection{Preliminaries}

We focus on four-point functions of (non-necessarily identical) scalar operators of equal conformal dimension $\Delta_\phi$. We write it as
\be
\left\langle\phi_1(x_1)\phi_2(x_2)\phi_3(x_3)\phi_4(x_4)\right\rangle = \frac{\cG(z,\zb)}{(x_{13}^2 x_{24}^2)^{\Df}}\, ,
\ee
with $z$  and $\zb$ the usual cross ratios
\be
u =  \frac{x_{12}^2 x_{34}^2}{x_{13}^2 x_{24}^2} = z \bar z \, , \quad v =  \frac{x_{14}^2 x_{23}^2}{x_{13}^2 x_{24}^2} = (1-z)(1- \bar z) \, . \label{cross-ratios}
\ee
The correlator  $\cG(z,\zb)$ admits  conformal block expansions in  three (a priori inequivalent) channels,
\begin{equation} \label{OPE}
{\cal G}(z,\zb) =  \sum\limits_{\cO}
a_{\cO}\,
G^{s}_{\Delta_{\cO},J_\mathcal{O}}(z,\bar{z})=
\sum\limits_{\mathcal{P}} a_{\mathcal{P}}\,
G^{t}_{\Delta_{\mathcal{P}},J_\mathcal{P}}(z,\bar{z}) =  \sum\limits_{\mathcal{Q}} a_{\mathcal{Q}}\,
G^{u}_{\Delta_{\mathcal{Q}},J_\mathcal{Q}}(z,\bar{z}) \, .
\end{equation}
We will use the following normalization,
\be \label{normalization}
G^{s}_{\Delta,J}(z,\zb) \sim (z\zb)^{-\Df} \times z^{\frac{\Delta-J}{2}}\zb^{\frac{\Delta+J}{2}}\quad\textrm{for}\quad 0<z  \ll   \zb  \ll 1\,. 
\ee

\begin{figure}
\begin{center}
\includegraphics[width=0.45\textwidth]{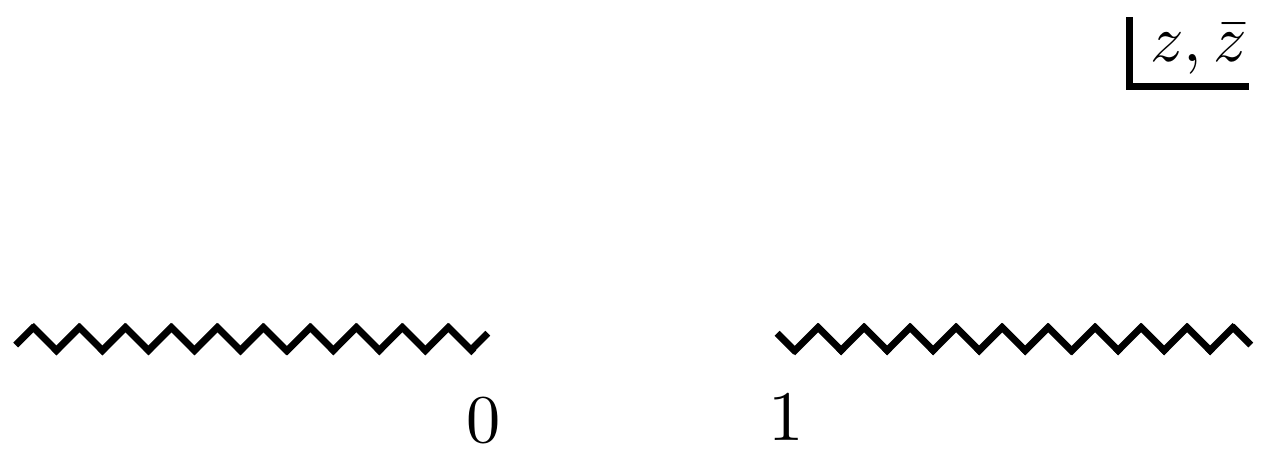}
\caption{ \label{fig:cutPlane} The region of the complex plane (for the independent complex variables  $z$ and $\bar z$) in which the s- and t- channel OPEs converge simultaneously.}
\end{center}
\end{figure}

Our analysis will treat the three OPE channels asymmetrically, singling out a pair.  With no loss of generality, we choose to study the $\mS = \mT$ crossing equation, i.e.~the equality of the s- and t-channel expansions above in their common region of convergence. 
The s-channel OPE converges when both $z$ and $\zb$ are away from the interval $[1, \infty)$, 
and the t-channel OPE converges away from the interval $(-\infty, 0]$ \cite{Pappadopulo:2012jk}.
The two OPEs thus simultaneously converge in the cut plane shown in Figure~\ref{fig:cutPlane}.
It is important that in unitary theories one can analytically continue $z$ and $\bar z$ to independent complex variables.
We will keep both within this cut plane.

\smallskip

Physical four-point functions are single-valued in Euclidean signature,
where $z$ and $\zb$ are complex conjugate of each other.
Within the cut plane, this amounts to a constraint on the boundary values on both sides of the cut.
Specifically, in Euclidean signature we can reach $z,\zb>1$ in two ways: either taking $z$ above the axis and $\zb$ below, or the other way round.  We say that a correlator single-valued around $(z,\zb)=(1,1)$
if these agree: $\cG(z{+}i0,\zb{-}i0)=\cG(z{-}i0,\zb{+}i0)$ for $z,\zb>1$. Similarly, we say that it is single-valued around $(0,0)$
if these agree for $z,\zb<0$.
Individual blocks do not satisfy this property: an s-channel block is not single valued around $(1,1)$.

\smallskip

When both $z$ and $\zb$ variables are on the same cut in Figure~\ref{fig:cutPlane}, we define
the {\it double discontinuity} by taking independent discontinuities in $z$ and $\zb$:
\ba
\dDisc_s[\cG(z,\zb)] &= \tfrac12\big(\cG(z_+,\zb_-)+\cG(z_-,\zb_+)-\cG(z_+,\zb_+)-\cG(z_-,\zb_-)\big) \qquad (z,\zb<0)\ ,
\\
\dDisc_t[\cG(z,\zb)] &= \tfrac12\big(\cG(z_+,\zb_-)+\cG(z_-,\zb_+)-\cG(z_+,\zb_+)-\cG(z_-,\zb_-)\big) \qquad (z,\zb>1)\ ,
\label{eq:dDisc}
\ea
where we use $z_{\pm}=z\pm i 0$ to denote which side of the cut we evaluate the correlator.
When $\cG$ is a single-valued function, the first two terms both reduce to the Euclidean correlator $\cG_E$,
and the others two give its analytic continuation. The definition is then equivalent to the one in~\cite{Caron-Huot:2017vep}:
\be
\dDisc_s[\cG(z,\zb)] = 
\cG_{\mathrm{E}}(z,\zb) - \tfrac{1}{2}\cG_{\mathrm{E}}(z,\zb\circlearrowright0)- \tfrac{1}{2}\cG_{\mathrm{E}}(z,\zb\circlearrowleft0) \qquad \mbox{($\cG$ single-valued)}\,. \label{dDisc SV}
\ee
The double discontinuity will be significant in this paper since it multiplies a block by a simple factor with double zeros on double-twist dimensions:
\be \dDisc_s[G^s_{\Delta,J}(z,\zb)] = 2\sin^2\left(\tfrac{\Delta-J-2\Df}{2} \, \pi \right)G^s_{\Delta,J}(z,\zb)\,.
\label{dDisc block}
\ee
As reviewed in Introduction, these double zeros play an in important role in analytic bootstrap applications.

%%%%%%%%
\subsection{Position-space dispersion relation}

In physics, dispersion relations are formulas expressing dynamics in terms of absorptive physics. 
Paradigmatic examples are the Kramers-Kronig dispersion relation in optics, which expresses the index of refraction in terms of its imaginary (absorptive) part, and dispersion relations for the relativistic S-matrix.
Both cases exploit analyticity properties that encode causality.
Dispersion relations are useful because the absorptive part is often both simpler and more directly accessible than the full quantity, while maintaining desirable properties such as positivity.

\smallskip

In the context of CFTs, the operation which correctly implements the notion of absorptive part is the double discontinuity just described.  A natural question is whether a (superbounded) correlator is uniquely determined by its absorptive part.
This was answered positively by the Lorentzian inversion formula, which reconstructs OPE data from the double discontinuity \cite{Caron-Huot:2017vep,Simmons-Duffin:2017nub,Kravchuk:2018htv}.
Resumming this formula gives the conformal dispersion relation of~\cite{Carmi:2019cub}, which we now briefly review.

\smallskip

\begin{figure}[ht]
\begin{center}
\includegraphics[width=0.45\textwidth]{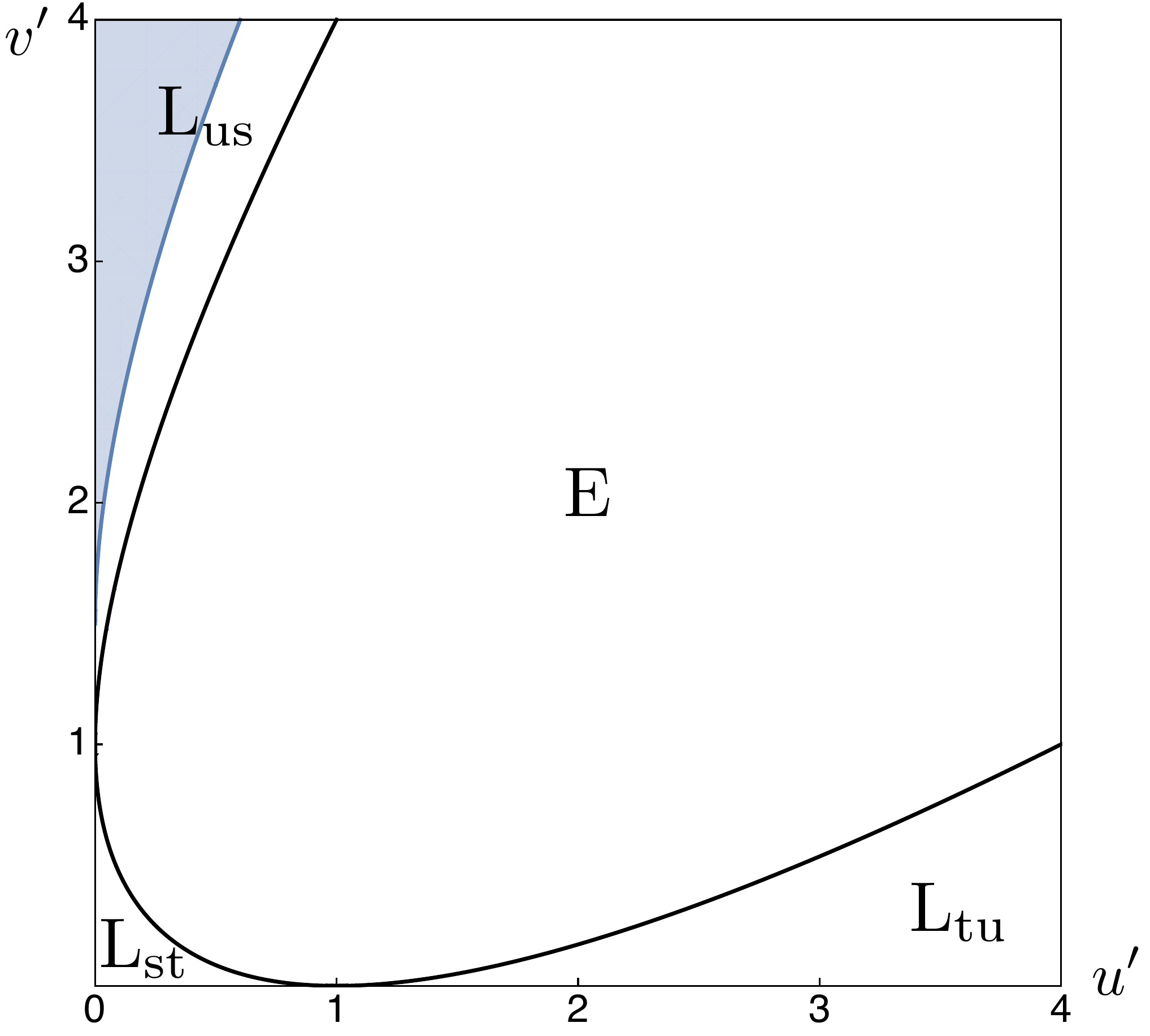}
\caption{Relevant regions in the $u',v'$ variables. The black curve is the image of $w=\wb$ and separates the Euclidean region $\mathrm{E}$, where $w$ and $\bar{w}$ are complex conjugate, from the three Lorentzian lightcones $\mathrm{L_{us}}$: $w,\wb<0$, $\mathrm{L_{st}}$: $0<w,\wb<1$ and $\mathrm{L_{tu}}$: $w,\wb>1$. The s-channel contribution to the dispersion relation \eqref{eq:Dispersion1} is an integral over the shaded region $\sqrt{v'}\geq\sqrt{u'}+\sqrt{u}+\sqrt{v}$. When $\sqrt{u}+\sqrt{v}\geq1$, the integration region lies inside $\mathrm{L_{us}}$. When $\sqrt{u}+\sqrt{v}<1$, it covers all of $\mathrm{L_{us}}$ and parts of $\mathrm{L_{st}}$ and $\mathrm{E}$.}
\label{fig:RegionsUV}
\end{center}
\end{figure}

The conformal dispersion relation derived in \cite{Carmi:2019cub} expresses the four-point function as an integral transform of its double discontinuities in two channels.
To write down the relation, we start by choosing one of the three channels s,t,u. For the purposes of this paper, we will focus on the fixed-u dispersion relation, which builds on discontinuities in the s- and t- channels.\footnote{We will see that this dispersion relation is indeed equivalent to a standard one in Mellin space with fixed Mellin variable $\mU$. Reference~\cite{Carmi:2019cub} worked with the fixed-s dispersion relation.}
For the relation to be valid, the correlator must satisfy a boundedness condition in the u-channel Regge limit.
In this section we focus on the unsubtracted dispersion relation, which requires that $\cG(z,\zb)$ is superbounded in the u-channel, i.e.
\be
 |\cG(z,\zb)|< C |z\zb|^{-\frac{1}{2}-\epsilon} \qquad \textrm{(superboundedness)}
\label{superbounded}
\ee
for some $C,\epsilon>0$ and all $|z|,|\zb|>R>1$. Superboundedness ensures that the Lorentzian inversion formula \cite{Caron-Huot:2017vep} for the u-channel OPE data holds for all spins. Since the u-channel Lorentzian inversion formula depends only on the s- and t-channel dDisc, we conclude that a u-channel superbounded correlator $\cG(z,\zb)$ is uniquely fixed by $\dDisc_s[\cG(z,\zb)]$ and $\dDisc_t[\cG(z,\zb)]$.
The dispersion relation makes this explicit and takes the form
\ba 
\cG(z,\zb) &= \cG^s(z,\zb) + \cG^t(z,\zb)\,,\;\textrm{where}\\
\cG^s(z,\zb) &= \iint\! du'dv'K(u,v;u',v')\dDisc_s[\cG(w,\wb)]\ ,\\
\cG^t(z,\zb) &= \iint\! du'dv'K(v,u;v',u')\dDisc_t[\cG(w,\wb)]\,.
\label{eq:Dispersion1}
\ea
Here we used the notation
$z,\zb$ (or $u,v$) for the free variables, and $w,\wb$ (or $u',v'$) the integrated variables:
\ba
u\phantom{'} &= z\zb\,,\quad\;\, v\phantom{'}=(1-z)(1-\zb)\ ,\\
u' &= w\wb\,,\quad v'=(1-w)(1-\wb)\ .
\ea
We wrote \eqref{eq:Dispersion1} using $u,v,u',v'$ since, as we will see soon, the dispersion kernel $K(u,v;u',v')$ looks particularly simple in these variables. What are the integration regions in \eqref{eq:Dispersion1}? $\cG^s(z,\zb)$ is an integral over $u',v'>0$ such that $\sqrt{v'}\geq\sqrt{u'}+\sqrt{u}+\sqrt{v}$. The integration region is shown in Figure \ref{fig:RegionsUV}. $\cG^t(z,\zb)$ is an integral over the region obtained by $\mS\leftrightarrow\mT$ crossing-symmetry, i.e. $\sqrt{u'}\geq\sqrt{v'}+\sqrt{u}+\sqrt{v}$. Note that when $\sqrt{u}+\sqrt{v}\geq 1$, the integrations lie inside the Lorentzian lightcones $\mathrm{L_{us}}$, $\mathrm{L_{tu}}$.

\smallskip

Let us describe the dispersion kernel $K(u,v;u',v')$. As explained in \cite{Carmi:2019cub},
it is a sum of a bulk term and a contact term
\ba
K(u,v;u',v') = K_B(u,v;u',v') \,&\theta(\sqrt{v'}>\sqrt{u'}+\sqrt{u}+\sqrt{v}) +\\
+ \,K_C(u,v;u') \,&\delta(\sqrt{v'}-\sqrt{u'}-\sqrt{u}-\sqrt{v})\,.
\label{eq:positionKernelFull}
\ea
The contact term is localized at the boundary of the bulk region. The kernels are respectively
\ba
K_B(u,v;u',v') &= \frac{u-v+u'-v'}{64 \pi (u v u' v')^{\frac{3}{4}}} x^{\frac{3}{2}}{}_2F_1\!\left(\tfrac{1}{2},\tfrac{3}{2};2;1-x\right)\,,\\
K_C(u,v;u') &=\frac{1}{4 \pi (u v u'v')^{\frac{1}{4}} (\sqrt{u}+\sqrt{u'})}\,,
\label{eq:kernelsUV}
\ea
where $x$ is the following combination of the four cross-ratios
\be
x = \frac{16 \sqrt{u v u' v'}}{[(\sqrt{u}+\sqrt{v})^2-(\sqrt{u'}+\sqrt{v'})^2][(\sqrt{u}-\sqrt{v})^2-(\sqrt{u'}-\sqrt{v'})^2]}\,.
\label{eq:xDefinition}
\ee
A few comments are in order. First, the dispersion kernel defining $\cG^t(z,\zb)$ in \eqref{eq:Dispersion1} is related to the one defining $\cG^s(z,\zb)$ by crossing symmetry $u\leftrightarrow v$, $u'\leftrightarrow v'$. Therefore, if $\cG(z,\zb)$ is s-t symmetric, i.e. $\cG(z,\zb)=\cG(1-z,1-\zb)$, then we have manifestly $\cG^t(z,\zb) = \cG^s(1-z,1-\zb)$.
Second, in the above formulas we have assumed $u,v$ are real and nonnegative. This is always true in Euclidean kinematics. In Lorentzian kinematics, it is true if the four operators are spacelike separated. In more general situations, \eqref{eq:Dispersion1} needs to be analytically continued in $u$ and $v$.
Third, although the bulk and contact term in \eqref{eq:kernelsUV} look rather different at first sight, they can be combined
\be
K(u,v;u',v') = K_B(u,v;u',v')\left[\theta(x<1) -4\delta(x-1)\right]\,,
\label{eq:kernelsCombined}
\ee
revealing their common origin.\footnote{A little care is needed in interpreting \eqref{eq:kernelsCombined} since besides $\sqrt{v'}=\sqrt{u'}+\sqrt{u}+\sqrt{v}$ the equation $x=1$ can have another solution inside $\mathrm{L_{us}}$, where $(\sqrt{u'}+\sqrt{v'})^2=(\sqrt{u}-\sqrt{v})^2$. In writing \eqref{eq:kernelsCombined}, we implicitly assume that we only keep the part of the theta-function term where $\sqrt{v'}>\sqrt{u'}+\sqrt{u}+\sqrt{v}$, and the part of the delta-function term where $\sqrt{v'}=\sqrt{u'}+\sqrt{u}+\sqrt{v}$.}

Finally, a remarkable feature of \eqref{eq:kernelsUV} (unnoticed in~\cite{Carmi:2019cub}!) is that $K(u,v;u',v')$ is a homogenous function of $u,v,u',v'$ of weight minus two, which implies that the dispersion relation commutes with $u\partial_u+v\partial_v$. This foreshadows a connection with a fixed-u Mellin dispersion relation, which we will confirm below.

%%%%%%%%
\subsection{A basis of analytic functionals}\label{ssec:FunctionalBasis}

A claim closely related to (\ref{eq:Dispersion1}) was made in  \cite{Mazac:2019shk}.
The authors proposed that any function $\cG(z,\zb)$ which is holomorphic in the cut plane and superbounded,\footnote{To avoid long sentences, we use ``superbounded" as shortcut for ``superbounded in the u-channel''. Also, by holomorphic, we will mean holomorphic in the cut plane from now on.}
can be written as a sum of two pieces
\ba
\cG(z,\zb) = \cG^{s}(z,\zb) + \cG^{t}(z,\zb)\,,
\label{eq:Dispersion2}
\ea
where $\cG^s(z,\zb)$ and $\cG^t(z,\zb)$ are both holomorphic and superbounded, and furthermore
\begin{enumerate}
\item $\cG^s(z,\zb)$ is single-valued around $(z,\zb)=(1,1)$ and satisfies $\dDisc_t[\cG^s(z,\zb)] = 0$\,,
\item $\cG^t(z,\zb)$ is single-valued around $(z,\zb)=(0,0)$ and satisfies $\dDisc_s[\cG^t(z,\zb)] = 0$\,.
\end{enumerate}
Unlike for the dispersion relation of the previous subsection, here we do {\it not} assume that $\cG(z,\zb)$ is single-valued around either channel. Correspondingly, $\cG^s(z,\zb)$ need not be single-valued around $(z,\zb)=(0,0)$ and $\cG^t(z,\zb)$ need not be single-valued around $(z,\zb)=(1,1)$. An important example of non-single-valued functions are the conformal blocks, and we would like to stress that the decomposition \eqref{eq:Dispersion2} applies even when $\cG(z,\zb)$ is a conformal block, which will soon play an important role. It turns out that when $\cG(z,\zb)$ \emph{is} single-valued around both s- and t-channel, the decomposition \eqref{eq:Dispersion2} agrees with the dispersion relation \eqref{eq:Dispersion1}. This is by no means obvious and we will provide a proof later in this subsection.  

\smallskip

We will give a constructive proof of  \eqref{eq:Dispersion2} below. It is easy to see that if such a decomposition exists, it is unique. Suppose there is an alternative decomposition
$\cG = \widetilde \cG^{s} + \widetilde \cG^{t}$, with  $\widetilde \cG^{s}$ and $\widetilde \cG^{t}$ obeying Properties 1 and 2. By our assumptions, the difference ${\cal H} \equiv  \widetilde \cG^{s} -  \cG^{s} =  -\widetilde \cG^{t} +  \cG^{t}$ is superbounded,  singled-valued around both (0, 0) and (1, 1), and with vanishing dDisc around both channels,  $\dDisc_s[{\cal H}] = \dDisc_t[{\cal H}] =0$. It immediately follows from the dispersion relation \eqref{eq:Dispersion1} that ${\cal H}$ is identically zero, and hence the decomposition is unique.

\smallskip

Before justifying \eqref{eq:Dispersion2}, let us review some of its consequences, following \cite{Mazac:2019shk}. 
Since $\cG^t(z,\zb)$ in \eqref{eq:Dispersion2} is single-valued around $(z,\zb)=(0,0)$ and satisfies $\dDisc_s[\cG^t(z,\zb)] = 0$, it can be decomposed into a basis of functions with these properties. A natural such basis is the set of all s-channel double-trace conformal blocks and their derivatives with respect to the exchanged dimension $\Delta$: $G^s_{\Delta_{n,\ell},\ell}(z,\zb)$, $\partial_{\Delta}G^s_{\Delta_{n,\ell},\ell}(z,\zb)$ where $\Delta_{n,\ell} = 2\Df+2n+\ell$. We have
\be
\cG^t(z,\zb) = \sum\limits_{n,\ell}\left\{\alpha^s_{n,\ell}[\cG]\,G^s_{\Delta_{n,\ell},\ell}(z,\zb)+\beta^s_{n,\ell}[\cG]\,\partial_{\Delta}G^s_{\Delta_{n,\ell},\ell}(z,\zb)\right\}\,,
\ee
where the sum over $n,\ell$ runs over all nonnegative integers, including both even and \emph{odd} $\ell$. Similarly, we can decompose $\cG^s(z,\zb)$ in t-channel double-traces
\be
\cG^s(z,\zb) = \sum\limits_{n,\ell}\left\{\alpha^t_{n,\ell}[\cG]\,G^t_{\Delta_{n,\ell},\ell}(z,\zb)+\beta^t_{n,\ell}[\cG]\,\partial_{\Delta}G^t_{\Delta_{n,\ell},\ell}(z,\zb)\right\}\,.
\ee
Here $\alpha^s_{n,\ell}$, $\beta^s_{n,\ell}$, $\alpha^t_{n,\ell}$ and $\beta^t_{n,\ell}$ are linear functionals acting on the space of holomorphic, superbounded functions, which extract the coefficient of the double-trace conformal blocks in the decomposition of $\cG(z,\zb)$. An equivalent way to state the proposal is that $G^s_{\Delta_{n,\ell},\ell}(z,\zb)$, $\partial_{\Delta}G^s_{\Delta_{n,\ell},\ell}(z,\zb)$, $G^t_{\Delta_{n,\ell},\ell}(z,\zb)$ and $\partial_{\Delta}G^t_{\Delta_{n,\ell},\ell}(z,\zb)$ form a basis for the space of holomorphic superbounded functions, and $\alpha^s_{n,\ell}$, $\beta^s_{n,\ell}$, $\alpha^t_{n,\ell}$ and $\beta^t_{n,\ell}$ is the corresponding dual basis. The decomposition \eqref{eq:Dispersion2}
is simply the completeness relation
\ba
\cG(z,\zb) =  &\sum\limits_{n,\ell=0}^{\infty}\left\{\alpha^t_{n,\ell}[\cG]\,G^t_{\Delta_{n,\ell},\ell}(z,\zb)+\beta^t_{n,\ell}[\cG]\,\partial_{\Delta}G^t_{\Delta_{n,\ell},\ell}(z,\zb)\right\} +\\
+&\sum\limits_{n,\ell=0}^{\infty}\left\{\alpha^s_{n,\ell}[\cG]\,G^s_{\Delta_{n,\ell},\ell}(z,\zb)+\beta^s_{n,\ell}[\cG]\,\partial_{\Delta}G^s_{\Delta_{n,\ell},\ell}(z,\zb)\right\}\,.
\ea
The statement that functionals $\alpha^{s,t}_{n,\ell}$, $\beta^{s,t}_{n,\ell}$ are dual to the primal basis is captured by their action on double-trace conformal blocks
\be
\begin{array}{l l}
\alpha^{s}_{n,\ell}[G^{s}_{\Delta_{n',\ell'},\ell'}] = \delta_{n n'}\delta_{\ell\ell'}\ ,
&\quad\alpha^{s}_{n,\ell}[\partial_{\Delta}G^{s}_{\Delta_{n',\ell'},\ell'}] = 0\ ,\vspace{.2cm}\\
\beta^{s}_{n,\ell}[G^{s}_{\Delta_{n',\ell'},\ell'}] = 0\ ,
&\quad\beta^{s}_{n,\ell}[\partial_{\Delta}G^{s}_{\Delta_{n',\ell'},\ell'}] = \delta_{n n'}\delta_{\ell\ell'}\ ,\vspace{.4cm}\\
\alpha^{s}_{n,\ell}[G^{t}_{\Delta_{n',\ell'},\ell'}] = 0\ ,
&\quad\alpha^{s}_{n,\ell}[\partial_{\Delta}G^{t}_{\Delta_{n',\ell'},\ell'}] = 0\ ,\vspace{.2cm}\\
\beta^{s}_{n,\ell}[G^{t}_{\Delta_{n',\ell'},\ell'}] = 0\ ,
&\quad\beta^{s}_{n,\ell}[\partial_{\Delta}G^{t}_{\Delta_{n',\ell'},\ell'}] = 0\ .
\end{array}
\label{eq:dualityC1}
\ee
and similarly with s and t exchanged. These relations in particular imply that each of the functionals in the dual basis has double zeros on essentially all s- and t-channel double trace conformal blocks.

\smallskip 
Reference \cite{Mazac:2019shk} explained how to construct the functionals $\alpha^{s,t}_{n,\ell}$, $\beta^{s,t}_{n,\ell}$ in terms of contour integrals within the cut plane. It also sketched how these functionals are related to the position-space dispersion relation \eqref{eq:Dispersion2}. We will now make the connection precise.

\smallskip

\begin{figure}[ht]
\begin{center}
\includegraphics[width=0.45\textwidth]{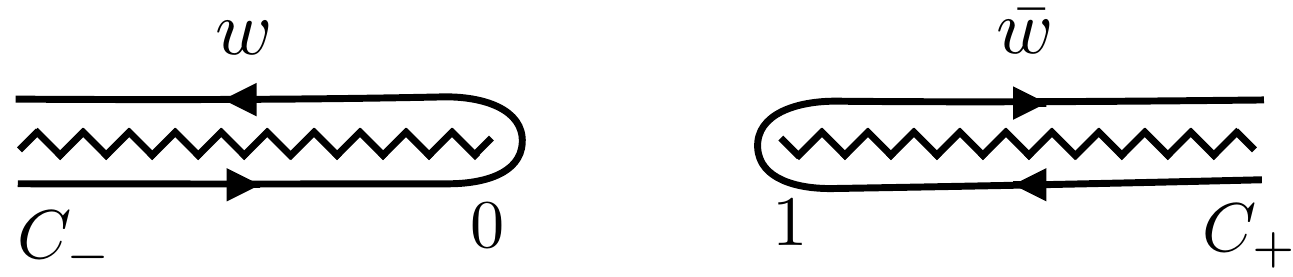}
\caption{The integration contour used to define the $\Omega$ transform \eqref{Omega-s}, \eqref{Omega-t}. Note that the contour is invariant under $(w,\wb)\mapsto (1-\wb,1-w)$, i.e.~it is crossing-symmetric.}
\label{fig:Contours}
\end{center}
\end{figure}

It turns out there are closed formed expressions for $\cG^s(z,\zb)$ and $\cG^t(z,\zb)$, analogous to the second and third line of the dispersion relation in \eqref{eq:Dispersion1}, but valid also when $\cG(z,\zb)$ is not single-valued. 
It will be useful to have a compact notation for these formulas. For a holomorphic superbounded function $f(z, \zb)$, we define its s-channel $\Omega$-transform by 
\be \label{Omega-s}
\Omega^{s | u }[f] (z, \zb) =\theta(v-u)f(z,\zb)+
\!\iint\limits_{C_-\times C_+}\!\!\frac{dwd\wb}{(2\pi i)^2} \pi^2(\wb-w)K_B(u,v;u',v')  f(w, \wb)
\ee
and similarly its t-channel transform by
\be \label{Omega-t}
\Omega^{t | u }[f] (z, \zb) =\theta(u-v)f(z,\zb)
-\!\!\iint\limits_{C_-\times C_+}\!\!\frac{dwd\wb}{(2\pi i)^2}\pi^2(\wb-w)K_B(u,v;u',v') f(w,\wb)\,,
\ee
where $K_B(u,v;u',v')$ is the bulk dispersion kernel defined in \eqref{eq:kernelsUV}.
The second superscript in $\Omega$ is a reminder that our whole discussion pertains to the space of u-channel superbounded functions. It is apparent that
\be
\Omega^{s | u } + \Omega^{t | u } = {\rm Identity} \,.
\ee
The integration contours are the same as in \cite{Mazac:2019shk} and are shown in Figure \ref{fig:Contours}: $w$ is integrated over the contour $C_{-}$ wrapping the left-hand branch cut $(-\infty,0)$, and $\wb$ over $C_{+}$, which wraps the right-hand branch cut $(1,\infty)$.

The piecewise definition of the $\Omega$ transforms in the regions $u<v$ and $u>v$ may appear awkward at first, but it is in fact dictated by analyticity -- the expressions in the two different domains are the analytic continuations of each other.\footnote{The theta function term in \eqref{Omega-s}, \eqref{Omega-t} only applies literally when $u,v$ are real and non-negative. As shown in Appendix \ref{app:Continuation}, for general $z,\zb$ in the cut plane, the argument $u-v$ should be replaced by $\Re(\sqrt{u}-\sqrt{v})$.} We show this in detail in Appendix \ref{app:Continuation}, and give the gist of the argument here.
Take for example \eqref{Omega-t} for $u<v$ and analytically continue it in $z,\zb$ outside of the region $u<v$. The singularities of the kernel $K_B(u,v;u',v')$ move in the space of $w,\wb$ and we are forced to deform the contour $C_{-}\times C_{+}$ to avoid the singularities. One can then deform the contour back to the original configuration $C_{-}\times C_{+}$, but only at the expense of picking the contribution of the singularities. It turns out that the only singularity which contributes is a simple pole at $(w,\wb)=(z,\zb)$, where the kernel has residue 1
\be
\pi^2(\wb-w)K_B(u,v;u',v') \sim \frac{1}{(w-z)(\wb-\zb)}\,.
\ee
In other words, the extra term that $\Omega^{t |  u}[f](z,\zb)$ picks in the process is precisely $f(z,\zb)$.
Therefore, the step function in (\ref{Omega-t}) precisely ensures that the $\Omega$-transform is analytic.

\smallskip

With these definitions in place,  we  claim that the decomposition (\ref{eq:Dispersion2}) is achieved by the two Omega-transforms of ${\cal G}$,
\ba \label{punchline}
\cG^s =  \Omega^{s | u }[\cG]  \, , \quad \cG^t =  \Omega^{t | u }[\cG]\,.
\ea
While the original dispersion relation \eqref{eq:Dispersion1} was written as an integral over a real contour in the $u',v'$ variables, the new version (\ref{eq:Dispersion2}, \ref{punchline}) 
entails integrals over a different, complex contour in the $w,\wb$ variables.  Note that the factor $\wb-w$ is simply the Jacobian of the coordinate change: $du'dv'=(\wb-w) dw d\wb$.

\smallskip

One of the virtues of the new dispersion relation is that Properties 1 and 2 are manifest, i.e. it is easy to show that  $\cG^t$ and $\cG^s$ are single-valued and have vanishing double discontinuities around the s- and t-channel respectively. To see that, let us focus on $\cG^t(z,\zb)$ and take $|z|$ and $|\zb|$ small, so that we can drop the step function term in \eqref{Omega-t}.
The $w,\wb$ integral converges uniformly in a neighborhood of $z=\zb=0$. This means that to study analytic properties of $\cG^t(z,\zb)$ around $z=\zb=0$, we can simply expand in $z,\zb$ under the integral sign. It is straightforward to check that $K_B(u,v;u',v')$ is single-valued around $z=\zb=0$ and satisfies $\dDisc_{s}[K_B(u,v;u',v')] = 0$. Indeed, the magic combination $x$ goes to zero as $u\rightarrow 0$, and the hypergeometric in $K_B$ leads to at most a single $\log(u)$, i.e.
\be
K_B(u,v;u',v') = K^{(0)}_B(u,v;u',v') + \log(u)K^{(1)}_B(u,v;u',v')\,,
\label{eq:dDiscKZero}
\ee
where $K^{(0)}_B(u,v;u',v')$ and $K^{(1)}_B(u,v;u',v')$ are holomorphic at $z=\zb=0$. The same comments apply to $\cG^s(z,\zb)$, showing that it is indeed single-valued around $z=\zb=1$ and satisfies $\dDisc_{t}[\cG^s(z,\zb)] = 0$. 

\smallskip

We can obtain explicit formulas for the functionals $\alpha^{s,t}_{n,\ell}$, $\beta^{s,t}_{n,\ell}$ by expanding (\ref{Omega-s}, \ref{Omega-t}) into double-trace blocks and their derivatives under the integral sign.
This immediately leads to the contour-integral description of $\alpha^{s,t}_{n,\ell}$, $\beta^{s,t}_{n,\ell}$ found in \cite{Mazac:2019shk}. Specifically, for each of $s$ and $t$:
\ba
\alpha^{s,t}_{n,\ell}[\cG(w,\wb)] &= \!\!\!\iint\limits_{C_-\times C_+}\!\!\!\frac{dwd\wb}{(2\pi i)^2} \,\mathcal{A}^{s,t}_{n,\ell}(w,\wb) \cG(w,\wb)\\
\beta^{s,t}_{n,\ell}[\cG(w,\wb)]&= \!\!\!\iint\limits_{C_-\times C_+}\!\!\frac{dwd\wb}{(2\pi i)^2} \,\mathcal{B}^{s,t}_{n,\ell}(w,\wb) \cG(w,\wb)\, .
\label{eq:DualBasisContours}
\ea
where the kernels $\mathcal{A}^{s,t}_{n,\ell}$, $\mathcal{B}^{s,t}_{n,\ell}$ are defined by expanding \eqref{Omega-s}, \eqref{Omega-t} in double-trace conformal blocks under the integral sign
\ba
\pi^2(\wb-w)K_B(v,u;v',u') &= \sum\limits_{n,\ell}\left[\mathcal{A}^s_{n,\ell}(w,\wb)G^s_{\Delta_{n,\ell},\ell}(z,\zb)+\mathcal{B}^s_{n,\ell}(w,\wb)\partial_{\Delta}G^s_{\Delta_{n,\ell},\ell}(z,\zb)\right]\\
\pi^2(\wb-w)K_B(u,v;u',v') &= \sum\limits_{n,\ell}\left[\mathcal{A}^t_{n,\ell}(w,\wb)G^t_{\Delta_{n,\ell},\ell}(z,\zb)+\mathcal{B}^t_{n,\ell}(w,\wb)\partial_{\Delta}G^t_{\Delta_{n,\ell},\ell}(z,\zb)\right]
\label{eq:kernelExpansion}
\ea
The resulting kernels agree with those found in \cite{Mazac:2019shk} using a different, indirect method.\footnote{Note that the expansion in \eqref{eq:kernelExpansion} must be done in the domain of the contours, i.e. for $\mathrm{Re}(\wb)>\mathrm{Re}(w)$. The nontrivial branch-cut structure of $K_B(u,v;u',v')$ then implies $\mathcal{A}^{s}_{n,\ell}(w,\wb) = \mathcal{A}^{t}_{n,\ell}(1-w,1-\wb)$, $\mathcal{B}^{s}_{n,\ell}(w,\wb) = \mathcal{B}^{t}_{n,\ell}(1-w,1-\wb)$ although naively \eqref{eq:kernelExpansion} seems to give an extra minus sign in the transformation.} For example,
\ba \label{eq:kernelExample}
\mathcal{A}^{s}_{0,0}(w,\wb) &= \frac{w+\wb-2}{(\wb-w)^2}\left\{4+\log\left[\frac{w\wb(1-w)(1-\wb)}{(\wb-w)^4}\right]\right\}\\
\mathcal{B}^{s}_{0,0}(w,\wb) &= \frac{2(w+\wb-2)}{(\wb-w)^2}\,.
\ea

\smallskip
It remains to be shown that when $\cG(z,\zb)$ is single-valued, the contour integrals \eqref{Omega-s}, \eqref{Omega-t} defining $\cG^{s,t}(z,\zb)$ through \eqref{punchline} reduce to the original dispersion relation \eqref{eq:Dispersion1} involving 
$\dDisc$.
We sketch the argument here, leaving the details to Appendix \ref{app:deformationToDDisc}. Let us focus on $\cG^{t}(z,\zb)$. The idea is to start from \eqref{Omega-t} in a region where the step function vanishes
and wrap both $w$ and $\wb$ contours on the right-hand branch cut, wrapping the $\wb$ contour first and $w$ second. The resulting integral comes in four pieces, containing $\cG(w{+}i\epsilon,\wb{+}i\epsilon)$, $\cG(w{+}i\epsilon,\wb{-}i\epsilon)$, $\cG(w{-}i\epsilon,\wb{+}i\epsilon)$ and $\cG(w{-}i\epsilon,\wb{-}i\epsilon)$ respectively. $\cG(w,\wb)$ being single-valued around the t-channel is equivalent to
\be
\cG(w+i\epsilon,\wb-i\epsilon)= \cG(w-i\epsilon,\wb+i\epsilon)=\cG_{\rm E}(w,\wb)\,.
\ee
Using this condition, together with nontrivial identities satisfied by the kernel $K_B$, the four pieces
nontrivially recombine to an integral depending only on the double-discontinuity defined in (\ref{dDisc SV}).
Furthermore, the antisymmetry of the integrand in \eqref{Omega-t} under $w\leftrightarrow\wb$ implies that many of the contributions cancel out and we are left only with the integral over the sliver $v'>0$, $u'\geq(\sqrt{v'}+\sqrt{u}+\sqrt{v})^2$,
which is precisely the support of $K$:
\be
 \Omega^{t | u }[f] (z, \zb) = \iint\! du'dv'K(v,u;v',u')\dDisc_t[f(w,\wb)] \quad \mbox{($f(z,\zb)$ single-valued)}\ .
\ee
Both the bulk term and the contact term in \eqref{eq:kernelsUV} are correctly reproduced by the argument.

\smallskip

To see how the decomposition $\cG=\cG^s+\cG^t$ works in simple cases, the reader can consult Appendix \ref{app:decompositions}. There, we consider the correlator $\langle\phi^2\bar{\phi}^2\phi^2\bar{\phi}^2\rangle$, where $\phi$ is a free complex scalar in $d=3$ and $d=4$, corresponding to $\cG(u,v) = 1/\sqrt{uv}$ and $\cG(u,v)=1/(uv)$.

\smallskip 
To summarize this subsection, we used the kernel of the dispersion relation
to define a transform $\Omega$ in (\ref{Omega-s}) which establishes in a constructive way
the decomposition (\ref{eq:Dispersion2}).
Generally, the $\Omega$-transform cannot be written in terms of dDisc alone, except when its acts
on single-valued correlators. In this case it reduces to the dispersion relation by a contour deformation.

\subsection{The Polyakov-Regge bootstrap}
\label{ssec:polyakovreggebootstrap}

We now introduce one of the unifying ideas of our whole discussion,
the Polyakov-Regge expansion. We start by defining the $s$- and $t$-channel {\it Polyakov-Regge blocks} as the $\Omega$-transforms of the conformal blocks in the same channel,
\be \label{Pdefinitions}
 P_{\Delta, J}^{s|u} \equiv  \Omega^{s|u} [G_{\Delta, J}^s] \, , \quad  P_{\Delta, J}^{t|u} \equiv  \Omega^{t|u} [G_{\Delta, J}^t] \,.
\ee
A Polyakov-Regge block obeys the following properties:
\begin{enumerate}
\item[(i)] It is  u-channel superbounded.
\item[(ii)] It is Euclidean singled-valued (in all channels).
\item[(iii)]  It has the same double discontinuity as the corresponding conformal block in its defining channel, and vanishing double discontinuity in the other channel. In formulas, 
\ba \label{dDiscP}
{\rm dDisc}_s[P_{\Delta, J}^{s|u}]  & = {\rm dDisc}_s[G^s_{\Delta, J}]\,, \quad {\rm dDisc}_t[P_{\Delta, J}^{s|u}]  = 0\,, \\
{\rm dDisc}_t[P_{\Delta, J}^{t|u}]  & = {\rm dDisc}_t[G^t_{\Delta, J}]\,, \quad {\rm dDisc}_s[P_{\Delta, J}^{t|u}]  = 0\,.
\ea
\end{enumerate}
These properties follow almost immediately from the general discussion in the previous subsection. Perhaps the least obvious is the statement of Euclidean singled-valuedness.
Focus for definiteness on  $P_{\Delta, J}^{s|u} (z, \zb)$.  Singled-valuedness around $(1,1)$ follows at once from the definition (\ref{Pdefinitions}) and from property 1 of the s-channel Omega transform
(as stated at the beginning of the previous subsection). To see singled-valuedness around $(0,0)$, we use the $\Omega$-decomposition of the conformal block,
\be
G^s_{\Delta, J} =  \Omega^{s|u} [G_{\Delta, J}^s] +  \Omega^{t|u} [G_{\Delta, J}^s]  =  P_{\Delta, J}^{s|u} +  \Omega^{t|u} [G_{\Delta, J}^s] \, , \label{Omega decomposition block}
\ee
and observe that both $G^s_{\Delta, J}$ and   $\Omega^{t|u} [G_{\Delta, J}^s]$ are singled-valued around $(0,0)$,
for integer spin $J$.
The same decompositions makes clear that 
${\rm dDisc}_s[P_{\Delta, J}^{s|u}]   = {\rm dDisc}_s[G^s_{\Delta, J}]$.

\smallskip

Polyakov-Regge blocks are in fact the {\it unique} functions satisfying properties (i)--(iii). This is clear from the fact that singled-valued, superbounded functions can be unambiguously reconstructed from their s- and t-channel double discontinuities -- most directly by applying the position-space dispersion relation~(\ref{eq:Dispersion1}). 
This uniqueness will be very helpful.

\smallskip

Since $G^s_{\Delta,J}$ is single-valued around $(0,0)$, we can write \eqref{Pdefinitions} as the dispersive transform of an individual block
\be
\label{P as dispersive integral}
 P^{s|u}_{\Delta,J}(z,\zb) = \iint du'dv'K(u,v;u',v')\dDisc_s[G^s_{\Delta,J}(w,\wb)]\,.
\ee

Polyakov-Regge blocks admit useful expansions in the  s- and the t-channel. The general structure of these expansions follows  from (\ref{dDiscP}).
From  ${\rm dDisc}_s[P_{\Delta, J}^{s|u}]   = {\rm dDisc}_s[G^s_{\Delta, J}]$ we deduce that $P_{\Delta, J}^{s|u}$ can be written in the s-channel as $G^s_{\Delta, J}$ plus an infinite sum
of double-twist conformal blocks and their derivatives; while ${\rm dDisc}_t[P_{\Delta, J}^{s|u}] =0$ implies a t-channel expansion just in terms of double-twists and their derivatives:
\ba \label{P_OPEs}
P_{\Delta, J}^{s|u} & = 
   G_{\Delta, J}^s -   \sum\limits_{n,\ell=0}^{\infty}\left\{\alpha^s_{n,\ell}[  G_{\Delta, J}^s  ]\,G^s_{\Delta_{n,\ell},\ell} +\beta^s_{n,\ell}[  G_{\Delta, J}^s]\,\partial_{\Delta}G^s_{\Delta_{n,\ell},\ell} \right\} \\
&= \sum\limits_{n,\ell=0}^{\infty}\left\{\alpha^t_{n,\ell}[ G_{\Delta, J}^s]\,G^t_{\Delta_{n,\ell},\ell} +\beta^t_{n,\ell}[ G_{\Delta, J}^s]\,\partial_{\Delta}G^t_{\Delta_{n,\ell},\ell} \right\}\,.
\ea
We remind the reader that the sum contains both even and odd $\ell$. The coefficients in these expansions are immediately fixed in terms of the dual basis of functionals by applying the orthonormality relations  (\ref{eq:dualityC1}).
In complete analogy,
\ba  \label{P_OPEt}
P_{\Delta, J}^{t|u} & = 
   G_{\Delta, J}^t -   \sum\limits_{n,\ell=0}^{\infty}\left\{\alpha^t_{n,\ell}[  G_{\Delta, J}^t  ]\,G^t_{\Delta_{n,\ell},\ell} +\beta^t_{n,\ell}[  G_{\Delta, J}^t]\,\partial_{\Delta}G^t_{\Delta_{n,\ell},\ell} \right\} \\
&= \sum\limits_{n,\ell=0}^{\infty}\left\{\alpha^s_{n,\ell}[ G_{\Delta, J}^t]\,G^s_{\Delta_{n,\ell},\ell} +\beta^s_{n,\ell}[ G_{\Delta, J}^t]\,\partial_{\Delta}G^s_{\Delta_{n,\ell},\ell} \right\}\,.
\ea

\begin{figure}
\centering{\def\svgwidth{10cm}%% Creator: Inkscape 1.0beta1 (32d4812, 2019-09-19), www.inkscape.org
%% Accompanies image file 'witten_diagram.pdf' (pdf, eps, ps)
\begingroup%
  \makeatletter%
  \providecommand\color[2][]{%
    \errmessage{(Inkscape) Color is used for the text in Inkscape, but the package 'color.sty' is not loaded}%
    \renewcommand\color[2][]{}%
  }%
  \providecommand\transparent[1]{%
    \errmessage{(Inkscape) Transparency is used (non-zero) for the text in Inkscape, but the package 'transparent.sty' is not loaded}%
    \renewcommand\transparent[1]{}%
  }%
  \providecommand\rotatebox[2]{#2}%
  \newcommand*\fsize{\dimexpr\f@size pt\relax}%
  \newcommand*\lineheight[1]{\fontsize{\fsize}{#1\fsize}\selectfont}%
  \ifx\svgwidth\undefined%
    \setlength{\unitlength}{333.02446333bp}%
    \ifx\svgscale\undefined%
      \relax%
    \else%
      \setlength{\unitlength}{\unitlength * \real{\svgscale}}%
    \fi%
  \else%
    \setlength{\unitlength}{\svgwidth}%
  \fi%
  \global\let\svgwidth\undefined%
  \global\let\svgscale\undefined%
  \makeatother%
  \begin{picture}(1,0.27450632)%
    \lineheight{1}%
    \setlength\tabcolsep{0pt}%
    \put(0,0){\includegraphics[width=\unitlength,page=1]{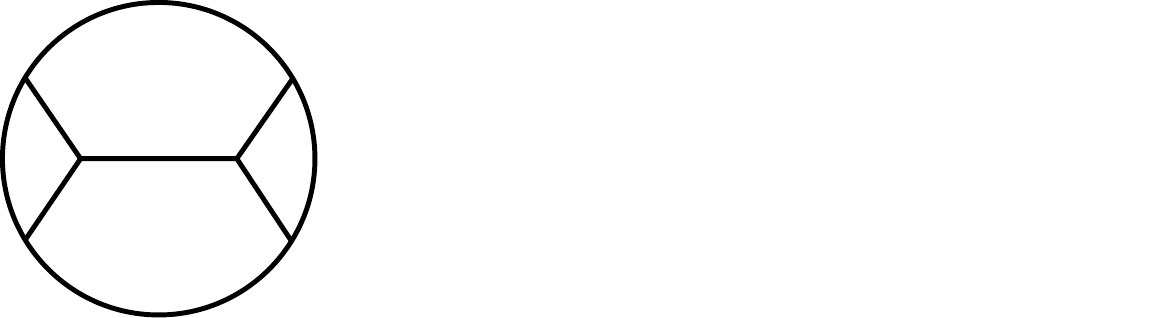}}%
    \put(0.10266751,0.14690494){\makebox(0,0)[lt]{\lineheight{1.25}\smash{\begin{tabular}[t]{l}$\Delta,J$\end{tabular}}}}%
    \put(0.22733662,0.01097539){\makebox(0,0)[lt]{\lineheight{1.25}\smash{\begin{tabular}[t]{l}u\end{tabular}}}}%
    \put(0,0){\includegraphics[width=\unitlength,page=2]{witten_diagram.pdf}}%
    \put(0.46219704,0.14770922){\makebox(0,0)[lt]{\lineheight{1.25}\smash{\begin{tabular}[t]{l}$\Delta,J$\end{tabular}}}}%
    \put(0.30535532,0.12197114){\makebox(0,0)[lt]{\lineheight{1.25}\smash{\begin{tabular}[t]{l}=\end{tabular}}}}%
    \put(0.66729783,0.1251884){\makebox(0,0)[lt]{\lineheight{1.25}\smash{\begin{tabular}[t]{l}+\end{tabular}}}}%
    \put(0,0){\includegraphics[width=\unitlength,page=3]{witten_diagram.pdf}}%
  \end{picture}%
\endgroup%
}
\caption{\label{fig:witten} The Polyakov-Regge block $P_{\Delta, J}^{s|u}$ is a combination of exchange and contact
Witten diagrams such that the $u$-channel Regge behavior is the best possible.}
\end{figure}

\smallskip
We will now give a different description of the same Polyakov-Regge blocks. There is in fact {\it another} familiar class of functions that are Euclidean single-valued and have a conformal block decomposition of the form (\ref{P_OPEs})-(\ref{P_OPEt}): Witten exchange diagrams in $AdS_{d+1}$. 
Consider an s-channel Witten exchange diagram $W_{\Delta, J}^s$ with
 internal propagator of dimension $\Delta$ and spin $J$. The function $W_{\Delta, J}^s$ shares with $P_{\Delta, J}^{s|u}$ properties (ii) and (iii), but
it is not guaranteed to be u-channel superbounded.
Note that there is an inherent freedom in defining exchange Witten diagrams with spin since
different ways of writing the vertices can change the result by contact diagrams.
Now it is always possible to ``improve'' an exchange diagram
by adding a finite number of contact diagrams (with up to $2(J-1)$ derivatives in the quartic vertices)
such that  the sum {\it is} superbounded. (This statement is particularly transparent in Mellin space, as we will review in the next subsection.) By uniqueness of  Polyakov-Regge blocks, it follows that
\be
P_{\Delta, J}^{s|u} = W_{\Delta, J}^s + \sum {\rm contacts}\,, \quad P_{\Delta, J}^{t|u} = W_{\Delta, J}^t +  \sum {\rm contacts}\,.
\ee
A Polyakov-Regge block is simply a Witten diagram, where contact ambiguities are
fixed by Regge superboundedness in the u-channel! This is depicted in Figure \ref{fig:witten}.

A comment is in order about crossing properties. While $W_{\Delta, J}^s$ can be defined to have a simple transformation property under the t$\leftrightarrow$u crossing transformation, namely it picks up an overall factor of $(-1)^J$,
this is not the case for $P_{\Delta, J}^{s|u}$, because the additional contact diagrams spoil this symmetry.  Similarly,  $W_{\Delta, J}^t \to (-1)^J  W_{\Delta, J}^t$ under  s$\leftrightarrow$u crossing, but
 $P_{\Delta, J}^{t|u}$ does not have such  property. This is of course not surprising, since the definition of a Polyakov-Regge blocks treats the three channels asymmetrically.

\smallskip

We are ready to derive the main statement.
Starting from the $\Omega$-decomposition of superbounded correlator ${\cal G}$,
\be
{\cal G} = \Omega^{s|u} [{\cal G}] +  \Omega^{t|u} [{\cal G}]\,,
\ee
we simply expand $\cG$ inside $\Omega^{s|u} [ {\cal G}]$ using the s-channel OPE and $\cG$ in $\Omega^{t|u} [ {\cal G}]$ using the t-channel OPE (recall our conventions (\ref{OPE})):
\be \label{PR expansion}
\cG(z,\zb) = \sum\limits_{\cO}a_{\cO} P^{s|u}_{\Delta_{\cO},J_{\cO}}(z,\zb)+\sum\limits_{\cP}a_{\cP} P^{t|u}_{\Delta_{\cP},J_{\cP}}(z,\zb)\,.
\ee
This is the Polyakov-Regge expansion of ${\cal G}$.
It is a fully non-perturbative statement and converges in the cut plane of Figure~\ref{fig:cutPlane}.
The OPE sum and $\Omega$ functionals (\ref{Omega-s})
commute because the OPE is absolutely convergent everywhere inside the integration contour,
and it does not diverge strongly enough near its boundaries
(assuming superboundedness in the Regge limit $w,\wb\to\infty$).

\smallskip

One could also derive (\ref{PR expansion}) starting with the dispersion relation in (\ref{eq:Dispersion1}),
by inserting the s- and t- channel OPEs to compute $\dDisc_s[\cG]$ and $\dDisc_t[\cG]$, respectively,
using \eqref{P as dispersive integral}.
The appearance of Witten diagrams and AdS space is purely kinematical and \emph{a priori} unrelated to holography:
a (Polyakov-Regge) Witten diagram is simply the dispersive transform of a conformal block.

\smallskip

In contrast with the usual conformal block expansion,
individual terms in the Polyakov-Regge expansion~(\ref{PR expansion}) are single-valued.
The price to pay is that we must sum over two channels.
Note that summing Witten diagrams in all three channels would be incorrect: in the best case scenario,
for example if one made such a sum
symmetrical by replacing $P^{s|u}$ by $\tfrac12(P^{s|u}+P^{s|t})$, one would find $\tfrac32\cG$.
In~(\ref{PR expansion}), u-channel singularities are generated by infinite towers of terms in the other channels. 
As usual with dispersion relations, two channels suffice.
 
 \smallskip
 
Compatibility of the Polyakov-Regge expansion (\ref{PR expansion})
with the usual Euclidean expansion is nontrivial: subtracting the s-channel OPE one must find
\be
 0 = \sum\limits_{\cO}a_{\cO} \Big(P^{s|u}_{\Delta_{\cO},J_{\cO}}(z,\zb)-G^s_{\Delta_{\cO},J_{\cO}}(z,\zb)\Big)+
 \sum\limits_{\cP}a_{\cP} P^{t|u}_{\Delta_{\cP},J_{\cP}}(z,\zb)\,. \label{PR minus Euclidean}
\ee
We may now use the first line of (\ref{P_OPEs}) and the second line of (\ref{P_OPEt}) to expand
both summands over s-channel double-twist blocks and their derivatives.
Compatibility thus give the following nontrivial sum rules which relate s- and t-channel data:
\be
 \sum\limits_{\cO}a_{\cO}\,\alpha_{n,\ell}^s[G^s_{\Delta_{\cO},J_{\cO}}]
= \sum\limits_{\cP}a_{\cP}\,\alpha_{n,\ell}^s[G^t_{\Delta_{\cP},J_{\cP}}] \qquad \forall n,\ell\geq 0\ , \label{first sum rule}
\ee
together with the same sums with $\alpha$ replaced by $\beta$. These sum rules can be derived more directly by applying the functionals \eqref{eq:DualBasisContours} to the standard s=t crossing equation. We stress that the sum rules are valid whether or not exact double-trace blocks appear in the OPE. They follow from nothing more than unitarity, crossing symmetry and (in this case) Regge superboundedness. The sum rules will be discussed at length in Section \ref{sec:Edges}.

%%%%%%%%
\subsection{Polyakov-Regge blocks in Mellin space and dispersion relation}\label{ssec:MellinDispersion}

Mellin space provides a representation of CFT correlators which shares many intuitive features
with gapped S-matrices. It is particularly effective in the context of holographic theories
since Witten diagrams admit convenient Mellin-space expressions.
In this subsection we use this to rewrite the Polyakov-Regge expansion (\ref{PR expansion}) in Mellin space.
We will find that the decomposition $\mathcal{G}=\mathcal{G}^s+\mathcal{G}^t$ is nothing but
the Mellin-space dispersion relation studied in \cite{Penedones:2019tng}!

\smallskip

The Mellin representation of a four-point function $\cG(u,v)$ takes the form\footnote{Recall that we are assuming the dimensions of the external operators are all equal to $\Df$, but the operators themselves are not necessarily identical.}
\be
\cG(u,v) = \iint\!\!\frac{d\mS\,d\mT}{(4\pi i)^2}\,
\Gamma\!\left(\Df-\tfrac{\mS}{2}\right)^2\Gamma\!\left(\Df-\tfrac{\mT}{2}\right)^2\Gamma\!\left(-\Df+\tfrac{\mS+\mT}{2}\right)^2 u^{\tfrac{\mS}{2}-\Df}v^{\tfrac{\mT}{2}-\Df} M(\mS,\mT)\,.
\label{eq:mellinRep}
\ee
Here $\mS,\mT$ are the Mellin variables and $M(\mS,\mT)$ is the Mellin amplitude. It is convenient to introduce a third Mellin variable $\mU$, related to the other two by
\be
\mS+\mT+\mU=4\Df\,.
\ee
We denote the Mellin variables by roman letters to avoid confusion with the cross-ratios $u,v$. Our notation for the Mellin variables is related to that of reference \cite{Penedones:2019tng} by
\be
\gamma_{12} = \Df-\frac{\mS}{2}\,,\qquad
\gamma_{13} = \Df-\frac{\mU}{2}\,,\qquad
\gamma_{14} = \Df-\frac{\mT}{2}\,.
\ee
The contours in the $\mS,\mT$ variables run in the imaginary direction from $-i\infty$ to $+i\infty$. As discussed in detail in \cite{Penedones:2019tng}, various subtleties enter in choosing the contour.

\smallskip

One of the main virtues of the Mellin representation is that operators present in the OPEs show up as simple poles of $M(\mS,\mT)$. Specifically, each primary $\cO$ present in the s-channel gives rise to an infinite sequence of simple poles of $M(\mS,\mT)$ at $\mS=\tau_{\cO}+2m$ with $m=0,1,\ldots$, where $\tau_{\cO} = \Delta_{\cO}-J_{\cO}$ is the twist. The s-channel OPE is then reproduced by \eqref{eq:mellinRep} by closing the $\mS$ contour to the right and picking the contribution of these poles. The residues at these poles are functions of the remaining Mellin variable $\mT$. Upon closing the $\mT$ contour, they must correctly reproduce the expansion of conformal blocks at small $u$. It can be shown that the residues are polynomials in the $\mT$ variable of degree $J_{\cO}$. These polynomials will be denoted $\mathcal{Q}^{m}_{\Delta_{\cO},J_{\cO}}(\mT)$. They are uniquely fixed by the conformal symmetry. Thus,
\be
\cG(z,\zb) \supset a_{\cO} \,G^s_{\Delta_{\cO},J_{\cO}}(z,\zb) \quad\Rightarrow\quad M(\mS,\mT) \sim\,a_{\cO}\frac{\mathcal{Q}^{m}_{\Delta_{\cO},J_{\cO}}(\mT)}{\mS-\tau_{\cO}-2m}\quad\textrm{as}\quad\mS\rightarrow \tau_{\cO}+2m
\label{eq:mellinPoles}
\ee
for $m=0,1,\ldots$.
Primary operators in the t-channel give rise to analogous poles with $\mS$ and $\mT$ swapped.

\smallskip

The kinematical polynomials $\mathcal{Q}^{m}_{\Delta,J}(\mT)$ depend additionally on $\Df$ and $d$.
They are typically written as
\be
\mathcal{Q}^{m}_{\Delta,J}(\mT)=-\sin ^2\left[\tfrac{\pi}{2}(\Delta-J -2 \Df)\right]
\frac{(\Delta -1)_J\,\Gamma (\Delta+J)\,\Gamma\!\left(\tfrac{\Delta-J-2 \Df+2 m+2}{2}\right)^2}{2^{J-1}\,\pi ^2\,m!\left(\Delta-\frac{d}{2}+1\right)_m\,\Gamma\!\left(\tfrac{\Delta+J}{2}\right)^4}\,Q^{m}_{\Delta,J}(\mT)
\label{eq:qCal}
\ee
where $Q^{m}_{\Delta,J}(\mT)$ are the so-called Mack polynomials. $Q^{m}_{\Delta,J}(\mT)$ is a polynomial in $\mT-2\Df$ of degree $J$ with coefficients depending only on $d$, $\Delta$ and $m$ (and not on $\Df$). For example,
\ba
Q^{m}_{\Delta,0}(\mT) &= 1\\
Q^{m}_{\Delta,1}(\mT) &= \mT-2\Df+\frac{\Delta+2m-1}{2}\,.
\ea
$Q^{m}_{\Delta,J}(\mT)$ can be computed for example using a recursion relation presented in Appendix A of reference \cite{Costa:2012cb}.
For our purposes, the most important feature of $\mathcal{Q}^{m}_{\Delta,J}(\mT)$ is that they all have double zeros as a function of $\Delta$ at $\Delta=2\Df+J+2n$ for $n=0,1,\ldots$, thanks to the factor $\sin ^2\left[\tfrac{\pi}{2}(\Delta-J -2 \Df)\right]$ in \eqref{eq:qCal}. The role of these double zeros is to cancel the double poles of the squared gamma function $\Gamma\!\left(\Df-\tfrac{\mS}{2}\right)^2$ in the Mellin representation \eqref{eq:mellinRep} and thus ensure the Mellin-space poles \eqref{eq:mellinPoles} contribute conformal blocks with the correct coefficients $a_{\cO}$.
It is expected that \eqref{eq:mellinPoles}, together with the t-channel poles, are the only singularities of $M(\mS,\mT)$ at fixed $\mU=4\Df-\mS-\mT$.

\smallskip

Nontrivially, the Mellin amplitude $M(\mS,\mT)$ is power-behaved at large complex
$\mS,\mT$ \cite{Mack:2009mi,Penedones:2019tng}.
Thanks to the gamma factors, the integrand in (\ref{eq:mellinRep}) is thus exponentially decaying in Euclidean signature.
The amplitude
$M(\mS,\mT$) corresponding to an individual conformal block is
however generally not power-behaved \cite{Mack:2009mi,Costa:2012cb}.
This is easy to see since for example an s-channel conformal block can only have the poles in (\ref{eq:mellinPoles}),
but it must have an infinite sequence of double zeros for each double-twist value $\mS=2\Df+2m$,
in order to cancel the gamma function poles from (\ref{eq:mellinRep}).
For this reason, blocks are not a very convenient basis for Mellin amplitudes,
and they are often replaced by Witten diagrams, which share the same poles but not the double zeros \cite{Penedones:2010ue}.

\smallskip

We are now ready to write down the Mellin-space form of
of the Polyakov-Regge blocks, $\mathcal{P}^{s|u}_{\Delta,J}(\mS,\mT)$.
The uniqueness property discussed below (\ref{Omega decomposition block})
will come in handy: we only need to cook up a function with the right properties. Superboundedness in the u-channel Regge limit means that
\be
 \lim_{|\mS|\to\infty} \mathcal{P}^{s|u}_{\Delta,J}(\mS,\mT)_{\mS+\mT={\rm fixed}} =0 \label{PR super bounded}
\ee
along any (non-real) direction in the complex plane.\footnote{Generally $M(\mS,\mT)= O(|s|^{J_{0}})$ in this limit, where $J_0$ is the u-channel Regge intercept \cite{Costa:2012cb,Penedones:2019tng}. Superboundedness implies $J_{0}<0$.}
The statement in (\ref{P_OPEs}) that the s-channel OPE of $\mathcal{P}^{s|u}_{\Delta,J}$ contains a single physical block,
plus double-twist and their derivatives, means that its only poles are at $s=\Delta-J+2m$ with residue given in (\ref{eq:mellinPoles}) and no poles in $\mT$. (Double-twist blocks and their derivatives are generated by the gamma factors in (\ref{eq:mellinRep}).)
Since $\mT=4\Df-\mS-\mU$, all singularities of $\mathcal{P}^{s|u}_{\Delta,J}(\mS,\mT)$ (at fixed $\mU$)
are the poles in $\mS$. There is a unique function with these properties:
\be \label{PR Mellin}
\mathcal{P}^{s|u}_{\Delta,J}(\mS,\mT) =
\sum\limits_{m=0}^{\infty}\frac{\mathcal{Q}^{m}_{\Delta,J}(\mT+\mS-\tau-2m)}{\mS-\tau-2m}\,.
\ee
The numerator is simply the residue in (\ref{eq:mellinPoles}) written as a function of $\mU$.
We define the $t$-channel Polyakov-Regge block $\mathcal{P}^{t|u}_{\Delta,J}(\mS,\mT)$
by the same expression with $\mS$ and $\mT$ interchanged.

Since the Mellin transform of $\mathcal{P}^{s|u}_{\Delta,J}(\mS,\mT)$ satisfies all the properties of the Polyakov-Regge block
listed below (\ref{Pdefinitions}), it follows, by uniqueness, that it must be the Polyakov-Regge block!
That is:
\ba
P^{s|u}_{\Delta,J}(z,\zb) = \iint\!\!\frac{d\mS\,d\mT}{(4\pi i)^2}\,
\Gamma\!\left(\Df-\tfrac{\mS}{2}\right)^2\Gamma\!\left(\Df-\tfrac{\mT}{2}\right)^2\Gamma\!\left(-\Df+\tfrac{\mS+\mT}{2}\right)^2\times\\
\times u^{\tfrac{\mS}{2}-\Df}v^{\tfrac{\mT}{2}-\Df} \mathcal{P}^{s|u}_{\Delta,J}(\mS,\mT)\,.
\label{eq:PRMellin2}
\ea
Recall the formula for the cross-ratios $u=z\zb$ and $v=(1-z)(1-\zb)$.

\smallskip

The formula (\ref{PR Mellin}) explicitly shows that the Polyakov-Regge block is equal to the s-channel
exchange Witten diagram for an operator of dimension $\Delta$ and spin $J$, supplemented by contact terms.
The contact terms are chosen precisely so that $\mathcal{P}^{s|u}_{\Delta,J}(\mS,\mT)$ vanishes in the u-channel Regge limit,
$\mS,\mT\rightarrow\infty$ with $\mS+\mT$ fixed. Note that in general $\mathcal{P}^{s|u}_{\Delta,J}(\mS,\mT) \neq \mathcal{P}^{s|t}_{\Delta,J}(\mS,\mT)$ since a different set of contact diagrams is needed to improve the u- and t-channel Regge behaviour of a given s-channel exchange. Let us check that $\mathcal{P}^{s|u}_{\Delta,J}(\mS,\mT)$ and $\mathcal{P}^{s|t}_{\Delta,J}(\mS,\mT)$ indeed differ by contact terms. We have
\be
\mathcal{P}^{s|t}_{\Delta,J}(\mS,\mT) = (-1)^{J}\mathcal{P}^{s|u}_{\Delta,J}(\mS,4\Df-\mS-\mT)=
\sum\limits_{m=0}^{\infty}\frac{\mathcal{Q}^{m}_{\Delta,J}(\mT)}{\mS-\tau-2m}\,.
\ee
The first equality is the definition which ensures $G^s_{\Delta,J}$ appears with unit coefficient in $P^{s|t}_{\Delta,J}$. The second equality uses symmetry of Mack polynomials $\mathcal{Q}^{m}_{\Delta,J}(\mT) = (-1)^{J}\mathcal{Q}^{m}_{\Delta,J}(4\Df-\mT-\tau-2m)$. Thus
\be
\mathcal{P}^{s|u}_{\Delta,J}(\mS,\mT) - \mathcal{P}^{s|t}_{\Delta,J}(\mS,\mT) =
\sum\limits_{m=0}^{\infty}\frac{\mathcal{Q}^{m}_{\Delta,J}(\mT+\mS-\tau-2m)-\mathcal{Q}^{m}_{\Delta,J}(\mT)}{\mS-\tau-2m}\,.
\ee
The numerator of each summand vanishes as $\mS\rightarrow\tau+2m$, cancelling the pole, which means the RHS is a polynomial in $\mS,\mT$ of degree $J-1$, i.e. a contact diagram \cite{Heemskerk:2009pn,Fitzpatrick:2011ia}.

\smallskip

Since the Mellin representation is unique \cite{Penedones:2019tng}, (\ref{eq:PRMellin2}) allows us to
write the Polyakov-Regge expansion (\ref{PR expansion}) directly in Mellin space:\footnote{We apologize for $\mathcal{P}$ denoting both the t-channel primary and Mellin-space Polyakov-Regge block.}
\be
M(\mS,\mT) = \sum_{\cO}a_{\cO}\,\mathcal{P}^{s|u}_{\Delta_{\cO},J_{\cO}}(\mS,\mT)
 +\sum_{\cP}a_{\cP}\,\mathcal{P}^{t|u}_{\Delta_{\cP},J_{\cP}}(\mS,\mT)\,.
\label{eq:mellinDispersion}
\ee
This will be an important formula. What is its physical interpretation?
We now show that this is nothing but the fixed-$\mU$ dispersion relation for the meromorphic function $M(\mS,\mT)$!

\smallskip

To simplify the discussion, let us introduce a notation for the fixed-$\mU$ amplitude:
\be
 M(\mS;\mU) \equiv M(\mS,4\Df-\mS-\mU)
\ee
where we have simply expressed $\mT$ as a function of $\mS$ and $\mU$.
In a (nonperturbative) unitary CFT, it is known that the correlator grows at most linearly in the Regge limit
$|\mS|\to\infty$.  This is equivalent to boundedness of $\cG(z,\zb)$ in the Regge limit \cite{Penedones:2019tng}.
In this section we restrict our attention to superbounded correlators, where the analogous condition is that
\be
\lim_{|\mS|\to\infty} |M(\mS;\mU)| = 0 \label{no arcs}
\ee
as was imposed already for individual Polyakov-Regge blocks in (\ref{PR super bounded}).
This condition ensures convergence of an unsubtracted dispersion relation.
We start from the Cauchy formula
\be
M(\mS;\mU) = \oint\limits_{C_\mS}\frac{d\mS'}{2\pi i}\frac{M(\mS';\mU)}{\mS'-\mS}\,,
\label{eq:mellinCauchy}
\ee
where $C_\mS$ is a tiny clockwise circle around $\mS$. We now expand the contour,
picking up contributions form the poles of $M(\mS';\mU)$, and dropping
arcs at infinity thanks to (\ref{no arcs}).
We are left with the s-channel poles listed in \eqref{eq:mellinPoles}, and t-channel poles
whose location (at fixed $\mU$) satisfy $\mT'=4\Df-\mS'-\mU=\tau_{\mathcal{P}}+2m$.
Thus each s- and t-channel primary contributes an infinite tower of residues:
\ba
M(\mS;\mU) &= 
\sum_{\cO}a_{\cO}\sum\limits_{m=0}^{\infty}\frac{\mathcal{Q}^{m}_{\Delta_{\cO},J_{\cO}}(4\Df-\mU-\tau_{\cO}-2m)}{\mS-\tau_{\cO}-2m} \\
&\quad 
+\sum_{\cP}a_{\cP}\sum\limits_{m=0}^{\infty}\frac{\mathcal{Q}^{m}_{\Delta_{\cP},J_{\cP}}(4\Df-\mU-\tau_{\cP}-2m)}{
4\Df-\mS-\mU-\tau_{\cP}-2m}\,.
\ea
Collecting the contribution of each operator, and eliminating $\mU$ in favor of $\mT$,
this is precisely (\ref{eq:mellinDispersion}) with the Polyakov-Regge block in (\ref{PR Mellin}).

\smallskip

To summarize this section so far, we started from the dispersion relation in position space in (\ref{eq:Dispersion1}),
which reconstructs correlators from their double discontinuities in two channels.
We showed that it expresses correlators as a sum of Polyakov-Regge blocks,
(\ref{PR expansion}), and we now found that this is equivalent to a dispersion relation in Mellin space.
This is a surprising and unexpected result: \emph{all CFT dispersion relations are equivalent}.

\subsection{Position-space kernel from Mellin space}\label{ssec:kernelFromMellin}

In light of this observation, and given the simplicity of the Mellin-space dispersion relation, it is rewarding to derive the relatively complex position-space kernel \eqref{eq:kernelsUV} starting from (\ref{eq:mellinCauchy}).
This will provide the last missing edge, the rightmost edge, to the triangle in Figure~\ref{fig:triangle}.
Our goal is not to be fully rigorous, but simply to present convincing evidence for the equivalence. Indeed, the point of view of this paper is that Mellin space can be used to motivate various position-space results. These results are in turn rigorously established by working directly in position space.

\begin{figure}[ht]
\begin{center}
\includegraphics[width=0.95\textwidth]{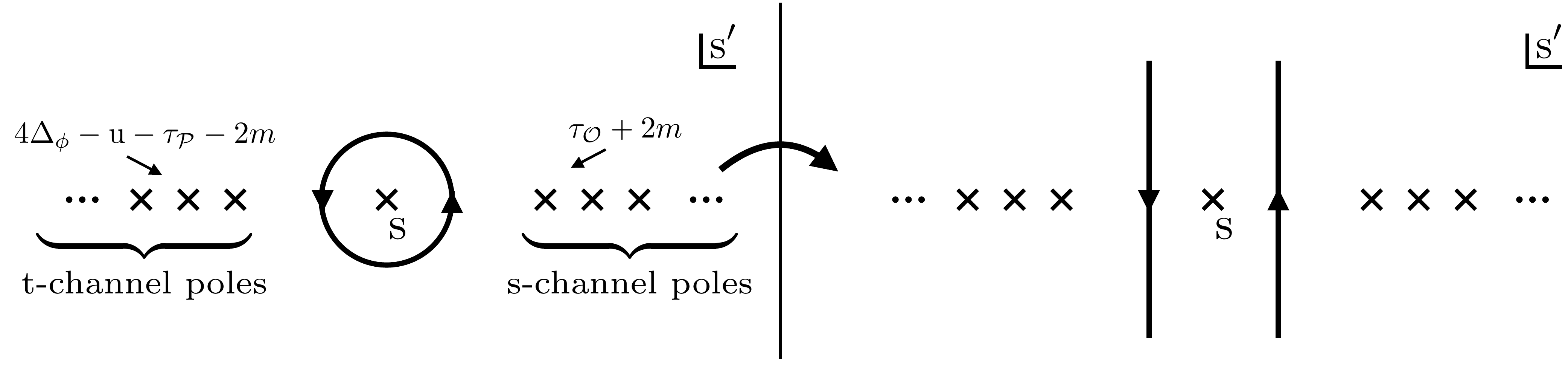}
\caption{Contour deformation used in deriving the position-space dispersion kernel from the Mellin-space dispersion relation. We start from a Cauchy integral formula for $M(\mS,\mT)$, \eqref{eq:mellinCauchy}. Deforming the circular contour to a pair of straight contours in the imaginary direction leads to the split representation \eqref{eq:mellinSplit}.}
\label{fig:MellinContours}
\end{center}
\end{figure}

The idea is simply to insert a suitable form of the Mellin-space dispersion relation into the Mellin representation \eqref{eq:mellinRep}. We start with the Cauchy integral formula at fixed $\mU$, as shown in \eqref{eq:mellinCauchy}. We then deform the circular contour to a pair of contours running in the imaginary direction, to the right and to the left of $\mS$, as shown in Figure \ref{fig:MellinContours}. This leads to a natural split of $M(\mS,\mT)$ into s- and t-channel parts
\be
M(\mS,\mT) = M^s(\mS,\mT) + M^t(\mS,\mT)\,,
\label{eq:mellinSplit}
\ee
where
\ba
M^s(\mS,\mT) &= \int\limits_{\mS+\epsilon-i\infty}^{\mS+\epsilon+i\infty}\frac{d\mS'}{2\pi i}\frac{M(\mS',\mT')}{\mS'-\mS}\\
M^t(\mS,\mT) &= \int\limits_{\mS-\epsilon-i\infty}^{\mS-\epsilon+i\infty}\frac{d\mS'}{2\pi i}\frac{M(\mS',\mT')}{\mS-\mS'} = 
\int\limits_{\mT+\epsilon-i\infty}^{\mT+\epsilon+i\infty}\frac{d\mT'}{2\pi i}\frac{M(\mS',\mT')}{\mT'-\mT}\,,
\label{eq:MsMt}
\ea
where we defined $\mT' = \mT+\mS-\mS'$. When we insert \eqref{eq:mellinSplit} into the Mellin representation \eqref{eq:mellinRep}, we obtain
\be
\cG(u,v) = \cG^s(u,v) + \cG^t(u,v)\,,
\ee
where
\ba
\cG^s(u,v) = \frac{1}{4}\iiint\!\frac{d\mS\,d\mT\,d\mS'}{(2\pi i)^3}\,
\Gamma\!\left(\Df-\tfrac{\mS}{2}\right)^2\Gamma\!\left(\Df-\tfrac{\mT}{2}\right)^2\Gamma\!\left(-\Df+\tfrac{\mS+\mT}{2}\right)^2\times\\
\times \,u^{\tfrac{\mS}{2}-\Df}v^{\tfrac{\mT}{2}-\Df}
\frac{M(\mS',\mT')}{\mS'-\mS}\,.
\label{eq:GsMellin} 
\ea
with the $\mS'$ contour the same as in \eqref{eq:MsMt}.
$\cG^t(u,v)$ is given by an identical formula with $\frac{d\mS'}{\mS'-\mS}$ replaced with $\frac{d\mT'}{\mT'-\mT}$.
We focus on $\cG^s(u,v)$.

\smallskip

At this point, it may be tempting to substitute for $M(\mS',\mT')$ in \eqref{eq:GsMellin} using the fact that it is the Mellin transform of $\cG(u,v)$, i.e. using the inverse of \eqref{eq:mellinRep}\footnote{Formula \eqref{eq:mellinRep} applies with straight vertical contours provided $\Df<(\tau^{\mS}_{\min}+\tau^{\mT}_{\min}+\tau^{\mU}_{\min})/4$, where $\tau^{\mS,\mT,\mU}_{\min}$ are the minimal twists in each channel. Similarly, formula \eqref{eq:mellinTransform} applies literally only if the same condition is satisfied, since otherwise the integral \eqref{eq:mellinTransform} does not converge for any $\mS,\mT$. The condition can be satisfied by subtracting low-twist conformal blocks from $\cG(u,v)$ \cite{Penedones:2019tng}.}
\be
M(\mS,\mT) =
\int\limits_{0}^{\infty}\!\int\limits_{0}^{\infty}\frac{du dv}{u v}\frac{u^{\Df-\tfrac{\mS}{2}}v^{\Df-\tfrac{\mT}{2}}}{ \Gamma\!\left(\Df-\tfrac{\mS}{2}\right)^{2}\Gamma\!\left(\Df-\tfrac{\mT}{2}\right)^{2}\Gamma\!\left(-\Df+\tfrac{\mS+\mT}{2}\right)^{2}}\,\cG(u,v)\,.
\label{eq:mellinTransform}
\ee
Naively, this leads to a representation of $\cG^s(u,v)$ as a double integral of $\cG(u',v')$, with the kernel given by
\be
\frac{1}{4u'v'}\iiint\!\frac{d\mS\,d\mT\,d\mS'}{(2\pi i)^3}
\frac{u^{\tfrac{\mS}{2}-\Df}v^{\tfrac{\mT}{2}-\Df}}{u'^{\tfrac{\mS'}{2}-\Df}v'^{\tfrac{\mT'}{2}-\Df}}
\frac{\Gamma(\Df-\tfrac{\mS}{2})^2\Gamma(\Df-\tfrac{\mT}{2})^2}{\Gamma(\Df-\tfrac{\mS'}{2})^2\Gamma(\Df-\tfrac{\mT'}{2})^2}\frac{1}{\mS'-\mS}\,.
\label{eq:kernelMellinNaive}
\ee
However, this idea does not work because the integrand \eqref{eq:kernelMellinNaive} diverges exponentially for large imaginary $\mS'$ because of the gamma functions in the denominator. Moreover, we want to express $\cG^s(u,v)$ as an integral of $\dDisc_s[\cG(u',v')]$ rather than $\cG(u',v')$ anyway. Thus, to recover the position-space dispersion relation \eqref{eq:Dispersion1}, it would be nice to have a formula for $M(\mS',\mT')$ given as an integral of $\dDisc_s[\cG(u',v')]$ and $\dDisc_t[\cG(u',v')]$, which could then be inserted into \eqref{eq:GsMellin}. We do not know  of such formula but fortunately it is not needed to finish the argument. The reason is that \eqref{eq:GsMellin} only receives contributions from singularities of $M(\mS',\mT')$ with $\mathrm{Re}(\mS')>\mathrm{Re}(\mS)$, since we can close the $\mS'$ contour to the right. All of these singularities come from s-channel operators. Thus $M(\mS',\mT')$ in \eqref{eq:GsMellin} can be replaced with any function with the same s-channel poles and residues. Thus, we can make the substitution
\ba
M(\mS',\mT')\big|_{\mS'\textrm{-channel poles}}=
\iint\frac{du' dv'}{u' v'}\frac{u'^{\Df-\tfrac{\mS'}{2}}v'^{\Df-\tfrac{\mT'}{2}}}{\Gamma(\Df-\tfrac{\mS'}{2})^{2}\Gamma(\Df-\tfrac{\mT'}{2})^{2}\Gamma(-\Df+\tfrac{\mS'+\mT'}{2})^{2}}\times\\
\times\frac{\dDisc_s[\cG(u',v')]}{2\sin ^2\!\left[\tfrac{\pi}{2}(\mS'-2\Df)\right]}
\,.
\label{eq:mPolarParts}
\ea
in \eqref{eq:GsMellin}.
Indeed, the s-channel poles come entirely from the small-$u$ region in the integral \eqref{eq:mellinTransform},
where replacing $\cG(u,v)$ with $\dDisc_s[\cG(u,v)]$ amounts to multiplying each block $G^s_{\Delta_{\cO},J_{\cO}}(u,v)$
by $2\sin ^2\!\left[\tfrac{\pi}{2}(\tau_{\cO}-2\Df)\right]$, see (\ref{dDisc block}).
This can be compensated by dividing 
the Mellin amplitude by $2\sin ^2\!\left[\tfrac{\pi}{2}(\mS'-2\Df)\right]$ as was done in \eqref{eq:mPolarParts}. This works because the poles corresponding to $\cO$ occur at $\mS' = \tau_{\cO}+2m$ and so $\sin ^2\!\left[\tfrac{\pi}{2}(\mS'-2\Df)\right]=\sin ^2\!\left[\tfrac{\pi}{2}(\tau_{\cO}-2\Df)\right]$ at the locations of the poles.
Furthermore, the division has formally not introduced any new singularity
since it only cancelled zeros from existing gamma functions. For $\cG^t(u,v)$ we use a similar substitution with the t-channel sine-squared factors.
We are unsure what the correct integration region is in \eqref{eq:mPolarParts}.
A natural guess is to use the maximal region where the OPE converges, i.e. $\sqrt{u'}<\sqrt{v'}+1$. As we will see soon, the precise shape of the region is not important for the purpose of reproducing the position-space dispersion relation.

\smallskip

When we use \eqref{eq:mPolarParts} in \eqref{eq:GsMellin} and swap the order of Mellin-space and position-space integrations, we arrive at
\be
\cG^s(u,v) = \iint\!\!du' dv' K_{\textrm{Mellin}}(u,v;u',v')\dDisc_{s}[\cG(u',v')]
\ee
where
\ba
K_{\textrm{Mellin}}(u,v;u',v') = 
\frac{1}{4u'v'}\iiint\!\frac{d\mS\,d\mT\,d\mS'}{(2\pi i)^3}
&\frac{\Gamma(\Df-\tfrac{\mS}{2})^2\Gamma(\Df-\tfrac{\mT}{2})^2}{2\sin ^2\!\left[\tfrac{\pi}{2}(\mS'-2\Df)\right]\Gamma(\Df-\tfrac{\mS'}{2})^2\Gamma(\Df-\tfrac{\mT'}{2})^2}\\
\times&\frac{u^{\tfrac{\mS}{2}-\Df}v^{\tfrac{\mT}{2}-\Df}}{u'^{\tfrac{\mS'}{2}-\Df}v'^{\tfrac{\mT'}{2}-\Df}}\frac{1}{\mS'-\mS}\,.
\label{eq:KMellin}
\ea
The extra factor $\sin^2\!\left[\tfrac{\pi}{2}(\mS'-2\Df)\right]$ in the denominator cancels the exponential growth of the gamma factors, resolving our earlier problem.  We have thus converted the Mellin-space dispersion relation to a position-space one,
and it remains only to verify that $K_{\textrm{Mellin}}(u,v;u',v')$ agrees with the position-space kernel $K(u,v;u',v')$ given in equations \eqref{eq:positionKernelFull}, \eqref{eq:kernelsUV}.

\smallskip

To evaluate (\ref{eq:KMellin}) concretely, let us first simplify the integrand by changing coordinates to $p,q,p'$ given by
\be
p=\frac{\mS}{2}-\Df\,,\qquad q=\frac{\mT}{2}-\Df\,,\qquad p'=\frac{\mS'}{2}-\Df\qquad\Rightarrow\qquad p+q-p' = \frac{\mT'}{2}-\Df\,.
\ee
We get
\ba
K_{\textrm{Mellin}}(u,v;u',v') = 
\frac{1}{2\pi^2u'v'}\iiint\!\frac{dp\,dq\,dp'}{(2\pi i)^3}
\frac{\Gamma(-p)^2\Gamma(-q)^2\Gamma(p'+1)^2}{(p'-p)\Gamma(p'-p-q)^2}\times\\
\times\left(\frac{u\phantom{'}}{v'}\right)^{p}\left(\frac{v\phantom{'}}{v'}\right)^{q}\left(\frac{v'}{u'}\right)^{p'}\,.
\label{eq:KMellin2}
\ea
Dependence on $\Df$ has dropped out as expected. Furthermore, $K_{\textrm{Mellin}}(u,v;u',v')$ is manifestly homogeneous in $u,v,u',v'$ of degree $-2$, just like $K(u,v;u',v')$. We show in Appendix \ref{app:Kmellin} that \eqref{eq:KMellin2} vanishes for $\sqrt{v'}<\sqrt{u'}+\sqrt{u}+\sqrt{v}$ as expected from \eqref{eq:positionKernelFull}, and that it contains precisely the right contact term, proportional to $\delta(\sqrt{v'}-\sqrt{u'}-\sqrt{u}-\sqrt{v})$.
In general, computing the triple integral looks like a formidable task.

\smallskip

For $\sqrt{v'}>\sqrt{u'}+\sqrt{u}+\sqrt{v}$ we could check that \eqref{eq:KMellin2} agrees with $K_B(u,v;u',v')$ in an expansion around $v=0$, $u'=0$. We start by closing the $q$ contour to the right, encountering double poles of $\Gamma(-q)^2$ at $q=n$, where $n=0,1,\ldots$. This gives an expansion of $K_{\textrm{Mellin}}(u,v;u',v')$ around $v=0$ containing terms $v^n\log(v)$ and $v^n$, which precisely agrees with the structure of $K_B(u,v;u',v')$. Let us focus here on the term proportional to $v^0\log(v)$
\ba
K_{\textrm{Mellin}}(u,v;u',v')|_{v^0\log(v)} =
-\frac{1}{2\pi^2u'v'}\iint\!\frac{dp\,dp'}{(2\pi i)^2}
\frac{\Gamma(-p)^2\Gamma(p'+1)^2}{\Gamma(p'-p)\Gamma(p'-p+1)}\times\\
\times\left(\frac{u\phantom{'}}{v'}\right)^{p}\left(\frac{v'}{u'}\right)^{p'}\,.
\ea
The $p'$ contour can now be closed to the left, encountering poles at $p'=-1,-2,\ldots$, which leads to a power-series expansion around $u'=0$. Focusing on the $u'^0$ term, we get
\be
K_{\textrm{Mellin}}(u,v;u',v')|_{v^0\log(v)\cdot u'^0} =
\int\!\frac{dp}{2\pi i}
\frac{ \left[(p+1) \left(-2 H_{-p-2}-\log u'+\log v'\right)+1\right]}{2 \pi ^2 v'^2}\left(\frac{u}{v'}\right)^p
\ee
where $H_{z}$ is the harmonic number. The contour of the $p$ integral can now be closed to the right, encountering the poles of the harmonic number at $p=0,1,\ldots$. The resulting sum over residues is equal to
\be
K_{\textrm{Mellin}}(u,v;u',v')|_{v^0\log(v)\cdot u'^0} = \frac{1}{\pi^2(u-v')^2}\,,
\ee
which precisely agrees with the $v^0\log(v)\cdot u'^0$ term in $K_B(u,v;u',v')$ (related to (\ref{eq:kernelExample})).
We have checked many further terms, finding exact agreement with $K_B(u,v;u',v')$. This makes us certain that $K_{\textrm{Mellin}}$ exactly agrees with the dispersion kernel \eqref{eq:positionKernelFull}.

\subsection{Formal properties of the dispersion kernel}

The equivalence between the position-space and Mellin-space dispersion relations is surprising,
because the different approaches make very different properties manifest.

\smallskip

As noticed in~\cite{Carmi:2019cub}, the kernel does not depend on the dimensionality of the space-time in which the CFT resides. This was nontrivial there since the kernel was computed as an infinite sum over conformal blocks, which depend on space-time dimension, and only the two cases $d=2$ and $d=4$ could be studied.
In contrast, in Mellin space, it is evident that the kernel in (\ref{eq:KMellin2}) is independent of space-time dimension.

\smallskip

On the other hand, it is far from obvious from Mellin space that the kernel has nice group-theoretic properties.
It was manifest from the starting point in~\cite{Carmi:2019cub} that the dispersion relation
commutes with the u-channel Casimir.\footnote{Reference~\cite{Carmi:2019cub} considered the fixed-s dispersion relation and (2.23) there manifestly commutes with the s-channel Casimir.}
Schematically, if $C_u$ denotes the u-channel quadratic Casimir invariant of the conformal group (see~\cite{Dolan:2003hv} for explicit expressions),
\be
 C_u\circ \int_{u',v'} K(u,v,u',v')\dDisc_s[\cG(u',v')] = \int_{u',v'} K(u,v,u',v')\dDisc_s[C_u\circ\cG(u',v')]\,.
\ee
This implies that $K(u,v,u',v')$ satisfies certain second-order differential equations discussed
in \cite{Carmi:2019cub}. In Mellin space, $C_u$ becomes the following finite-difference operator:
\ba
 C_u\circ M(\mS,\mT) &= d(\tfrac{\mS+\mT}{2}-2\Df) M(\mS,\mT)+
 \Bigg[(\tfrac{\mS}{2}-\Df)^2\big(M(\mS-2,\mT+2) -M(\mS,\mT)\big)+
\\ &\hspace{10mm}+ (\tfrac{\mS+\mT}{2}-2\Df)^2M(\mS,\mT)-(\tfrac{\mS+\mT}{2}-\Df)^2M(\mS,\mT{+}2)+ (\mS{\leftrightarrow}\mT)\Bigg]\ .
\ea
In the notation ($\mS{\leftrightarrow}\mT$) the two arguments of $M$ should be permuted
so that the first argument remains $\mS$ plus an integer shift.
The Casimir is a linear polynomial in $d$.
The fact that its $d$-dependent part commutes with the dispersion relation in any $d$ is simply the fact that $\mU$ is fixed.
It is nontrivial, but also true, that the $d$-independent part also commutes. That is,
on any contour that is invariant under the shift of $\mS$ and $\mT$ by integers, one can check the following surprising identity:
\ba
 C_u\circ \int \frac{d\mS'}{\mS'-\mS} M(\mS'; \mU) =\int \frac{d\mS'}{\mS'-\mS} C_u\circ M(\mS'; \mU).
\ea
Thus the Mellin dispersion relation commutes with the Casimir equation.\footnote{This is analogous to the fact
that $\int \frac{ds'}{s'-s} P_J(s')$ commutes with the Legendre differential equation, where $s=\cos\theta$.
The polynomial solution $P_J(s)$ turns into the non-polynomial solution $Q_J(s)$ to the same equation.}

\smallskip

Besides the Casimir differential equations, an \emph{additional} first-order equation
was also derived in~\cite{Carmi:2019cub} encoding the equality of $d=2$ and $d=4$ kernels
((3.3) there). It would be interesting to find the Mellin interpretation of that equation.
As shown in section 3.4 of that paper, this set of differential equations uniquely determines the kernel
$K(u,v,u',v')$. Without such differential equations
it would seem virtually impossible to directly compute the triple integral in (\ref{eq:KMellin2})
(without already knowing the answer to compare to).

% !TEX root = ../main.tex

\section{Sum rules for superbounded correlators}\label{sec:Edges}
The logic of Section \ref{sec:Vertices} implies the existence of infinitely many sum rules satisfied by the OPE data in superbounded correlators. There are several complementary ways of understanding the sum rules, corresponding to the vertices of the triangle in Figure \ref{fig:triangle}. The sum rules arise by using the OPE on both sides of the dispersion relation (the first line of \eqref{eq:Dispersion1}). This leads to the Polyakov-Regge sum rules \eqref{PR minus Euclidean} labelled by position-space variables $z,\zb$. It proves useful to expand these at small $z,\zb$, where one finds that the sum rules can be labelled by double-trace operators \eqref{first sum rule}. We will see that if we work in Mellin space, the same set of sum rules can be formally derived by combining the Mellin-space dispersion relation with the `Polyakov conditions' which say $M(\mS,\mT) = \partial M(\mS,\mT) = 0$ at double-trace locations. We would like to stress that all of these sum rules are a consequence of nothing more than crossing symmetry and unitarity.\footnote{This section also assumes superboundedness. Unitarity guarantees just boundedness, used in Section~\ref{sec:Subtractions}.} Indeed, they can all be derived by starting from the standard crossing equation in position space and applying suitable linear functionals to it.
In other words, one should not think of the Polyakov condition as a new constraint on conformal field theories but rather as a useful way of organizing the existing constraints of crossing symmetry.

\smallskip

We will now discuss the sum rules in more detail, and use them to further illustrate the equivalence of the above approaches.

\subsection{Generalities}

To simplify the discussion, we will assume that $\cG(z,\zb)$ is u-channel superbounded in this entire section. The assumption of superboundedness is not entirely
unphysical. Perhaps the simplest example is the four-point function of a complex scalar field $\langle\phi\bar{\phi}\phi\bar{\phi}\rangle$. This correlator is u-channel superbounded provided the lightest operator in the $\phi\times\phi$ OPE has $\Delta_{0}>1$.\footnote{The correlator goes as $(z\zb)^{-\Delta_0/2}$ in the u-channel Euclidean OPE limit, and the u-channel Regge limit is bounded in the OPE limit by positivity of the s-channel OPE.} Supersymmetry also provides many examples \cite{Beem:2013qxa,Caron-Huot:2018kta}.

\smallskip

All sum rules covered in this paper can be obtained by acting with suitable linear functionals on the position-space crossing equation
\be
\sum\limits_{\cO}a_{\cO}\,G^s_{\Delta_{\cO},J_{\cO}}(z,\zb) = \sum\limits_{\cP}a_{\cP}\,G^t_{\Delta_{\cP},J_{\cP}}(z,\zb)\,.
\label{eq:crossing3}
\ee
Given a functional $\omega$ which commutes with the OPE sums, we obtain the sum rule
\be
\sum\limits_{\cO}a_{\cO}\,\omega[G^s_{\Delta_{\cO},J_{\cO}}] = \sum\limits_{\cP}a_{\cP}\,\omega[G^t_{\Delta_{\cP},J_{\cP}}]\,.
\label{eq:sumRule3}
\ee
As discussed in \cite{Mazac:2019shk} and reviewed around (\ref{first sum rule}), 
the functionals $\alpha^s_{n,\ell}$, $\beta^s_{n,\ell}$, $\alpha^t_{n,\ell}$ and $\beta^t_{n,\ell}$ are a basis for the space of functionals which commute with OPE sums in superbounded correlators. The sum rules arising from these basis functionals are particularly interesting because they automatically suppress the contribution of double-twist operators. Indeed, the duality equations \eqref{eq:dualityC1} imply that each of the basis functionals has double zeros as a function of $\Delta$ on almost all s- and t-channel double-twist blocks. This property makes these functionals ideally suited for studying holographic CFTs since in the large-$c$ limit, they directly constrain single-trace operators. We will discuss applications to holographic CFTs in a companion paper.

\smallskip

We will often be interested in correlators invariant under switching the s- and t-channels (those where the u-channel OPE only contains primaries of even spin). In that case, the crossing equation takes the form
\be
\sum\limits_{\cO}a_{\cO}\,[G^s_{\Delta_{\cO},J_{\cO}}(z,\zb) -G^t_{\Delta_{\cO},J_{\cO}}(z,\zb)]=0\,,
\ee
and the complete set of sum rules consists of
\be
\sum\limits_{\cO}a_{\cO}\,\alpha_{n,\ell}[G^s_{\Delta_{\cO},J_{\cO}}] = 0 \,,\qquad
\sum\limits_{\cO}a_{\cO}\,\beta_{n,\ell}[G^s_{\Delta_{\cO},J_{\cO}}] = 0\,,
\ee
for $n,\ell\in\mathbb{N}$, where we introduced the notations
\be
\alpha_{n,\ell} = \alpha^s_{n,\ell} - \alpha^t_{n,\ell}\,,\qquad
\beta_{n,\ell} = \beta^s_{n,\ell} - \beta^t_{n,\ell}\,.
\ee

\subsection{Free theory examples}
To make the discussion more concrete, let us apply the basis functionals to some simple correlators. Arguably the simplest correlator is the s-channel identity
\be
\cG(z,\zb) = G^s_{0,0}(z,\zb)= (z\zb)^{-\Df}\,.
\ee
This is realized as the correlator $\langle\phi_1\bar{\phi}_1\phi_2\bar{\phi}_2\rangle$ in the theory of two copies of complex scalar mean field theory. $\cG(z,\zb)$ is superbounded provided $\Df>1/2$. The s-channel OPE contains only the identity block, and the t-channel OPE only double-twist blocks
\be
G^s_{0,0}(z,\zb) = \sum\limits_{n,\ell=0}^{\infty}q^{\textrm{MFT}}_{n,\ell}G^{t}_{\Delta_{n,\ell},\ell}(z,\zb)\,,
\label{eq:QMFT}
\ee
where $q^{\textrm{MFT}}_{n,\ell}$ are known functions of $d$ and $\Df$ \cite{Fitzpatrick:2011dm}. Let us apply the basis functionals to this equation and use \eqref{eq:dualityC1}. The validity of the resulting sum rules is equivalent to the following action of the basis on the identity block
\be
\alpha^s_{n,\ell}[G^s_{0,0}] = 0\,,\qquad\beta^s_{n,\ell}[G^s_{0,0}] = 0\,,\qquad
\alpha^t_{n,\ell}[G^s_{0,0}] = q^{\textrm{MFT}}_{n,\ell}\,,\qquad\beta^t_{n,\ell}[G^s_{0,0}] = 0\,.
\ee
The actions on $G^t_{0,0}$ follow by switching $s\leftrightarrow t$ everywhere. These formulas can be explicitly verified from the description of the basis functionals as contour integrals \eqref{eq:DualBasisContours}. For example $\beta^s_{n,\ell}[G^s_{0,0}] = \beta^t_{n,\ell}[G^s_{0,0}] = 0$ follows immediately by deforming the $\wb$ contour in \eqref{eq:DualBasisContours} to the right, where we encounter no singularity and the integral thus vanishes. It follows that the Polyakov-Regge block of the identity operator is equal to the identity conformal block
\be
P^{s|u}_{0,0}(z,\zb) = G^{s}_{0,0}(z,\zb)\,,\qquad P^{t|u}_{0,0}(z,\zb) = G^{t}_{0,0}(z,\zb)\,.
\ee
We note in passing that it also immediately follows from \eqref{eq:dualityC1} that the Polyakov-Regge blocks identically vanish to the second order around double-twist dimensions
\be
P^{s|u}_{\Delta_{n,\ell},\ell}(z,\zb) = \partial_{\Delta}P^{s|u}_{\Delta_{n,\ell},\ell}(z,\zb) = 0\,,\qquad 
P^{t|u}_{\Delta_{n,\ell},\ell}(z,\zb) = \partial_{\Delta}P^{t|u}_{\Delta_{n,\ell},\ell}(z,\zb) = 0
\ee
for all $n,\ell\in\mathbb{N}$.

\smallskip

For a less trivial example, we will now consider the four-point function $\langle\cO\bar{\cO}\cO\bar{\cO}\rangle$, where $\cO = \phi^2$ with $\phi$ a free complex scalar in $d=4$, so that $\Delta_{\cO}=2$. This correlator has a disconnected part, equal to $G^s_{0,0}+G^t_{0,0}$, and a connected part. We will focus on the connected part, which is proportional to
\be
\cG(z,\zb) = \frac{1}{u v}\,.
\ee
$\cG(z,\zb)$ is superbounded and $\text{s}\leftrightarrow\text{t}$ symmetric. The conformal block expansion consists of a single family of twist-two primaries
\be
\cG(z,\zb) = \sum\limits_{J=0}^{\infty}a_J G^s_{J+2,J}(z,\zb)
= \sum\limits_{J=0}^{\infty}a_J G^t_{J+2,J}(z,\zb)\,,
\ee
where $a_J=\frac{(J!)^2}{(2J)!}$. Note that the whole family lies below the double-twist threshold.
The $J=2$ primary is the stress-tensor. If we apply the basis functionals to this equation, we find the following sum rules satisfied by the coefficients $a_{J}$
\be
\sum\limits_{J=0}^{\infty}a_J\,\alpha_{n,\ell}[G^s_{J+2,J}] = 0\,,\qquad \sum\limits_{J=0}^{\infty}a_J\,\beta_{n,\ell}[G^s_{J+2,J}] = 0\,.
\ee
This is a toy version of the idea that the basis functionals directly constrain the single-trace operators in holographic CFTs.\footnote{We can think of $\cO$ as a single-trace operator. After we apply any of the basis functionals to the crossing equation for $\langle\cO\bar{\cO}\cO\bar{\cO}\rangle$, the disconnected part drops out by the above arguments, and we are left with a sum rule for the `single-trace' operators of twist two.} Let us consider the sum rule coming from $\beta_{0,0}$. This functional has the following integral representation
\be
\beta_{0,0}[\cG] =
\!\iint\limits_{C_-\times C_+}\!\!\frac{dwd\wb}{(2\pi i)^2}\frac{4(w+\wb-1)}{(\wb-w)^2}\cG(w,\wb)\,.
\ee
We can explicitly evaluate $\beta_{0,0}[G^s_{J+2,J}]$ by picking the residues at $w=\wb=0$, finding
\be
\beta_{0,0}[G^s_{J+2,J}] =
\begin{cases}
\;\;\,\frac{2}{3} &\text{for }J=0\\
-4 &\text{for }J=2\\
\;\;\;0 &\text{for }J\neq0,2\,,
\end{cases}
\ee
so the sum rule is satisfied since $a_{0}=1$ and $a_{2}=1/6$. Similarly, we consider $\alpha_{0,0}$, which takes the form
\be
\alpha_{0,0}[\cG] =
\!\iint\limits_{C_-\times C_+}\!\!\frac{dwd\wb}{(2\pi i)^2}\frac{2(w+\wb-1)}{(\wb-w)^2}\left\{4+\log\left[\frac{w\wb(1-w)(1-\wb)}{(\wb-w)^4}\right]\right\}\cG(w,\wb)\,.
\ee
With some more work, we can evaluate
\be
\alpha_{0,0}[G^s_{J+2,J}] =
\begin{cases}-\frac{2}{9} &\text{for }J=0\\
\;\;\;2 &\text{for }J=1\\
-\frac{43}{15} &\text{for }J=2\\
\frac{2 (-1)^J(2 J+2)!}{(J-2) J (J+3) ((J+1)!)^2}&\text{for }J>2\,,
\end{cases}
\ee
so that the $\alpha_{0,0}$ sum rule holds thanks to the identity
\be
\sum\limits_{J=3}^{\infty}\frac{(-1)^J(2J+1)}{(J-2)J(J+1)(J+3)} = -\frac{3}{40}\,.
\ee
This example illustrates the general phenomenon that the sum rules considered in this paper converge as inverse power-laws at large $\Delta$ and $J$. This is in contrast with the sum rules coming from equating a pair of Euclidean OPEs at a fixed $z,\zb$, which converge exponentially fast. This difference is due to the fact that functionals such as $\alpha_{n,\ell}$, $\beta_{n,\ell}$ probe the u-channel OPE and Regge limits, where the exponential convergence of the s- and t-channel OPEs becomes arbitrarily slow.

\subsection{Generating functionals}\label{ssec:GeneratingFunctionals}
The discussion of Section \ref{ssec:FunctionalBasis} makes it clear that it is natural to combine the basis functionals $\alpha^s_{n,\ell}$, $\beta^s_{n,\ell}$, $\alpha^s_{n,\ell}$ and $\beta^s_{n,\ell}$ to the following `generating' functionals
\ba
\Omega^{s|u}_{z,\zb} &=
 \sum\limits_{n,\ell}\left[G^t_{\Delta_{n,\ell},\ell}(z,\zb)\alpha^{t}_{n,\ell}+
 \partial_{\Delta}G^t_{\Delta_{n,\ell},\ell}(z,\zb)\beta^{t}_{n,\ell}\right]\\
\Omega^{t|u}_{z,\zb} &=
 \sum\limits_{n,\ell}\left[G^s_{\Delta_{n,\ell},\ell}(z,\zb)\alpha^{s}_{n,\ell}+
 \partial_{\Delta}G^s_{\Delta_{n,\ell},\ell}(z,\zb)\beta^{s}_{n,\ell}\right]\,.
\label{eq:OmegaFunctionals}
\ea
$\Omega^{s|u}_{z,\zb}$ is the linear functional obtained by applying the integral transform $\Omega^{s|u}$ defined in \eqref{Omega-s}, followed by evaluating the result at $z,\zb$. Similarly for $\Omega^{t|u}_{z,\zb}$. Equation \eqref{punchline} then implies
\be
\Omega^{s|u}_{z,\zb} + \Omega^{t|u}_{z,\zb} = \mathrm{ev}_{z,\zb}\,,
\ee
where $\mathrm{ev}_{z,\zb}$ is the functional which evaluates a superbounded function at $z,\zb$. This identity demonstrates the functionals $\alpha^{s,t}_{n,\ell}$, $\beta^{s,t}_{n,\ell}$ lead to a complete set of sum rules in the following sense. Suppose we are given a putative set of s-channel and t-channel OPE data $\{(\Delta_{\cO},a_{\cO})\}_{\cO}$, $\{(\Delta_{\cP},a_{\cP})\}_{\cP}$, both of which define a superbounded correlator. Then this OPE data satisfies the crossing equation \eqref{eq:crossing3} for all $z,\zb$ if and only if it satisfies the sum rules \eqref{eq:sumRule3}, where $\omega$ ranges over all $\alpha^{s,t}_{n,\ell}$, $\beta^{s,t}_{n,\ell}$.

\smallskip

Consider now the sum rule obtained by applying $\Omega^{t|u}_{z,\zb}$
\be
\sum\limits_{\cO}a_{\cO}\,\Omega^{t|u}_{z,\zb}[G^s_{\Delta_{\cO},J_{\cO}}] = \sum\limits_{\cP}a_{\cP}\,\Omega^{t|u}_{z,\zb}[G^t_{\Delta_{\cP},J_{\cP}}]\,.
\label{eq:OmegaTSumRule}
\ee
This sum rule suppresses all the t-channel double-twist operators because from \eqref{eq:OmegaFunctionals} we see that $\Omega^{t|u}_{z,\zb}[G^t_{\Delta,J}]$ has double zeros as a function of $\Delta$ at $\Delta = 2\Df+2n+J$. On the other hand, the s-channel double-twists are not suppressed because $\Omega^{t|u}_{z,\zb}[G^s_{2\Df+2n+J,J}] =G^s_{2\Df+2n+J,J}(z,\zb)\neq 0$. We can suppress also the s-channel double-twists with $n>0$ by taking the $z\rightarrow 0$ limit. This is allowed because the sums over $\cO,\cP$ in \eqref{eq:OmegaTSumRule} converge uniformly in $z$ in a neighbourhood of $z=0$, thanks to the discussion around equation \eqref{eq:dDiscKZero}.

\smallskip

It can be helpful to give an alternative definition of the sum rule (\ref{eq:OmegaTSumRule}).
Recall from eqs.~\eqref{P_OPEs}-\eqref{P_OPEt} that the $\alpha,\beta$ functionals describe the series expansion
of Polyakov-Regge blocks, which are the dispersive transform \eqref{P as dispersive integral} of blocks.
Thus the $\Omega^{t|u}_{z,\zb}$ sum rule \eqref{eq:OmegaTSumRule} states the equality between dispersion relations and s-channel OPE (see also
\eqref{PR minus Euclidean})
\be\begin{aligned} \label{PR minus Euclidean 2}
&\sum\limits_{\cO}a_{\cO}\left[G^s_{\Delta_{\cO},J_{\cO}}(z,\zb)-\iint du'dv'K(u,v;u',v')\dDisc_s[G^s_{\Delta_{\cO},J_{\cO}}(w,\wb)]
\right]
\\
&=\sum\limits_{\cP}a_{\cP} \iint du'dv' K(v,u;v',u')\dDisc_t[G^t_{\Delta_{\cP},J_{\cP}}(w,\wb)]\ .
\end{aligned}\ee
From this perspective, the reason why each summand admits an expansion in $z^n$ and $z^n\log(z)$ is because the dispersion integral cancels all the non-integer powers
from $G^s_{\Delta,J}(z,\zb)$.
In practice, since the functional action is analytic in $\Delta$ (this is clear from the $C_-\times C_+$ contour integral),
the coefficient of a given integer power of $z$ may be computed by starting with $\Delta$ large enough
that $G^s_{\Delta,J}(z,\zb)$ can be neglected, and then analytically continuing in $\Delta$. This is further described in Appendix \ref{app:numericalbtwov}.

\smallskip

Thus, let us define the functional $B^s_{v}$ as the coefficient of $\log(z)$ in the small-$z$ expansion of $\Omega^{t|u}_{z,\zb}$, with $\zb=1-v$ fixed
\be
B^s_{v} = \,\Omega^{t|u}_{z,\zb}|_{(\log z)z^0}\,. \label{def B}
\ee
We can use \eqref{eq:OmegaFunctionals} to give an expansion of $B^s_{v}$ in the basis functionals
\be
B^s_{v} = \frac{1}{2}\sum\limits_{\ell=0}^{\infty}\,(1-v)^{-\Df}k_{\Df+\ell}(1-v)\,\beta^s_{0,\ell}\,,
\label{eq:BsbExpansion}
\ee
where $k_h(z)$ is the 1D block
\be
k_h(z) = z^h{}_2F_1(h,h;2h;z)\,,
\ee
and we used
\be
G^s_{\Delta,J}(z,\zb)\sim z^{\frac{\Delta-J}{2}-\Df}\zb^{-\Df}k_{\frac{\Delta+J}{2}}(\zb)\quad\text{as}\quad z\rightarrow 0\,.
\ee
The superscript $s$ reminds us that $B^s_{v}$ is a linear combination of $\beta^s$ functionals. It follows that $B^s_{v}[G^s_{\Delta,J}]$ has double zeros on all $n>0$ double-twist dimensions and simple zeros on the $n=0$ ones, with slope given by $(1-v)^{-\Df}k_{\Df+\ell}(1-v)$. $B^s_{v}[G^t_{\Delta,J}]$ has double zeros on all double-twist dimensions. By expanding $K_B(u,v;u',v')$ in \eqref{Omega-t} at small $u$, we get a representation of $B^s_{v}$ acting on a general superbounded function $\cG(w,\wb)$ as a contour integral
\be
B^s_{v}[\cG] = \!\iint\limits_{C_-\times C_+}\!\!\frac{dwd\wb}{(2\pi i)^2}
\frac{(\wb-w) (u'-v'-v)}{\left[v^2-2(u'+v')v +(u'-v')^2\right]^{\frac{3}{2}}}\cG(w,\wb)\,,
\label{eq:BsbContours}
\ee
where as usual $u'=w\wb$, $v'=(1-w)(1-\wb)$ and $C_{-}$, $C_{+}$ wrap the left, right branch cut respectively. More precisely, the oriented simple curve $C_{-}$ must encircle the cut $(-\infty,0]$ to its left but not the roots
of the quadratic polynomial $p=v^2-2(u'+v')v +(u'-v')^2$ (as a function of $w$), which should be on its right.
Similarly, $C_+$ must wrap the cut $[1,\infty)$ to its right but not the roots of $p$ as a function of $\wb$. This can be achieved by wrapping both contours within a distance $\epsilon>0$ from the cuts, as long as $v$ is away from the negative real axis.\footnote{To see this, it is enough to check that if $w,\wb$ are both real such that $w<0$ and $\wb>1$, then both roots of $p$ as a function of $v$ are real and nonpositive. Denoting the roots $v_{1,2}$, we have $v_1+v_2 = 2(u'+v')<0$ and $v_1v_2 = (u'-v')^2\geq 0$, so $v_1,v_2\leq 0$.}

\smallskip

This argument shows that $B^s_{v}$ is a swappable functional (for superbounded correlators) as long as $v$ is on the first sheet and away from $v\in(-\infty,0]$. In fact, $B^s_{v}[\cG]$ as a function of $v$ is holomorphic away from $v\in(-\infty,0]$ for any superbounded $\cG$. Expansion of $B^s_{v}$ in the basis functionals in equation \eqref{eq:BsbExpansion} supports these conclusions, since the expansion coefficients $(1-v)^{-\Df}k_{\Df+\ell}(1-v)$ are indeed holomorphic for $v\in\mathbb{C}\backslash(-\infty,0]$.

\smallskip

It is useful to have an integral representation of $B^s_{v}[\cG]$ which manifests the double zeros on s- and t-channel double-twist conformal blocks. Such representation can be derived either by expanding (\ref{PR minus Euclidean 2}) or by
closing both $w$ and $\wb$ contours in \eqref{eq:BsbContours} onto the same branch cut of $\cG(w,\wb)$. This argument is just a special case of the contour deformation explained at the end of Section \ref{ssec:FunctionalBasis}, and in more detail in Appendix \ref{app:deformationToDDisc}. Suppose $\cG(w,\wb)$ is single-valued around s-channel. Closing both contours onto the branch cut $w,\wb\in(-\infty,0]$, we find, using special properties of the kernel,
that the integral depends only on $\dDisc_{s}[\cG(w,\wb)]$ and only the region $\sqrt{v'}\geq\sqrt{u'}+\sqrt{v}$ contributes
\be
B^s_{v}[\cG] = \int\limits_{v}^{\infty}dv'\!\!\!\!\!\!\!\!\int\limits_{0}^{(\sqrt{v'}-\sqrt{v})^2}\!\!\!\!\!\!\!\!\!du'
\frac{u'-v'-v}{\pi^2\left[v^2-2(u'+v')v +(u'-v')^2\right]^{\frac{3}{2}}}
\dDisc_{s}[\cG(u',v')]\,.
\label{eq:BsbdDiscS}
\ee
Similarly, if $\cG(w,\wb)$ is single-valued around t-channel, we can close both contours onto the cut $w,\wb\in[1,\infty)$ to find an integral depending on $\dDisc_t[\cG(w,\wb)]$ over the region $\sqrt{u'}\geq\sqrt{v'}+\sqrt{v}$
\be
B^s_{v}[\cG] = \int\limits_{v}^{\infty}du'\!\!\!\!\!\!\!\!\int\limits_{0}^{(\sqrt{u'}-\sqrt{v})^2}\!\!\!\!\!\!\!\!\!dv'
\frac{u'-v'-v}{\pi^2\left[v^2-2(u'+v')v +(u'-v')^2\right]^{\frac{3}{2}}}
\dDisc_{t}[\cG(u',v')]\,.
\label{eq:BsbdDiscT}
\ee
A few comments are in order. Firstly, the polynomial in the denumerator factors as
\ba
v^2-2(u'+v')v +(u'-v')^2
&= [u'-(\sqrt{v}+\sqrt{v'})^2][u'-(\sqrt{v}-\sqrt{v'})^2] =\\
&= [v'-(\sqrt{v}+\sqrt{u'})^2][v'-(\sqrt{v}-\sqrt{u'})^2]\,.
\ea
Therefore, the polynomial is manifestly positive in the integration regions above. The polynomial vanishes on the integration boundaries at $\sqrt{v'}=\sqrt{u'}+\sqrt{v}$ and $\sqrt{u'}=\sqrt{v'}+\sqrt{v}$. Thus in fact the integrals as written are divergent at this boundary and need to be regularized, since $\int_0x^{-3/2}dx=\infty$. The original definition \eqref{eq:BsbContours} is equivalent to cutting off the $u'$ integral in \eqref{eq:BsbdDiscS} at $u' = (\sqrt{v'}-\sqrt{v})^2-\epsilon$, and throwing away the $\epsilon^{-1/2}$ divergent term as $\epsilon\rightarrow0$, and similarly for \eqref{eq:BsbdDiscT}. The integrals in (\ref{eq:BsbdDiscS}) and (\ref{eq:BsbdDiscT}) should be interpreted as being regularized in that fashion.

\smallskip

Equations \eqref{eq:BsbdDiscS} and \eqref{eq:BsbdDiscT} make the structure of double zeros of $B^s_{v}[G^s_{\Delta,J}]$ and $B^s_{v}[G^t_{\Delta,J}]$ manifest since
\ba
\dDisc_s[G^s_{\Delta,J}(u',v')] &= 2\sin^2[\tfrac{\pi}{2}(\Delta-J-2\Df)]\,G^s_{\Delta,J}(u',v')\\
\dDisc_t[G^t_{\Delta,J}(u',v')] &= 2\sin^2[\tfrac{\pi}{2}(\Delta-J-2\Df)]\,G^t_{\Delta,J}(u',v')\,.
\ea
 To evaluate $B^s_{v}[G^s_{\Delta,J}]$, we can use \eqref{eq:BsbdDiscS}. The integral converges as long as $\Delta>2\Df+J$ and diverges at $u'=0$ otherwise. Thus, the integration leads to a simple pole $(\Delta-J-2\Df)^{-1}$, which combines with the double zero of $\sin^2[\tfrac{\pi}{2}(\Delta-J-2\Df)]$ to make a simple zero, in agreement with the expansion \eqref{eq:BsbExpansion}, which only contains $\beta^s_{0,\ell}$. Similarly, we use \eqref{eq:BsbdDiscT} to evaluate $B^s_{v}[G^t_{\Delta,J}]$. The argument is the same except in this case the factor $u'-v'-v$ in the numerator suppresses the divergence at $\Delta=2\Df+J$ and we find double zeros also on the $n=0$ double-twist family, as expected. We stress that while \eqref{eq:BsbdDiscS} and \eqref{eq:BsbdDiscT} diverge on conformal blocks with small $\Delta<2\Df$, the original definition \eqref{eq:BsbContours} is finite on all physical conformal blocks. Since \eqref{eq:BsbContours} is analytic in $\Delta$, we can compute the functional in general by analytically continuing \eqref{eq:BsbdDiscS} and \eqref{eq:BsbdDiscT}.
 
 \smallskip
 
Expressions such as \eqref{eq:BsbdDiscS} and \eqref{eq:BsbdDiscT} will be very useful later to diagnose positivity of the functional actions on conformal blocks. Positivity will follow from positivity of the integrand. In the present case, positivity is obscured by the need to regularize the integrals, but luckily this issue will go away when we deal with physical (non-superbounded) correlators.

\subsection{The Polyakov condition follows from crossing symmetry}\label{ssec:Polyakov conditions}
Reference \cite{Penedones:2019tng} obtained interesting sum rules on OPE data by imposing the `Polyakov condition' in Mellin space. After quickly reviewing their argument, we will show that the resulting sum rule is equivalent to the sum rules arising from applying $B^s_{v}$ to the crossing equation.

\smallskip

The argument of \cite{Penedones:2019tng} starts by explaining that if $\cG(z,\zb)$ is a nonpeturbative CFT correlator, then we expect the Mellin amplitude $M(\mS,\mT)$ to vanish at $\mS = 2\Df$. Indeed, if $M(2\Df,\mT)\neq0$, the
Mellin representation \eqref{eq:mellinRep} would cause the s-channel OPE of $\cG(z,\zb)$ to contain terms $\partial_{\Delta}G^s_{2\Df+\ell,\ell}(z,\zb)$, due to the double pole of $\Gamma(\Df-\tfrac{\mS}{2})^2$. Nonperturbative correlators can only contain genuine conformal blocks and not their $\Delta$-derivatives, hence $M(2\Df,\mT)=0$.\footnote{The full argument is more involved since $\mS=2\Df$ is an essential singularity of $M(\mS,\mT)$, but the conclusion remains unchanged.} The next step is to use this constraint on the LHS of the Mellin-space dispersion relation, as written in \eqref{eq:mellinDispersion}.
This gives the following family of sum rules on the OPE data,\footnote{We denote the t-channel primaries by $\cO'$ rather than $\cP$ to avoid the clash of notation with $\cP_{\Delta,J}$.} parametrized by $\mT$,
\be
 -\sum_{\cO}a_{\cO}\,\cP_{\Delta_{\cO},J_{\cO}}(2\Df,\mT)
= \sum_{\cO'}a_{\cO'}\,\cP_{\Delta_{\cO'},J_{O'}}(\mT,2\Df)\,,
\label{eq:PolyakovSumRule}
\ee
where $\cP_{\Delta,J}(\mS,\mT)=\cP^{s|u}_{\Delta,J}(\mS,\mT) = \cP^{t|u}_{\Delta,J}(\mT,\mS)$
is the Polyakov-Regge block given by equation \eqref{PR Mellin}, which we reproduce here:
\be
\cP_{\Delta,J}(\mS,\mT)=\sum\limits_{m=0}^{\infty}\frac{\mathcal{Q}^{m}_{\Delta,J}(\mT+\mS-\tau-2m)}{\mS-\tau-2m}\,.
\ee
We claim that the sum rule \eqref{eq:PolyakovSumRule} is nothing but the Mellin transform (with respect to $v$) of the sum rules arising from $B^s_{v}$
\be
\sum\limits_{\cO}a_{\cO}\,B^s_{v}[G^s_{\Delta_{\cO},J_{\cO}}] = \sum\limits_{\cO'}a_{\cO'}\,B^s_{v}[G^t_{\Delta_{\cO'},J_{\cO'}}]\,.
\label{eq:BsbSumRule}
\ee
Intuitively, it is clear that the Polyakov condition and $B^s_{v}$ encode the same constraint. Indeed, $B^s_{v}$ is obtained by expanding the position-space dispersion relation around $u=0$ and keeping the leading term proportional to $\log u$. This term arises precisely from the double-twist terms $\partial_{\Delta}G^s_{2\Df+\ell,\ell}(z,\zb)$ in the Polyakov-Regge blocks.

\smallskip

To genuinely derive \eqref{eq:PolyakovSumRule} from \eqref{eq:BsbSumRule}, let us define the functional $\widehat{B}^s_{\mT}$ as the Mellin transform of functional $B^s_{v}$
\be
\widehat{B}^s_{\mT} = \Gamma\!\left(\Df-\tfrac{\mT}{2}\right)^{-2}\Gamma\!\left(\tfrac{\mT}{2}\right)^{-2}\int\limits_{0}^{\infty}\!\frac{dv}{v}\,v^{\Df-\tfrac{\mT}{2}}B^s_{v}\,.
\label{eq:BHatDef}
\ee
We claim that
\ba
\widehat{B}^s_{\mT}[G^s_{\Delta_{\cO},J_{\cO}}] &= \cP_{\Delta_{\cO},J_{\cO}}(2\Df,\mT)\\
\widehat{B}^s_{\mT}[G^t_{\Delta_{\cO'},J_{\cO'}}] &= -\cP_{\Delta_{\cO'},J_{\cO'}}(\mT,2\Df)\,,
\label{eq:BHatActions}
\ea
{\it i.e.} the Polyakov condition \eqref{eq:PolyakovSumRule} is precisely the action of the $\widehat{B}^s_{\mT}$ functional
on the position-space crossing equation.

\smallskip

Let us illustrate the equivalence \eqref{eq:BHatActions} by comparing the structure of zeros on double trace blocks. Recall from \eqref{eq:qCal} that $\mathcal{Q}^{m}_{\Delta,J}(\mT)$ has double zeros as a function of $\Delta$ at all double-twist dimensions $\Delta=2\Df+2n+J$. It follows that for generic $\mT$, the term $\cP_{\Delta_{\cO'},J_{\cO'}}(\mT,2\Df)$ on the RHS of \eqref{eq:PolyakovSumRule} has double zeros on all double-trace dimensions $\Delta_{\cO'}=2\Df+2n+J_{\cO'}$. On the other hand, the term $\cP_{\Delta_{\cO},J_{\cO}}(2\Df,\mT)$ on the LHS contains factor $2\Df-\tau_{\cO}-2m$ in the denominator. For $m=0$, this extra pole combines with the double zero of $\mathcal{Q}^{m}_{\Delta_{\cO},J_{\cO}}$ in the numerator to give a simple zero on the $n=0$ double-twist. In other words, the functional which leads to \eqref{eq:PolyakovSumRule} is a linear combination of $\beta^s_{0,\ell}$ for $\ell\in\mathbb{N}$, which is also what we got for $B^s_{v}$. We can apply the Mellin transform to \eqref{eq:BsbExpansion} term by term to find the expansion of $\widehat{B}^s_{\mT}$ in the dual basis
\be
\widehat{B}^s_{\mT} = \sum\limits_{\ell=0}^{\infty}
b_{\ell}(\mT)\,\beta^s_{0,\ell}\,,
\label{eq:BHatExpansion}
\ee
where
\ba
b_{\ell}(\mT) &=
\frac{1}{2}\Gamma\!\left(\Df-\tfrac{\mT}{2}\right)^{-2}\Gamma\!\left(\tfrac{\mT}{2}\right)^{-2}
\int\limits_{0}^{\infty}\!\frac{dv}{v}\,v^{\Df-\tfrac{\mT}{2}}(1-v)^{-\Df}k_{\Df+\ell}(1-v)=\\
&= \frac{\Gamma (2\Df +2\ell)}{2\Gamma (\Df )^2 \Gamma (\Df +\ell)^2}
\,{}_3F_2\!\left(-\ell,\Df -\tfrac{\mT}{2},\ell+2 \Df -1;\Df ,\Df ;1\right)\,.
\ea

It is now easy to check \eqref{eq:BHatActions} to the second order around all double-trace dimensions using \eqref{eq:BHatExpansion}. Indeed, we have \cite{Costa:2012cb}
\ba
\mathcal{Q}^{0}_{\Delta,\ell}(\mT) =
-\sin ^2\!\left[\tfrac{\pi}{2}  (\tau -2 \Df )\right]&\frac{2 \left(\frac{\tau }{2}\right)^{2}_{\ell}\Gamma \left(\tfrac{\tau}{2}-\Df+1\right)^2 \Gamma (\tau+2\ell)}{\pi ^2 \Gamma \left(\frac{\tau }{2}+\ell\right)^4}\times\\
&\times{}_3F_2\left(-\ell,\Df -\tfrac{\mT}{2},\ell+\tau -1;\tfrac{\tau }{2},\tfrac{\tau }{2};1\right)
\ea
and so $\partial^2_{\Delta}\mathcal{Q}^{0}_{\Delta,\ell}(\mT)|_{\Delta=2\Df+\ell} = - 2b_{\ell}(\mT)$, which implies
\be
\cP_{\Delta,\ell}(2\Df,\mT) =
b_{\ell}(\mT)(\Delta-2\Df-\ell) + O((\Delta-2\Df-\ell)^2)\,,
\label{eq:PolyakovSimpleZero}
\ee
in agreement with \eqref{eq:BHatExpansion}. Since \eqref{eq:PolyakovSumRule} and $-\widehat{B}^s_{\mT}$ have the same expansions into the functional basis, they are identical.

\smallskip

We can now adress the puzzle stated at the head of this subsection.
If \cite{Penedones:2019tng} derived the sum rule (\ref{eq:PolyakovSumRule}) by assuming that $M(\mS,\mT)$ vanishes at $\mS=2\Df$, and we derived the sum rule \eqref{eq:BsbSumRule} using only crossing symmetry. How can the resulting sum rules be the same?

\smallskip

The resolution is that the sum rules \eqref{eq:PolyakovSumRule} and \eqref{eq:BsbSumRule} hold
whether $M(\mS,\mT)$ vanishes at $\mS=2\Df$ or not.
This must be true because, as we have shown, the sum rules follow by applying appropriate linear functionals to the position-space crossing equation, without assuming anything about the OPE spectrum. This may sound surprising: wasn't the sum rule \eqref{eq:PolyakovSumRule} derived precisely from assuming $M(2\Df,\mT)=0$?
To understand what is going on, let us derive the sum rule \eqref{eq:PolyakovSumRule} in the case that
$M(2\Df,\mT)$ does not vanish. $M(2\Df,\mT)\neq 0$ is equivalent to saying that the s-channel OPE contains some terms of the form $\partial_{\Delta}G^s_{2\Df+\ell,\ell}$. So let us write the s-channel OPE of $\cG$ as follows
\be
\cG(z,\zb) = \sum\limits_{\ell=0}^{\infty}c_{\ell}\,\partial_{\Delta}G^s_{2\Df+\ell,\ell}(z,\zb) +\sideset{}{'}\sum\limits_{\cO}a_{\cO}\,G^s_{\Delta_{\cO},J_{\cO}}(z,\zb)\,,
\ee
where the primed sum runs over all OPE contributions other than $\partial_{\Delta}G^s_{2\Df+\ell,\ell}(z,\zb)$. For the Mellin representation \eqref{eq:mellinRep} to correctly reproduce this OPE, we must have
\be
M(2\Df,\mT) = \sum\limits_{\ell=0}^{\infty}\frac{c_{\ell}}{2}\partial^2_{\Delta}\mathcal{Q}^{0}_{\Delta,\ell}(\mT)|_{\Delta=2\Df+\ell} =- \sum\limits_{\ell=0}^{\infty}c_{\ell}\,b_{\ell}(\mT)\,.
\label{eq:MPolyakov1}
\ee
On the other hand, if we repeat the derivation of the Mellin-space dispersion relation of section \ref{ssec:MellinDispersion}, we find
\be
M(2\Df,\mT) = \sideset{}{'}\sum_{\cO}a_{\cO}\,\cP_{\Delta_{\cO},J_{\cO}}(2\Df,\mT) + \sum_{\cO'}a_{\cO'}\,\cP_{\Delta_{\cO'},J_{\cO'}}(\mT,2\Df)\,,
\label{eq:MPolyakov2}
\ee
where the s-channel sum does not include the $\partial_{\Delta}G^s_{2\Df+\ell,\ell}(z,\zb)$ contributions since those are on the inside of the contour in \eqref{eq:mellinCauchy}. By equating the RHS of \eqref{eq:MPolyakov1} and \eqref{eq:MPolyakov2}, we arrive at the sum rule
\be
 -\sum\limits_{\ell=0}^{\infty}c_{\ell}\,b_{\ell}(\mT)-\sideset{}{'}\sum_{\cO}a_{\cO}\,\cP_{\Delta_{\cO},J_{\cO}}(2\Df,\mT)
= \sum_{\cO'}a_{\cO'}\,\cP_{\Delta_{\cO'},J_{\cO'}}(\mT,2\Df)\,.
\ee
But this is equivalent to the original sum rule \eqref{eq:PolyakovSumRule} thanks to the simple zeros of $\cP_{\Delta_{\cO},J_{\cO}}(2\Df,\mT)$ on the leading-twist family, i.e. eq. \eqref{eq:PolyakovSimpleZero}. In summary, the real meaning of the Polyakov condition is not the constraint $M(2\Df,\mT) = 0$ but rather the equality of two independent ways of evaluating $M(2\Df,\mT)$, one using the dispersion relation (eq. \eqref{eq:MPolyakov2}) and the other directly the s-channel OPE
(\eqref{eq:MPolyakov1}). This is manifest in \eqref{PR minus Euclidean 2}.

\smallskip

Note that a simple object with $M(2\Df,\mT)\neq 0$ for which the sum rule \eqref{eq:PolyakovSumRule} holds
is the Polyakov-Regge block of an individual `single-trace' operator.
In theories that are close to mean field theory, the sum rule thus relates
anomalous dimensions of $n=0$ double-trace operators to the single-trace primaries that can be exchanged, which is
one way that the Polyakov conditions are often (correctly) used, see for example \cite{Gopakumar:2016wkt,Gopakumar:2016cpb,Sleight:2018ryu}.  Of course, to \emph{interpret} the coefficient of
$\partial_\Delta G^s_{2\Df+\ell,\ell}(z,\zb)$ as the anomalous dimension of a double-trace operator, it is necessary to assume that the spectrum contains no unexpected near-degenerate operators (or that we know the full set, in case we have a mixing matrix). One may refer to this assumption as elastic unitarity.\footnote{We thank the authors of \cite{Penedones:2019tng} for suggesting this name.} Our point here is that the sum rule \eqref{eq:PolyakovSumRule}
holds regardless of that assumption.

\smallskip

A technical comment is in order. We did not in fact demonstrate that $\widehat{B}^s_{\mT}$ are valid functionals, {\it i.e.} that they satisfy the swapping condition with OPE in superbounded correlators.
They are infinite combinations of the functionals $B^s_{v}$, which, as explained above, are valid
for $v\in\mathbb{C}\backslash(-\infty,0]$. This is made explicit by the position-space formula \eqref{eq:BsbContours} since the contours can be placed in a region where both s- and t-channel OPEs converge exponentially fast.\footnote{Except for $z,\zb\rightarrow\infty$, where the swapping condition holds thanks to superboundedness of $\cG(z,\zb)$.} On the other hand, to define $\widehat{B}^s_{\mT}$ through \eqref{eq:BHatDef}, we need to integrate $v$ all the way to $v=0$, i.e. arbitrarily close to the forbidden region $v\leq 0$, which is potentially problematic because it forces the $C_{-}\times C_{+}$ contour to the boundary of the region of OPE convergence. In the rest of this paper, we will mostly use the position-space functionals such as $B^s_{v}$, and so will not need to worry about this issue. It would be interesting to study the swappability of $\widehat{B}_{\mT}$ directly by using its representation as a position-space integral using the methods of \cite{Kravchuk:2020scc}.\footnote{See Appendix \ref{app:mellinFunctionals} for the position-space representation of a close cousin of $\widehat{B}^s_{\mT}$.}

\subsection{General Mellin-space sum rules}
The previous subsection discussed the simplest example of the Polyakov condition, associated with $M(2\Df,\mT)$.
The full set of Polyakov conditions similarly comes from\footnote{These are all the s-channel Polyakov conditions.
Non s$\leftrightarrow$t symmetric correlators also have an independent set of t-channel Polyakov conditions associated with $M(\mS,2\Df+2n)$ and $\partial_{\mT}M(\mS,2\Df+2n)$.}
\be
M(2\Df+2n,\mT)\quad\text{and}\quad\partial_{\mS}M(2\Df+2n,\mT)\,,\quad\text{where}\quad n\in\mathbb{N}\,.
\ee
Thanks to the factor $\Gamma(\Df-\tfrac{\mS}{2})^2$ in the Mellin representation, these quantities are directly computable from the s-channel OPE. Indeed, $M(2\Df+2n,\mT)$ only receives contributions from terms $\partial_{\Delta}G^s_{2\Df+2m+\ell,\ell}$ with $m\leq n$. Similarly, $\partial_{\mS}M(2\Df+2n,\mT)$ only comes from terms $G^s_{2\Df+2m+\ell,\ell}$ with $m\leq n$ and $\partial_{\Delta}G^s_{2\Df+2m+\ell,\ell}$ with $m< n$.

\smallskip

Exactly as in the previous subsection, the corresponding Polyakov-Mellin sum rules can be derived by equating the prediction of the s-channel OPE with the prediction of the dispersion relation \eqref{eq:mellinDispersion}, for each of the above quantities. As in the above, we get correct sum rules by setting $M(2\Df+2n,\mT) = 0$ and $\partial_{\mS}M(2\Df+2n,\mT)=0$ on the LHS of the dispersion relation, provided we include exact double-twist operators (if present in the theory) on the RHS.
As manifest in \eqref{PR minus Euclidean 2}, this is because the $\Omega$
sum rules automatically subtract the Euclidean OPE. Specifically, to derive the sum rules corresponding to all Polyakov conditions, we expand the generating functional $\Omega^{t|u}_{z,\zb}$ at small $u=z \zb$ and keep the coefficient of $u^n$ or $u^n\log(u)$. This yields a functional whose Mellin transform with respect to $v$ gives the Polyakov-Mellin sum rule when applied to the crossing equation.
The main conclusion remains the same. All of the stated sum rules follow from crossing symmetry and superboundedness alone, irrespective of the detailed structure of the double-trace spectrum of the OPEs.

\smallskip

A general Mellin-space sum rule can be obtained by formally taking the two-variable Mellin transform of the position-space equality between the Euclidean OPE and Polyakov-Regge expansion, i.e. \eqref{PR minus Euclidean 2}. The sum rule thus looks as follows
\be
\sum\limits_{\cO} a_{\cO}\,\left[G^s_{\Delta_{\cO},J_{\cO}}(\mS,\mT)-\cP_{\Delta_{\cO},J_{\cO}}(\mS,\mT)\right] = 
\sum\limits_{\cO'} a_{\cO'} \cP_{\Delta_{\cO'},J_{\cO'}}(\mT,\mS)\,,
\label{eq:generalMellinSumRule}
\ee
where $G^s_{\Delta_{\cO},J_{\cO}}(\mS,\mT)$ stands for the Mellin transform of the s-channel conformal block.\footnote{Explicit expressions for $G^s_{\Delta_{\cO},J_{\cO}}(\mS,\mT)$ appear in \cite{Mack:2009mi,Costa:2012cb}.} For $\Delta$ generic, $G^s_{\Delta,J}(\mS,\mT)$ has double zeros as a function of $\mS$ at $\mS=2\Df+2n$. The Polyakov conditions are precisely the sum rule \eqref{eq:generalMellinSumRule} expanded to the second order around each $\mS=2\Df+2n$. It would be interesting to study for what values of $\mS$ and $\mT$ \eqref{eq:generalMellinSumRule} is a convergent sum rule (i.e. comes from a swappable functional).

\subsection{Symmetry under $\text{s}\leftrightarrow\text{t}$ and $\text{t}\leftrightarrow\text{u}$}\label{ssec:stuSymmetry}

Finally, let us comment on situations where some of the external operators in $\langle\phi_1\phi_2\phi_3\phi_4\rangle$ are identical. If $\phi_1=\phi_3$ or $\phi_2=\phi_4$, the u-channel OPE contains only primaries of even spin, and thus the correlator is $\text{s}\leftrightarrow\text{t}$ symmetric
\be
\cG(z,\zb) = \cG(1-z,1-\zb)\,.
\ee
In this case, the complete set of sum rules in the superbounded setting consists of $\alpha_{n,\ell}=\alpha^s_{n,\ell}-\alpha^t_{n,\ell}$ and $\beta_{n,\ell}=\beta^s_{n,\ell}-\beta^t_{n,\ell}$.
The sum rule coming from applying $B^s_{v}$ to the s=t crossing equation takes the form
\be
\sum\limits_{\cO} a_{\cO}\,B_{v}[G^s_{\Delta_{\cO},J_{\cO}}] = 0\,, \label{Bv zero}
\ee
where the functional $B_{v}$ is defined by
\be
B_{v}[\cG(w,\wb)] = B^s_{v}[\cG(w,\wb)] - B^s_{v}[\cG(1-w,1-\wb)]\,.
\ee
From \eqref{eq:BsbExpansion}, we find the expansion of $B_{v}$ in the functional basis
\be
B_{v} = \frac{1}{2}\sum\limits_{\ell=0}^{\infty}\,(1-v)^{-\Df}k_{\Df+\ell}(1-v)\,(\beta^s_{0,\ell}-\beta^t_{0,\ell})\,.
\ee
The contour representation of $B_{v}$ follows from \eqref{eq:BsbContours}
\ba
B_{v}[\cG] &= \!\iint\limits_{C_-\times C_+}\!\!\frac{dwd\wb}{(2\pi i)^2}
\frac{2(\wb-w) (u'-v')}{\left[v^2-2(u'+v')v +(u'-v')^2\right]^{\frac{3}{2}}}\cG(w,\wb) \\
&= 
\int\limits_{v}^{\infty}du'\!\!\!\!\!\!\!\!\int\limits_{0}^{(\sqrt{u'}-\sqrt{v})^2}\!\!\!\!\!\!\!\!\!dv'
\frac{2(u'-v')}{\pi^2\left[v^2-2(u'+v')v +(u'-v')^2\right]^{\frac{3}{2}}}
\dDisc_{s}[\cG(u',v')]\,,
\ea
where we used the fact that the contour $C_{-}\times C_{+}$ is invariant under $(w,\wb)\mapsto(1-\wb,1-w)$. 
The vanishing of \eqref{Bv zero} is immediate on that contour due to symmetry reasons.

\smallskip

Now suppose that $\phi_1=\phi_2$ or $\phi_3=\phi_4$, so that the s-channel OPE contains only primaries of even spin and thus the correlator is $\text{t}\leftrightarrow\text{u}$ symmetric. The main difference between this and the $\text{s}\leftrightarrow\text{t}$ symmetric situation is that now the basis functionals $\alpha^s_{n,\ell}$, $\beta^s_{n,\ell}$, $\alpha^t_{n,\ell}$, $\beta^t_{n,\ell}$ do not satisfy any obvious relations.\footnote{In the $\text{s}\leftrightarrow\text{t}$-symmetric case, there are relations because we have $\alpha^s_{n,\ell}\sim-\alpha^t_{n,\ell}$ and $\beta^s_{n,\ell}\sim-\beta^t_{n,\ell}$, where two functionals are equivalent if they lead to identical sum rules for $s\leftrightarrow t$ symmetric correlators.} It may be tempting to throw out $\alpha^s_{n,\ell}$ and $\beta^s_{n,\ell}$ with odd $\ell$ since only even spin appears in the s-channel OPE. However, this would be a mistake because $\alpha^s_{n,\ell}[G^s_{\Delta,J}]$, $\beta^s_{n,\ell}[G^s_{\Delta,J}]$ are generally nonvanishing even if $\ell$ is odd and $J$ is even, and thus these functionals lead to nontrivial sum rules even if only even $J$ appears in the s-channel.

\smallskip

Finally, suppose $\phi_1=\phi_2=\phi_3=\phi_4$ and the correlator is thus fully symmetric. From the discussion so far, we know that in this case a complete set of functionals consists of $\alpha_{n,\ell}$ and $\beta_{n,\ell}$ with $\ell$ \emph{both even and odd}. In particular, $\alpha_{n,\ell}$ and $\beta_{n,\ell}$ with $\ell$ odd lead to sum rules with double zeros on all double-traces which can appear in the OPE. At the level of $B_{v}$, symmetry between the t- and u-channels is implemented as $v\leftrightarrow1/v$, but $B_{v}$ and $B_{1/v}$ still lead to independent sum rules. We can project onto the odd-spin basis functionals by defining
\be
\widetilde{B}_v =v^{\frac{\Df}{2}}B_{v} - v^{-\frac{\Df}{2}}B_{1/v}\,.
\ee
$\widetilde{B}_v$ has the expansion
\be
\widetilde{B}_v = \sum\limits_{\ell\text{ odd}}^{\infty}\,v^{\frac{\Df}{2}}(1-v)^{-\Df}k_{\Df+\ell}(1-v)\,\beta_{0,\ell}\,,
\ee
where we used
\be
(1-1/v)^{-\Df}k_{\Df+\ell}(1-1/v) = (-1)^{\ell}v^{\Df}(1-v)^{-\Df}k_{\Df+\ell}(1-v)\,.
\ee
When we consider subtractions, the analogue of $\widetilde{B}_v$ will be used to prove that mean field theory saturates the upper bound on the twist gap. In Mellin space, t-u crossing symmetry is the statement $M(\mS,\mT) = M(\mS,4\Df-\mS-\mT)$. We can get the corresponding functionals by antisymmetrizing \eqref{eq:generalMellinSumRule} under $\mT\leftrightarrow 4\Df-\mS-\mT$. The conformal blocks $G^s_{\Delta_{\cO},J_{\cO}}(\mS,\mT)$ for even $J_{\cO}$ then drop out from the LHS. However, if we want to decompose the corresponding functional into $\alpha_{n,\ell}$ and $\beta_{n,\ell}$, we also need to keep track of the action on odd-spin blocks, for which the term $G^s_{\Delta_{\cO},J_{\cO}}(\mS,\mT)$ does not drop out. In particular, this logic allows one to decompose the twist gap functional of \cite{Penedones:2019tng} (their eq. (139)) into $\alpha_{n,\ell}$ and $\beta_{n,\ell}$. One finds that their functional contains $\alpha_{n,\ell}$ and $\beta_{n,\ell}$ for all $n$ and all \emph{odd} $\ell$, as well as $\beta_{0,\ell}$ with $\ell=2,4,6,\ldots$. We omit the details. In Section \ref{ssec:Btilde21}, we will exhibit a candidate for a twist gap extremal functional valid for any $d$ and $\Df$ which only contains $\alpha_{n,\ell}$ and $\beta_{n,\ell}$ with $n=0$.

\smallskip

Finally, we should mention that imposing symmetry between channels in the superbounded case is rather unphysical. Indeed, we are not aware of any superbounded, $\text{s}\leftrightarrow\text{t}\leftrightarrow\text{u}$-symmetric correlator. The main purpose of this subsection was to set the stage for analogous constructions
including subtractions in the following sections.

% !TEX root = ../main.tex

\section{Subtractions}\label{sec:Subtractions}

In the rest of this paper, we overcome the remaining obstacle allowing us to apply analytic functionals
to generic correlators: the fact that they are generally ``only'' bounded, rather than superbounded, in the 
u-channel Regge limit.  This will force us to introduce subtractions.
For conciseness of notation,
we will focus on four-point functions of identical scalars, the generalization being straightforward.

\smallskip

The basic idea of subtraction is to replace a correlator with a related object that is better-behaved in the Regge limit.
For example, replacing the Mellin amplitude (or a flat-space $S$-matrix) in the following way:
\be
 M(\mS,\mT) \mapsto \frac{M(\mS,\mT)}{(\mS-\mS_1)(\mS-\mS_2)} \label{mellin sub}
\ee
improves its fixed-$\mU$ Regge behavior by $1/\mS^2$, i.e. two units of spin.
Here $\mS_{1,2}$ are arbitrary subtraction points.
Since $M(\mS,\mT)$ grows at most linearly in a unitary CFT,
the quantity on the right-hand-side vanishes in the Regge limit and
we can then apply dispersion relation logic to it, leading to what one would call a
twice-subtracted dispersion relation.

\smallskip

\begin{figure}
\centering{\def\svgwidth{8.2cm}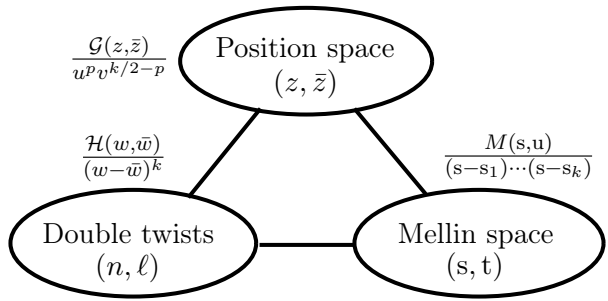}
\caption{\label{fig:triangle sub}
In each space a natural rescaling can be used to improve convergence
in the u-channel Regge limit. Each of the shown rescalings
leads to a $k$-times subtracted dispersion relation.
Approaches can be combined.
}
\end{figure}

As summarized in Figure~\ref{fig:triangle sub}, analogous constructions exist in the three spaces considered in the paper.
In position space, one can simply divide the correlator by powers of $1/u$ or $1/v$,
or, alternatively, one may
divide the kernel of the analytic functionals in \eqref{eq:kernelExample} by powers of $1/(w-\wb)^2$:
\be
\cG(z,\zb)\mapsto \frac{\cG(z,\zb)}{u}
\quad\mbox{or}\quad
\mathcal{A}^s_{n,\ell}(w,\wb)\mapsto \frac{\mathcal{A}^s_{n,\ell}(w,\wb)}{(w-\wb)^2}\,. \label{other subs}
\ee
Either of these three substitutions converts a superbounded sum rule into one which is valid in an arbitrary CFT. Every sum rule can be expanded in a unique way in the dual basis consisting of $\alpha_{n,\ell}$ and $\beta_{n,\ell}$. Demanding that the sum rule is subtracted, i.e. valid in arbitrary CFT, restricts the allowed linear combinations.

In this section we discuss respective advantages and interrelations between these approaches.
We begin by introducing notation for the Regge behavior of a given functional.

\subsection{Function spaces and $u$-channel growth}
\label{ssec:growth}

Let us denote the doubly cut plane where the cross-ratios live as
$\cR=\C\backslash ((-\oo,0]\cup[1,\oo))$.
It will be important to classify functions on $\cR\x\cR$ according to their growth (or decay)
in the $u$-channel Regge and OPE limits $z,\bar z\to\oo$. 
We say that a function $F(z,\bar z)$ is ``bounded by spin-$J$'' in the u-channel if there exist positive constants $R$ and $A$ such that for all $(z,\bar z) \in \cR\x\cR$ satisfying $|z|>R, |\bar z|>R$, we have
\begin{align}
\label{eq:uchannelgrowthcondition}
|F(z,\bar z)| &\leq A|z \bar z|^{\frac{J-1}{2}}.
\end{align}
This terminology comes from the fact that a u-channel block with spin-$J$, analytically continued to the $u$-channel Regge limit, behaves according to (\ref{eq:uchannelgrowthcondition}). For reference, let us recall that in Mellin space the u-channel Regge limit corresponds to $\mS,\mT$ large with $\mS+\mT$ fixed. Spin-$J$ behaviour translates to $M(\mS,\mT) = O(|s|^{J})$ in this limit.

\smallskip

Let $\cV_J$ be the space of holomorphic functions on $\cR\x\cR$ that are bounded by spin-$J$ in the $u$-channel.\footnote{The spaces $\cV,\cU$ defined in \cite{Mazac:2019shk} can be written in terms of $\cV_J$. The space of functions bounded by a constant in the $u$-channel is $\cV=\cV_1$. The space of Regge super-bounded functions is $\cU=\cup_{J<0} \cV_J$.} Let us describe some example elements of $\cV_J$ for various $J$. Firstly, a physical four-point function $\cG$ is bounded in the $u$-channel Regge limit, so satisfies $\cG\in \cV_1$. Individual $s$- and $t$-channel blocks satisfy $G^s_{\De,J},G^t_{\De,J} \in \cV_{\frac d 2-2\De_\f+\e}$ for any $\e>0$. In particular, in unitary theories, conformal blocks decay faster in the Regge limit than physical four-point functions. Note that this fails for u-channel conformal blocks since we have $G^u_{\De,J}(z,\zb)=O(|z|^{J-1})$ in the u-channel Regge limit. U-channel exchange Witten diagrams of spin $J$ belong to $\cV_{J}$. On the other hand, s- and t-channel exchange diagrams for a generic choice of the 3-point couplings belong to $\cV_{J-1}$. As we saw however, they can always be improved by contact diagrams to  belong to $\cV_{-\epsilon}$.

\smallskip

The space of functionals is dual to the space of functions.
We say that a functional $\omega$ has spin-$J$ decay or spin-$J$ convergence (in the u-channel)
if its action is well-defined on all functions in $\cV_{J-\e}$ with $\e>0$. 

\smallskip

An important class of functionals (including all functionals considered in this paper) can be
obtained via a contour integral against a kernel $\cH(w,\bar w)$ in cross-ratio space
\begin{align}
\label{eq:contourexample}
\w[\cG(w,\bar w)] &= \iint\limits_{C_{-}\times C_{+}}\frac{dw d\wb}{(2\pi i)^2} \,\cH(w,\bar w) \cG(w,\bar w).
\end{align}
where the contours $C_{-}$ and $C_{+}$ wrap the left and right branch cuts respectively,
see Figure \ref{fig:Contours}.
A functional of this form has spin-$J$ convergence if it obeys the scaling
\begin{align}
\label{eq:functionaldecay}
\cH(w,\bar w) & = O(w^{-J-1}) \qquad \textrm{as $w,\bar w\to \oo$ with fixed $w/\wb$},
\end{align}
and if, furthermore, it is a sufficiently nice function of $w/\bar w$. In practice, we want the functionals to be sufficiently nice that their action commutes with the s- and t-channel OPEs.\footnote{We will not attempt to characterize the conditions on the function of $w/\bar w$ in full generality. Instead, we will check that the integral (\ref{eq:contourexample}) is well-defined in specific examples.}

\smallskip

The functionals $\a_{n,\ell}$, $\b_{n,\ell}$ defined in \cite{Mazac:2019shk} all have spin-0 decay in the u-channel. For example, the kernel for $\b_{0,0}$ is
\begin{align}
\cH_{\b_{0,0}}(w,\bar w) &= -\frac{4(1-w-\wb)}{(\wb- w)^2} = O(w^{-1}) \qquad \textrm{with fixed $w/\bar w$}.
\end{align}
This means that individual $\a_{n,\ell}$, $\b_{n,\ell}$ do not give valid sum rules for generic (non-superbounded)
physical correlators of identical scalars.
To obtain functionals which do lead to valid sum rules, we need kernels that decay with spin-$J$ where $J>1$.
In practice, the ``subtraction" constructions
exemplified by \eqref{mellin sub}-\eqref{other subs} can only shift the spin by even integers,
and so our functionals will have spin-2 convergence: we need two subtractions, but not more.

\smallskip

We will exhibit several classes of such functionals.

\subsection{A simple positive sum rule}\label{ssec:B21}

Independent of the subtraction logic, it is rather easy to write down examples of kernels $\cH$
that create spin-2 functionals.
A particularly simple one is:
\be
\cH(w,\wb) = \frac{1-w-\wb}{w\wb(1-w)(1-\wb)}\,, \label{simple H}
\ee
which satisfies $\cH(w,\wb) = \cH(\wb,w)$ and $\cH(w,\wb) = -\cH(1-w,1-\wb)$ and decays as spin-2 in the Regge limit.
As we will now show, the corresponding functional is positive on blocks with $\Delta-J>2\Df$!
Furthermore, it exhibits double zeros on each double-twist above the leading one, $\tau = 2\Df+2n$ with $n>0$.

\smallskip

Some additional care is needed in deriving the sum rule coming from this functional. In the u-channel lightcone limit $w\rightarrow \infty$ with $\wb$ fixed, the kernel goes as $\cH(w,\wb) \sim w^{-1}$, in contrast with the $\alpha_{n,\ell}$ and $\beta_{n,\ell}$ kernels which are all $O(w^{-3})$. A physical correlator $\cG(w,\wb)$ is bounded by $1$ in this limit, but this is not enough to make the $w$ integral converge -- it goes as $\int dw/w$. The solution is to subtract the u-channel identity and apply the functional to $\cG(w,\wb) - 1$. If $w$ and $\wb$ are in opposite half-planes, the limit $w\rightarrow\infty$ with $\wb$ fixed is controlled by the u-channel OPE and we have $\cG(w,\wb)-1 = O(|w|^{-\tau_{0}/2})$, where $\tau_0>0$ is the u-channel twist gap. If $w$ and $\wb$ are in the same half-plane, we are dealing with the light-cone limit on the second sheet. It is expected that this limit is controlled by the usual OPE, analytically continued to the second sheet, for operators up to the first twist accumulation point (see \cite{Hartman:2015lfa} for a heuristic argument). Under this technical assumption we have $\cG(w,\wb)-1 = O(|w|^{-\tau_{0}/2})$ also on the second sheet. We will make this assumption in the rest of this paper.

\smallskip

To derive the sum rule, we thus start from the identity  
\be \label{simple H identity}
\iint\limits_{C_{-}\times C_{+}}\frac{dw d\wb}{(2\pi i)^2} \frac{1-w-\wb}{w\wb(1-w)(1-\wb)} [\cG(w,\bar w)-1] = 0\,,
\ee
which follows since the integrand is antisymmetric under $(w,\wb)\mapsto(1-\wb,1-w)$, while the contour is invariant under it. 
This will give us a sum rule upon inserting the s-channel OPE for $\cG$.

\smallskip

To study positivity properties of this sum rule,
it will be useful to first deform the contours to make manifest the double zeros on double-twist conformal blocks.
We keep this discussion general since it will apply to other functionals defined by similar contour integrals.
The idea is to wrap both $w$ and $\wb$ contours tightly on the left branch cut, similar to the relation between
the $\Omega$ transform and dispersion relation in Section \ref{ssec:FunctionalBasis}.
Let us temporarily ignore the identity subtraction $\cG-1$ and consider \eqref{eq:contourexample}, which becomes:
\be
\w[\cG(w,\bar w)] = 
 \iint\limits_{\!\!\!\!\!\!-\infty}^{\;\;\;\;\;0}\frac{dw d\wb}{(2\pi i)^2} \Disc_{\wb}[\Disc_{w}[\cH(w,\bar w)\cG(w,\bar w)]]\,.
\ee
In general, both $\cH(w,\bar w)$ and $\cG(w,\bar w)$ can have a nonzero discontinuity. To evaluate the discontinuity of a product, note that
\ba
\Disc_{w}[f(w)g(w)] &= f(w+i\epsilon)g(w+i\epsilon) - f(w-i\epsilon)g(w-i\epsilon) =\\
&= \Disc_{w}[f(w)] g(w+i\epsilon) + f(w-i\epsilon) \Disc_{w}[g(w)] =\\
&= \Disc_{w}[f(w)] g(w-i\epsilon) + f(w+i\epsilon) \Disc_{w}[g(w)]\,.
\ea
Taking the average of the last two lines, we get
\be
\Disc_{w}[f(w)g(w)] = \Disc_{w}[f(w)]\,\mathrm{P}_{w}[g(w)] + \mathrm{P}_{w}[f(w)]\,\Disc_{w}[g(w)]\,,
\ee
where we defined
\be
\mathrm{P}_w[f(w)] = \frac{f(w+i\epsilon)+f(w-i\epsilon)}{2}\,.
\ee
The integral against $\mathrm{P}_w[f(w)]$ is the principal value distribution, hence the notation $\mathrm{P}$. Thus
\ba
\w[\cG(w,\bar w)] = 
 \iint\limits_{\!\!\!\!\!\!-\infty}^{\;\;\;\;\;0}\frac{dw d\wb}{(2\pi i)^2}
 \{
 &\mathrm{P}_{\wb}[\mathrm{P}_{w}[\cH(w,\bar w)]]\;\Disc_{\wb}[\Disc_{w}[\cG(w,\bar w)]] +\\
 +&\Disc_{\wb}[\mathrm{P}_{w}[\cH(w,\bar w)]]\;\mathrm{P}_{\wb}[\Disc_{w}[\cG(w,\bar w)]] +\\
+&\mathrm{P}_{\wb}[\Disc_{w}[\cH(w,\bar w)]]\;\Disc_{\wb}[\mathrm{P}_{w}[\cG(w,\bar w)]] +\\
 +&\Disc_{\wb}[\Disc_{w}[\cH(w,\bar w)]]\;\mathrm{P}_{\wb}[\mathrm{P}_{w}[\cG(w,\bar w)]]
 \}\,. \label{Disc and P}
\ea
Note that we assumed that $\cH(w,\wb)$ has no $\wb$-singularities in $\cR$ when $w$ is on the left-cut;
this is clearly the case for $\cH$ in \eqref{simple H}.
The first line of \eqref{Disc and P} is what we're looking for: the double-discontinuity of $\cG$.
To get functionals with double-zeros on s-channel double-twists, we need all the other lines to cancel out.

\smallskip

Let us briefly recall how this happens for the $\alpha_{n,\ell}$ or $\beta_{n,\ell}$ functionals
exemplified in \eqref{eq:kernelExample} \cite{Mazac:2019shk}. 
Consider the action on one block, $\cG(w,\wb) \mapsto G^s_{\Delta,J}(w,\wb)$.
It turns out that for sufficiently large $\Delta$, the last three terms in the curly bracket do not contribute to the integral. The reason is that they can be reduced to a total derivative, which evaluates to a contact term at $z=\zb=0$, where the integrand vanishes for sufficiently large $\Delta$. How this works for the $\beta_{n,\ell}$ functionals was explained in detail in Section 5 of \cite{Mazac:2019shk}. \eqref{Disc and P} then reduces to
\be
\w[G^s_{\Delta,J}] =
\frac{\sin^2\left[\frac{\pi}{2}(\Delta-J-2\Df)\right]}{\pi^2} \iint\limits_{\!\!\!\!\!\!-\infty}^{\;\;\;\;\;0}\!dw d\wb\,
\mathrm{P}_{\wb}[\mathrm{P}_{w}[\cH(w,\bar w)]]\,G^s_{\Delta,J}(w,\bar w)\,,
\ee
where the conformal block in the integrand is evaluated in Euclidean kinematics, meaning $G^s_{\Delta,J}(w\pm i\epsilon,\wb\mp i\epsilon)$. The formula is valid as long as the integral converges, which is always true for sufficiently large $\Delta$. The analytic continuation in $\Delta$ gives the correct answer for general $\Delta$. The formula makes manifest that $\w[G^s_{\Delta,J}]$ has double zeros on all double-twist dimensions for which the integral converges. Simple zeros and non-zero values of $\w[G^s_{\Delta,J}]$ on low-lying double traces arise when the integral develops a simple or double pole, thus cancelling the double zero of the prefactor.
Note that in the integration region, $G^s_{\Delta,J}(w,\bar w)$ has definite sign $(-1)^{J}$.
The kernel of the $\alpha_{n,\ell}$ or $\beta_{n,\ell}$ functionals, however, have a distributional nature at $w=\wb$,
which makes it hard to understand their positivity properties.

\smallskip

We can now see why the functional from \eqref{simple H} enjoys simple positivity properties.
Consider its action on a block $G^s_{\Delta,J}$.
Since the kernel has no discontinuity, we can immediately discard the last three lines of \eqref{Disc and P}.
The first line is then the integral over a positive-definite function for even $J$, QED.
This argument holds for $\Delta$ large enough that the singularity at $\wb=0$ can be neglected.
This is the case for $\Delta-J>2\Df$. Thus we have shown that the functional is positive for
all operators above the double-twist threshold.

\smallskip

Let us return to the sum rule resulting from \eqref{simple H identity}, now evaluated on the contour $C_-\times C_+$.
We need to pay attention to the u-channel identity subtraction in $\cG-1$.
On this contour we can insert the s-channel OPE for $\cG$, and calculate the integral directly for the $-1$ term.
Due to cancellations between the upper and lower branches of the contour, the latter simply
gives $-1$ from a residue at $(w,\wb)=(0,0)$.
The resulting sum rule is thus:
\be
\sum\limits_{\cO} a_{\cO}\,B_{2,1}[G^s_{\Delta_{\cO},J_{\cO}}] = 1\,,
\label{eq:B1SumRule}
\ee
where
\ba
B_{2,1}[G^s_{\Delta,J}] &= 
\iint\limits_{C_{-}\times C_{+}}\frac{dw d\wb}{(2\pi i)^2} \frac{1-w-\wb}{w\wb(1-w)(1-\wb)} G^s_{\Delta,J}(w,\bar w)=\\
&=\frac{\sin^2\left[\frac{\pi}{2}(\Delta-J-2\Df)\right]}{\pi^2} \iint\limits_{\!\!\!\!\!\!-\infty}^{\;\;\;\;\;0}\!dw d\wb\,
\frac{1-w-\wb}{w\wb(1-w)(1-\wb)}\,G^s_{\Delta,J}(w,\bar w)\,.
\label{eq:OmegadDisc}
\ea
We denote this functional as $B_{2,1}$ because it will turn out to be a special case of a family introduced below.
The sum on the left-hand of \eqref{eq:B1SumRule} side is over all primaries in the s-channel OPE, including the identity. As we already explained, we have $B_{2,1}[G^s_{\Delta,J}] \geq 0$ for $\Delta>2\Df+J$ and even $J$.
For $\Delta<2\Df+J$ the integral can be evaluated by analytic continuation in $\Delta$, but need not be positive.

\smallskip

The second line of \eqref{eq:OmegadDisc} also makes it manifest that $B_{2,1}[G^s_{\Delta,J}]$ has double zeros on all $n>0$ double-twists. Therefore, $B_{2,1}$ should be a linear combination of
$\alpha_{0,\ell}$ and $\beta_{0,\ell}$ with $\ell=0,1,\ldots$. We can compute
\be
B_{2,1}[G^s_{2\Df+\ell,\ell}] = \delta_{\ell,0}
\ee
and
\ba
&B_{2,1}[\partial_{\Delta}G^s_{2\Df,0}] = H_{2\Df-1}-2 H_{\Df-1}\\
&B_{2,1}[\partial_{\Delta}G^s_{2\Df+\ell,\ell}] =
(-1)^\ell \frac{(\ell-1)!(2 \Df+\ell)_{\ell+1}}{2(\Df )_\ell (\Df )_{\ell+1}}\qquad\textrm{for }\ell\geq 1\,,
\ea
where $H_{z}$ is the harmonic number. In other words,
\be
B_{2,1} = \alpha_{0,0} + (H_{2\Df-1}-2 H_{\Df-1})\beta_{0,0} +
\sum\limits_{\ell=1}^{\infty}(-1)^\ell \frac{(\ell-1)!(2 \Df+\ell)_{\ell+1}}{2(\Df )_\ell (\Df )_{\ell+1}}\beta_{0,\ell}\,.
\label{B21 from alphas}
\ee
This expression is meaningful if both sides act on functions in the domain of the individual $\beta_{0,\ell}$ functionals, for example the conformal blocks $G^s_{\Delta,J}(w,\wb)$.

\smallskip

A lesson which will prove to be more general is that while the individual $\alpha_{n,\ell}$, $\beta_{n,\ell}$ are not positive, 
and only have spin-0 convergence,
their infinite sums over $\ell$ can be positive and enjoy spin-2 convergence.
Subtracted dispersion relations will provide a natural mechanism for producing such positive functionals, $B_{2,1}$ being a special case.

\subsection{Mellin-inspired subtracted dispersion relation}

In \eqref{def B}, we introduced a family of sum rules $B_v$ with spin-0 decay
by evaluating the dispersion relation (minus Euclidean OPE) in the limit $u\to 0$.
Convergence may be improved by acting with the same kernel from \eqref{eq:BsbdDiscS}
on a rescaled correlator $\cG\mapsto \cG/u$, which is the improvement scheme used in \cite{Carmi:2019cub}.
However, these sum rules are not sign-definite due to a singular denominator entering with power $3/2$.

\smallskip

We will now derive a distinct position-space dispersion relation,
valid for any $\langle\phi\phi\phi\phi\rangle$ correlator in a unitary theory, and which will enjoy nice positivity properties.

\smallskip

It will be based on the \emph{Mellin} form of the subtraction shown in \eqref{mellin sub}:
$M(\mS,\mT) \mapsto \frac{M(\mS,\mT)}{(\mS-\mS_1)(\mS-\mS_2)}$.
As in Section \ref{ssec:kernelFromMellin}, we will use Mellin space to motivate the formula, which can then be checked by working directly in position space.
The nonperturbative Mellin amplitude is bounded by spin-1 growth in the u-channel Regge limit, meaning
\be
M(\mS,4\Df-\mS-\mU)=O(|\mS|)\qquad\textrm{as}\quad |\mS|\rightarrow\infty\quad\textrm{at fixed }\mU\,.
\ee
Therefore, the arc at infinity in the dispersion relation vanishes if we apply Cauchy's integral formula to
\be
\frac{M(\mS,4\Df-\mS-\mU)}{(\mS-\mS_1)(\mS-\mS_2)}\,,
\label{eq:MellinSubtracted}
\ee
where $\mS_{1,2}$ are arbitrary subtraction points.
It is convenient to maintain $\mS\leftrightarrow\mT$ symmetry by choosing $\mS_2=4\Df-\mU-\mS_1$. Furthermore, we will set $\mS_1=2\Df$, since this choice puts the extra poles of \eqref{eq:MellinSubtracted} to double-twist locations and leads to simpler formulas in position space. In other words, we will apply Cauchy's formula to $M(\mS,\mT)/((\mS-2\Df)(\mT-2\Df))$.

\smallskip

Going through the same steps as in Section \ref{ssec:kernelFromMellin}, we find that the dispersion relation in Mellin space leads to the following position-space kernel
\ba
K_{2}(u,v;u',v') = 
\frac{1}{4u'v'}\iiint\!\frac{d\mS\,d\mT\,d\mS'}{(2\pi i)^3}
&\frac{\Gamma(\Df-\tfrac{\mS}{2})^2\Gamma(\Df-\tfrac{\mT}{2})^2}{2\sin ^2\!\left[\tfrac{\pi}{2}(\mS'-2\Df)\right]\Gamma(\Df-\tfrac{\mS'}{2})^2\Gamma(\Df-\tfrac{\mT'}{2})^2}\times\\
\times&\frac{1}{\mS'-\mS}
\frac{u^{\tfrac{\mS}{2}-\Df}v^{\tfrac{\mT}{2}-\Df}}{u'^{\tfrac{\mS'}{2}-\Df}v'^{\tfrac{\mT'}{2}-\Df}}
\times \frac{(\Df-\tfrac{\mS}{2})(\Df-\tfrac{\mT}{2})}{(\Df-\tfrac{\mS'}{2})(\Df-\tfrac{\mT'}{2})}\,.
\label{eq:KMellinSubtracted}
\ea
The superscript of $K_{2}(u,v;u',v')$ means it corresponds to the spin-2 subtracted dispersion relation; the only difference is the last factor.
Below, we will generalize it to multiple subtractions. Quite remarkably, $K_{2}(u,v;u',v')$ admits a concise closed form, which is very similar to the unsubtracted case \eqref{eq:kernelsUV}:
\ba
K_{2}(u,v;u',v') = K_{2;B}(u,v;u',v') \,&\theta(\sqrt{v'}>\sqrt{u'}+\sqrt{u}+\sqrt{v}) +\\
+ \,K_{2;C}(u,v;u') \,&\delta(\sqrt{v'}-\sqrt{u'}-\sqrt{u}-\sqrt{v})\,,
\ea
where
\ba
K_{2;B}(u,v;u',v') &=
-\frac{3}{64 \pi}\frac{v+v'-u-u'}{(u v u' v')^{\frac{3}{4}}}\left(\frac{uv}{u'v'}\right)^{\frac{1}{2}}
x^{\frac{5}{2}}{}_2F_1\!\left(\tfrac{3}{2},\tfrac{5}{2};2;1-x\right)
\\
K_{2;C}(u,v;u') &=-\frac{1}{4 \pi}\frac{1}{(u v u'v')^{\frac{1}{4}} (\sqrt{u}+\sqrt{u'})}\left(\frac{uv}{u'v'}\right)^{\frac{1}{2}}\,.
\label{eq:kernelsK2}
\ea
The contact term can be found by explicitly evaluating the Mellin integrals, as in Appendix \ref{app:Kmellin}. We obtained the bulk term by computing a few terms in its series expansion as in Section \ref{ssec:kernelFromMellin} and matching with an educated guess consisting of powers of $u,v$ times a function of the magic variable $x$, given in \eqref{eq:xDefinition}. We then verified the result to high order against the series expansion of \eqref{eq:KMellinSubtracted}.

\smallskip

The formulas in \eqref{eq:kernelsK2} look deceptively similar to their unsubtracted cousin in \eqref{eq:kernelsUV}, but there are deep differences.  For one thing, the relative sign between the bulk and contact term has changed,
and they are now both \emph{negative}.  Second, the behavior as $x\to 0$ now gives
$x^{\frac{5}{2}}{}_2F_1\!\left(\tfrac{3}{2},\tfrac{5}{2};2;1-x\right)\sim x^{1/2}$ as opposed to $x^{3/2}$.
As we will see, this will cure the $u\to 0$ sum rules from their distributional nature.
These will be crucial for applications.

\smallskip

We can now forget where the kernel came from and prove that it does the job by working directly in position space. We claim that the correct subtracted dispersion relation corresponding to kernels \eqref{eq:kernelsK2} takes the form
\ba 
\cG(z,\zb)&=1_{\rm u}+\cG^s(z,\zb) + \cG^t(z,\zb)\\
\cG^s(z,\zb) &= \iint\! du'dv'K_{2}(u,v;u',v')\dDisc_s[\cG(w,\wb)]\\
\cG^t(z,\zb) &= \iint\! du'dv'K_{2}(v,u;v',u')\dDisc_t[\cG(w,\wb)]\,.
\label{eq:DispersionSubtracted}
\ea
Note in particular the appearance of u-channel identity on the RHS of the first line (nonzero only if identity is exchanged in the u-channel).
To derive \eqref{eq:DispersionSubtracted}, we write down the corresponding primitive dispersion relation \eqref{punchline} with the $C_{-}\times C_{+}$ contour and run the same contour deformation argument as in Section \ref{ssec:FunctionalBasis}. Thus, in analogy with \eqref{Omega-s}, \eqref{Omega-t}, define the generating functionals 
\ba
\Omega^{s | u }_{2;z,\zb}[\cF] &=\theta(v-u)\cF(z,\zb)+
\!\iint\limits_{C_-\times C_+}\!\!\frac{dwd\wb}{(2\pi i)^2} \pi^2(\wb-w)K_{2;B}(u,v;u',v')  \cF(w, \wb)\\
\Omega^{t | u }_{2;z,\zb}[\cF] &=\theta(u-v)\cF(z,\zb)
-\!\!\iint\limits_{C_-\times C_+}\!\!\frac{dwd\wb}{(2\pi i)^2}\pi^2(\wb-w)K_{2;B}(u,v;u',v') \cF(w,\wb)\,,
\label{eq:OmegaSubtracted}
\ea
and set
\ba
\cG^s(z,\zb) &=\Omega^{s | u }_{2;z,\zb}[\cG(w,\wb)-1_{\rm u}]\\
\cG^t(z,\zb) &= \Omega^{t | u }_{2;z,\zb}[\cG(w,\wb)-1_{\rm u}]\,.
\label{eq:DispersionSubtractedPrimitive}
\ea
It immediately follows from this definition that
\be
\cG(z,\zb)  =1_{\rm u} + \cG^s(z,\zb) + \cG^t(z,\zb).
\ee
The subtraction of the u-channel identity is necessary to make the integrals converge in the lightcone limit $w\rightarrow\infty$ with $\wb$ fixed. Indeed,
\be
(\wb-w)K_{2;B}(u,v;u',v') \sim w^{-1}\quad\textrm{as }w\rightarrow\infty\textrm{ with fixed }\wb\,.
\ee
Recall that we are assuming $\cG(w,\wb) = 1+O(w^{-\tau_0/2})$ for $\tau_{0}>0$ in the u-channel lightcone limit on both sheets. The integrals in \eqref{eq:OmegaSubtracted} are automatically convergent in the u-channel Regge limit since
\be
(\wb-w)K_{2;B}(u,v;u',v') \sim w^{-3}\quad\textrm{as }w\rightarrow\infty\textrm{ with fixed }\wb/w\,.
\ee
In other words, the Mellin subtraction enhances the decay of the position-space kernel by $w^{-2}$ in the u-channel Regge limit, i.e. by two units of spin, as expected from \eqref{eq:MellinSubtracted}.

\smallskip

Exactly as we saw for $K_{B}(u,v;u',v')$, also $K_{2;B}(u,v;u',v')$ has a simultaneous pole at $(w,\wb)=(z,\zb)$ with a unit residue
\be
\pi^2(\wb-w)K_{2;B}(u,v;u',v') \sim \frac{1}{(w-z)(\wb-\zb)}\,.
\ee
The residue combines with the theta functions in \eqref{eq:OmegaSubtracted} to ensure that $\cG^s(z,\zb)$ and $\cG^t(z,\zb)$ are analytic in the cut plane.

\smallskip

Finally, we can deform the $C_{-}\times C_{+}$ to the left or to the right to get the second and third line of \eqref{eq:DispersionSubtracted}. Note that the subtracted u-channel identity does not appear since $\dDisc_{s,t}[1] = 0$. All the steps are covered in detail in Appendix \ref{app:contours} in the unsubtracted case.

\subsection{The $B_{2,v}$ family of sum rules}

As an application, we can derive the spin-2 subtracted version of the functional $B_{v}$ of Section \ref{ssec:GeneratingFunctionals}.
We start from the subtracted dispersion relation \eqref{eq:DispersionSubtracted}. We expand $\cG(z,\zb)$ using the s-channel OPE, and $\dDisc_{s,t}[\cG(w,\wb)]$ under the integral sign using respectively the s-,t-channel OPE. The resulting sum rule is the same as what we get by applying the crossing-antisymmetric functional
 \be
\Omega_{2;z,\zb} = \Omega^{t|u}_{2;z,\zb} - \Omega^{s|u}_{2;1-z,1-\zb}\,.
\ee
to the OPE of crossing-symmetric $\cG(w,\wb)-1_{\rm u}$. The sum rules read
\be
\sum\limits_{\cO} a_{\cO}\,\Omega_{2;z,\zb} [G^s_{\Delta_{\cO},J_{\cO}}] = 1_{\rm u}\,.
\label{eq:sumRuleOmega2}
\ee
We will define $B_{2,v}$ as $\Omega_{2;z,\zb}$ at the leading order as $z\rightarrow 0$ with fixed $\zb=1-v$.
\be
B_{2,v} = \Omega_{2;z=0,\zb=1-v}\,.
\ee
Note that there are no logarithms at this order in the subtracted kernel. This definition and the formula for the dispersion kernel \eqref{eq:kernelsK2} give us the explicit form
\be
B_{2,v}[\mathcal{F}] =
\!\iint\limits_{C_-\times C_+}\!\!\frac{dwd\wb}{(2\pi i)^2}\,
\frac{(\wb-w)(v'-u')}{u'v'\sqrt{v^2-2(u'+v')v+(u'-v')^2}}
\mathcal{F}(w,\wb)\,.
\label{eq:B2VDefinition}
\ee
The difference from $B_{v}$ is the extra factor $u'v'$ in the denominator, and also the power of the quadratic polynomial: 1/2 compared to 3/2. Overall, these lead to a faster decay of the functional in the Regge limit by a factor $w^{-2}$ as required. It is easy to see that $B_{2,v}$ is s-t antisymmetric when acting on functions satisfying $\cF(w,\wb)=\cF(\wb,w)$
\be
B_{2,v}[\cF(w,\wb)] = - B_{2,v}[\cF(1-\wb,1-w)]\,.
\ee
Therefore, for any s-t symmetric correlator $\cG(w,\wb)$, we have\footnote{
For non s-t symmetrical correlators, the functionals $B_{2,v}^{s,t}$
would be given by similar formula with $v'-u'\mapsto \frac{v \pm(v'-u')}{2}$, respectively, such that
$B_{2,v}=B_{2,v}^s-B_{2,v}^t$.
}
\be
B_{2,v}[\cG(w,\wb)-1_{\rm u}]=0.
\ee
The corresponding non-perturbative sum rule is derived by expanding $\cG(w,\wb)-1_{\rm u}$ using the s-channel OPE, or equivalently as the $z\rightarrow 0$ limit of \eqref{eq:sumRuleOmega2}
\be
\sum\limits_{\cO} a_{\cO}\,B_{2,v} [G^s_{\Delta_{\cO},J_{\cO}}] = 1_u\,,
\label{eq:B2vSumrule}
\ee
where $B_{2,v} [G^s_{\Delta_{\cO},J_{\cO}}]$ can be computed using the $\dDisc_{s}$ form of \eqref{eq:B2VDefinition}:
\be
B_{2,v} [G^s_{\Delta,J}] = 
 \int\limits_{v}^{\infty}dv'\!\!\!\!\!\!\!\!\int\limits_{0}^{(\sqrt{v'}-\sqrt{v})^2}\!\!\!\!\!\!\!\!\!du'
\frac{v'-u'}{\pi^2 u' v' \sqrt{v^2-2(u'+v')v+(u'-v')^2}}\,
\dDisc_{s}[G^s_{\Delta,J}(u',v')]\,. \label{eq:B2v dDisc}
\ee
This integral will be useful for practical evaluation.
It converges for all $\Delta>2\Df+J$. Therefore, $B_{2,v} [G^s_{\Delta,J}]$ has double zeros on all $n>0$ double-twists. Note that the factor multiplying $\dDisc_{s}[G^s_{\Delta,J}(u',v')]$ is always positive inside the integration region. On the other hand, $\dDisc_{s}[G^s_{\Delta,J}(u',v')]$ is only guaranteed to be positive in the Lorentzian lightcone, i.e. for $v'\geq1$, $u'\leq(\sqrt{v'}-1)^2$. If $v\geq 1$, then the integration region is a subset of the Lorentzian lightcone. Therefore, we can conclude
\be
B_{2,v} [G^s_{\Delta,J}] \geq 0 \quad\textrm{for}\quad\Delta\geq 2\Df+J\,,\quad\textrm{assuming}\quad v\geq 1\,.
\ee
Note that for $v=1$, the square root becomes simply $(\wb-w)$ and the
$B_{2,v}$ functional reduces precisely to the simple positive functional
introduced in section \ref{ssec:B21}.

\smallskip

It is instructive to understand the relation between the $C_-\times C_+$ and $C_-\times C_-$ forms of the sum rule
in \eqref{eq:B2VDefinition}-\eqref{eq:B2v dDisc}, in particular the appearance of dDisc.
A physical interpretation in terms of commutators of detectors is given in subsection \ref{ssec:superconvergence} below.

\smallskip

We thus have an infinite family of functionals labelled by $v$,
each of which is positive above the double-twist threshold.
To understand their physical implications, let us discuss the expansion of $B_{2,v}$ in the basis functionals $\alpha_{n,\ell}$, $\beta_{n,\ell}$. Since $B_{2,v} [G^s_{\Delta,J}]$ has double zeros on all $n>0$ double-twists,
all the $n>0$ basis functionals are absent and we can write:
\be \label{B2v alpha beta}
B_{2,v} = \sum\limits_{\ell=0}^{\infty}\left[a_{\ell}(v)\,\alpha_{0,\ell} + b_{\ell}(v)\,\beta_{0,\ell}\right]\,.
\ee
The coefficients can be computed as
\be
a_{\ell}(v) = B_{2,v}[G^s_{2\Df+\ell,\ell}]\,,\qquad b_{\ell}(v) = B_{2,v}[\partial_{\Delta}G^s_{2\Df+\ell,\ell}]\,,
\ee
starting from \eqref{eq:B2VDefinition}. The computation of $a_{\ell}(v)$ reduces to a residue at $(w,\wb)=(0,1-v)$, the result being
\be
a_{\ell}(v) = (1-v)^{-\Df}k_{\Df+\ell}(1-v)\,.
\label{eq:a2V}
\ee
A similar but more involved calculation gives
\ba
b_{\ell}(v) &= \frac{1}{2}(1-v)^{\ell}v^{\Df+\ell}\partial_{h}[v^{-h}{}_2F_1(h,h;2h;1-v)]_{h=\Df+\ell}
-\\
&-\int\limits_{v}^{\infty}dz\log\!\left(z-v\right)
\partial_z[(1-z)^{\ell}{}_2F_1(\Df+\ell,\Df+\ell;2\Df+2\ell;1-z)]\,.
\label{eq:b2V}
\ea
The $a_{\ell}(v)$ coefficients are trivial to diagonalize by doing a series expansion around $v=1$.
The interpretation of the generating function
\eqref{B2v alpha beta} is thus that any $\alpha_{0,\ell}$ can be promoted
to a spin-2 convergent functional by adding a unique (infinite) combination of $\beta_{0,\ell}$'s.
One can check that $b_{\ell}(v)$ admits a more explicit presentation as a linear combination of $a_{j}(v)$ with $0\leq j\leq \ell$
\ba
b_{\ell}&(v) = (H_{2 \Df+2\ell-1}+H_{2 \Df+2\ell-2}-2 H_{\Df+\ell -1}-H_{2 \Df+\ell-2})a_{\ell}(v)+\\
&+\sum\limits_{j=0}^{\ell-1}
\frac{(-1)^{j+\ell}\,\ell!\,(2 \Df+\ell-1)_{\ell+2}}{2 (\ell-j)j!\, (2 \Df +j+\ell-1) (2 \Df+j-1)_j (\Df+j)_{\ell-j} (\Df+j)_{\ell-j+1}}a_{j}(v)\,.
\label{eq:b2vExplicit}
\ea
We will soon provide a justification for this using Mellin space.
Note that $a_{\ell}(v)$, $b_{\ell}(v)$ are holomorphic in
$v\in\mathbb{C}\backslash(-\infty,0]$, as is $B_{2,v}[\mathcal{F}]$ for any $\mathcal{F}(w,\wb)$ in the domain of $B_{2,v}$.

\smallskip

Let us check that the $B_{2,v}$ sum rule \eqref{eq:B2vSumrule} holds for the $\langle\phi\phi\phi\phi\rangle$ correlator in mean field theory. The only contributing primaries are the identity and the double traces with $n=0$ and $\ell$ even. We can evaluate $B_{2,v}[G^s_{0,0}]$ most easily
by closing both contours in \eqref{eq:B2VDefinition} to the right, picking the residue at $(w,\wb)=(v,1)$
\be \label{B2v identity}
B_{2,v}[G^s_{0,0}] = -v^{-\Df}\,.
\ee
Thus
\ba
\sum\limits_{\cO} a_{\cO}\,B_{2;v} [G^s_{\Delta_{\cO},J_{\cO}}] &=
-v^{-\Df} + \sum\limits_{\ell=0}^{\infty}[1+(-1)^{\ell}]q^{\textrm{MFT}}_{0,\ell}(1-v)^{-\Df}k_{\Df+\ell}(1-v) =\\
&= - v^{-\Df} + v^{-\Df} + 1 = 1\,,
\ea
in agreement with the RHS of \eqref{eq:B2VDefinition}.

\smallskip

We see from \eqref{eq:a2V} that each $\alpha_{0,\ell}$ occurs with positive coefficient in $B_{2,v}$ for $v<1$,
and with alternating signs proportional to $(-1)^\ell$ for $v>1$. By antisymmetrizing under the switch of t- and u-channels,
implemented by $v\leftrightarrow 1/v$ with suitable powers of $v$, it is
thus possible to cancel the even-spin $\alpha$'s:
\be
\widetilde{B}_{2,v} \equiv v^{\frac{\Df}{2}} B_{2,v} - v^{-\frac{\Df}{2}} B_{2,1/v}\,. \label{B2 tilde}
\ee
Since $a_{\ell}(v) = (-1)^{\ell}v^{-\Df}a_{\ell}(1/v)$, it follows that $\widetilde{B}_{2,v}$ vanishes on the $n=0$, $\ell$ even double-twist conformal blocks, while having double zeros on all $n>0$ double traces. Relatedly, the contribution of the s-channel identity to the $\widetilde{B}_{2,v}$ sum rule precisely cancels the inhomogenous term coming from the subtracted u-channel identity, giving the sum rule
\be
\sideset{}{'}\sum\limits_{\cO}a_{\cO}\, \widetilde{B}_{2,v}[G^s_{\Delta_{\cO},J_{\cO}}] = 0\,,
\ee
where the sum runs over all non-identity primaries in the OPE.
Since it is a difference between two terms, positivity properties of this sum rule are not obvious;
they are studied in Section \ref{sec:MFTbounds}. The first derivative of $\widetilde{B}_{2,v}$ with respect to $v$ at $v=1$, denoted $\widetilde{B}'_{2,1}$, will turn out to enjoy nice properties.
We can see hints of this by using \eqref{eq:b2V} to compute the slope of the single zero around $\Delta=2\Df+\ell$.
By series-expanding the hypergeometric functions and integrating
term-by-term (or using \eqref{eq:b2vExplicit}), we obtain the following simple formula:
\be \label{slope B21}
 \widetilde{B}_{2,1}'[G^s_{2\Df+\ell+\gamma,\ell}]  = 
\frac{\ell (\ell-2)!\Df(2 \Df+\ell -1)_{\ell+1}}{(2 \Df +\ell)(\Df)^{2}_{\ell}} \gamma + O(\gamma^2)\qquad
 \mbox{($\ell\geq 2$ even)}\,,
\ee
whereas for $\ell=0$ the slope vanishes. 
Remarkably, the slopes in \eqref{slope B21} are positive! 
Thus the $\widetilde{B}_{2,1}'$ functional has double zeros for each double twist with $n>0$, and has
a single zero with positive slope on the leading twist (and only double zeros for spin 0).
Positivity above all double-twists is a desired property
which would allow the functional to establish the existence of spinning operators below
the double-twist thresholds, as will be further investigated below.\footnote{This fact has long been known form the lightcone bootstrap, in particular convexity property \cite{Komargodski:2012ek}, but an extremal functional which shows this was not known until \cite{Penedones:2019tng} found such functional for certain range of $\Df$.}

\smallskip

Finally, it is instructive to transform the functionals $B_{2,v}$ to Mellin space. Let us define
\be
\widehat{B}_{2,\mT} = \Gamma(\Df-\tfrac{\mT}{2})^{-2}\Gamma(\tfrac{\mT}{2})^{-2}
\int\limits_{0}^{\infty}\frac{dv}{v}v^{\Df-\tfrac{\mT}{2}} B_{2,v}\,,
\ee
which has the inverse
\be
B_{2,v} = \!\!\!\int\limits_{\Df-i\infty}^{\Df+i\infty}\!\!\!\frac{d\mT}{4\pi i}
v^{\tfrac{\mT}{2}-\Df}\Gamma(\Df-\tfrac{\mT}{2})^{2}\Gamma(\tfrac{\mT}{2})^{2} \widehat{B}_{2,\mT}\,.
\label{eq:B2VMellinRep}
\ee
Equivalently, $\widehat{B}_{2,\mT}$ is the dispersion relation for $M(\mS,\mT)/[(\mS-2\Df)(\mT-2\Df)]$ at the leading order around $\mS=2\Df$, that is, it implements the sum rule that the following integral along the arc at infinity vanishes:
\be \label{vanishing id}
 0 = \oint \frac{d\mS'}{2\pi i} \frac{M(\mS',\mT')}{(\mS'-2\Df)^2(\mT'-2\Df)}
\ee
with $\mT'=\mT+2\Df-\mS'$.
This identification leads to the following closed formula for the action of $\widehat{B}_{2,\mT}$ on arbitrary conformal blocks
\be
\widehat{B}_{2,\mT}[G^s_{\Delta,J}] = 
\sum\limits_{m=0}^{\infty}
\left(\frac{1}{\tau-\mT+2 m}+\frac{1}{\tau-2 \Df+2 m}\right)
\frac{2 (\mT-2 \Df ) (-1)^J \mathcal{Q}^{m}_{\Delta,J}(2 \Df -\mT)}{(\tau-2 \Df+2 m) (\tau-\mT+2 m)}\,.
\label{eq:B2HatAction}
\ee
The fact that the sum rule adds up to the u-channel identity, clear from the position space derivation of \eqref{eq:B2vSumrule}, is likely related to the order-of-limit issues elucidated in Section 5 of \cite{Penedones:2019tng}.

\smallskip

This formula combined with the Mellin representation \eqref{eq:B2VMellinRep} in principle gives an independent way to compute the action of $B_{2,v}$ on arbitrary conformal blocks. The formula makes it manifest that $\widehat{B}_{2,\mT}[G^s_{\Delta,J}]$ has double zeros on the $n>0$ double traces, so that we can write
\be
\widehat{B}_{2,\mT} = \sum\limits_{\ell=0}^{\infty}\left[\widehat{a}_{\ell}(\mT)\,\alpha_{0,\ell} + \widehat{b}_{\ell}(\mT)\,\beta_{0,\ell}\right]\,.
\label{eq:B2HatExpansion}
\ee
$\widehat{a}_{\ell}(\mT)$, $\widehat{b}_{\ell}(\mT)$ can be computed in two ways: either as the Mellin transform of $a_{\ell}(v)$, $b_{\ell}(v)$ or by expanding \eqref{eq:B2HatAction} around $\tau=2\Df$. The result
\ba
\widehat{a}_{\ell}&(\mT)=
\frac{\Gamma (2\Df+2\ell)}{\Gamma (\Df)^2 \Gamma (\Df+\ell)^2}
{}_3F_2\left(-\ell,\Df -\tfrac{\mT}{2},\ell+2 \Df -1;\Df ,\Df ;1\right)\\
\widehat{b}_{\ell}&(\mT) = (H_{2 \Df+2\ell-1}+H_{2 \Df+2\ell-2}-2 H_{\Df+\ell -1}-H_{2 \Df+\ell-2})\widehat{a}_{\ell}(\mT)+\\
&+\sum\limits_{j=0}^{\ell-1}
\frac{(-1)^{j+\ell}\,\ell!\,(2 \Df+\ell-1)_{\ell+2}}{2 (\ell-j)j!\, (2 \Df +j+\ell-1) (2 \Df+j-1)_j (\Df+j)_{\ell-j} (\Df+j)_{\ell-j+1}}\widehat{a}_{j}(\mT)\,.
\label{eq:abHatFormulas}
\ea
Note that the Mellin transform of $\widetilde{B}_{2,v}$ gives $\widehat{B}_{2,\mT}-\widehat{B}_{2,2\Df-\mT}$.
It is not clear what positivity properties are enjoyed in the Mellin basis $\widehat{B}_{2,\mT}$; it would be nice to clarify the range of $\mT$ for which the Mellin-space functionals are valid. Appendix \ref{app:mellinFunctionals} contains an explicit formula for the position-space kernel defining $\widehat{B}_{2,\mT}$. In this work we focus on the position-space basis $B_{2,v}$.

\newcommand\wL{\mathbf{L}}
\newcommand\R{\mathbb{R}}
\newcommand\cA{\mathcal{A}}

\subsection{A physical interpretation of $B_{2,v}$ and subtracted superconvergence}
\label{ssec:superconvergence}

In this section, we explain a simple physical interpretation of the $B_{2,v}$ sum rule: it is a subtracted version of the ``superconvergence'' sum rules of \cite{Kologlu:2019bco}.  Superconvergence is the statement that light-transformed (i.e.\ null-integrated) operators on the same null plane commute. For example,  consider null integrated scalars $\phi_1$ and $\phi_3$  in states created by two other scalars $\f_2$ and $\f_4$. We will refer to $\f_1$ and $\f_3$ as ``detectors.'' An example superconvergence sum rule is
\ba
\label{eq:superconvergencescalars}
0 &= \<\Omega|\f_4(x_4)\left[\int_{-\oo}^\oo dx^+ \f_1(x^+,x^-=0,\vec x_1),\int_{-\oo}^\oo dx^+ \f_3(x^+,x^-=0,\vec x_3)\right]\f_2(x_2)|\Omega\> \\
&\quad \textrm{if $J_0<-1$}.
\ea
Here, we have written the positions of $\f_1,\f_3$ in lightcone coordinates. The statement (\ref{eq:superconvergencescalars}) becomes a nontrivial constraint on CFT data when we evaluate $\<\Omega|\f_4 \f_1 \f_3 \f_2|\Omega\>$ using the $t$-channel OPE and $\<\Omega|\f_4 \f_3 \f_1 \f_2|\Omega\>$ using the $s$-channel OPE.

\smallskip

Let us explain the condition $J_0<-1$.
The na\"ive argument for (\ref{eq:superconvergencescalars}) is that $\f_1$ and $\f_3$ are spacelike-separated everywhere along their integration contours, and thus they commute. However, as explained in \cite{Kologlu:2019bco}, one must also study the ends of the integration contours $x_1^+,x_3^+\to\pm \oo$. The result is that (\ref{eq:superconvergencescalars}) holds if $\<\f_4\f_1\f_3\f_2\>$ decays in the $u$-channel Regge limit with spin $J_0 < -1$. This condition may be violated in physical correlators, so it is not immediately obvious how (\ref{eq:superconvergencescalars}) is useful for bounding physical  correlators.\footnote{The conditions for superconvergence sum rules of spinning operators are not as stringent, and they are easily satisfied in physical correlators.} Nevertheless, by studying (\ref{eq:superconvergencescalars}) in simpler kinematics we will see how it is related to $B_{2,v}$.

\smallskip

Let us choose a conformal frame such that $\f_1$ and $\f_3$ lie at future null infinity. Suppose that $x_{24}$ is future-pointing and timelike.  We can use translations to set $x_4=0$ and boosts and dilatations to set $x=x_{24}=(1,0,\dots,0)$. Finally, suppose $\f_1$ and $\f_3$ lie at positions $\vec n_1,\vec n_3\in S^{d-2}$ on the celestial sphere and are integrated over retarded time. In the language of \cite{Kravchuk:2018htv}, we are studying
\begin{align}
\label{eq:lighttransform}
\<\Omega|\f_4(0) [\wL[\f_1](\oo,z_1), \wL[\f_3](\oo,z_3)] \f_2(x)|\Omega\>,
\end{align}
where $\wL$ is the light-transform, and $z_1=(1,\vec n_1)$, $z_3=(1,\vec n_3)$ are future-pointing null vectors.

\smallskip

An event shape in conformal collider physics is defined by fourier-transforming (\ref{eq:lighttransform}) with respect to $x$ \cite{Hofman:2008ar}. This would give a matrix element of $[\wL[\f_1], \wL[\f_3]]$ in a momentum eigenstate. Here, we do {\it not} pass to momentum space. Thus, (\ref{eq:lighttransform}) could be called a ``position-space event shape.''

\smallskip

\begin{figure}[ht!]
	\centering
	\begin{tikzpicture}

\draw[fill=black] (0.98,0.4) circle (0.07);
\draw[fill=black] (-0.73,-0.4) circle (0.07);

\draw[] (-2.9,1.3) -- (-2.9,0.8) -- (-4,0.8);

\draw[->,line width=0.7] (-4,-0.1) -- (0,-0.1);
\draw[line width=0.7] (0,-0.1) -- (4,-0.1);
\draw[->,line width=0.7] (-4,0.1) -- (0,0.1);
\draw[line width=0.7] (0,0.1) -- (4,0.1);

\draw[blue,->,line width=0.7] (-4,-0.25) -- (-2.35,-0.25);
\draw[blue,line width=0.7] (-2.35,-0.25) -- (-0.7,-0.25) to[out=0,in=0,distance=0.2cm] (-0.7,-0.55);
\draw[blue,->,line width=0.7] (-0.7,-0.55) -- (-2.35,-0.55);
\draw[blue,line width=0.7] (-2.35,-0.55) -- (-4,-0.55);

\draw[red,line width=0.7] (2.35,0.25) -- (4,0.25);
\draw[red,->,line width=0.7]  (0.96,0.55) to[out=180,in=180,distance=0.2cm] (0.96,0.25) -- (2.35,0.25);
\draw[red,line width=0.7] (0.96,0.55)-- (2.35,0.55);
\draw[red,->,line width=0.7]  (4,0.55)-- (2.35,0.55);

	\node[above] at (-3.5,0.8) {$\a_1,\a_3$}; 
	\node[below] at (-0.73,-0.5) {$1,3\sim 4$};
	\node[above] at (0.96,0.5) {$1,3\sim 2$};
	\end{tikzpicture}
	\caption{Contour prescriptions for the $\a_1$ and $\a_3$ integrals in (\ref{eq:positionevent}). The singularities in $\a_1$ occur when $\f_1$ is lightlike to $\f_4$ at $\a_1=0$, and when $\f_1$ is lightlike to $\f_2$ at $\a_1=2$. The singularities in $\a_3$ are in the same places. There is no $1\sim 3$ lightcone singularity because $\f_1$ and $\f_3$ are not lightlike separated. The contours for $\a_1,\a_3$ in (\ref{eq:positionevent}) are in black. The contours that give rise to $\dDisc_s$ and $\dDisc_t$ are shown in red and blue, as described in the main text.}
	\label{fig:alphacontours}
\end{figure}
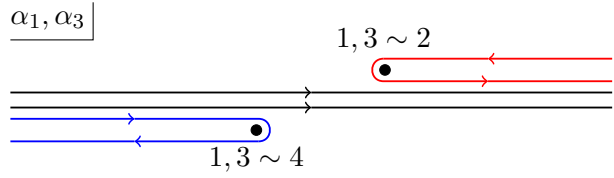

It is convenient to compute (\ref{eq:lighttransform}) using embedding-space coordinates $X=(X^+,X^-,X^\mu)\in \R^{d,2}$. The embedding-space positions of the points are
\begin{align}
X_1 &= (0,-\a_1,z_1),&
X_2 &= (1,x^2,x),\nn\\
X_3 &= (0,-\a_3,z_3),&
X_4 &= (1,0,0),
\end{align}
where we must integrate $\a_1,\a_3$ along the real line. Let us write
\begin{align}
\<\f_1\f_2\f_3\f_4\> &= \frac{\cA(u,v)}{(x_{13}^2 x_{24}^2)^{\De_\f}}.
\end{align}
Then we have
\begin{align}
\<\Omega|\f_4(0) \wL[\f_1](\oo,z_1)\wL[\f_3](\oo,z_3) \f_2(x)|\Omega\> &= 
\frac{1}{(x^2)^{\De_\f}z_{13}^{\De_\f}}\int_{-\oo}^\oo d\a_1 \int_{-\oo}^\oo d\a_3 \cA(u',v'),
\label{eq:positionevent}
\end{align}
where 
\begin{align}
z_{13}=-2z_1\.z_3,
\qquad
u' = \frac{\a_3(2-\a_1)}{z_{13}},
\qquad
v' = \frac{\a_1(2-\a_3)}{z_{13}},
\end{align}
and the phase of $(x^2)^{\De_\f}$ is defined by analytically continuing to timelike $x$ using the appropriate $i\e$ prescription. The $\a_1$ and $\a_3$ contours run along the real axis, below the $\f_2$ lightcone singularity at $\a_1,\a_3=2$, and above the $\f_4$ lightcone singularity at $\a_1,\a_3=0$, see Figure~\ref{fig:alphacontours}.

 Changing variables to cross-ratio space, we find
\begin{align}
\label{eq:caintegrand}
\int_{-\oo}^\oo d\a_1 \int_{-\oo}^\oo d\a_3 \cA(u',v') &= z_{13} \int du' dv' \frac{\cA(u',v')}{\sqrt{(u'-v')^2-2(u'+v')v+v^2}},
\end{align}
where the integration contour is the same as in \eqref{eq:B2VDefinition} and
\begin{align}
v &= \frac{4}{z_{13}} = \frac{2}{1-\vec n_1\.\vec n_3}.
\end{align}
Note that the Jacobian for the change of variables from $\a_1,\a_3$ to $u',v'$ is precisely the square-root factor in the kernel for $B_{2,v}$. To compute the commutator $[\wL[\f_1],\wL[\f_3]]$, we must subtract the same expression with $1\leftrightarrow 3$.

\smallskip

Comparing the integrands (\ref{eq:caintegrand}) and (\ref{eq:B2VDefinition}), we see that the $B_{2,v}$ sum rule is (up to a constant) a superconvergence sum rule for the subtracted correlator
\begin{align}
\label{eq:subtractionforA}
\cA(u',v') &= \frac{(v'-u')}{u'v'}\cF(u',v'),
\end{align}
where $\cF$ is a physical correlation function. The case $v=1$ corresponds to detectors that are ``back-to-back'' $\vec n_1\.\vec n_3=-1$, while $v\to \oo$ corresponds to nearly coincident detectors $\vec n_1\.\vec n_3\to 1$.

\smallskip

Let us describe some checks of this result. Firstly, a scalar superconvergence sum rule has spin $J=-1$ decay in the $u$-channel Regge limit \cite{Kologlu:2019bco,Kologlu:2019mfz}. However, the factor 
\begin{align}
\frac{v'-u'}{u'v'}&=\frac{1-w-\bar w}{w\bar w(1-w)(1-\bar w)}
\label{eq:subtractionfactor}
\end{align}
 adds 3 to that decay, resulting in a kernel with spin-2 decay, which matches the decay of $B_{2,v}$. Furthermore, scalar superconvergence sum rules only get contributions from odd-spin Regge trajectories in the $u$-channel, and  are thus trivial if applied to $s$-$t$ symmetric four-point functions. However, the factor (\ref{eq:subtractionfactor}) is $s$-$t$ antisymmetric, so that superconvergence applied to $\cA$ is nontrivial even if $\cF$ is $s$-$t$ symmetric.
 
 \smallskip

The superconvergence point of view gives a simple way to understand the contour manipulations in appendix~\ref{app:deformationToDDisc} that allow one to write $B_{2,v}$ and its cousins in terms of $\dDisc$. In $\a_1,\a_3$-space these manipulations are much simpler because there is no square-root factor $\sqrt{(u'-v')^2-2(u'+v')v+v^2}$ in the denominator. They are the same manipulations used to relate superconvergence sum rules to the $\dDisc$ in \cite{Kologlu:2019bco,Kologlu:2019mfz}, and also to prove the generalized Lorentzian inversion formula in \cite{Kravchuk:2018htv}. (It also appears to
have been used in a recent discussion of energy-energy correlators \cite{Henn:2019gkr}.)
To obtain the $s$-channel dDisc, we deform $\a_1$ into the upper half-plane so that it wraps the $\f_2$ lightcone singularity (red curve in Figure~\ref{fig:alphacontours}), and we deform $\a_3$ into the lower half-plane so that it wraps the $\f_4$ lightcone singularity (blue curve). Similarly, to obtain the $t$-channel dDisc, we deform $\a_1$ into the lower half-plane so that it wraps the $\f_4$ lightcone singularity (blue curve), and we deform $\a_3$ into the upper half-plane so that it wraps the $\f_2$ lightcone singularity (red curve).

\smallskip

A virtue of unsubtracted superconvergence sum rules is that, when they are well-defined, they are directly related to the $\dDisc$ by the fact that
\begin{align}
\<\Omega|\f_4 \wL[\f_1] \wL[\f_3] \f_2(0)|\Omega\> &= \<\Omega|[\f_4, \wL[\f_1]][\wL[\f_3], \f_2]|\Omega\>,\nn\\
\<\Omega|\f_4 \wL[\f_3] \wL[\f_1] \f_2(0)|\Omega\> &= \<\Omega|[\f_4, \wL[\f_3]][\wL[\f_1], \f_2]|\Omega\>,
\end{align}
since $\wL[\f_i]$ kills the vacuum. In particular, double-twist operators in the $s$- and $t$-channels do not contribute. The subtraction (\ref{eq:subtractionforA}) {\it almost} preserves this property, while improving the $u$-channel Regge decay. Because of the poles in (\ref{eq:subtractionfactor}), it gives a sum rule with support only on the leading tower of double-twists.

\smallskip

We see that there is a hierarchy between the naive crossing equation and superconvergence sum rules, with general dispersive sum rules in between. Non-dispersive sum rules come from the condition that spacelike-separated operators at fixed positions commute. This is just the usual crossing equation. We can obtain dispersive sum rules by integrating the operators along null lines weighted by meromorphic functions of $\a_1,\a_3$. Finally, when the weighting functions are constant, we obtain superconvergence sum rules.

\subsection{Finite sums of $\alpha_{n,\ell}$ and $\beta_{n,\ell}$}

The $B_{2,v}$ functionals provide a generating function for
spin-two convergent combinations of $\alpha_{0,\ell}$ and $\beta_{0,\ell}$ functionals.
We now discuss an alternative method to produce spin-two convergent combinations,
now involving only $\beta_{n,\ell}$'s but for different twists $n$.

\smallskip

Recall from \cite{Mazac:2019shk} that the kernels for $\beta_{n,\ell}$ take the form
\be
\mathcal{B}_{n,\ell}(w,\wb) = \frac{p_{n,\ell}(w,\wb)}{(\wb-w)^{4n+2\ell+2}}\,,
\label{eq:betaKernelGeneral 0}
\ee
where $p_{n,\ell}(w,\wb)$ is a polynomial of total degree $4n+2\ell+1$, meaning that it is a linear combination of monomials $w^a\wb^b$ with $a+b\leq 4n+2\ell+1$. The polynomials satisfy
\be
p_{n,\ell}(w,\wb) = p_{n,\ell}(\wb,w)\,,\qquad p_{n,\ell}(w,\wb) = -p_{n,\ell}(1-w,1-\wb)\,.
\label{eq:pConstraints}
\ee
Conversely, suppose we are given a kernel
\be
\cH(w,\wb) = \frac{p(w,\wb)}{(\wb-w)^{2N+2}}\,,
\label{eq:betaKernelGeneral}
\ee
where $N\in\mathbb{Z}_{\geq0}$ and $p(w,\wb)$ is a polynomial of total degree at most $2N+1$, satisfying \eqref{eq:pConstraints}. It follows from the analysis of Section 5.4 of \cite{Mazac:2019shk} that $\cH(w,\wb)$ can be written as a finite linear combination of $\cB_{n,\ell}(w,\wb)$ with $2n+\ell \leq 2N$. In other words, the space of finite linear combinations of $\beta_{n,\ell}$ is the same as the space of kernels of the form \eqref{eq:betaKernelGeneral} with the stated constraints on $p(w,\wb)$. Every finite linear combination of $\beta_{n,\ell}$ decays at least with spin zero.

\smallskip

How do we construct finite linear combinations of $\beta_{n,\ell}$ which decay at least with spin two and thus apply to general unitary correlators? The solution is to divide \eqref{eq:betaKernelGeneral} by $(\wb-w)^2$. Indeed, every finite combination of $\beta_{n,\ell}$ which decays at least with spin two has a kernel of the form
\be
\cH(w,\wb) = \frac{p(w,\wb)}{(\wb-w)^{2N+4}}\,,
\label{eq:hKernelSpinTwo}
\ee
where $p(w,\wb)$ satisfies \eqref{eq:pConstraints} and has total degree at most $2N+1$. Conversely, every such kernel gives rise to a combination of $\beta_{n,\ell}$ with at least spin two decay. The simplest example is the functional with kernel
\be
\cH(w,\wb) = \frac{1-w-\wb}{(\wb-w)^{4}}\,.
\label{eq:cHNu01}
\ee
One can check that it decomposes as
\be
\frac{1}{24} \beta_{1,0}
-\frac{d+2}{12 d}\beta_{0,2}
-\frac{\Df}{24} \beta_{0,1}-\frac{\Df ^3 (4 \Df -d)}{48 (2 \Df +1)(4 \Df-d +2)}\beta_{0,0}\,.
\label{eq:cHNu01Decomposition}
\ee
Every functional with kernel of the form \eqref{eq:hKernelSpinTwo} gives rise to a nonperturbative sum rule for the OPE of the $\langle\phi\phi\phi\phi\rangle$ four-point function. To show this, we need to check that the functional action can be swapped with the OPE. The swapping property holds if the tail of the OPE, i.e. primaries $\cO$ with $\Delta_{\cO}>\Delta_0$, give a vanishing contribution to the sum rule as $\Delta_{0}\rightarrow\infty$. In practice, we only need to check that the functional gives a finite answer when acting on any nonperturbative four-point function in a unitary theory. The only potentially problematic region of the double integral \eqref{eq:contourexample} occurs when $w$ and/or $\wb$ approach complex infinity. Unitarity guarantees that $\cG(w,\wb)$ is bounded by a constant for $|w|,|\wb|> R$. Thus we only need to check that the integral \eqref{eq:contourexample} contains no divergence at infinity when we replace $\cG(w,\wb)$ with a constant, which is indeed true.

\smallskip

Let us give an explicit basis for the space of finite combinations of $\beta_{n,\ell}$ which decay with spin two. Firstly, let us recall from \cite{Mazac:2019shk} the definition of $\widehat{\beta}_{i,j}$ functionals: they are dual to monomials $z^i\zb^{j}$, and can be recovered from the generating function
\be
\sum\limits_{n,\ell}\,G^s_{\Delta_{n,\ell},\ell}(z,\zb) \beta_{n,\ell}= \sum\limits_{i,j=0}^{\infty}z^{i}\zb^{j}\widehat{\beta}_{i,j}\,.
\ee
Each $\widehat{\beta}_{i,j}$ is a finite combination of $\beta_{n,\ell}$ and vice versa. Let us introduce the following combinations of $\widehat{\beta}_{i,j}$
\be
\label{eq:definitionofnuij}
\nu_{i,j} = (i+1)^2 \,\widehat{\beta}_{i+1,j} - (j+1)^2 \,\widehat{\beta}_{i,j+1} - (i-j)(i+j+1)\,\widehat{\beta}_{i,j}\,.
\ee
$\nu_{i,j}$ is defined for $i,j\geq 0$ and satisfies $\nu_{i,j} = -\nu_{j,i}$. We have found experimentally (and checked extensively) that $\nu_{i,j}$ with $0\leq i<j$ are a basis for the space of finite combinations of $\beta_{n,\ell}$ which decay with spin two. For example, the functional with kernel \eqref{eq:cHNu01} is $\nu_{0,1}/24$.

\smallskip

In order to have a complete basis for the space of functionals with spin-2 decay and with double zeros on all but finitely many double-twists, it remains to incorporate the $\alpha$-type functionals into the picture. Recall that the kernels defining $\alpha_{n,\ell}$ functionals take the form
\be
\mathcal{A}_{n,\ell}(w,\wb) = \frac{p_{n,\ell}(w,\wb)}{2(\wb-w)^{4n+2\ell+2}}\log\!\left[\tfrac{w\wb(1-w)(1-\wb)}{(\wb-w)^4}\right] + \frac{q_{n,\ell}(w,\wb)}{(\wb-w)^{4n+2\ell+2}}\,,
\label{eq:alphaKernelGeneral}
\ee
where $p_{n,\ell}(w,\wb)$ is the same polynomial which appears in \eqref{eq:betaKernelGeneral} and $q_{n,\ell}(w,\wb)$ is another polynomial of total degree bounded by $4n+2\ell+1$ satisfying the same constraints as $p_{n,\ell}(w,\wb)$, i.e. \eqref{eq:pConstraints}. The second term in \eqref{eq:alphaKernelGeneral} can be eliminated by adding a finite linear combination of $\beta$ functionals. This means that we can get a basis for functionals with spin-two decay as follows. Let $\mathcal{N}_{i,j}(w,\wb)$ be the kernel defining $\nu_{i,j}$. A simple basis of functionals is
\be
\label{eq:nukernels}
\begin{tabular}{c|c}
functional & kernel \\
\hline
$\nu_{i,j}$ & $\mathcal{N}_{i,j}(w,\wb)$ \\
$\mu_{i,j}$ & $\mathcal{N}_{i,j}(w,\wb)\log\!\left[\tfrac{w\wb(1-w)(1-\wb)}{(\wb-w)^4}\right]$
\end{tabular}
\qquad\textrm{with }0\leq i<j\,.
\ee
The $\nu_{i,j}$ functionals are finite sums of $\beta$-type functionals with spin-two decay, while the
 $\mu_{i,j}$ functionals are finite sums of $\alpha$-type and $\beta$-type functionals with spin-two decay.

\smallskip

Interestingly, the functional $B_{2,v=1}$ from \eqref{B21 from alphas}
is \emph{not} in the finite span of $\mu$ or $\nu$-type functionals.
The loophole is that it is a sum of $\alpha_{0,0}$ and infinitely many $\beta_{0,\ell}$'s:
this is what enables the new $1/(w\wb)$ singularity to appear.

\subsection{Polyakov-Regge expansion with subtraction}
Just like in the unsubtracted case, we can use the subtracted dispersion relation \eqref{eq:DispersionSubtracted} to exhibit a variant of the Polyakov-Regge expansion which now applies to every $\langle\phi\phi\phi\phi\rangle$ correlator in a unitary CFT. First, define the spin-2 subtracted Polyakov-Regge blocks
\ba
P^{s|u}_{2;\Delta,J}(z,\zb) &= 
\iint\! du'dv'K_{2}(u,v;u',v')\dDisc_s[G^s_{\Delta,J}(w,\wb)]\\
P^{t|u}_{2;\Delta,J}(z,\zb) &= 
\iint\! du'dv'K_{2}(v,u;v',u')\dDisc_t[G^t_{\Delta,J}(w,\wb)] = P^{s|u}_{2;\Delta,J}(1-z,1-\zb)\,.
\ea
Inserting the OPE into the dispersion relation \eqref{eq:DispersionSubtracted} leads to the Polyakov-Regge expansion
\be
\cG(z,\zb) -1 = \sum\limits_{\cO} a_{\cO} \left[P^{s|u}_{2;\Delta_{\cO},J_{\cO}}(z,\zb) + P^{t|u}_{2;\Delta_{\cO},J_{\cO}}(z,\zb)\right]\,.
\label{eq:PolyakovReggeSubtracted}
\ee
The subtracted Polyakov-Regge blocks admit a simple Mellin representation
\ba
P^{s|u}_{2;\Delta,J}(z,\zb) = \iint\!\!\frac{d\mS\,d\mT}{(4\pi i)^2}\,
\Gamma\!\left(\Df-\tfrac{\mS}{2}\right)^2\Gamma\!\left(\Df-\tfrac{\mT}{2}\right)^2\Gamma\!\left(-\Df+\tfrac{\mS+\mT}{2}\right)^2\times\\
\times u^{\tfrac{\mS}{2}-\Df}v^{\tfrac{\mT}{2}-\Df}\cP^{s|u}_{2;\Delta,J}(\mS,\mT)\,,
\ea
where $\cP^{s|u}_{2;\Delta,J}(\mS,\mT)$ is the subtracted Polyakov-Regge block in Mellin-space
\be
\cP^{s|u}_{2;\Delta,J}(\mS,\mT)=
\sum\limits_{m=0}^{\infty}
\frac{(\Df-\tfrac{\mS}{2})(\Df-\tfrac{\mT}{2})}
{(\Df-\tfrac{\tau}{2}-m)(\Df-\tfrac{\mS+\mT-\tau}{2}+m)}
\frac{\mathcal{Q}^{m}_{\Delta,J}(\mT+\mS-\tau-2m)}{\mS-\tau-2m}\,.
\ee
$P^{s|u}_{2;\Delta,J}(z,\zb)$ is the unique single-valued function with the following properties
\begin{enumerate}
\item $\dDisc_{s}[P^{s|u}_{2;\Delta,J}] = \dDisc_{s}[G^{s}_{\Delta,J}]$, $\dDisc_{t}[P^{s|u}_{2;\Delta,J}] = 0$.
\item $P^{s|u}_{2;\Delta,J}$ is bounded by spin $J<2$ in the u-channel Regge limit.
\item The $n=0$ double-twist terms $\partial_{\Delta}G^s_{2\Df+\ell,\ell}$, $\partial_{\Delta}G^{t}_{2\Df+\ell,\ell}$ are absent in the s- and t-channel OPEs.
\end{enumerate}
Properties 1 and 2 are true for a general spin-2 subtraction, and property 3 is specific to the choice of subtraction scheme
made above. Property 3 is related to the observation made below \eqref{eq:b2V} that,
starting from a given $\alpha_{0,\ell}$ functional, it is always possible to add a (unique)
combination of $\beta_{0,\ell}$'s to make it spin-2 convergent.
The combination is found by series-expanding \eqref{eq:a2V} around $v=1$ and diagonalizing in $\ell$.
Thus, in this subtraction scheme, we lose the $\beta_{0,\ell}$'s but get to keep all the individual $\alpha_{0,\ell}$'s (supplemented by an infinite sum of $\beta_{0,\ell}$).

\smallskip

It is clear from these properties that the Polyakov-Regge block for s-channel identity is simply the identity:
$P_{2;0,0}^{s|u}=u^{-\Df}$. For identical external operators, the Polyakov-Regge expansion can thus be written as
mean field theory plus non-gaussianity:
\be \label{PR minus MFT}
\cG(z,\zb) = 1+ u^{-\Df} + v^{-\Df} + \sideset{}{'}\sum\limits_{\cO} a_{\cO}
\left[P^{s|u}_{2;\Delta_{\cO},J_{\cO}}(z,\zb)+P^{t|u}_{2;\Delta_{\cO},J_{\cO}}(z,\zb)\right]\,,
\ee
where the sum $\sideset{}{'}\sum$ runs over non-identity primaries.
It is worth mentioning that the bracket is negative definite for all $\tau_{\cO}> 2\Df$ and $z,\zb$ Euclidean, due
to the minus signs in \eqref{eq:kernelsK2}.  In \cite{Rychkov:2016mrc} it was observed numerically
that non-gaussianity is negative-definite in the critical 3d Ising model, in agreement with a theorem relying on the lattice formulation.  Here we expressed non-gaussianity as a sum over mostly negative-definite terms;
it would be interesting to further investigate the overall sign of the sum.

\smallskip

For general correlators,  we can derive, as before, nonperturbative OPE sum rules by replacing $\cG(z,\zb)$ in \eqref{eq:PolyakovReggeSubtracted} by its Euclidean OPE:
\be
\sum\limits_{\cO} a_{\cO} \left[G^s_{\Delta_{\cO},J_{\cO}}(z,\zb) - P^{s|u}_{2;\Delta_{\cO},J_{\cO}}(z,\zb)\right] -1 = \sum\limits_{\cO} a_{\cO}P^{t|u}_{2;\Delta_{\cO},J_{\cO}}(z,\zb)\,,
\label{eq:PRSubtracted}
\ee
which is a rewriting of \eqref{eq:sumRuleOmega2}.
Both sides of this equation have an expansion in $G^s_{2\Df+2n+\ell,\ell}$ with $n\geq 0$ and $\partial_{\Delta}G^s_{2\Df+2n+\ell,\ell}$ with $n\geq 1$. The expansion coefficients are nonperturbative sum rules valid for the OPE in an arbitrary unitary correlator $\langle\phi\phi\phi\phi\rangle$.

\smallskip

The leading coefficient as $z\to 0$ is noting but $B_{2,v}$. The higher coefficients have both $z^n$ and $z^n\log z$ components and (when expanded also around $\zb=0$) generate finite sums of $\alpha_{0,\ell}$, $\alpha_{n,\ell}$ and $\beta_{n,\ell}$'s, supplemented (in a unique way) by infinite sums of $\beta_{0,\ell}$ to make them spin-2 convergent.

\smallskip

As we explain next, there exist other convenient generating functions for the higher-twist $\alpha$'s and $\beta$'s,
which enjoy better and better Regge convergence.

\subsection{Multiple subtractions}

It is easy to further improve convergence in the Regge limit by adding
powers of $\mS$ and $\mT$ to the denominator of Cauchy's integral formula in Mellin space, \eqref{eq:mellinCauchy}.
Given the (naive) vanishing of the Mellin amplitude on double-twist families, we find it
natural to put the zeros on the first few double-twist families, i.e.\ we apply Cauchy to the sequence:
\be
M(\mS,\mT), \quad \frac{M(\mS,\mT)}{(\mS-2\Delta_\phi)(\mT-2\Delta_\phi)},\quad
\frac{M(\mS,\mT)}{(\mS-2\Delta_\phi)(\mS-2\Delta_\phi-2)(\mT-2\Delta_\phi)(\mT-2\Delta_\phi-2)},\ldots
\ee
to obtain dispersion relations with spin-0,2,4\ldots convergence, respectively.
The resulting dispersion relation is then identical to (\ref{eq:mellinDispersion}) but with each residue rescaled
by a factor
\be
\cP^{s|u}_{k;\Delta,J}(\mS,\mT)=
\sum\limits_{m=0}^{\infty}
\frac{(\Df-\tfrac{\mS}{2})_{\frac{k}{2}}(\Df-\tfrac{\mT}{2})_{\frac{k}{2}}}
{(\Df-\tfrac{\tau}{2}-m)_{\frac{k}{2}}(\Df-\tfrac{\mS+\mT-\tau}{2}+m)_{\frac{k}{2}}}
\frac{\mathcal{Q}^{m}_{\Delta,J}(\mT+\mS-\tau-2m)}{\mS-\tau-2m}\,.
\label{eq:PRKMellin}
\ee
This may be interpreted physically as a spin-$k$-subtracted Polyakov-Regge block ($k$ must be even).
Its defining property is that
double-twist $\partial_\Delta G_\Delta$ are absent for the first $k/2$ double-twist trajectories in both the $s$- and $t$-channels, due to the zeros on the numerator.

\smallskip

Quite remarkably, this subtraction leads to natural formulas in position space as well.
Repeating the steps leading to (\ref{eq:KMellin}), we obtain a Mellin representation of
the spin-$k$-convergent kernel, where $k$ is any positive even integer:
\ba
K_k(u,v;u',v') = 
\frac{1}{4u'v'}\iiint\!\frac{d\mS\,d\mT\,d\mS'}{(2\pi i)^3}
&\frac{\Gamma(\Df-\tfrac{\mS}{2})^2\Gamma(\Df-\tfrac{\mT}{2})^2}{2\sin ^2\!\left[\tfrac{\pi}{2}(\mS'-2\Df)\right]\Gamma(\Df-\tfrac{\mS'}{2})^2\Gamma(\Df-\tfrac{\mT'}{2})^2}
\times\\
\times& \frac{1}{\mS'-\mS}
\frac{u^{\tfrac{\mS}{2}-\Df}v^{\tfrac{\mT}{2}-\Df}}{u'^{\tfrac{\mS'}{2}-\Df}v'^{\tfrac{\mT'}{2}-\Df}}
\times \frac{(\Df-\tfrac{\mS}{2})_{\frac{k}{2}}(\Df-\tfrac{\mT}{2})_{\frac{k}{2}}}
{(\Df-\tfrac{\mS'}{2})_{\frac{k}{2}}(\Df-\tfrac{\mT'}{2})_{\frac{k}{2}}}\,.\label{eq:KMellin k2}
\ea
This admits a concise closed form (obtained by the combination of educated guesswork and extensive checks)
\ba
K_{k}(u,v;u',v') &= K_{k;B}(u,v;u',v') \,\theta(\sqrt{v'}>\sqrt{u'}+\sqrt{u}+\sqrt{v}) +\\
&+ \,K_{k;C}(u,v;u') \,\delta(\sqrt{v'}-\sqrt{u'}-\sqrt{u}-\sqrt{v})\,,
\ea
where
\ba
K_{k;B}(u,v;u',v') &=
(-1)^{\frac{k}{2}}(k^2-1)\frac{v\!+\!v'\!-\!u\!-\!u'}{64 \pi (u v u' v')^{\frac{3}{4}}}\left(\frac{uv}{u'v'}\right)^{\frac{k}{4}}
x^{\frac{k+3}{2}}{}_2F_1\!\left(\tfrac{k+1}{2},\tfrac{k+3}{2};2;1-x\right)
\\
K_{k;C}(u,v;u') &=\frac{(-1)^{\frac{k}{2}}}{4 \pi (u v u'v')^{\frac{1}{4}} (\sqrt{u}+\sqrt{u'})}\left(\frac{uv}{u'v'}\right)^{\frac{k}{4}}\,.
\label{eq:kernels k}
\ea
The improved convergence in the Regge limit is due to the extra powers of $1/(u'v')$.
The behavior in the lightcone limit $w\to\infty$ remains the same for all $k\geq 2$.

\smallskip

Analogously to the case $k=2$ discussed in the preceding sections, the correct position-space dispersion relation takes the form
\ba 
\cG(z,\zb)&=1+\cG^s(z,\zb) + \cG^t(z,\zb)\\
\cG^s(z,\zb) &= \iint\! du'dv'K_{k}(u,v;u',v')\dDisc_s[\cG(w,\wb)]\\
\cG^t(z,\zb) &= \iint\! du'dv'K_{k}(v,u;v',u')\dDisc_t[\cG(w,\wb)]\,.
\label{eq:DispersionKSubtracted}
\ea
It can be derived by applying the following identity between functionals
\be\
\Omega^{s|u}_{k;z,\zb} + \Omega^{t|u}_{k;z,\zb} = \mathrm{ev}_{z,\zb}
\ee
to $\cG(w,\wb)-1$, where
\ba
\Omega^{s | u }_{2;z,\zb}[\cF] &=\theta(v-u)\cF(z,\zb)+
\!\iint\limits_{C_-\times C_+}\!\!\frac{dwd\wb}{(2\pi i)^2} \pi^2(\wb-w)K_{2;B}(u,v;u',v')  \cF(w, \wb)\\
\Omega^{t | u }_{2;z,\zb}[\cF] &=\theta(u-v)\cF(z,\zb)
-\!\!\iint\limits_{C_-\times C_+}\!\!\frac{dwd\wb}{(2\pi i)^2}\pi^2(\wb-w)K_{2;B}(u,v;u',v') \cF(w,\wb)\,.
\label{eq:OmegaSTK}
\ea
The Polyakov-Regge sum rules are equivalent to applying the crossing-antisymmetric functional $\Omega_{k;z,\zb}$ to the OPE of crossing-symmetric $\cG(w,\wb)-1$, where
\be
\Omega_{k;z,\zb} = \Omega^{t|u}_{k;z,\zb} - \Omega^{s|u}_{k;1-z,1-\zb}\,.
\ee
The sum rules read
\be
\sum\limits_{\cO} a_{\cO}\,\Omega_{k;z,\zb} [G^s_{\Delta_{\cO},J_{\cO}}] = 1\,.
\label{eq:sumRuleOmegak}
\ee
$\Omega_{k;z,\zb}$ admits an expansion in double-twist blocks $G^s_{2\Df+2n+\ell,\ell}(z,\zb)$ with $n\geq 0$ and their derivatives $\partial_{\Delta}G^s_{2\Df+2n+\ell,\ell}(z,\zb)$ with $n\geq k/2$. We will define the higher-$k$ analogue of $B_{2,v}$ by taking the leading term as $z\rightarrow 0$ 
\be
B_{k,v} = (-1)^{\frac{k}{2}-1}\Omega_{k;0,1-v}\,.
\ee
It follows $B_{k,v}$ takes the form of the integral
\be
B_{k,v}[\mathcal{F}] =
\frac{\Gamma\big(\tfrac{k}{2}\big)^2}{\Gamma(k-1)}\!\iint\limits_{C_-\times C_+}\!\!\frac{dwd\wb}{(2\pi i)^2}\,
\frac{(\wb-w)(v'-u')}{(u'v')^{\frac{k}{2}}[v^2-2(u'+v')v+(u'-v')^2]^{\frac{3-k}{2}}}
\mathcal{F}(w,\wb)\,.
\label{eq:BKVDefinition}
\ee
When $\mathcal{F}(w,\wb)$ is single-valued around the s-channel, e.g. for $\mathcal{F}(w,\wb) = G^s_{\Delta,J}(w,\wb)$, we have the $\dDisc_{s}$-manifesting formula
\be
B_{k,v} [\mathcal{F}] = 
\frac{\Gamma\big(\tfrac{k}{2}\big)^2}{\pi^2\Gamma(k-1)}\int\limits_{v}^{\infty}dv'\!\!\!\!\!\!\!\!\int\limits_{0}^{(\sqrt{v'}-\sqrt{v})^2}\!\!\!\!\!\!\!\!\!du'
\frac{(v'-u')\,\dDisc_{s}[\mathcal{F}(u',v')]}{(u'v')^{\frac{k}{2}}[v^2-2(u'+v')v+(u'-v')^2]^{\frac{3-k}{2}}}\,.
\ee
Finally, the sum rule corresponding to $B_{k,v}$ follows by taking $z\rightarrow 0$ in \eqref{eq:sumRuleOmegak}
\be
\sum\limits_{\cO} a_{\cO}\,B_{k,v} [G^s_{\Delta_{\cO},J_{\cO}}] = (-1)^{\frac{k}{2}-1}\,.
\ee
The sum runs over all primaries in the OPE including identity and the sum rules holds
for all $v\in\mathbb{C}\backslash(-\infty,0]$. We note in passing that the the $\nu_{i,j}$ functionals defined in \eqref{eq:definitionofnuij} can be obtained from the $B_{k,v}$ functionals. Indeed, one can check that $B_{k,v}+(-1)^{\frac{k}{2}}B_{2,v}$ contains only $\beta_{n,\ell}$ in its expansion in $\alpha_{n,\ell}$ and $\beta_{n,\ell}$. The $\nu_{i,j}$ functionals are finite combinations of derivatives of $B_{k,v}+(-1)^{\frac{k}{2}}B_{2,v}$  with respect to $v$ at $v=1$.

\smallskip

A simple physical interpretation of the $B_{k,v}$ sum rule can be given in Mellin space
following \eqref{vanishing id}: they implement the vanishing of the following integral on an arc at infinity
\be \label{vanishing id k}
 0 = \oint\limits_{\mS'\to\infty} \frac{d\mS'}{2\pi i} \frac{M(\mS',\mT')}{\big(\tfrac{2\Df-\mS'}{2}\big)
 \big(\tfrac{2\Df-\mS'}{2}\big)_{k/2}\big(\tfrac{2\Df-\mT'}{2}\big)_{k/2}}\,.
\ee
Similar sum rules are widely used in S-matrix theory to study the implications
of causal and unitary short-distance physics on low-energy effective theories \cite{Adams:2006sv}.
The $B_{k,v}$ sum rules will be analyzed in the context of holographic theories
in a forthcoming publication \cite{CHMRSD}.

% !TEX root = ../main.tex
\newpage
\section{Intermezzo: Regge boundedness versus full crossing symmetry}\label{sec:AppliedSubtractions}

In this section we take stock of the formal developments described so far and tie a few conceptual loose ends.  

\smallskip

 We begin in Section~\ref{ssec:contacts} by showing how our functional formalism leads to a simple way to classify AdS contact diagrams and derive their conformal block expansion. The idea is to define
contact diagrams as functions with vanishing dDisc, bounded by some power in the Regge limit. The prescribed Regge behavior dictates that they must be annhilated by certain finite linear combinations of the double-twist functionals; this is enough to determine them completely.\footnote{Reference \cite{Paulos:2019gtx} used a  different set of functionals, with finite support in twist rather than spin and twist, to arrive at the same conclusions.}
As a byproduct of this analysis, we find that for any $n$ and any even $\ell$ it is possible (generally in more than one way) to construct a ``Regge improved'' version $\beta_{n, \ell}^{\rm imp}$ of the $\beta_{n, \ell}$ functional, which decays with spin $\ell$ and differs from the ordinary $\beta_{n, \ell}$ by the addition of  {odd}-spin basis functionals.

\smallskip

In Section~\ref{sec:lightconerelationship}, we use this new class of functionals to explain the connection between our formalism and the lightcone bootstrap \cite{Fitzpatrick:2012yx, Komargodski:2012ek},
concisely  implemented by  the LIF~\cite{Caron-Huot:2017vep}. When acting on a four-point function of identical scalars $\phi$,
 $\beta_{n, \ell}^{\rm imp}$ yields a convergent sum rule that expresses the anomalous  dimensions $\gamma_{n , \ell}$ of the double-twist family $[\phi \phi]_{n, \ell}$ in terms of the conformal data of all the other operators. This is to be contrasted with applying the LIF block by block in the cross channel: $\gamma_{n , \ell}$ is  expressed as divergent series, asymptotic for large $\ell$.  The two formulas are  related: our dispersive sum rule agrees with the LIF for the contribution to $\gamma_{n, \ell}$ from a cross channel primary of spin $J < \ell$, but the two sums are cut off differently for large $J$. We hope that our formalism will provide the basis for a version of the lightcone bootstrap with rigorous error bars.
 
 \smallskip
 
In  Section~\ref{ssec:symmetricPolyakov} we comment on the fully crossing symmetric Polyakov expansion put forward in \cite{Gopakumar:2016wkt,Gopakumar:2016cpb,Gopakumar:2018xqi}.
Our version of the Polyakov expansion (which we  call the ``Polyakov-Regge expansion'') only maintains crossing symmetry between two channels (s and t), while manifesting good Regge behavior in the third (the u channel).
This asymmetry is very natural from the viewpoint of dispersion relations, and we  have proven the full nonperturbative validity of the Polyakov-Regge expansion. There is currently no complete proposal for 
 an s-t-u symmetric version of the Polyakov expansion, because of intrinsic ambiguities in the higher spin  s-t-u symmetric Polyakov blocks. We will comment on some of the difficulties in deriving such an expansion.

\subsection{AdS contact diagrams and finite sums of $\beta_{n,\ell}$}\label{ssec:contacts}

Our analysis  has relied  crucially on  nonperturbative boundedness of correlators in the Regge limit: in the terminology introduced in Section~\ref{ssec:growth}, a nonperturbative
 correlator is bounded by spin one (it belongs to ${\cal V}_{J=1}$).
Perturbative expansions of CFTs 
generally have worse Regge behavior order by order in the expansion parameter. An important class of examples are holographic CFTs,
understood as  asymptotic expansions around mean field theory in inverse powers of the central charge $c$.  At the leading non-trivial order, $O(1/c)$, 
tree level Witten exchange diagrams in AdS  have Regge behavior controlled by the spin of the exchanged bulk field. To wit, an u-channel exchange of spin $J$ behaves as spin $J$ in the u-channel Regge limit.
Thus the tree-level (u-channel) graviton exchange belongs to ${\cal V}_2$, exhibiting a worse Regge behavior than the true nonperturbative bound. 
What's more, tree level contact diagrams  diverge faster and faster in the Regge limit as one increases the number of derivatives in the quartic vertex.  If the theory has a large gap $\Delta_{\rm gap}$, 
we  should be able write an effective field theory in AdS, with higher derivative vertices weighted by inverse powers of $\Delta_{\rm gap}$. Resumming this derivative expansion is expected to restore good Regge behavior. A case is point is  string theory, where $\Delta_{\rm gap} \sim  M_{\rm string} R_{\rm AdS}$, and the  Regge behavior of tree-level four-point amplitudes is strictly better than spin two.

\smallskip

We will now explain how one can use finite combinations of $\beta_{n,\ell}$ functionals to recover the classification and OPE decomposition of contact Witten diagrams \cite{Heemskerk:2009pn, Penedones:2010ue}.\footnote{See also  e.g.~\cite{Hoffmann:2000mx, Dolan:2000ut} for some early references on the OPE decomposition of Witten diagrams and \cite{Sleight:2017fpc, Sleight:2018epi, Sleight:2018ryu,  Liu:2018jhs, Zhou:2018sfz, Sleight:2019hfp, Sleight:2020obc}  for recent analyses.}
For simplicity, we will focus on the case of identical external scalars. In this case, the contact diagrams are precisely the four-point functions $\cG(z,\zb)$ which are s-t-u symmetric, have vanishing double discontinuity and are bounded by some power in the Regge limit. We take this as the definition of contact diagrams and attempt to classify them. We will not assume that the OPE has finite support in spin -- that will be a result rather than an assumption in the ensuing analysis.

\smallskip

It follows from t-u symmetry and $\dDisc_s[\cG(z,\zb)] = 0$ that the OPE only contains even-spin double-trace blocks and their derivatives
\be
\cG(z,\zb) = \!\! \sum\limits_{\ell=0,2,\ldots}\,\sum\limits_{n=0}^{\infty}
\left[
a_{n,\ell}\,G^s_{\Delta_{n,\ell},\ell}(z,\zb) + 2q^{\textrm{MFT}}_{n,\ell}\gamma_{n,\ell}\,\partial_{\Delta}G^s_{\Delta_{n,\ell},\ell}(z,\zb)
\right]\,,
\ee
where $a_{n,\ell}$ and $\gamma_{n,\ell}$ are the anomalous OPE coefficients and anomalous dimensions, and $q^{\textrm{MFT}}_{n, \ell}$ was defined in (\ref{eq:QMFT}). Let us assume that $\cG(z,\zb)$ is bounded by spin $J$ in the Regge limit (as defined in Section \ref{ssec:growth}). This means that any functional which decays with spin $>J$ will give a valid sum rule satisfied by $a_{n,\ell}$ and $\gamma_{n,\ell}$. In particular, functionals which are finite linear combinations of $\alpha_{n,\ell}$ and $\beta_{n,\ell}$ that decay with spin $>J$ lead to finite linear relations among $a_{n,\ell}$ and $\gamma_{n,\ell}$. Since each individual $\alpha_{n,\ell}$ and $\beta_{n,\ell}$ decays with spin zero, we immediately learn that there are no contact diagrams which are bounded by negative spin in the Regge limit.

\smallskip

Next, consider finite combinations of $\beta_{n,\ell}$ which decay with spin two. The simplest example is \eqref{eq:cHNu01} with decomposition \eqref{eq:cHNu01Decomposition}. This example implies the following relation among anomalous dimensions in every contact diagram Regge-bounded by spin $<2$,
\be
\widehat{\gamma}_{1,0}
-\frac{2(d+2)}{d}\,\widehat{\gamma}_{0,2}
-\frac{\Df ^3 (4 \Df -d)}{2 (2 \Df +1)(4 \Df-d +2)}\,\widehat{\gamma}_{0,0} = 0\,,
\ee
where we defined $\widehat{\gamma}_{n,\ell} = q^{\textrm{MFT}}_{n,\ell}\gamma_{n,\ell}$. Note that $\beta_{n,\ell}$ with odd $\ell$ appearing in the functional drop out from the relation since the OPE only contains even spin. With some more work, one can also check
\be
-\frac{1}{2 \Df +1}\nu_{0,2}+\frac{\Df +1}{2 (2 \Df +1)}\nu_{0,1} = \beta_{0,2} + [\textrm{odd spin}]\,,
\label{eq:beta02functional}
\ee
where here and in the following $[\textrm{odd spin}]$ stands for an arbitrary linear combination of $\beta_{n,\ell}$ with $\ell$ odd. Since \eqref{eq:beta02functional} decays with spin two, we learn that
\be
\widehat{\gamma}_{0,2} = 0\qquad\textrm{and}\qquad
\widehat{\gamma}_{1,0}=\frac{\Df ^3 (4 \Df -d)}{2 (2 \Df +1)(4 \Df-d +2)}\,\widehat{\gamma}_{0,0}
\ee
in all contact diagrams bounded by spin $<2$. By exploring the space of finite combinations of the $\nu_{i,j}$ functionals, we convinced ourselves that their span includes functionals of the form
\be
\beta_{n,0}-\frac{4^{-n}(\Df )^3_n (2 \Df -\tfrac{d}{2})_n}{\Gamma (n+1)^2 (\Df +\tfrac{1}{2})_n(-\tfrac{d}{2}+n+2 \Df)_n}\beta_{0,0} + [\textrm{odd spin}]\quad\textrm{for all }n>0
\ee
and of the form
\be
\beta_{n,\ell}+ [\textrm{odd spin}]\quad\textrm{for all }\ell=2,4,\ldots\,.
\ee
However, it does not include functionals of the form
\be
\beta_{n,0} + [\textrm{odd spin}]\,.
\ee
This implies that the solution space for $\widehat{\gamma}_{n,\ell}$ is one-dimensional, satisfying
\be
\widehat{\gamma}_{n,\ell} = 0\quad\textrm{for all }\ell=2,4,\ldots\;\;\;\textrm{and}\;\;\;
\widehat{\gamma}_{n,0} =
\frac{4^{-n}(\Df )^3_n (2 \Df -\tfrac{d}{2})_n}{\Gamma (n+1)^2 (\Df +\tfrac{1}{2})_n(-\tfrac{d}{2}+n+2 \Df)_n}\,\widehat{\gamma}_{0,0}
\ee
These are precisely the relation satisfied by the $\Phi^4$ contact Witten diagram \cite{Heemskerk:2009pn}.
In other words, we have recovered the well-known fact that this is the only contact diagram bounded by spin $<2$ in the Regge limit. Note that the fact that the anomalous dimensions are supported in $\ell=0$ was a result rather than an assumption in our analysis. This fact is obvious if we work in Mellin space.

\smallskip

Contact diagrams with derivatives are recovered by relaxing the Regge bound. To constrain the anomalous dimensions in contact diagrams Regge-bounded by spin $<J$, we consider linear combinations of $\beta_{n,\ell}$ which decay with spin $J$. The decomposition of all such functionals into $\beta_{n,\ell}$ is a technically challenging exercise which we have not solved in full generality. Nevertheless, by studying numerous examples we convinced ourselves that for all even $J$ and any $n$, we can improve the Regge behavior of $\beta_{n,J}$ by adding a suitable combination of odd spin $\beta$ functionals,
\be
\beta^{\rm imp}_{n,J} = \beta_{n, J}  + [\textrm{odd spin}]=  [\textrm{spin-$J$ decay}]\,, \label{beta plus odd}
\ee
where $[\textrm{spin-$J$ decay}]$ is a finite combination of $\beta_{n,\ell}$ which decays as spin $J$ in the Regge limit. This observation implies the (true) fact that $\widehat{\gamma}_{n,J} = 0$ in all contact diagrams Regge-bounded by spin $<J$. More generally, we expect that for any finite relation
\be
\sum\limits_{n,\ell} b_{n,\ell}\,\widehat{\gamma}_{n,\ell} = 0
\ee
valid in all contact diagrams bounded by spin $<J$, there exists a functional $[\textrm{spin-$J$ decay}]$ such that
\be
\sum\limits_{n,\ell} b_{n,\ell}\,\beta_{n,\ell} = [\textrm{spin-$J$ decay}] + [\textrm{odd spin}]\,.
\ee
It would be interesting to prove this claim explicitly.

\smallskip

Finally, let us comment on functionals whose expansion in $\alpha_{n,\ell}$ and $\beta_{n,\ell}$ is supported in $\ell$ odd. We will call them odd-spin functionals. Odd-spin functionals do not lead to constraints on contact diagrams but \emph{do} lead to non-perturbative sum rules since they are non-vanishing on generic even-spin conformal blocks. They are interesting because they suppress the contribution of all approximate double-traces, allowing to reach into the ultraviolet.
An interesting question is if there exist odd-spin functionals which decay with spin $>1$. The answer is yes, the simplest example which decays with spin 2 being
\be
2 \nu_{1,3}-(\Df+2) \nu_{1,2}
-16 \nu_{0,4}+(11 \Df +30) \nu_{0,3}-4 (\Df +2)^2 \nu_{0,2}+(\Df +2) (\Df +1)^2 \nu_{0,1}\,.
\label{eq:oddSpinFunctional}
\ee
Indeed, one can check that this functional is a linear combination of
\ba
&\beta_{0,5}\\
&\beta_{0,3}\quad \beta_{1,3}\\
&\beta_{0,1}\quad \beta_{1,1}\quad \beta_{2,1}\,.
\ea
We have also found examples which decay with spin 4 and spin 6 and believe there are infinitely many odd-spin functionals which decay with arbitrarily large fixed spin. It will be important to understand the space and consequences of odd-spin functionals systematically. We conclude with an interesting vanishing property of odd-spin functionals. Suppose $\omega$ is an odd-spin functional which decays with spin $J$, where $J$ is even. Then
\be
\label{eq:vanishingproperty}
\omega[G^s_{\Delta,J'}] = 0 \quad\textrm{for all even }J'<J \qquad\mbox{($\omega$ odd-spin)}\,,
\ee
where $\Delta$ is any dimension allowed by unitarity bounds. For example \eqref{eq:oddSpinFunctional} identically vanishes on scalar conformal blocks. The proof is as follows. Consider a fully crossing-symmetric sum of exchange Witten diagrams
\be
\cG(z,\zb) = W^s_{\Delta,J'}(z,\zb) + W^t_{\Delta,J'}(z,\zb) + W^u_{\Delta,J'}(z,\zb)\,.
\ee
$\cG(z,\zb)$ is bounded by spin $J'$ in the Regge limit. Its s-channel OPE contains only the single-trace block $G^s_{\Delta,J'}$ together with even-spin double-trace contributions. Since $\omega$ decays with $J>J'$, it leads to a valid sum rule for $\cG(z,\zb)$. Since it is an odd-spin functional, all the double-trace contributions drop out and we find $\omega[G^s_{\Delta,J'}]=0$.

\subsection{Relationship with the lightcone bootstrap}
\label{sec:lightconerelationship}
The logic of the previous subsection serves as a good preparation for understanding the relationship between dispersive sum rules and the lightcone bootstrap \cite{Komargodski:2012ek,Fitzpatrick:2012yx}. Again, let us focus on the case of identical scalars $\phi$ in a nonperturbative CFT. The lightcone bootstrap allows us to estimate anomalous dimensions and OPEs of the double-twist operators $[\phi\phi]_{n,\ell}$ in terms of the data of primaries exchanged in the $\phi\times\phi$ OPE. Let us focus on the anomalous dimensions
\be
\gamma_{n,\ell} = \Delta_{[\phi\phi]_{n,\ell}}-2\Df-\ell\,.
\ee
The lightcone bootstrap can be implemented using the Lorentzian inversion formula (LIF) \cite{Caron-Huot:2017vep} as follows. We consider $[\phi\phi]_{n,\ell}$ exchanged in the s-channel. $\Delta_{[\phi\phi]_{n,\ell}}$ is found as the location of a simple pole of $c^s(\Delta,\ell)$, which is in turn fixed by the LIF in terms of $\dDisc_{t}\cG$ and $\dDisc_{u}\cG$ (the latter two giving the same contribution in the case of identical scalars). Now, if we apply the LIF block-by-block to the t/u-channel OPE, we find that each primary contributes a double pole to $c^s(\Delta,\ell)$ at $\tau=\Delta-\ell=2\Df+2n$. The double pole can be interpreted as a contribution to the anomalous dimension because
\be
\frac{1}{\tau-2\Df-2n-\gamma_{n,\ell}} = 
\frac{1}{\tau-2\Df-2n}+
\frac{\gamma_{n,\ell}}{(\tau-2\Df-2n)^2} + [\text{higher-order poles}]\,.
\label{eq:simpledoublepole}
\ee
Summing the contributions of all primaries to $\gamma_{n,\ell}$ then leads to the formula
\be
\gamma_{n,\ell} \,\stackrel{\text{asymp.}}{=}\, \sum\limits_{\cO} a_{\cO}\,\gamma_{n,\ell}(\Delta_{\cO},J_{\cO})\,,
\label{eq:gammaFromLIF}
\ee
where $\gamma_{n,\ell}(\Delta_{\cO},J_{\cO})$ is the function giving the contribution of primary $\cO$ through the LIF. However, \eqref{eq:gammaFromLIF} is \emph{not} an exact sum rule for $\gamma_{n,\ell}$. In fact, the sum over $\cO$ on the RHS does not even converge. The correct interpretation of \eqref{eq:gammaFromLIF} is that it gives the asymptotic expansion of $\gamma_{n,\ell}$ as $\ell\rightarrow\infty$. Why does \eqref{eq:gammaFromLIF} fail to be a nonperturbative sum rule? The reason is that the procedure of extracting the anomalous dimension from the LIF does not commute with expanding $\dDisc_{t,u}\cG$ in conformal blocks. Indeed, individual blocks only lead to double poles of $c^s(\Delta,\ell)$ at $\Delta=2\Df+2n+\ell$, whereas their total sum leads to a shifted simple pole at $\Delta = 2\Df +2n+ \ell + \gamma_{n,\ell}$. As explained in \cite{Simmons-Duffin:2016wlq}, this means that the OPE must contain infinite towers of multitwist operators which give rise to the higher-order poles in \eqref{eq:simpledoublepole}. It is precisely the tail of the infinite sum over the spin of these multitwist towers which makes the RHS of \eqref{eq:gammaFromLIF} diverge.

\smallskip

Let us explain how these problems are avoided using the dispersive sum rules. In order to study the anomalous dimension $\gamma_{n,\ell}$, we will use a functional $\beta^{\text{imp}}_{n,\ell}$, which is $\beta_{n,\ell}$ improved by an odd-spin functional so that it decays with spin $\ell$ in the u-channel Regge limit,
\be
\beta^{\text{imp}}_{n,\ell} = \beta_{n,\ell} + [\textrm{odd spin}] = [\text{spin-$\ell$ decay}]\,.
\label{eq:betaIMP}
\ee
As explained in the previous subsection, we expect that such an improvement  always exists, but is not unique. See \eqref{eq:beta02functional} for an example of $\beta^{\text{imp}}_{0,2}$. The ambiguity of $\beta^{\text{imp}}_{n,\ell}$ is parametrized by all odd-spin functionals that decay with spin $\ell$. When we apply $\beta^{\text{imp}}_{n,\ell}$ to the $\langle\phi\phi\phi\phi\rangle$ crossing equation, we get the nonperturbative sum rule
\be
a_{[\phi\phi]_{n,\ell}}\,\beta^{\text{imp}}_{n,\ell} [G^s_{[\phi\phi]_{n,\ell}}] = -\!\!\sum\limits_{\cO\neq [\phi\phi]_{n,\ell}} a_{\cO}\, \beta^{\text{imp}}_{n,\ell} [G^s_{\cO}]\,,
\label{eq:gammaFromFunctionalsExact}
\ee
where we separated the contribution of the double-twist operator $[\phi\phi]_{n,\ell}$ from the rest. Since $\beta^{\text{imp}}_{n,\ell}$ is a swappable functional, the sum on the RHS of \eqref{eq:gammaFromFunctionalsExact} converges. We also know from \eqref{eq:betaIMP} that
\be
\beta^{\text{imp}}_{n,\ell} [G^s_{[\phi\phi]_{n,\ell}}  ] = \gamma_{n,\ell} + O(\gamma_{n,\ell}^2)\,.
\ee
We can thus think of \eqref{eq:gammaFromFunctionalsExact} as an implicit equation determining $\gamma_{n,\ell}$ from the rest of the CFT data. We can play the same game with the $\alpha_{n,\ell}$ functional, which combined with \eqref{eq:gammaFromFunctionalsExact} gives a coupled pair of nonlinear equations for $a_{[\phi\phi]_{n,\ell}}$ and $\gamma_{n,\ell}$ valid nonperturbatively.

\smallskip

When we are close to mean field theory, such as for $\ell\gg 1$ or in the presence of a small coupling, we have
\be
\left|\frac{a_{[\phi\phi]_{n,\ell}}}{2 q^{\text{MFT}}_{n,\ell}}-1\right| \ll 1\quad\text{and}\quad |\gamma_{n,\ell}|\ll 1\,.
\ee
Equation \eqref{eq:gammaFromFunctionalsExact} can then be solved perturbatively, and gives at the leading order
\be
\gamma_{n,\ell} =  - \sum\limits_{\cO\neq [\phi\phi]_{n,\ell}} a_{\cO}\, \frac{\beta^{\text{imp}}_{n,\ell} [G^s_{\Delta_{\cO},J_{\cO}}]}{2 q^{\text{MFT}}_{n,\ell}} + O(\gamma^2_{n,\ell})\,.
\label{eq:gammaFromFunctionals}
\ee
Let us compare this equation for $\gamma_{n,\ell}$ with the result of the LIF \eqref{eq:gammaFromLIF}. One obvious difference is that the sum over $\cO$ in \eqref{eq:gammaFromLIF} is divergent whereas the sum in \eqref{eq:gammaFromFunctionals} converges. This is because the large-spin tail of the sum is cut off differently in the two cases, as we explain below. This feature makes \eqref{eq:gammaFromFunctionalsExact}, \eqref{eq:gammaFromFunctionals} more suitable as a basis for rigorous estimates of $\gamma_{n,\ell}$ with controlled errors, and for carrying out the analytic bootstrap to higher orders.

\smallskip

On the other hand, the two formulas are closely related. In fact, the dispersive sum rule \eqref{eq:gammaFromFunctionals} agrees with the LIF about the contribution of primaries $\cO$ with small spin, namely for $J_{\cO}<\ell$. In other words, we have
\be
\beta^{\text{imp}}_{n,\ell} [G^s_{\Delta_{\cO},J_{\cO}}] = - 2 q^{\text{MFT}}_{n,\ell}\gamma_{n,\ell}(\Delta_{\cO},J_{\cO})\qquad\text{for all }\Delta_{\cO}\text{ and all }\ell>J_{\cO}\,.
\label{eq:betagammaArgree}
\ee
This is the best we could hope for because as already stated $\beta^{\text{imp}}_{n,\ell}$ has an additive ambiguity in the form of odd-spin functionals with spin-$\ell$ decay. Such modifications will in general change the action of $\beta^{\text{imp}}_{n,\ell}$ on $G^s_{\Delta,J}$ with $J\geq\ell$. On the other hand, as proved at the end of the previous subsection, the action of $\beta^{\text{imp}}_{n,\ell}$ on $G^s_{\Delta,J}$ with $J<\ell$ is unambiguous. Another obvious reason why \eqref{eq:gammaFromLIF} disagrees with \eqref{eq:gammaFromFunctionals} for $J_{\cO}\geq\ell$ is that the $J_{\cO}=\ell$ double trace operator $\cO=[\phi\phi]_{n,\ell}$ itself appears on the RHS of the former but not of the latter.

\smallskip

To see that \eqref{eq:betagammaArgree} holds, we will relate both sides to the OPE of exchange Witten diagrams. Consider the s-t-u symmetric correlator
\be
\cG(z,\zb) = W^s_{\Delta,J}(z,\zb) + W^t_{\Delta,J}(z,\zb) + W^u_{\Delta,J}(z,\zb)\,,
\ee
where $W^{s,t,u}_{\Delta,J}$ are exchange Witten diagrams (with an arbitrary choice of three-point couplings). $\cG(z,\zb)$ is bounded by spin $J$ in the Regge limit. Its OPE takes the form
\be
\cG =G^s_{\Delta,J}+ \sum\limits_{n=0}^{\infty}\sum\limits_{\ell=0,2,\ldots}
\left[q^{(1)}_{n,\ell}(\Delta,J)\,G^s_{2\Df+2n+\ell,\ell} + q^{(2)}_{n,\ell}(\Delta,J) \,\partial_{\Delta}G^s_{2\Df+2n+\ell,\ell}\right]\,.
\ee
Since $\beta^{\text{imp}}_{n,\ell}$ decays with spin $\ell>J$, we can apply it to the s-t crossing equation of $\cG$, which leads to the sum rule
\ba
&\beta^{\text{imp}}_{n,\ell}[G^s_{\Delta,J}] = \\&- \sum\limits_{n'=0}^{\infty}\sum\limits_{\ell'=0,2,\ldots}\!\!\!\!
\left[q^{(1)}_{n',\ell'}(\Delta,J)\,\beta^{\text{imp}}_{n,\ell}[G^s_{2\Df+2n'+\ell',\ell'}] + q^{(2)}_{n',\ell'}(\Delta,J) \,\beta^{\text{imp}}_{n,\ell}[\partial_{\Delta}G^s_{2\Df+2n'+\ell',\ell'}]\right]\,.
\ea
By definition of $\beta^{\text{imp}}_{n,\ell}$ in \eqref{eq:betaIMP}, only the second term in the square bracket with $n'=n$, $\ell'=\ell$ survives, and we get
\be
\beta^{\text{imp}}_{n,\ell}[G^s_{\Delta,J}] = -q^{(2)}_{n,\ell}(\Delta,J)\quad\text{for }\ell>J\,.
\label{eq:betaImpAction}
\ee
Note that $q^{(2)}_{n,\ell}(\Delta,J)$ for $\ell>J$ are independent of the contact ambiguity in $\cG$.

\smallskip

In order to relate $q^{(2)}_{n,\ell}(\Delta,J)$ to the function $\gamma_{n,\ell}(\Delta,J)$ appearing in \eqref{eq:gammaFromLIF}, we apply the LIF to $\cG$.\footnote{This argument appeared in \cite{Liu:2018jhs}, see also~\cite{Sleight:2018ryu}.} We have
\be
\dDisc_{t}[\cG] = \dDisc_{t}[G^t_{\Delta,J}]\quad\text{and}\quad \dDisc_{u}[\cG] = \dDisc_{u}[G^u_{\Delta,J}]\,.
\ee
Thus we are effectively inverting a single conformal block. Since $\cG$ is bounded by spin $J$, we can trust the LIF for $c^s(\Delta',\ell)$ in the range $\ell>J$. It then follows by definition of $\gamma_{n,\ell}(\Delta,J)$ that
\be
q^{(2)}_{n,\ell}(\Delta,J) = 2\,q^{\text{MFT}}_{n,\ell} \gamma_{n,\ell}(\Delta,J)\quad\text{for }\ell>J\,
\ee
which concludes the proof of \eqref{eq:betagammaArgree}.

\smallskip

Note that for \eqref{eq:betaImpAction} to hold, it is crucial that $[\text{odd spin}]$ on the RHS of \eqref{eq:betaIMP} is nonzero. This is because $\beta_{n,\ell}[G^s_{\Delta,J}]$ computes minus the coefficient of $\partial_{\Delta}G^s_{2\Df+2n+\ell,\ell}$ in $P^{s|u}_{\Delta,J}+P^{t|u}_{\Delta,J}$, which differs from $W^{s}_{\Delta,J}+W^{t}_{\Delta,J}+W^u_{\Delta,J}$ at the very least by the absence of $W^u_{\Delta,J}$. Of course, we already know that $[\text{odd spin}]$ in \eqref{eq:betaIMP} is nonzero because $\beta_{n,\ell}$ only decays with spin $0$ whereas $\beta^{\text{imp}}_{n,\ell}$ decays with spin $\ell > 0$.

\smallskip

The summary of this subsection is as follows. There exist dispersive sum rules \eqref{eq:gammaFromFunctionalsExact} which provide a rigorous, convergent alternative to the lightcone bootstrap \eqref{eq:gammaFromLIF}. The ``contribution of a conformal block of spin $J_{\cO}$ to the anomalous dimension of a spin-$\ell$ double-twist operator'' is only well-defined if $\ell>J_{\cO}$, where it agrees with the Lorentzian inverse of a single block. In particular, \eqref{eq:gammaFromFunctionalsExact} correctly reproduces the asymptotic large-$\ell$ tails in $\gamma_{n,\ell}$ coming from individual blocks. The contributions to \eqref{eq:gammaFromFunctionalsExact} with $J_{\cO}\geq \ell$ differ from the Lorentzian inverse of a single block. This is what cuts off the large-$J_{\cO}$ tails in \eqref{eq:gammaFromFunctionalsExact} differently from \eqref{eq:gammaFromLIF}, ultimately leading to a convergent formula. The contribution of all double-twist operators to the dispersive sum rule for $\gamma_{n,\ell}$ is suppressed thanks to the double zeros of $\beta^{\text{imp}}_{n,\ell}[G^s_{\Delta_{\cO},J_{\cO}}]$. The ambiguity of the $J_{\cO}\geq\ell$ contributions is parametrized by odd-spin functionals. These lead to sum rules having double zeros on all double-trace operators. Further examples of $\beta^{\text{imp}}_{0,\ell}$ will be discussed in Section \ref{ssec:fixed spin}.

\subsection{Comments on the s-t-u symmetric Polyakov bootstrap}

\label{ssec:symmetricPolyakov}

A key feature of the Polyakov-Regge expansion is that it only maintains crossing symmetry between two channels (s and t), while treating the third channel asymmetrically. Physical primaries in the s- and t-channel are present term-by-term in the expansion, while physical primaries in the u-channel only arise from the infinite sum over Polyakov-Regge blocks. In other words, taking the s- and t-channel OPE limits commutes with the sum over Polyakov-Regge blocks, while taking the u-channel OPE limit does not. The latter fact prefectly agrees with expectation from the u-channel Lorentzian inversion formula: Inverting individual s- and t-channel conformal blocks only produces poles in $c^u(\Delta,J)$ at double-trace locations. Poles at locations of actual u-channel primaries arise by performing the sum over s- and t-channel blocks first, and inverting afterwards.

\smallskip

In that light, it is interesting to examine the proposal for an s-t-u symmetric Polyakov expansion, put forward in \cite{Gopakumar:2016wkt,Gopakumar:2016cpb,Gopakumar:2018xqi}. These references postulated the existence of s-t-u symmetric Polyakov blocks $P^{\text{sym}}_{\Delta,J}(z,\zb)$. The main defining property of these hypothetical objects is that any crossing-symmetric four-point function can be expaded in them as follows
\be
\cG(z,\zb) = \sum\limits_{\cO} a_{\cO}\,G^{s}_{\Delta_{\cO},J_{\cO}}(z,\zb) =
\sum\limits_{\cO} a_{\cO}\,P^{\text{sym}}_{\Delta_{\cO},J_{\cO}}(z,\zb)\,.
\label{eq:PolyakovSym}
\ee
Furthermore, $P^{\text{sym}}_{\Delta,J}(z,\zb)$ were proposed to be a sum of exchange Witten diagrams in the three channels
\be
P^{\text{sym}}_{\Delta,J}(z,\zb) = W^s_{\Delta,J}(z,\zb) + W^t_{\Delta,J}(z,\zb) + W^u_{\Delta,J}(z,\zb)\,,
\ee
where $W^s_{\Delta,J}(z,\zb)$ is normalized so that the single-trace block $G^s_{\Delta,J}$ appears with a unit coefficient
\be
P^{\text{sym}}_{\Delta,J} = G^s_{\Delta,J}+ \sum\limits_{n=0}^{\infty}\sum\limits_{\ell=0,2,\ldots}
\left[q^{(1)}_{n,\ell}(\Delta,J)\,G^s_{2\Df+2n+\ell,\ell} + q^{(2)}_{n,\ell}(\Delta,J) \,\partial_{\Delta}G^s_{2\Df+2n+\ell,\ell}\right]\,.
\ee
The validity of \eqref{eq:PolyakovSym} would then be equivalent to the cancellation of the double-traces, which amounts to the sum rules
\be
\sum\limits_{\cO} a_{\cO}\,q^{(1,2)}_{n,\ell}(\Delta_{\cO},J_{\cO}) = 0\quad\text{for }n\in\mathbb{N},\,\ell\in2\mathbb{N}\,.
\label{eq:stuSumRules}
\ee
$W^{s,t,u}_{\Delta,J}(z,\zb)$ carry an ambiguity which leads to a contact-term ambiguity of $P^{\text{sym}}_{\Delta,J}(z,\zb)$ supported in spins $0,2,\ldots,J-2$. However, a choice of this ambiguity which makes \eqref{eq:PolyakovSym} hold has not been found and thus \eqref{eq:PolyakovSym} has remained speculative as a nonperturbative statement about CFT correlators. On the other hand, the proposal has been successfully used to compute some double-trace data of the $\langle\phi\phi\phi\phi\rangle$ correlator in the Wilson-Fischer fixed point up to $O(\epsilon^3)$. This was possible because up to this order the only relevant Polyakov block exchanged on the RHS of \eqref{eq:PolyakovSym} is that of the $\phi^2$ operator, which is a scalar and thus carries no contact ambiguity.

\smallskip

We are thus faced with the question whether \eqref{eq:PolyakovSym} holds nonperturbatively for a suitable choice of $P^{\text{sym}}_{\Delta,J}(z,\zb)$. Let us explain why one can \emph{not} arrive at \eqref{eq:PolyakovSym} by symmetrizing the Polyakov-Regge expansion over the three channels. Assuming $\cG(z,\zb)$ is s-t-u symmetric and superbounded, we have the rigorous expansions
\ba
\cG(z,\zb) &= \sum\limits_{\cO} a_{\cO} \left[P^{s|u}_{\Delta_{\cO},J_{\cO}}(z,\zb) + P^{t|u}_{\Delta_{\cO},J_{\cO}}(z,\zb)\right] = \\
&= \sum\limits_{\cO} a_{\cO} \left[P^{t|s}_{\Delta_{\cO},J_{\cO}}(z,\zb) + P^{u|s}_{\Delta_{\cO},J_{\cO}}(z,\zb)\right] = \\
&= \sum\limits_{\cO} a_{\cO} \left[P^{u|t}_{\Delta_{\cO},J_{\cO}}(z,\zb) + P^{s|t}_{\Delta_{\cO},J_{\cO}}(z,\zb)\right]\,.
\label{eq:PRSTU}
\ea
We can then fix the contact ambiguity by setting $W^s_{\Delta,J}$ to be the average between the two distinct s-channel Polyakov-Regge blocks
\be
W^s_{\Delta,J} = \frac{P^{s|u}_{\Delta,J}+P^{s|t}_{\Delta,J}}{2}\,,\qquad
W^t_{\Delta,J} = \frac{P^{t|s}_{\Delta,J}+P^{t|u}_{\Delta,J}}{2}\,,\qquad
W^u_{\Delta,J} = \frac{P^{u|t}_{\Delta,J}+P^{u|s}_{\Delta,J}}{2}\,,
\ee
where the factor of a half is fixed by requiring that $W^s_{\Delta,J}$ contains $G^s_{\Delta,J}$ with unit coefficient. This definition ensures that $W^{s,t,u}_{\Delta,J}$ are exchange diagrams in their respective channels, supplemented by a contact term. However, when we now take the average of the three equations \eqref{eq:PRSTU}, we find
\be
\sum\limits_{\cO}\left[W^s_{\Delta,J}(z,\zb)+W^t_{\Delta,J}(z,\zb)+W^u_{\Delta,J}(z,\zb)\right] = \frac{3}{2}\cG(z,\zb)\,,
\ee
which is in contradiction with \eqref{eq:PolyakovSym}. The additional $\cG(z,\zb)/2$ on the RHS comes from the primaries which do not appear in the individual terms but are correctly reconstructed by the infinite sums in \eqref{eq:PRSTU} (u-channel primaries on the first line of \eqref{eq:PRSTU}). Since this is a general feature of the Polyakov-Regge expansion, there is no simple way to fix this problem and recover \eqref{eq:PolyakovSym}. This means that if \eqref{eq:PolyakovSym} can be given nonperturbative meaning, its conceptual starting point should be different from the dispersion relations used in the present work.

\smallskip

We would also like to stress that the breaking of the t-u symmetry by the fixed-u Polyakov-Regge expansion is an important feature, rather than a bug. In particular, it implies that even in the case of identical external operators both \emph{even-} and \emph{odd}-spin double-traces appear in the Polyakov-Regge blocks. As explained in Section \ref{ssec:stuSymmetry}, the cancellation of odd-spin double-traces leads to infinitely-many nontrivial sum-rules and as explained in the previous two subsections, these ``odd-spin'' sum rules exist also in the subtracted case. It appears challenging to derive such sum rules from the s-t-u symmetric Polyakov expansion, where we only get \eqref{eq:stuSumRules} for even $\ell$. One possibility compatible with everything said so far is that \eqref{eq:stuSumRules} hold nonperturbatively for some choice of the contact ambiguity, but simply do not give a complete set of sum rules (since they miss the odd-spin sum rules). It is worth pointing out that the s-t-u symmetric Polyakov bootstrap has been rigorous derived in the context of 1D CFTs using the analytic functionals point of view in \cite{Mazac:2018ycv}, which served as another inspiration for the present work. In 1D, there is no spin and therefore no mismatch due to the odd-spin sum rules, which is what allows us to seamlessly connect the functional basis with the s-t-u symmetric Polyakov sum rules in 1D.

\bigskip
\noindent
\textbf{Note added:} Since the original publication of this work, these issues have been clarified in references \cite{Sinha:2020win,Gopakumar:2021dvg}. Their starting point is a dispersion relation \cite{Auberson:1973pf} which manifests the full s-t-u symmetry. Expanding the dispersion using the OPE leads to an expansion of the correlator in terms of crossing-symmetric objects. These objects only differ from a crossing-symmetric sum of exchange Witten diagrams by the presence of certain unphysical singularities in Mellin space. Demanding that these singularities are absent gives rise to the odd-spin sum rules. Demanding that the resulting expansion agrees with the usual OPE gives rise to the usual `Polyakov' sum rules associated with even-spin double traces.

% !TEX root = ../main.tex

\section{Extremal functionals for mean field theory}\label{sec:MFTbounds}

In this section we apply our functionals to the spin-$\ell$ \emph{gap-maximization problem}:
for a fixed external scaling dimension $\Df$, find a unitary solution to crossing
where the lightest spin-$\ell$ operator has the largest possible scaling dimension.
Throughout, we assume that $\cG$ is an s-t-u symmetric correlator.

\smallskip

The maximal value for the twist gap is easily found from
large-spin perturbation theory: since infinitely many operators must have twists accumulating to $2\Df$,
by negativity of their anomalous dimensions, the twist gap cannot exceed $2\Df$ \cite{Komargodski:2012ek,Fitzpatrick:2012yx}.
This value is saturated by mean field theory.
It is natural to expect that, if a twist gap of $2\Df$
is imposed for any particular spin $\ell\geq 2$, the only solution to crossing is the mean field theory correlator.

\smallskip

This can be argued directly using the Lorentzian inversion formula. The LIF
predicts the existence of double-twist operators of twist $\tau_{0,\ell}=2\Df+\gamma_{0,\ell}$ and $\ell\geq 2$.
Furthermore, the contribution of each crossed-channel primary to anomalous dimensions
$\gamma_{0,\ell}$ is nonpositive, so $\gamma_{0,\ell}$ is nonpositive.
It can only vanish if dDisc of the correlator is saturated by identity exchange, meaning that the spectrum contains only identity and double-twists. The only such unitary correlator is mean field theory.
A drawback of this argument is that it does not lead to a rigorous sum rule for $\gamma_{0,\ell}$ with controlled errors
when the theory is not mean field theory.
In fact, simply summing over all crossed-channel contributions to $\gamma_{0,\ell}$ predicted by the LIF gives infinity \cite{Alday:2016mxe,Simmons-Duffin:2016wlq,Caron-Huot:2017vep}, see also Section \ref{sec:lightconerelationship}.

\smallskip

For this reason, it would be preferable to prove the result using an extremal functional. By the virtue of being a nonperturbative sum rule, such a functional would provide rigorous estimates for $\gamma_{0,J}$ with controlled errors.
If such a functional exists, it must have the following properties
\begin{enumerate}
\item It is nonnegative on all conformal blocks with spin $\ell$ and $\tau\geq 2\Df$, and on all conformal blocks with spin $\neq \ell$ and $\tau$ allowed by unitarity bounds.
\item It has double zeros on all double-twist conformal blocks $G_{2\Df+2n+J,J}$ with
the exception of $n=0$, $J=\ell$, where it has a simple zero.
\end{enumerate}
We expect that such functional exists for general $d$, $\Df$ and $\ell\geq 2$. Numerical evidence for this claim in $d=3$ appears in Section 5.4 of \cite{ElShowk:2012ht}. In the language of our Section \ref{sec:lightconerelationship}, this functional is precisely $\beta^{\text{imp}}_{0,\ell}$ of equation \eqref{eq:betaIMP}, provided we can choose the odd-spin ambiguity to satisfy Property 1 above.

\smallskip

The playground in this section will be the $\widetilde{B}_{2,v}$ family
of sum rules introduced in \eqref{B2 tilde}:
\be
 \sideset{}{'}\sum\limits_{\cO} a_{\cO} \widetilde{B}_{2,v}[G^s_{\Delta_\cO,J_\cO}] = 0\,,
 \label{B2 tilde sum rule}
\ee
where the prime indicates that the sum runs over non-identity operators.
We explain in Section \ref{ssec:Btilde21} that the first derivative of $\widetilde{B}_{2,v}$ with respect to $v$ at $v=1$, denoted $\widetilde{B}'_{2,1}$ satisfies a weaker version of Properties 1 and 2: For any $d$ and $\Df$,
this sum rule enjoys double zeros on all double-twists, except for simple zeros on all $n=0$ operators with $\ell\geq 2$. Furthermore, we will show numerically that $\widetilde{B}'_{2,1}$ is nonnegative above $\tau=2\Df$ for $\ell\geq 2$ and above the unitarity bound for $\ell=0$.
This suffices to show that the only correlator with no spinning operator (of \emph{any} spin) below twist gap $2\Df$ is mean-field-theory.
This is refined in subsection \ref{ssec:fixed spin}, where we construct combinations of
$\widetilde{B}_{2,v}$'s, called  $\Phi_\ell$,  which have a unique single zero, at spin-$\ell$. We show
numerically that $\Phi_2$ satisfies Property 1 for a range of $\Df$.

\smallskip

Since the results of this section may be of broader interest, we attempt to make it self-contained
and summarize all definitions in \ref{ssec:recap}. The section concludes with a discussion of possible application to the numerical bootstrap.

\subsection{Self-contained recap and definitions of sum rules}\label{ssec:recap}

The $B_{2,v}$ family of sum rules, for any real $v>0$, can be defined by the following integral:
\be
B_{2,v}[\mathcal{F}] =
\!\iint\limits_{C_-\times C_+}\!\!\frac{dwd\wb}{(2\pi i)^2}\,
\frac{(\wb-w)(v'-u')}{u'v'\sqrt{v^2-2(u'+v')v+(u'-v')^2}}
\mathcal{F}(w,\wb)\,.
\label{eq:B2VDefinition a}
\ee
where the contour is depicted in Figure \ref{fig:Contours}, and $u'=w \wb$, $v'=(1-w)(1-\wb)$.
It is manifest by symmetry that the integral vanishes if $\mathcal{F}$ is any s $\leftrightarrow$ t
crossing-symmetric function.
Specifically, the integrand is odd under the transformation
$w \leftrightarrow 1-\wb$, which preserves the contour.
Another important feature is spin-2 decay in the Regge limit $w,\wb\to \infty$,
where the integral scales as $\frac{d^2w}{w^3}$.
This ensures the the integral actually converges for any bounded $\mathcal{F}$ (see Section \ref{ssec:growth});
recall that correlators in any unitary CFT satisfy this condition.
To avoid a logarithmic divergence in the collinear limit $w\to \infty$, however, we must subtract the u-channel identity.
Our sum rules thus originate from the following identity, valid for any physical crossing-symmetric correlator:
\be
  0 = B_{2,v}[\cG-1]\,. \label{B2v starting point}
\ee
$\cG-1$ can be expanded using the s-channel conformal blocks. The sum over said blocks can be swapped with the action of $B_{2,v}$. The tail of the sum over conformal blocks does not spoil swappability since the integral in \eqref{eq:B2VDefinition a} converges for any bounded $\mathcal{F}$.

Let us start by studying the expansion around $v=1$:
\be
 B_{2,v} = B_{2,1} + (v-1) B_{2,1}' + \ldots
\ee
One may obtain corresponding sum rules by expanding \eqref{eq:B2VDefinition a}
under the integration sign:
\ba \label{around 1}
B_{2,1}[\mathcal{F}] &=
\!\iint\limits_{C_-\times C_+}\!\!\frac{dwd\wb}{(2\pi i)^2}\,
\frac{1-w-\wb}{w\wb(1-w)(1-\wb)}\mathcal{F}(w,\wb)\,, \\
B_{2,1}'[\mathcal{F}] &=
\!\iint\limits_{C_-\times C_+}\!\!\frac{dwd\wb}{(2\pi i)^2}\,
\frac{1-w-\wb}{w\wb(1-w)(1-\wb)}\frac{2w\wb-w-\wb}{(\wb-w)^2}\mathcal{F}(w,\wb)\,,
\ea
and so on.
Again it is easy to see directly that these converge and annihilate $\cG-1$ by symmetry;
it is also straightforward to write down generalizations.\footnote{
For example, the sum rule with kernel $\frac{dwd\wb}{(2\pi i)^2(\wb-w)^{2q}}\,
\left(\frac{1}{u'^a v'^b}-\frac{1}{u'^b v'^a}\right)$ converges in the Regge limit provided  $2q+2a+2b>1$
and is collinear-safe (identity does not need to be subtracted) if $2q+a+b>1$.
}

\smallskip

A crucial fact, not evident from the above definition, is
that all sum rules stemming from $B_{2,v}$ enjoy double-zeros on $n>0$ double-twists.
This is shown by deforming the contour to a new contour $C_-\times C_-$ where both variables wrap the left-cut.
In principle this produces multiple combinations of principal values and discontinuities of the kernel and correlator
(see \eqref{Disc and P}),
however, thanks to special properties of the square root and overall antisymmetry in $\wb\leftrightarrow w$,
the result involves a clean dDisc:
\be
B_{2,v} [G^s_{\Delta_\cO,J_\cO}]= 
 \int\limits_{v}^{\infty}dv'\!\!\!\!\!\!\int\limits_{0}^{(\sqrt{v'}-\sqrt{v})^2}\!\!\!\!\!\!\!du'
\frac{v'-u'}{\pi^2 u' v' \sqrt{v^2-2(u'+v')v+(u'-v')^2}}\,
\dDisc_{s}[G^s_{\Delta_\cO,J_\cO}]\,. \label{eq:B2v dDisc ch5}
\ee
The dDisc operation produces the desired double zeros, see \eqref{dDisc block}.

\smallskip

In this form, we can now can separate the u-channel identity and
substitute in the s-channel OPE for $\cG$, integrating term by term.
As shown shortly, the u- and s- channel identities integrate
to $1$ and $-v^{-\Df}$ respectively on this contour, hence the identity \eqref{B2v starting point} may be written as:
\be
\sideset{}{'}\sum\limits_{\cO} a_{\cO} B_{2,v}[G^s_{\Delta_\cO,J_\cO}] = 1+v^{-\Df}\ . \label{B2v primmed sum}
\ee
We have only used s $\leftrightarrow$ t symmetry so far.
The idea of the $\widetilde{B}_{2,v}$ sum rules is to further use t $\leftrightarrow$ u symmetry,
by antisymmetrizing:
\be
 \widetilde{B}_{2,v} \equiv v^{\frac{\Df}{2}} B_{2,v} - v^{-\frac{\Df}{2}} B_{2,1/v}\ . \label{Btilde 1}
\ee
This cancels the right-hand-side and gives the sum rule advertised above in \eqref{B2 tilde sum rule}.
There would be many other
ways to achieve the same cancellation, but the $\widetilde{B}$ construct
is unique in that the resulting functional vanishes on all double-twists.
This follows from the action on $n=0$ double-twists recorded in \eqref{eq:a2V}:
\be
 B_{2,v}[G^s_{2\Df+\ell,\ell}] =  (1-v)^{-\Df}k_{\Df+\ell}(1-v) = G^s_{2\Df+\ell,\ell}(u=0,v)\,. \label{B2v n=0 ch5}
\ee
Loosely, the reason why \eqref{eq:B2v dDisc ch5} does not vanish on $n=0$ double-twists,
despite the double zero from dDisc, is that the zeros get canceled by divergences
which effectively localize the integral to $u'=0$ and $v'=v$.
For any correlator with t $\leftrightarrow$ u symmetry, only even spins $\ell$ contribute
to the s-channel OPE and the combination $\widetilde{B}_{2,v}$
vanishes thanks to the symmetries of the block in \eqref{B2v n=0 ch5}.
Thus, all $\widetilde{B}_{2,v}$ have double zeros on double-twists with $n>0$, and at least
single zeros on even-spin $n=0$ double-twists.

\smallskip

The result for u-channel identity (which is a sum of s-channel double-twists)
quoted above \eqref{B2v primmed sum} follows from \eqref{B2v n=0 ch5},
and the similar claim for s-channel identity follows using that $B_{2,v}[u^{-\Df}]=-B_{2,v}[v^{-\Df}]$.

\smallskip

The integral representation in \eqref{eq:B2v dDisc ch5} converges when $\Delta>2\Df+J$,
and can be correctly computed outside that range by analytic continuation.
In Appendices \ref{app:numericalbtwov} and \ref{app:wsrecursion} we detail two concrete methods to numerically
compute the $B_{2,v}$ sum rules, one based on direct numerical integration of dDisc
(both at generic $v$ and for series expanding around $v=1$),
and the other based on weight-shifting operators.
Another method is based on evaluating the Mellin space sum in \eqref{eq:B2HatAction}; all methods agree.

\subsection{Twist-gap bound: $\widetilde{B}_{2,1}'$ is positive above $2\Df$} \label{ssec:Btilde21}

\begin{figure}[t]
\begin{center}
\includegraphics[width=0.7\textwidth]{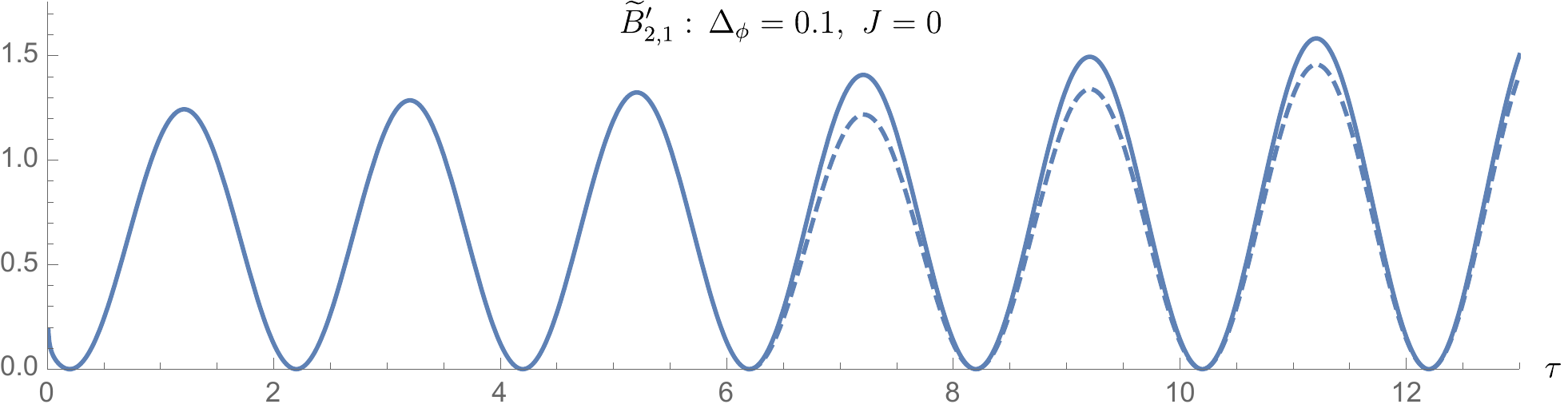}
\\\includegraphics[width=0.7\textwidth]{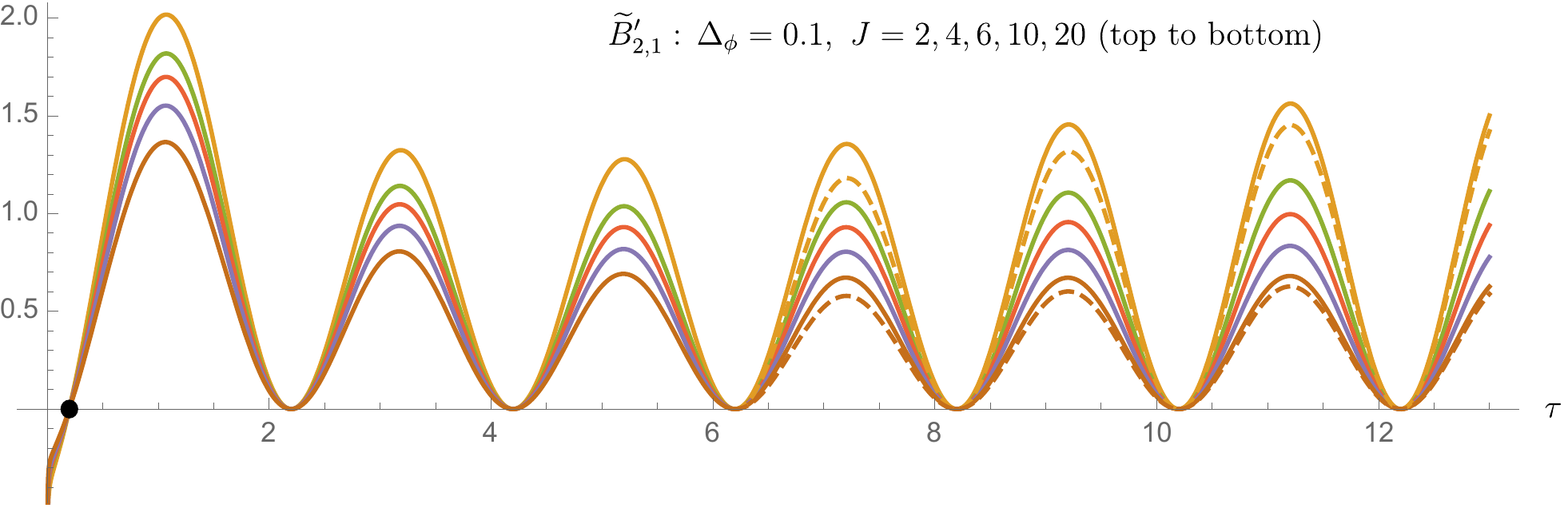}
\caption{Action of the functional $\widetilde{B}_{2,1}'$ on conformal blocks $G^s_{\Delta,J}$ for $d=2$,
$\Df=0.1$. The function being plotted is $(J+1)^{-\frac12} (\tau +1)^2 4^{-\Delta}\,\widetilde{B}_{2,1}'[G^s_{\Delta,J}]$.
Dashed lines show the asymptotic formula \eqref{2d heavy} for $J=0,2,20$.
}
\label{fig:Btilde21}
\end{center}
\end{figure}

We report here on the positivity properties of $\widetilde{B}_{2,1}'$.
While $B_{2,v}$ (for $v\geq 1$) is trivially positive above all double-twists, 
positivity of $\widetilde{B}_{2,1}'$ is less obvious since it is a difference between two terms.
We already showed in \eqref{slope B21} that $\widetilde{B}_{2,1}'$ has single zeros with positive slope
on each $n=0$ double twists, and double zeros on all $n>0$ double-twists.
This suggests positivity for $\Delta>2\Df+J$ generally, which we now demonstrate numerically.

\smallskip

Since higher-dimensional conformal blocks (above the unitarity bound)
are positive sums of two-dimensional blocks,
it suffices to show positivity in $d=2$ to establish it in any (integer) spacetime dimension.

\smallskip

\begin{figure}[t]
\begin{center}
\includegraphics[width=0.7\textwidth]{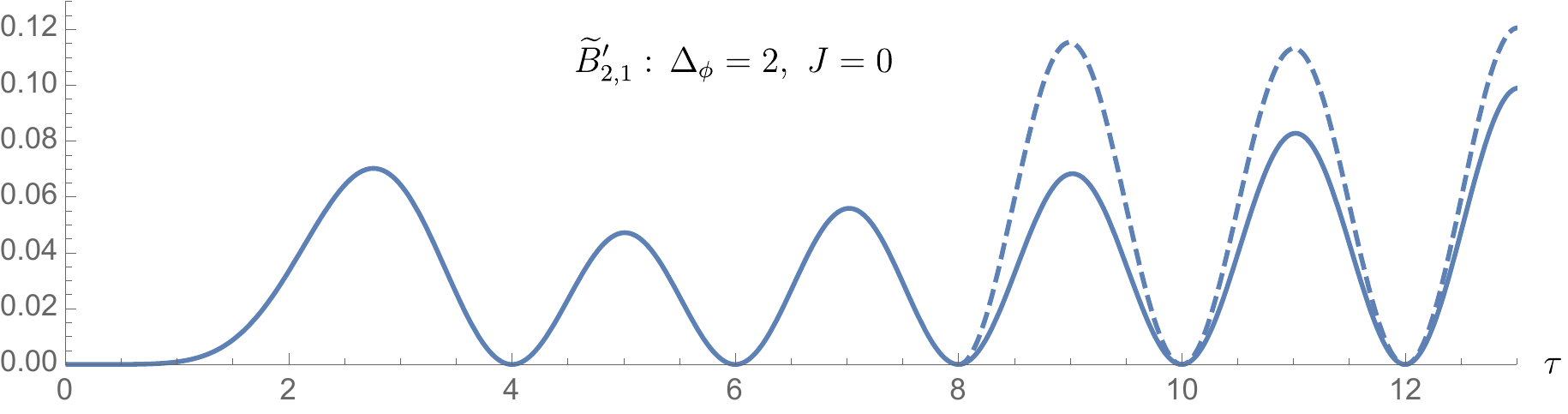}
\\\includegraphics[width=0.7\textwidth]{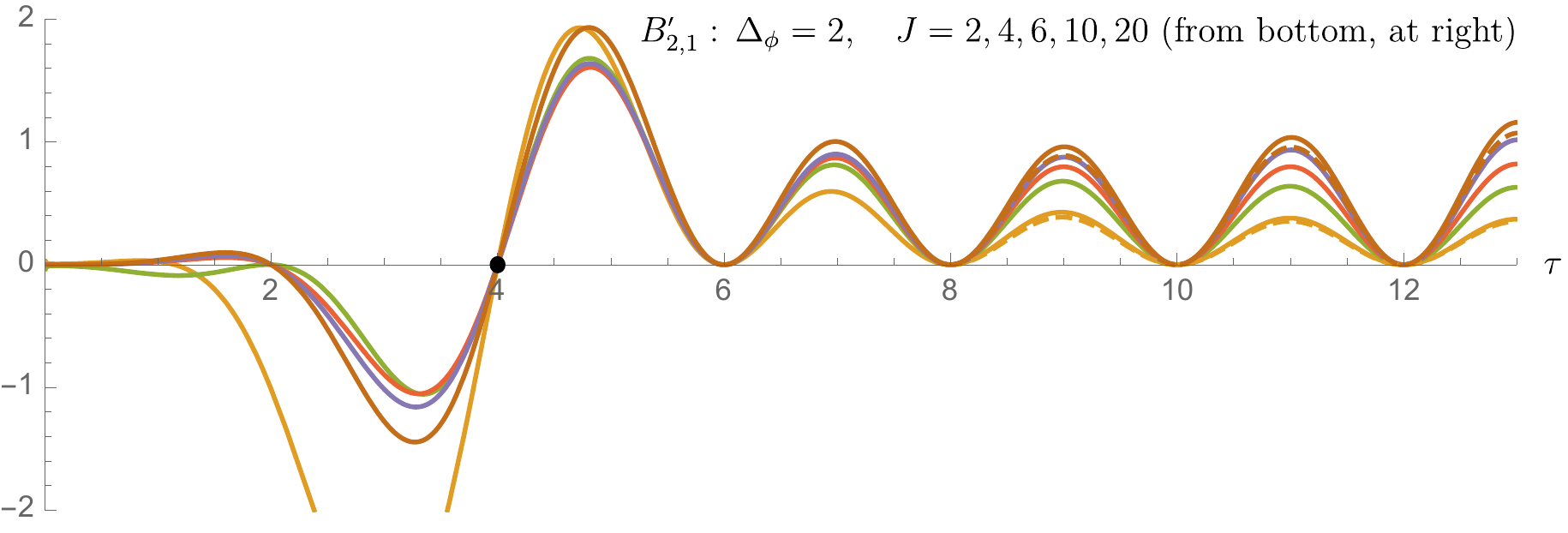}
\caption{Similar content to Figure \ref{fig:Btilde21} but for $\Delta_\phi=2$: the functional $\widetilde{B}_{2,1}'$
is again positive for $\tau>2\Df$.
The function being plotted is $(\Delta+J)^4 (\tau+1)^74^{-\Delta-8}\,\widetilde{B}_{2,1}'[G^s_{\Delta,J}]$.
\label{fig:Btilde21 2}
}
\end{center}
\end{figure}

The functional action for $\Df=0.1$ in $d=2$ is shown in Figure \ref{fig:Btilde21}.
The spin-0 action is clearly positive for all $\tau\geq 0$, while the higher-spin functionals exhibit the expected
sign-change at $\tau=2\Df$ and are positive above that.
We explored a range of values of $\Df$ between 0 and 3 and always found qualitatively similar plots.
A second example, with $\Df=2$, is shown in Figure \ref{fig:Btilde21 2}.

To convince ourselves that the functional remains positive at larger values of $\Delta$,
we computed its asymptotic behavior in appendix \ref{app:heavy}.
For generic $v$ the functional is given as \eqref{2d heavy}, which we recopy here:
\ba
\label{2d heavy text}
\lim\limits_{\Delta-J\gg 1} B_{2,v}[G^{s(d=2)}_{\Delta,J}] &=
\frac{2}{C^{\Df}_{\Delta,J}}
\frac{\Gamma(p)^4}{\Gamma(2p)}
\int_0^{t_{\rm max}}
\frac{dt(1- t)}{t\sqrt{(1+t)^2-4t v}}
\frac{1+t}{\sqrt{t}}t^{\frac{p}{2}}\times
\\
&\qquad \times\left[\xi^p \ {}_2F_1(p,p,2p,1-t\xi^2) + (\xi\mapsto \xi^{-1})\right]
\ea
with $p=2\Df+1$ and $\xi=\frac{\Delta-J}{\Delta+J}$, and the normalization
factor $C^{\Df}_{\Delta,J}$ is in \eqref{C factor}.
Specializations relevant to the $v\to 1$ limit can be found below \eqref{2d heavy}.

\smallskip

As is visible in Figure \ref{fig:Btilde21}, we
find that the asymptotic behavior (dashed curves) is rapidly approached with increasing twist;
the error is found to be at the percent level already at twist 20. 
We are therefore confident that the plots indicate the genuine behavior of the functionals.

\smallskip

\begin{figure}[t]
\begin{center}
\includegraphics[width=0.5\textwidth]{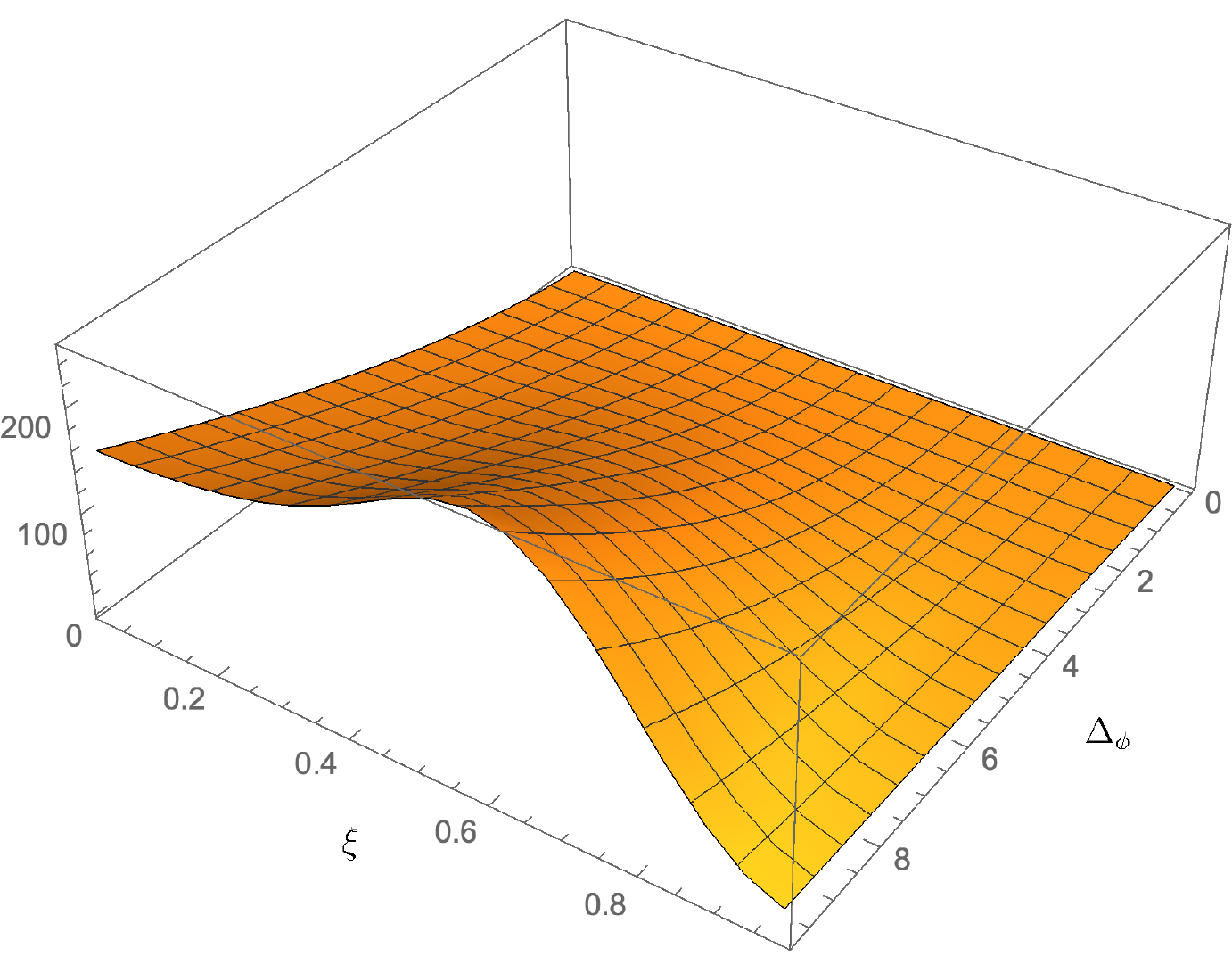}
\caption{The functional $\widetilde{B}_{2,1}'$ is positive on large-twist conformal blocks $G^s_{\Delta,J}$ in $d=2$
for all values of $\Df$ and $\xi=\frac{\Delta-J}{\Delta+J}$. 
The plot shows $\frac{C^{\Df}_{\Delta,J}}{2} \frac{\Gamma(4\Df+2)}{\Gamma(2\Df+1)^4} \times \xi\Df\widetilde{B}_{2,1}'[G^s_{\Delta,J}]$.
\label{fig:Btilde21 heavy}}
\end{center}
\end{figure}

A noteworthy feature of the asymptotic formula \eqref{2d heavy text} is that it depends only
on $\Df$ and on the ratio $\xi=\frac{\Delta-J}{\Delta+J}$ (up to the overall positive factor $C^{\Df}$).
This ratio was interpreted holographically as an impact parameter in AdS space in \cite{Cornalba:2006xm}.
For our current application, the upshot is that to check that a combination of $B_{2,v}$ functionals is
positive on heavy states, it suffices to scan over a single variable $\xi$.
In Figure \ref{fig:Btilde21 heavy} we show the result of this exercise for different values of $\Df$: we find
that the functional $\widetilde{B}_{2,1}'$ is always positive on high-twist states!

\subsection{Optimal bounds at fixed spin} \label{ssec:fixed spin}

It is expected that the result of the previous subsection can be significantly strengthened.
Not only there has to be a primary in the $\phi\times\phi$ OPE with $\tau\leq 2\Df$ and \textit{some} $\ell\geq 2$, but in fact there should be a primary with $\tau\leq 2\Df$ for \textit{each} even $\ell\geq 2$. 

\smallskip

How can we approach the task of constructing these functionals?
First of all, Property 2 stated at the top of this section
implies that the functionals must have the form (up to an overall positive rescaling)
\be
\Phi_{\ell} = \beta_{0,\ell} + [\textrm{odd spin}]\,,
\label{eq:FixedSpinBoundGeneral}
\ee
where [odd spin] stands for a (possibly infinite) linear combination of odd-spin $\alpha$s and $\beta$s. Furthermore, the functional must decay with spin $>1$ in the u-channel Regge limit. We did not find \textit{finite} linear combinations of the form \eqref{eq:FixedSpinBoundGeneral} which also satisfy Property 1. Indeed as noted above, finite linear combinations of $\alpha$s and $\beta$s generally have additional simple zeros away from double-traces.

\smallskip

We have seen previously that the $B_{2,v}$ sum rules, which are infinite linear combinations of the basis functionals, have good positivity properties. So let us guess that $\Phi_{\ell}$ arises as a projection of $B_{2,v}$. The good news is that there is a unique projection of $B_{2,v}$ of the form \eqref{eq:FixedSpinBoundGeneral}. Furthermore, we will check that for some values of $d$ and $\Df$, the resulting sum rule has the correct positivity Property 2.

\smallskip

To construct the projection, we will work in Mellin space and use orthogonality of Mack polynomials. A general projection takes the form of an integral against a kernel $f_{\ell}(\mT)$ over a vertical $\mT$ contour
\be
\Phi_{\ell} = \!\!\!\int\limits_{\Df-i\infty}^{\Df+i\infty}\!\!\!\frac{d\mT}{2\pi i}\,f_{\ell}(\mT)\,\widehat{B}_{2,\mT}\,.
\ee
Any such $\Phi_{\ell}$ has double zeros on all $n>0$ double traces. We will fix $f_{\ell}(\mT)$ by requiring the correct structure of zeros on the $n=0$ family. $\Phi_{\ell}$ vanishes on the $n=0$, even $\ell$ double traces if and only if $f_{\ell}(\mT)$ is an odd function
\be
f_{\ell}(\mT) = -f_{\ell}(2\Df-\mT)\,.
\label{eq:fIsOdd}
\ee
This is the condition that the Mellin integral yields a superposition of $\widetilde{B}_{2,v}$ functionals.
In order to have \textit{double} zeros on all $n=0$ double traces with even spin $\neq\ell$, we must impose
\be
\int\limits_{\Df-i\infty}^{\Df+i\infty}\!\!\!\frac{d\mT}{2\pi i}\,f_{\ell}(\mT)\,\widehat{b}_{\ell'}(\mT) = \delta_{\ell,\ell'}\quad\textrm{for}\quad \ell'=0,2,4,\ldots\,,
\label{eq:fIsOrthogonal}
\ee
where $\widehat{b}_{\ell'}(\mT)$ is the coefficients of $\beta_{0,\ell'}$ in the decomposition of $\widehat{B}_{2,\mT}$, see \eqref{eq:B2HatExpansion} and \eqref{eq:abHatFormulas}.

\smallskip

In order to fix $f_{\ell}(\mT)$ using these constraints, we use completeness and orthogonality of the Mack polynomials $\widehat{a}_{\ell}(\mT)$
\be
\int\frac{d\mT}{2\pi i}
\Gamma(\Df-\tfrac{\mT}{2})^{2}\Gamma(\tfrac{\mT}{2})^{2} \,\,\widehat{a}_{\ell}(\mT)\,\,\widehat{a}_{\ell'}(\mT) = 
\frac{2(-1)^\ell \Gamma (\ell+1) \Gamma (2\Df+2\ell)^2}{(2 \Df +2\ell-1) \Gamma (\Df+\ell)^4 \Gamma (2 \Df+\ell -1)}\delta_{\ell\ell'}\,.
\label{eq:aOrthogonality}
\ee
The $\mT$ contour runs from $\mT_{0}-i\infty$ to $\mT_{0}+i\infty$ with $0<\mT_{0}<2\Df$. Since $f_{\ell}(\mT)$ satisfies \eqref{eq:fIsOdd}, it can be expanded in $\widehat{a}_{\ell'}(\mT)$ with $\ell'$ odd. Equations \eqref{eq:fIsOrthogonal} then uniquely fix the expansion coefficients since $\widehat{b}_{\ell'}(\mT)$ are related to $\widehat{a}_{\ell'}(\mT)$ by a lower-triangular change of basis, see \eqref{eq:abHatFormulas}. Note that the construction breaks down for $\ell=0$ since $\widehat{b}_{0}(\mT) \sim \widehat{a}_{0}(\mT)$ and thus there is no $f_{0}(\mT)$ satisfying both \eqref{eq:fIsOdd} and \eqref{eq:fIsOrthogonal}.

\smallskip

We can give a closed formula for $f_{\ell}(\mT)$. Firstly, one checks that for even $\ell$, we have the relationship
\be
\widehat{b}_{\ell}(\mT) - \widehat{b}_{\ell}(2\Df-\mT) = -\frac{d\,\widehat{a}_{\ell}(\mT)}{d\mT}\,.
\label{eq:bda}
\ee
The formula for $f_{\ell}(\mT)$ is
\be
f_{\ell}(\mT) =
c_{\ell}\int\limits_{\Df}^{\mT}\!d\mT' \,\Gamma(\Df-\tfrac{\mT'}{2})^{2}\Gamma(\tfrac{\mT'}{2})^{2}
\,\widehat{a}_{\ell}(\mT')\,,
\ee
for some normalization $c_{\ell}$ to be determined. Why does this formula work? Firstly, note that \eqref{eq:fIsOdd} holds because $\widehat{a}_{\ell}(\mT) = (-1)^{\ell}\widehat{a}_{\ell}(2\Df-\mT)$. To see that \eqref{eq:fIsOrthogonal} holds, compute
\ba
\int\limits_{\Df-i\infty}^{\Df+i\infty}\!\!\!\frac{d\mT}{2\pi i}\,f_{\ell}(\mT)\,\widehat{b}_{\ell'}(\mT) &= 
c_{\ell}\int\limits_{\Df-i\infty}^{\Df+i\infty}\!\!\!\frac{d\mT}{2\pi i}
\,\widehat{b}_{\ell'}(\mT)\!
\int\limits_{\Df}^{\mT}\!d\mT' \,\Gamma(\Df-\tfrac{\mT'}{2})^{2}\Gamma(\tfrac{\mT'}{2})^{2}
\,\widehat{a}_{\ell}(\mT')
=\\
&=-\frac{c_{\ell}}{2}\int\limits_{\Df-i\infty}^{\Df+i\infty}\!\!\!\frac{d\mT}{2\pi i}
\frac{d\,\widehat{a}_{\ell'}(\mT)}{d\mT}
\int\limits_{\Df}^{\mT}\!d\mT' \,\Gamma(\Df-\tfrac{\mT'}{2})^{2}\Gamma(\tfrac{\mT'}{2})^{2}
\,\widehat{a}_{\ell}(\mT') =\\
&=\frac{c_{\ell}}{2}\int\limits_{\Df-i\infty}^{\Df+i\infty}\!\!\!\frac{d\mT}{2\pi i}
\,\widehat{a}_{\ell'}(\mT)\widehat{a}_{\ell}(\mT)
\Gamma(\Df-\tfrac{\mT}{2})^{2}\Gamma(\tfrac{\mT}{2})^{2} =\\
&=
c_{\ell}
\frac{\Gamma (\ell+1) \Gamma (2\Df+2\ell)^2}{(2 \Df +2\ell-1) \Gamma (\Df+\ell)^4 \Gamma (2 \Df+\ell -1)}\delta_{\ell\ell'}\,.
\ea
The second equality follows from antisymmetry of $\widehat{f}_{\ell}(\mT)$ and \eqref{eq:bda}. The third equality follows by integrating by parts. The boundary term vanishes for $\ell>0$ thanks to the orthogonality \eqref{eq:aOrthogonality} between $a_{\ell}(\mT)$ and $a_{0}(\mT) = \mathrm{const}$. We conclude
\be
c_{\ell}=
\frac{(2 \Df +2\ell-1) \Gamma (\Df+\ell)^4 \Gamma (2 \Df+\ell -1)}{\Gamma (\ell+1) \Gamma (2\Df+2\ell)^2}\,.
\ee
The formula for $\Phi_{\ell}$ therefore is
\ba
\Phi_{\ell} = &c_{\ell}\!\!\!\int\limits_{\Df-i\infty}^{\Df+i\infty}\!\!\!\frac{d\mT}{2\pi i}\,\widehat{B}_{2,\mT}
\int\limits_{\Df}^{\mT}\!d\mT' \,\Gamma(\Df-\tfrac{\mT'}{2})^{2}\Gamma(\tfrac{\mT'}{2})^{2}
\,\widehat{a}_{\ell}(\mT')=\\
=-&c_{\ell}\!\!\!\int\limits_{\Df-i\infty}^{\Df+i\infty}\!\!\!\frac{d\mT}{2\pi i}
\,\Gamma(\Df-\tfrac{\mT}{2})^{2}\Gamma(\tfrac{\mT}{2})^{2}
\,\widehat{a}_{\ell}(\mT)
\int\limits_{\Df}^{\mT}\!d\mT' 
\widehat{B}_{2,\mT'}\,.
\label{eq:PhiLFormula}
\ea
We believe that this formula defines a swappable functional which gives rise to a nonperturbative sum rule with vanishing RHS
\be
\sum\limits_{\cO}a_{\cO}\,\Phi_{\ell}[G^s_{\Delta_{\cO},J_{\cO}}] = 0\,.
\label{eq:PhiSumRule}
\ee
Equation \eqref{eq:PhiLFormula} gives $\Phi_{\ell}$ as a weighted integral of $\widehat{B}_{2,\mT}$. We can write $\widehat{B}_{2,\mT}$ as the Mellin transform of $B_{2,v}$, which gives $\Phi_{\ell}$ as a weighted integral of $B_{2,v}$. In particular, for $\ell=2$ and $\Df=1$ we were able to find the closed form
\ba
\Phi_2 = &\landupint\limits_{0}^{\infty}dv
\frac{v+1}{2 \pi ^2 (v-1)^2} \left[\frac{1}{\log (v)-2 i \pi }-\frac{1}{\log (v)}\right] B_{2,v} +\\
+&\landdownint\limits_{0}^{\infty}dv
\frac{v+1}{2 \pi ^2 (v-1)^2} \left[\frac{1}{\log (v)+2 i \pi }-\frac{1}{\log (v)}\right] B_{2,v}\,, \label{messy Phi2}
\ea
where the integral symbol shows how the pole at $v=1$ is avoided. Curiously, the kernel is essentially $\dDisc_{v=0}[(v+1)/((v-1)^2\log(v))]$. One could presumably prove swappability of $\Phi_{2}$ using this formula by bounding the contribution of the $v\rightarrow 0$ and $v\rightarrow\infty$ limits. We leave a detailed proof to future work and content ourselves with checking \eqref{eq:PhiSumRule} in examples.

\smallskip

We can compute $\Phi_{\ell}[G^s_{\Delta,J}]$ starting from the definition \eqref{eq:PhiLFormula} and using the known expression for $\widehat{B}_{2,\mT}[G^s_{\Delta,J}]$, equation \eqref{eq:B2HatAction}. In practice, the integral over $\mT'$ in \eqref{eq:PhiLFormula} can be done analytically. We did the remaining integral over $\mT$ numerically and approximated the sum over $m$ in \eqref{eq:B2HatAction} by a large but finite number of terms. We found that $\Phi_{\ell}[G^s_{\Delta,J}]$ is finite for all $(\Delta,J)$ allowed by unitarity bounds.\footnote{This is unlike $\widehat{B}_{2,\mT}[G^s_{\Delta,J}]$, which has a double pole at $\tau = \mT$.} In that region, it is a smooth function of $\Delta$ satisfying Property 1. We checked that $\Phi_{\ell}[G^s_{0,0}] = 0$, which implies \eqref{eq:PhiSumRule} is satisfied by the $\langle\phi\phi\phi\phi\rangle$ MFT correlator. Furthermore, we checked that the $\ell=2$ sum rule \eqref{eq:PhiSumRule} is satisfied for the $d=4$, $\Df=2$ four-point function $\langle\varphi^2\varphi^2\varphi^2\varphi^2\rangle$, where $\varphi$ is a free real scalar field.

\smallskip

\begin{figure}[ht]
\begin{center}
\includegraphics[width=0.7\textwidth]{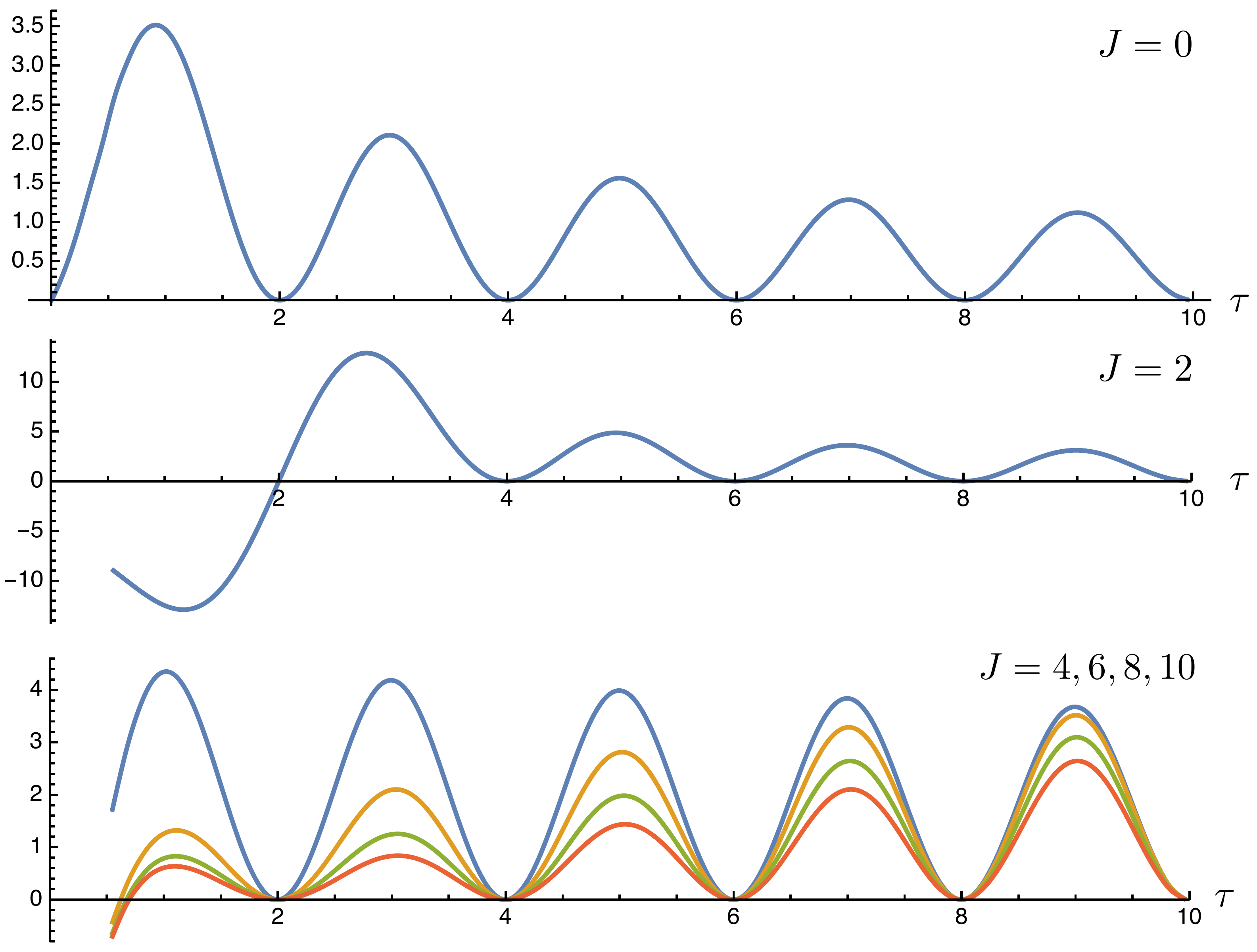}
\caption{The action of the extremal functional $\Phi_{2}$ on conformal blocks $G^s_{\Delta,J}$ for $d=3$, $\Df=1$. $\Phi_{2}$ has double zeros on all even-spin double traces apart from $n=0$, $\ell=2$, where it has a simple zero. It is nonnegative above the unitarity bound for all spins $\neq 2$, and also for spin 2 provided $\tau\geq 2\Df$. In order to emphasize the key features, the function being plotted is $\sqrt{J+1} (\tau +1)^5 4^{2-J-\tau}\,\Phi_{2}[G^s_{\Delta,J}]$.}
\label{fig:Phi2Action}
\end{center}
\end{figure}

In assessing the positivity properties of $\Phi_{\ell}[G^s_{\Delta,J}]$, we focused on $\Phi_2$. The following facts were observed experimentally.
\begin{enumerate}
\item $\Phi_{2}[G^s_{0,0}]= 0$ and $\Phi_{2}[G^s_{\Delta,0}]\geq 0$ for $\Delta\geq \frac{d-2}{2}$.
\item $\Phi_{2}[G^s_{\Delta,2}]\geq 0$ for $\Delta\geq2\Df+2$.
\item $\Phi_{2}[G^s_{\Delta,J}]\geq 0$ for even $J>2$ and $\Delta\geq \tau_{0}(d,\Df)+J$.
\end{enumerate}
$\tau_{0}(d,\Df)$ is a critical value of the twist below which $\Phi_{2}[G^s_{\Delta,J}]$ can be negative. If $\tau_{0}(d,\Df)<d-2$, then $\Phi_{2}[G^s_{\Delta,J}]$ satisfies all the properties needed to be an extremal functional. We found that $\tau_{0}(d,\Df)$ increases roughly linearly with $\Df$ and crosses the unitarity bound for some $\Df$. For $d=3$ the crossing happens for some $\Df\in(1,2)$. For $d=4$, it happens for $\Df\in(2,3)$. In particular, $\Phi_2$ is an extremal functionals for $d=3$ provided $\Df\in[1/2,1]$ and for $d=4$ provided $\Df\in[1,2]$. Figure \ref{fig:Phi2Action} shows the action $\Phi_{2}[G^s_{\Delta,J}]$ in the case $d=3$ $\Df=1$.

It would be interesting to study $\Phi_\ell$ for $\ell>2$ and also understand how to construct the extremal functional for gap maximization at $\ell>0$ for general $d$ and $\Df$ since our construction has the right positivity properties only in a certain range of $\Df$. In general the extremal functional with correct positivity properties must differ from $\Phi_{\ell}$ by a functional which has support in odd-spin $\alpha_{n,\ell}$ and $\beta_{n,\ell}$, and which itself decays with spin $>1$. As discussed in Section \ref{ssec:contacts}, such functionals certainly exist. It was also shown there that such odd-spin functionals identically annihilate all scalar conformal blocks. It follows that the action of the extremal functional on scalar blocks agrees with that of $\Phi_{\ell}$. In that light, it is encouraging that $\Phi_{2}[G^s_{\Delta,0}]\geq 0$ above the unitarity bound for general $d$ and $\Df$. In fact, since $\Phi_{2}$ is a special case of $\beta^{\text{imp}}_{0,\ell}$ of Section \ref{sec:lightconerelationship}, we know that $\Phi_{2}[G^s_{\Delta,0}] = - 2 q^{\text{MFT}}_{0,2}\gamma_{0,2}(\Delta,0)$ (the result of applying LIF to scalar blocks), which is known to be positive.

\smallskip

Finally, note that mean field theory does not saturate the bound on the gap above identity for $\ell=0$. Indeed, the scalar bound is an interesting function of $\Df$, conjecturally passing through the 3D Ising CFT in $d=3$ \cite{ElShowk:2012ht}. It is clear from our perspective why the case $\ell=0$ is special. As noted below equation \eqref{eq:aOrthogonality}, our construction breaks down for $\ell=0$. More generally, there can not exist a functional of the form
\be
\beta_{0,0} + [\textrm{odd spin}]
\ee
which decays with spin $>1$ in the Regge limit. Such a functional would be incompatible with the scalar contact diagram of the AdS $\Phi^4$ interaction. The latter grows with spin 0 in the Regge limit and only contains terms $G^s_{2\Df+2n,0}$ and $\partial_{\Delta}G^s_{2\Df+2n,0}$. On the other hand, the existence of this contact diagram is perfectly compatible with functionals of the form \eqref{eq:FixedSpinBoundGeneral} with $\ell=2,4,\ldots$.

\subsection{$\Phi_2$ sum rule in the 3D Ising CFT}
\begin{figure}[ht]
\begin{center}
\includegraphics[width=0.7\textwidth]{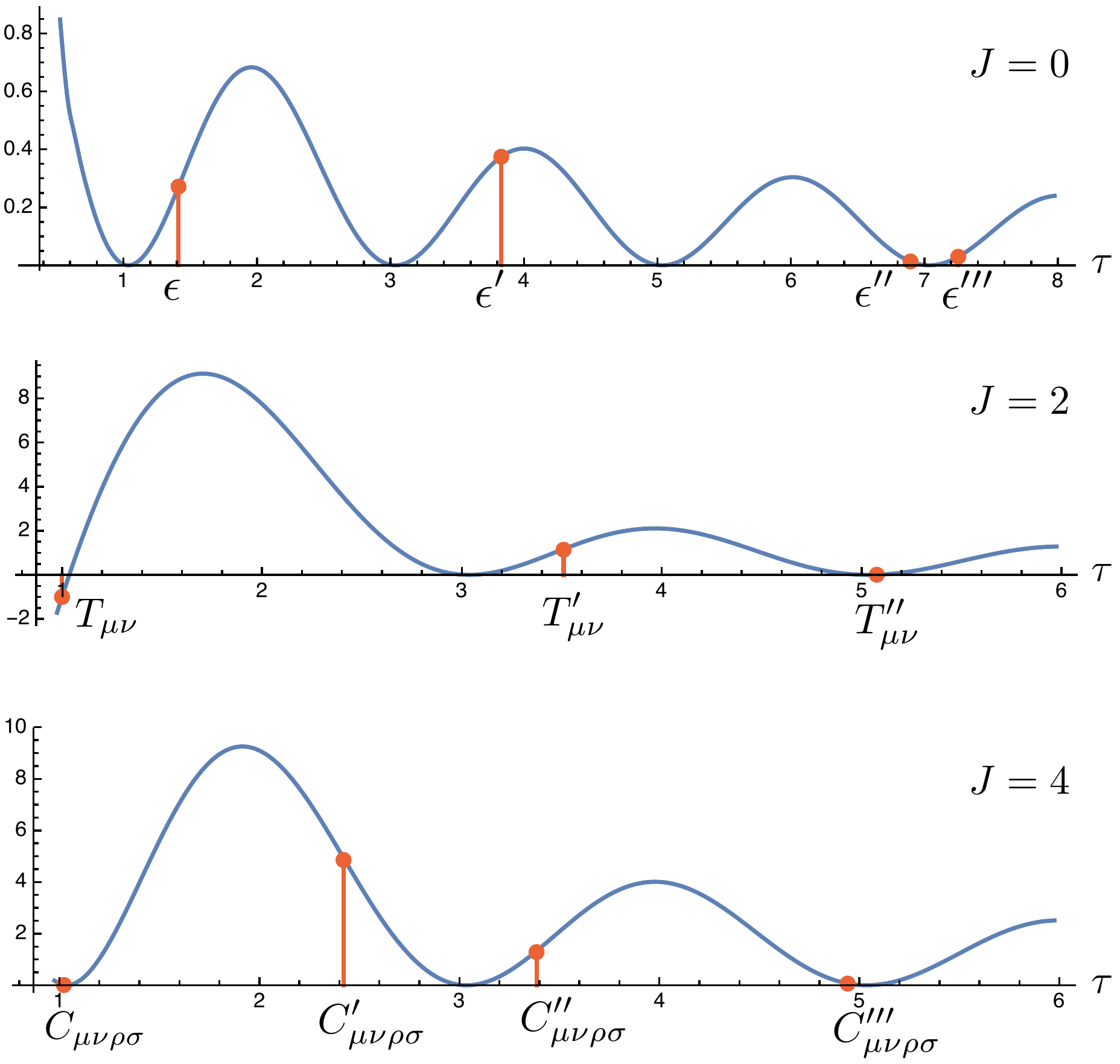}
\caption{The action of functional $\Phi_2$ on conformal blocks of spin $J=0,2,4$ for $d=3$ and $\Df = 0.51815$, i.e. in the 3D Ising CFT. Also shown are the locations of several low-lying primaries in the 3D Ising CFT, as predicted in \cite{Simmons-Duffin:2016wlq}. The only negative contribution to the sum rule \eqref{eq:3DIsingSumRule} comes from the stress tensor. The breakdown of contributions is shown in Table \ref{tab:3DIsingOperators}. For visual clarity, the plotted function is $y (\Delta +5)^5 4^{-\Delta}\,\Phi_{2}[G^s_{\Delta,J}]$, where the constant $y$ is chosen so that the action on the stress tensor is $-1$. To get the contributions of primaries to the sum rule, we must multiply the plotted function by the OPE coefficient squared and divide by the rescaling factor.}
\label{fig:Phi2_3DIsing}
\end{center}
\end{figure}

Let us apply $\Phi_{2}$ to the $\langle\sigma\sigma\sigma\sigma\rangle$ four-point function in the 3D Ising CFT. We have $\Delta_{\sigma} \approx 0.51815$ and so the only negative region of $\Phi_{2}$ is the small window $1\leq \tau <2\Delta_\sigma$ for $J=2$. There is precisely one operator in this window: the stress tensor, at $\tau = 1$. Thus $\Phi_{2}$ equates minus the contribution of $T_{\mu\nu}$ with a positive sum over all $\mathbb{Z}_{2}$-even primaries in the 3D Ising CFT
\be
-(f_{\sigma\sigma T_{\mu\nu}})^2 \Phi_{2}[G_{3,0}] = \sum\limits_{\cO\neq T_{\mu\nu}} (f_{\sigma\sigma\cO})^2 \Phi_{2}[G_{\Delta_{\cO},J_{\cO}}]\,.
\label{eq:3DIsingSumRule} 
\ee
Since $\Phi_{2}$ has double zeros at $\tau = 2\Delta_{\sigma}+2n$, the contribution of approximate double-twist operators at large spin is suppressed. Therefore, we can expect that the sum rule is nearly saturated by a handful of low-lying operators. We used the OPE data provided in \cite{Simmons-Duffin:2016wlq} to test the validity of the sum rule,
computed numerically using the method below \eqref{messy Phi2}.
Remarkably, merely including the $\epsilon$ operator on the RHS of \eqref{eq:3DIsingSumRule} accounts for $95\%$ of the LHS. In other words, the sum rule provides an approximate equation satisfied by $\Delta_{\sigma}$, $\Delta_{\epsilon}$, $c_{T}$ and $f_{\sigma\sigma\epsilon}$ to $95\%$ accuracy. Adding also the next-to-leading primaries with $J=0$ and $J=2$, i.e. $\epsilon'$ and $T'_{\mu\nu}$, brings the agreement to $98\%$. Thanks to positivity, this puts a stringent upper bound on the contribution of the remaining operators. Detailed results are shown in Figure \ref{fig:Phi2_3DIsing} and Table \ref{tab:3DIsingOperators}.

\begin{table}
\centering
  \begin{tabular}{| c | c | c | c | c |}
    \hline
$\cO$ & $J_{\cO}$ & $\tau_{\cO}$ & $(f_{\sigma\sigma \cO})^2\Phi_{2}[G_{\cO}]$\\[1pt]
\hline
$T_{\mu\nu}$ & 2& 1 &  $- 1$\\[2pt]
$\epsilon$ & 0& 1.412 &  0.950 \\[2pt]
$\epsilon'$ & 0 & 3.830 &  0.020\\[2pt]
$T_{\mu\nu}'$ & 2 & 3.509  & 0.011\\[2pt]
$C'_{\mu\nu\rho\sigma}$ & 4 & 2.420 & 0.004\\[2pt]
$C''_{\mu\nu\rho\sigma}$ & 4 & 3.385 & 0.004\\[2pt]
$C_{\mu\nu\rho\sigma}$ & 4 & 1.023 & 0.002\\[2pt]
$\epsilon''$ & 0 & 6.896 &$<10^{-5}$\\[2pt]
$C'''_{\mu\nu\rho\sigma}$ & 4 & 4.941 &  $<10^{-5}$\\[2pt]
$T_{\mu\nu}''$ & 2 & 5.076 & $<10^{-6}$\\[2pt]
identity & 0 & 0 & $ 0$\\[2pt]
    \hline
  \end{tabular}
\caption{\label{tab:3DIsingOperators}The low-lying primaries of the 3D Ising CFT ordered according to the size of their contribution to the $\Phi_2$ sum rule \eqref{eq:3DIsingSumRule}. The normalization is chosen so that $T_{\mu\nu}$ contributes $-1$, and thus all $\mathbb{Z}_2$-even operators must add up to $+1$. We see that $\epsilon$ alone gives $95\%$ accuracy, and the primaries in the table together give $99\%$ accuracy. This means that the remaining operators either have twists very close to $2\Delta_{\sigma}+2n$, or have very small OPE coefficients.}
\end{table}

\subsection{Comparison of $\Phi_2$ with numerical functionals}
\label{sec:numericaltwistgap}

We have presented strong evidence that $\Phi_2$ is an extremal functional for the twist gap problem at $\ell=2$. It is interesting to compare it to extremal functionals obtained from the numerical bootstrap \cite{Poland:2010wg,ElShowk:2012hu,El-Showk:2014dwa}. An important feature of $\Phi_2$ is that it is an example of an ``improved'' $\beta_{0,2}$:
\be
\label{eq:phi2isimproved}
\Phi_2 = \beta_{0,2} + [\textrm{odd spin}] = [\textrm{spin-$2$ decay}].
\ee
The condition (\ref{eq:phi2isimproved}) has an ambiguity parametrized by odd-spin functionals that decay with spin $2$. In other words, any positive functional of the form
\be
\label{eq:phi2ambiguity}
\Phi_2 + [\textrm{odd spin with spin-$2$ decay}]
\ee
would be another extremal functional for the $\ell=2$ twist gap problem. Given that $\Phi_2$ is conjecturally positive (for appropriate values of $\De_\phi$), we should ask: can other functionals of the type (\ref{eq:phi2ambiguity}) be positive as well? We expect that the answer is: yes, the solution to the spin-2 twist gap problem is {\it not unique}, and is given by a nontrivial convex subspace of the space (\ref{eq:phi2ambiguity}). We will obtain evidence for this claim using numerics.

\smallskip

We can approximate a solution to the spin-2 twist gap problem using the numerical bootstrap. Following \cite{Rattazzi:2008pe}, we study linear combinations of ``derivative'' functionals\footnote{We expect the analysis of this section could be performed with any basis of functionals satisfying the conditions in section~\ref{sec:numericalcomments}.}
\be
\label{eq:derivativebasis}
\left.\w^\mathrm{deriv}_{mn}[\cG] \equiv \ptl_z^m \ptl_{\bar z}^n \cG(z,\bar z) \right|_{z=\bar z = \frac 1 2}.
\ee
By searching over a subspace of s-t antisymmetric functionals with derivative order up to $\Lambda$
\be
\label{eq:limitderivativebasis}
\mathrm{Span}\{\w^\mathrm{deriv}_{mn} \textrm{ such that }m+n \leq \Lambda\textrm{ and $m+n$ odd}\},
\ee
we obtain an upper bound on the twist gap at spin $\ell=2$, witnessed by a functional $\Phi^\mathrm{deriv}_\mathrm{2,\Lambda}$. Taking $\Lambda$ larger and larger, this bound becomes closer and closer to optimal. We claim that if the limit $\lim_{\Lambda\to\oo}\Phi^\mathrm{deriv}_\mathrm{2,\Lambda}$ exists, it should live in the space (\ref{eq:phi2ambiguity}). 

\begin{figure}[t!]
\begin{center}
\includegraphics[width=\textwidth]{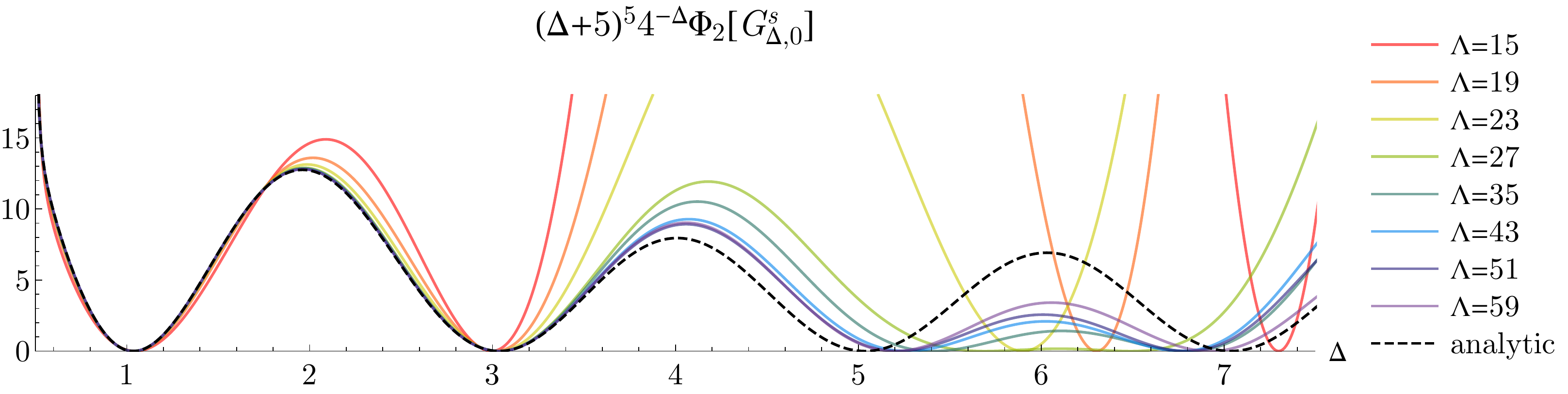}
\caption{The action of numerical functionals $\Phi_{2,\Lambda}^\mathrm{deriv}$ and the analytic functional $\Phi_2$ on scalar blocks $G^s_{\De,0}$, as a function of $\De$. We show the case $\De_\f=0.5181489$ and $d=3$. For visual clarity, the plotted function is $(\Delta +5)^5 4^{-\Delta}\,\Phi[G^s_{\Delta,0}]$. The colored curves show numerical functionals with derivative orders $\Lambda\in\{15,19,23,27,35,43,51,59\}$. The black dashed curve is the analytic functional $\Phi_2$. As $\Lambda$ increases, the numerical functionals converge to the analytic one, in accordance with (\ref{eq:expectscalars}). See appendix~\ref{app:extremalfunctionalnumerics} for details of our numerical implementation.}
\label{fig:spin0ActionPlot}
\end{center}
\end{figure}

\begin{figure}[t!]
\begin{center}
\includegraphics[width=0.8\textwidth]{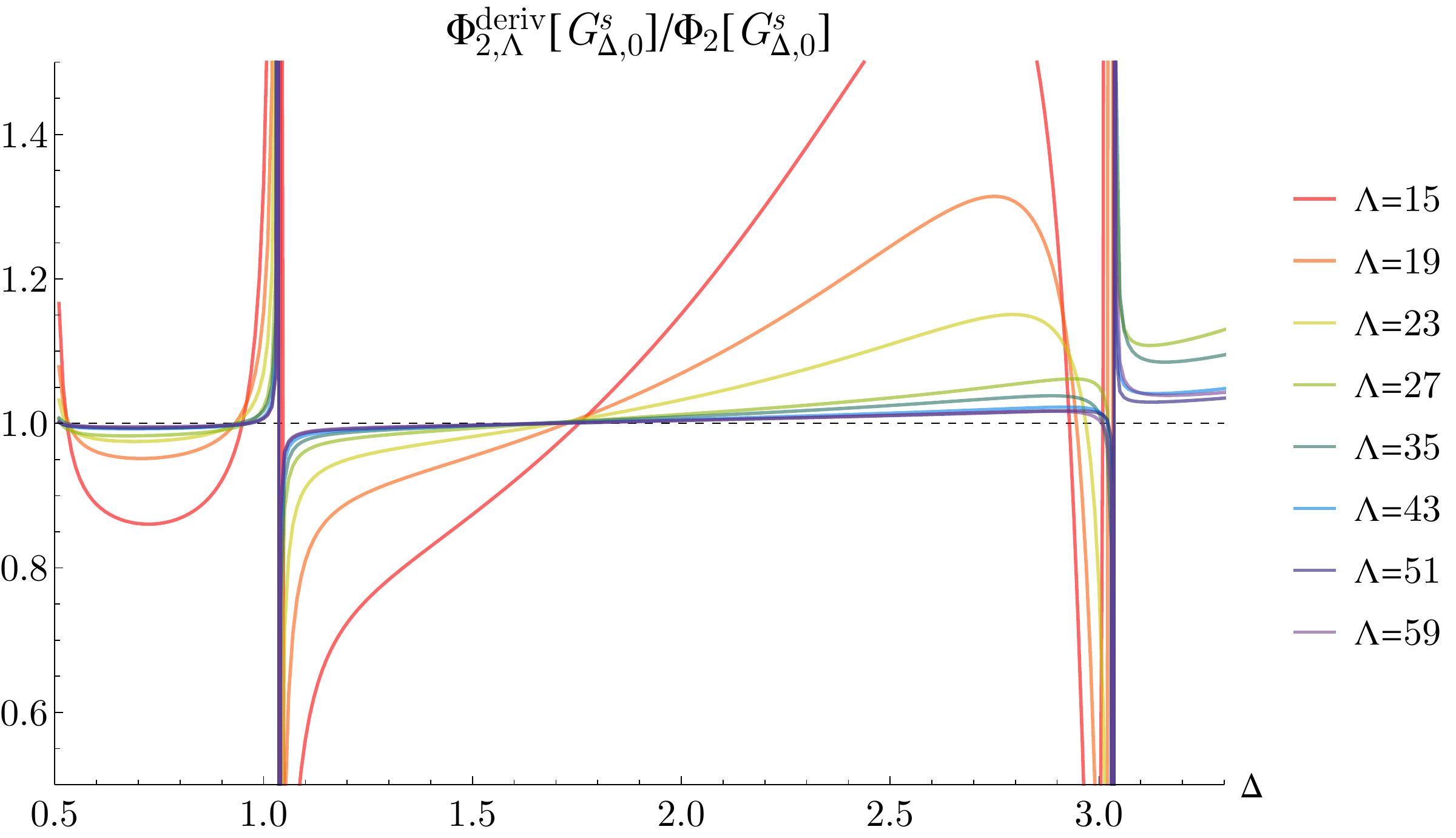}
\caption{The ratio $\Phi_{2,\Lambda}^\mathrm{deriv}[G^s_{\De,0}]/\Phi_2[G^s_{\De,0}]$ of numerical functionals to the analytic functional $\Phi_2$, acting on scalar blocks, as a function of $\De$.  We show the case $\De_\f=0.5181489$ and $d=3$. The numerical functionals are computed with derivative orders $\Lambda\in\{15,19,23,27,35,43,51,59\}$. As $\Lambda$ increases, the ratio approaches 1, in accordance with (\ref{eq:expectscalars}).}
\label{fig:ratioScalarAcionPlot}
\end{center}
\end{figure}

Property (\ref{eq:betagammaArgree}) gives a way to check this claim, despite the ambiguity in (\ref{eq:phi2ambiguity}). We can compare $\Phi^\mathrm{deriv}_\mathrm{2,\Lambda}$ and $\Phi_2$ acting on scalar blocks $G^s_{\De,0}$. Equation~(\ref{eq:betagammaArgree}) implies
\be
\Big(\Phi_2 + [\textrm{odd spin with spin-$2$ decay}]\Big)[G^s_{\De,0}] = -2 q^\mathrm{MFT}_{0,2} \g_{0,2}(\De,0).
\ee
Thus, we expect
\be
\label{eq:expectscalars}
\lim_{\Lambda\to\oo} \Phi^\mathrm{deriv}_\mathrm{2,\Lambda}[G^s_{\De,0}] = \Phi_2[G^s_{\De,0}] = -2 q^\mathrm{MFT}_{0,2} \g_{0,2}(\De,0).
\ee
In figure~\ref{fig:spin0ActionPlot}, we plot the action of $\Phi_{2,\Lambda}^\mathrm{deriv}$ on scalar blocks  for various $\Lambda$, along with the action of $\Phi_2$, for $\De_\f=0.5181489$ in $d=3$ dimensions. Details of our numerical implementation are described in appendix~\ref{app:extremalfunctionalnumerics}. As $\Lambda$ gets larger, the numerical functionals indeed approach $\Phi_2$. We give further detail in figure~\ref{fig:ratioScalarAcionPlot}, where we plot the ratio of numerical functionals to the analytic functional. The ratio clearly approaches $1$ as $\Lambda$ increases. This is a spectacular check of our claim that $\Phi_2$ is an extremal functional for the spin-2 twist-gap problem!

\begin{figure}[t!]
\begin{center}
\includegraphics[width=\textwidth]{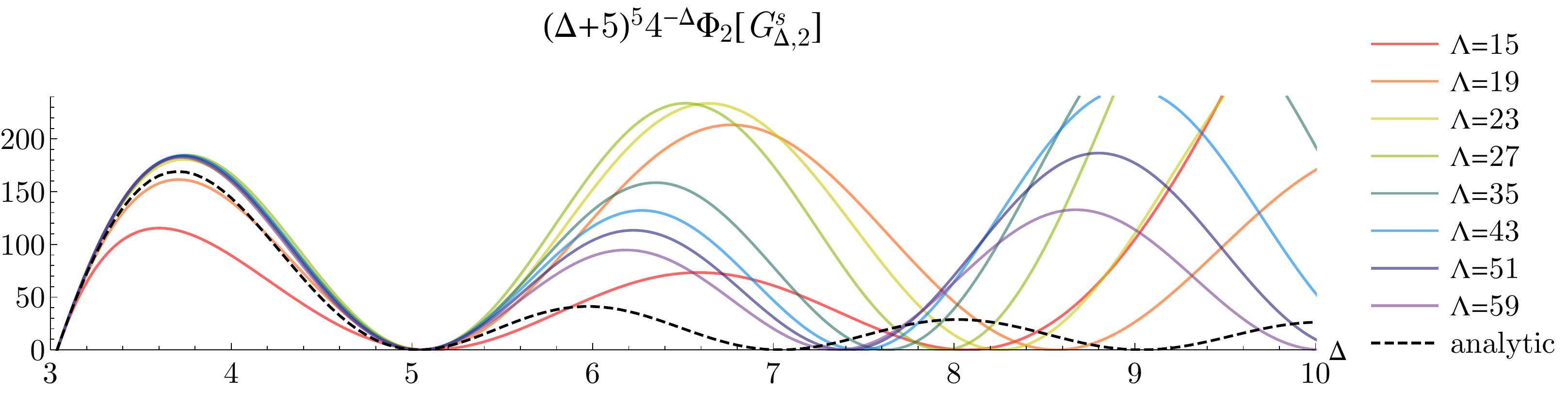}\\
\vspace{0.2in}
\includegraphics[width=\textwidth]{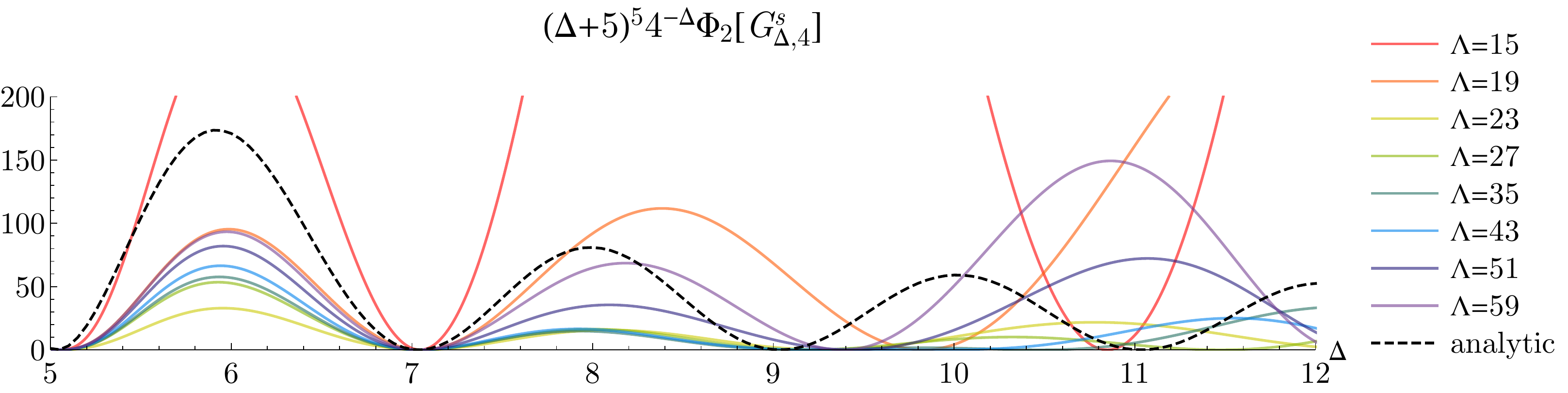}
\caption{The action of numerical functionals $\Phi_{2,\Lambda}^\mathrm{deriv}$ (colored curves) and the analytic functional $\Phi_2$ (dashed curve) on blocks $G^s_{\De,J}$ with spins $J=2$ and $J=4$, as a function of $\De$. The setup is the same as in figure~\ref{fig:spin0ActionPlot}. In the case $J=2$, when $\De\lesssim 5$ the numerical functionals appear to converge to a different value from the analytic functional $\Phi_2$. In the case $J=4$, the numerical functionals do not appear to converge (at this derivative order). Both behaviors are consistent with non-uniqueness of the form (\ref{eq:phi2ambiguity}) in the solution of the spin-2 twist gap problem.}
\label{fig:spin2and4ActionPlot}
\end{center}
\end{figure}

On the other hand, we do not necessarily expect the action of $\lim_{\Lambda\to\oo} \Phi^\mathrm{deriv}_\mathrm{2,\Lambda}$ on higher-spin blocks $G^s_{\De,J\geq 2}$ to match that of $\Phi_2$, because of the ambiguity in (\ref{eq:phi2ambiguity}). In fact, it is not obvious whether $\Phi^\mathrm{deriv}_\mathrm{2,\Lambda}[G^s_{\De,J\geq 2}]$ should converge at all.  Whether $\Phi^\mathrm{deriv}_\mathrm{2,\Lambda}$ converges, along with its limiting value,  could depend on details of the numerical implementation such as the objective function used to test feasibility and the choice of which derivatives to include. In figure~\ref{fig:spin2and4ActionPlot}, we plot the action of $\Phi_{2,\Lambda}^\mathrm{deriv}$ on blocks with $J=2$ and $J=4$. For $J=2$, the functionals appear to converge to a different value from $\Phi_2$. For $J=4$, it is unclear whether they are converging at the given derivative order. It would be interesting to explore the dependence of numerical functional actions for $J\geq 2$ on different implementation choices.

\subsection{Comments on numerical bootstrap applications}\label{sec:numericalcomments}

Numerical bootstrap computations require a basis of s-t antisymmetric functionals. The most commonly-used basis is the derivative basis (\ref{eq:limitderivativebasis}), first proposed in the original numerical bootstrap work \cite{Rattazzi:2008pe}. Although simple, the derivative basis has important properties that make it suitable for numerical applications:
\begin{enumerate}
\item Swappability.
\item Completeness in the limit $\Lambda \to \oo$.
\item Asymptotic positivity of finite linear combinations.
\end{enumerate}

``Swappability'' means that the action of a functional commutes with the infinite sum over conformal blocks inside a physical four-point function. Its importance was emphasized in \cite{Rychkov:2017tpc}. For $\w^\mathrm{deriv}_{mn}$, swappability follows readily from the fact that the conformal block expansion converges exponentially near the point $z=\bar z= \frac 1 2$ \cite{Pappadopulo:2012jk}. By ``completeness'' of the derivative basis, we mean that a sum of conformal blocks $\cG$ satisfying $\w^\mathrm{deriv}_{mn}[\cG]=0$ for all $m,n$ with $m+n$ odd is exactly crossing symmetric.

\smallskip

Finally, we call a functional $\w$ ``asymptotically positive'' if $\w[G_{\De,J}]$ is positive for all $(\De,J)$ above the unitarity bound, except for a compact subset of $(\De,J)$-space which we call the ``compact negative region.'' An asymptotically positive functional is equivalent to a proof that the spectrum of a CFT must contain an operator inside the compact negative region. Many of the most interesting applications of the numerical bootstrap require such proofs.

\smallskip

On the other hand, the derivative basis may not be the most efficient for obtaining certain bounds on CFT data. For example, extremal functionals computed numerically using the derivative basis can recover the existence of double-twist operators: they often exhibit double-zeros close to the values $\De=2\De_\f+2n+\ell$. However, to obtain these double-zeros, one must use a relatively high derivative order $\Lambda$. For example, the studies \cite{Simmons-Duffin:2016wlq,Liu:2020tpf} found about 100 double-twist operators using linear combinations of approximately 1000 functionals. The problem is that double-twist operators are needed to solve crossing symmetry in the lightcone regime, and this is far from the point $z=\bar z=\frac 1 2$ around which the derivative basis is defined. By contrast, dispersive functionals automatically have double-zeros at most double-twist locations. In a precise sense, they automatically account for the unit operator contributions to the crossing equation. It is conceivable that fewer of them will be needed to obtain a realistic spectrum.

\smallskip

Thus, it is interesting to identify candidate bases of dispersive functionals satisfying the three properties of swappability, completeness, and asymptotic positivity of finite linear combinations. For the functionals considered in this work, swappability is equivalent to spin-2 (or faster) decay as defined in (\ref{eq:functionaldecay}). Any of our subtracted (or multiply-subtracted) functionals have this property.

\smallskip

We do not have a systematic understanding of completeness for dispersive functionals. In \cite{Mazac:2019shk}, it was conjectured that $\a_{n,\ell},\b_{n,\ell}$ form a complete set of $s$-$t$ antisymmetric functionals with spin-0 decay. Assuming this, we furthermore conjecture that $\nu_{i,j}$, $\mu_{i,j}$ defined in (\ref{eq:definitionofnuij}) and (\ref{eq:nukernels}) provide a complete set of $s$-$t$ antisymmetric functionals with spin-2 decay.

\smallskip

Unfortunately, finite linear combinations of $\nu_{i,j}$, $\mu_{i,j}$ do not appear to be asymptotically positive. We have checked this claim numerically as follows. First, we evaluated the $\nu_{i,j}$, $\mu_{i,j}$ for several $i,j$ acting on $G_{\De,J}$ in the limit $\De,J\to \oo$ with fixed $\xi=\frac{\De-J}{\De+J}$, using the methods described in section~\ref{ssec:Btilde21} and appendix~\ref{app:heavy}. Using {\tt SDPB} \cite{Simmons-Duffin:2015qma,Landry:2019qug}, we searched numerically for finite linear combinations whose leading behavior at large $\De$ is positive as a function of $\xi\in[0,1]$. We did not find any such linear combinations. 

\smallskip

Consequently, it appears necessary to include at least one functional that is an infinite sum of $\a_{n,\ell}$'s and $\beta_{n,\ell}$'s, possessing a compact negative region. One possible candidate is $\Phi_2$ defined in (\ref{eq:PhiLFormula}). Specifically, $\Phi_2$ has a compact negative region if $\tau_0(d,\De_\f)<d-2$, which is true for sufficiently small $\De_\f$. In this case, a possible basis of dispersive functionals for the numerical bootstrap --- conjecturally satisfying all three properties above --- is $\{\Phi_2,\nu_{ij},\mu_{ij}\}$. This is not necessarily the only or optimal choice. It will be important to identify other examples of functionals with compact negative regions and find efficient methods for computing them.

\smallskip

Positivity properties of dispersive functionals are simpler to analyze in the context of 1D CFTs. Reference \cite{Paulos:2019fkw} applied a basis of  dispersive functionals to numerical 1D bootstrap, finding improvement over the derivative basis.

% !TEX root = ../main.tex

\section{Conclusions}\label{sec:Conclusions}

In this paper we have studied ``dispersive'' sum rules,  a class of constraints on conformal field theories
embodying crossing symmetry of four-point correlation functions.
The defining property of dispersive sum rules is that they possess double zeros which make them insensitive
to operators with scaling dimensions $\Delta=2\Df+2n+J$, commonly called double-twist operators.
These zeros are highly desirable for analytic purposes
and enable, physically, to probe highly boosted Lorentzian configurations. 

\smallskip

At a technical level,  our sum rules are double
integrals over cross-ratios, where one variable runs over the s-channel and the other over the t-channel cut (see Figure~\ref{fig:Contours}).
After deforming the contour so both variables wrap the same cut, the integral can be most naturally
computed as a sum of either s-channel or t-channel primaries, and by equating these, one relates the OPE data in the two channels.
When the correlator is crossing symmetric, the same manipulations yield sum rules on the common data.  Special properties of the integration kernel ensure the appearance of the desired double zeros. The ingredients are highly versatile: the quadratic polynomial raised to half-integer power in $B_v$ in eq.~\eqref{eq:B2VDefinition}, or its $v\to 1$ limit which gives $1/(w-\wb)$ in eq.~\eqref{around 1}, all times powers of cross-ratios and possible single logarithms. These can be combined to give sum rules with various physical properties.

\smallskip

More conceptually, all such sum rules originate from dispersion relations.
A remarkable finding of this paper is that \emph{all CFT dispersion relations are the same}.
One can reconstruct a Mellin amplitude from its poles, or reconstruct a position space
correlators from its dDisc, and these two processes are simply Mellin transform of each other.
Expanding either in the OPE limit, one obtains a basis of analytic functionals.
Physically, convergence of dispersive sum rules exploits causality and unitarity,
which ensure that correlators remain analytic and bounded 
in the Regge limit, where points approach null infinity.
Our sum rules, as discussed in Section \ref{ssec:superconvergence}, are precisely the statement that detectors
placed at null infinity commute with each other.

\smallskip

Different spaces manifest different properties.
The Mellin representation offers great insight and computability but
tends to obscure positivity and swappability properties;
position space clarifies both, but computations can be technically difficult; finally
the double-twist basis $\alpha_{n,\ell}$ and $\beta_{n,\ell}$ diagonalizes the OPE data but finite combinations 
are not positive.
The concept of \emph{Polyakov-Regge block} $P^{s|u}_{\Delta,J}$
allows to seamlessly translate between all spaces. A Polyakov-Regge block can be defined equivalently
as the dispersive transform of a conformal block, as a Witten exchange diagram with improved Regge behavior, and more abstractly as the unique solution to physical conditions of single-valuedness and Regge (super)boundedness.  
Uniqueness allows to directly translate results obtained in different spaces, bypassing difficult integral transforms.

\smallskip

Compared with the usual OPE, the Polyakov-Regge expansion is term-by-term single-valued,
at the price of using OPE data in two channels.  The expansion also introduces spurious double-twist
operators, and our sum rules can be understood as the requirement that spurious terms cancel out.
Such constraints are often referred to as Polyakov conditions.
Our approach makes clear, as discussed in Section  \ref{ssec:Polyakov conditions},
that the Polyakov conditions are independent of any spectral assumption and are purely consequences of crossing:
they hold whether or not physical double-twist operators are present in the spectrum.

\smallskip

An important step taken in this paper was to define \emph{subtracted} Polyakov-Regge blocks, which
have improved Regge behavior. These are necessary for applying the formalism to generic unitary theories.
Various natural subtraction schemes can be used, amounting to dividing by powers of Mellin variable
or position-space cross-ratios.  A certain subtraction scheme which enhances the leading double-twist trajectory
(rescaling the Mellin amplitude as $M(\mS,\mT)/((\mS-2\Df)(\mT-2\Df))$) enjoys an
interesting sign property: for Euclidean cross-ratios it is negative definite above the double-twist threshold.
The subtracted Polyakov-Regge expansion in~\eqref{PR minus MFT}
expresses the correlator as mean-field-theory plus a mostly-negative non-gaussianity.

\smallskip

The $B_{2,v}$ family of sum rules, obtained from the twice-subtracted dispersion relation (whence the $2$ subscript),
turns out to also enjoy interesting sign properties.
A certain derivative around $v=1$, $\widetilde{B}'_{2,1}$, has a strictly positive slope on all double-twist operators.
The parameter $v$ offers enough freedom to diagonalize in spin, creating functionals with
double-zeros on all but one double-twists operator.
The functional called $\Phi_2$, with a single zero on the first spin-2 double-twist,
appears to be an extremal functional for the spin-2 gap-maximization problem: it proves that mean field theory correlator
is the unique solution to crossing with no spin-2 operator of $\Delta<2\Df$.  An important open problem is to construct large classes of dispersive functionals with manifest positivity above a fixed twist. Such functionals can be used to constrain effective field theories in AdS. We plan to return to this problem in the near future.

\smallskip

The $B_{2,v}$ sum rules have a physical interpretation as subtracted versions of superconvergence sum rules~\cite{Kologlu:2019bco}. To see this connection in Section~\ref{ssec:superconvergence}, we studied a ``position space event shape,'' where initial and final states are created by position-space local operators, and we place detectors at future null infinity.  The subtractions that give rise to $B_{2,v}$ are easily generalized: we include a general weighting function of retarded time when integrating the detectors along null infinity. When the weighting function is meromorphic with poles at the retarded times of the source and sink operators, we obtain a dispersive sum rule from the condition that the commutator of detectors should vanish.  In conformal collider physics, it is more natural to study event shapes where the initial and final states are momentum eigenstates. It would be interesting to study the Fourier transform of $B_{2,v}$ and its generalizations and understand whether momentum space makes manifest any representation-theoretic or positivity properties. In general, Fourier space offers other approaches to deriving dispersion relations, see e.g.\ \cite{Polyakov:1974gs,Gillioz:2018kwh,Gillioz:2019iye,Gillioz:2020mdd,CutkoskyFuture}. It would be interesting to understand the relationship between the resulting sum rules and the sum rules identified in this work.

\smallskip

As explained in Section~\ref{sec:lightconerelationship}, dispersive functionals yield a new approach to the lightcone bootstrap \cite{Komargodski:2012ek,Fitzpatrick:2012yx} with the possibility of rigorously-controlled errors. Applying the Lorentzian inversion formula (LIF) to study double-twist operators in nonperturbative theories involves several subtleties. Because of the presence of multi-twist operators in the other channel, the LIF, na\"ively applied, yields only an asymptotic expansion for anomalous dimensions at large spin. To understand finite spin, one can introduce a generating function $C(z,\bar h)$ that is well-defined at finite spin and perform numerical fits to extract anomalous dimensions, as in \cite{Simmons-Duffin:2016wlq,Iliesiu:2018zlz,Liu:2020tpf,Caron-Huot:2020ouj}. Dispersive functionals give an alternative way to control the sum over multi-twists and obtain well-defined results at finite spin. In pursuing this direction, it will be important to understand the space of odd-spin functionals with spin-$J$ Regge decay, which provide the ambiguities in $\b_{n,\ell}^\mathrm{imp}$. In particular, it will be interesting to explore the possible positivity properties that $\b_{n,\ell}^\mathrm{imp}$ can have.

\smallskip

As stated above, all of our sum rules come from integrating the crossing equation over the two cross-ratios. The integration contour $C_{-}\times C_{+}$ naturally lives on the boundary of the common region of convergence of the pair of OPEs. In other words, the sum rules come from distributions on this boundary. The recent article \cite{Kravchuk:2020scc} developed an understanding of CFT correlators and OPE convergence in the language of such distributions. It would be interesting to use their work as a foundation for a systematic treatment of dispersive sum rules, addressing questions of swappability and completeness in a unified manner.

\smallskip

We explained how some dispersive sum rules with good positivity properties lead to rigorous results concerning the distribution of operators in $(\Delta,J)$-space. There exists another approach which has been successfully used to address this class of questions, based on Tauberian theorems \cite{Qiao:2017xif, Mukhametzhanov:2018zja, Mukhametzhanov:2019pzy, Pal:2019zzr, Mukhametzhanov:2020swe}. We expect that Tauberian theorems and dispersive sum rules are closely related and it would be rewarding to make the connection sharp.

\smallskip

One of the original motivations for the development of analytic extremal functionals was the observation that very similar functionals lead to the optimal bounds in the numerical bootstrap. There is numerical evidence that the 3D Ising CFT saturates the upper bound on dimension gap in the scalar sector \cite{ElShowk:2012ht}. Assuming it does, there must exist a corresponding extremal functional, which exhibits double zeros on the primaries present in the $\sigma\times\sigma$ OPE.\footnote{The numerical bootstrap works by constructing successively better approximations of this functional.} Since most double-twist operators of the 3D Ising CFT have small anomalous dimensions, $\alpha_{n,\ell}$, $\beta_{n,\ell}$ with even $\ell>0$ will appear in the extremal functional with small coefficients. It would be very interesting to learn more about the exact 3D Ising functional from this viewpoint.

%%%%%%%%%%%%%%%%%%%%%%%%%%%%%%%%

\section*{Acknowledgments}
It is a pleasure to thank David Meltzer, Eric Perlmutter, Anh-Khoi Trinh, Sasha Zhiboedov, and Xinan Zhou for useful conversations. The work of D.M. and L.R. is supported in part by NSF grant \# PHY-1915093.
The work of S.C.-H. is supported by the National Science and Engineering Council of Canada, the Canada Research Chair program, the Fonds de Recherche du Qu\'ebec - Nature et Technologies,
and the Simons Collaboration on the Nonperturbative Bootstrap. D.S.-D. is supported by Simons Foundation grant 488657 (Simons Collaboration on the Nonperturbative Bootstrap), a Sloan Research Fellowship, and a DOE Early Career Award under grant no. DE-SC0019085. Some of the computations in this work were performed on the Caltech High-Performance Cluster, partially supported by a grant from the Gordon and Betty Moore Foundation.

\appendix

% !TEX root = ../main.tex

\section{Details on the dispersion relation}\label{app:contours}
\subsection{Proof that $\cG=\cG^s+\cG^t$}\label{app:Continuation}
In this appendix, we will show that the piecewise definitions of $\cG^s(z,\zb)$ and $\cG^t(z,\zb)$ in equations \eqref{Omega-s}, \eqref{Omega-t} are analytic, i.e. the discontinuity of the $\theta$ function cancels the discontinuity of the integral. This is equivalent to showing that if we define
\ba
\cG^s(z,\zb) &= \!\!\iint\limits_{C_-\times C_+}\!\!\frac{dwd\wb}{(2\pi i)^2}\pi^2(\wb-w)K_B(u,v;u',v') \cG(w,\wb)\quad\textrm{for}\quad\mathrm{Re}(\sqrt{u})>\mathrm{Re}(\sqrt{v})\,,\\
\cG^t(z,\zb) &= \!\!\iint\limits_{C_-\times C_+}\!\!\frac{dwd\wb}{(2\pi i)^2}\pi^2(\wb-w)K_B(v,u;v',u') \cG(w,\wb)
\quad\textrm{for}\quad\mathrm{Re}(\sqrt{u})<\mathrm{Re}(\sqrt{v})
\,,
\label{eq:Dispersion3App}
\ea
then the analytic continuations satisfy $\cG=\cG^s+\cG^t$. We will focus on $\cG^t(z,\zb)$. We will start in the region $\mathrm{Re}(\sqrt{u})<\mathrm{Re}(\sqrt{v})$, where the contour definition \eqref{eq:Dispersion3App} is valid, and analytically continue in $z,\zb$ to the region $\mathrm{Re}(\sqrt{u})>\mathrm{Re}(\sqrt{v})$. It will be very convenient to switch to the variables $a,b,a',b'$, defined by
\ba
a&=\sqrt{u}\,,\qquad b=\sqrt{v}\,,\\
a'&=\sqrt{u}\,,\qquad b'=\sqrt{v'}\,.
\ea
For simplicity, we will restrict to Lorentzian kinematics $0<z,\zb<1$, which is equivalent to $0<a,b<1$, $0<a+b\leq1$. Inside this region, the contour definition of $\cG^t(z,\zb)$ in \eqref{eq:Dispersion3App} is valid for $a<b$, i.e. $z+\zb<1$, while the definition of $\cG^s(z,\zb)$ in \eqref{eq:Dispersion3App} is valid for $a>b$, i.e. $z+\zb>1$. Let us start by transforming the contour $C_{-}\times C_{+}$ in $w,\wb$ variables to the $a',b'$ space. They both run parallel to the imaginary axis and have a small positive real part $\epsilon>0$. The Jacobian of the transformation is $(\wb-w)dw d\wb = -4a'b'da'db'$, so that we find
\ba
\cG^s(z,\zb) &= \int\limits_{\epsilon-i\infty}^{\epsilon+i\infty}\frac{da'}{2\pi i}\int\limits_{\epsilon-i\infty}^{\epsilon+i\infty}\frac{db'}{2\pi i}\,4\pi^2a'b'K_B(a^2,b^2;a'^2,b'^2) \cG(w,\wb)\quad\textrm{for}\quad a>b\\
\cG^t(z,\zb) &= \int\limits_{\epsilon-i\infty}^{\epsilon+i\infty}\frac{da'}{2\pi i}\int\limits_{\epsilon-i\infty}^{\epsilon+i\infty}\frac{db'}{2\pi i}\,4\pi^2a'b'K_B(b^2,a^2;b'^2,a'^2) \cG(w,\wb)\quad\textrm{for}\quad a<b\,.
\ea
The branch points of $\cG(w,\wb)$ are at $w=0,1$ and $\wb=0,1$, and these map to $a'=0$ and $b'=0$. The corresponding branch cuts of $\cG(w,\wb)$ can be chosen to coincide with $a',b'$ real and negative. Thus $\epsilon>0$ ensures that the contour stays inside $\mathcal{R}\times\mathcal{R}$ in the $w,\wb$ variables.

\smallskip

Next, we need to analyze singularities of the kernel $K_B(v,u;v',u')$. These can only occur when $x=0$, $x=1$ or $x=\infty$, with $x$ given by \eqref{eq:xDefinition}. We reproduce $x$ here in the $a,b$ variables
\ba
x &= \frac{16 a b a' b'}{(a+b+a'+b')(a+b-a'-b')(a-b+a'-b')(a-b-a'+b')}\\[8pt]
1-x &= \frac{(a+b+a'-b')(a+b-a'+b')(a-b+a'+b')(a-b-a'-b')}{(a+b+a'+b')(a+b-a'-b')(a-b+a'-b')(a-b-a'+b')}\,.
\ea
This means that for fixed $a,b,b'$, the singularities can occur only at the following values of $a'$
\ba
x=0\,:\quad &a'=0,\quad a'=\infty\\
x=1\,:\quad &a'=b'+a+b,\quad a'=b'-a-b,\quad a'=-b'+a-b,\quad a'=-b'-a+b\\
x=\infty\,:\quad &a'=b'+a-b,\quad a'=b'-a+b,\quad a'=-b'+a+b,\quad a'=-b'-a-b\,.
\label{eq:xSingularities}
\ea
As long as $a<b$, the contour $\mathrm{Re}(a')=\mathrm{Re}(b')=\epsilon$ avoids all singularities. What happens as we increase $a-b$ from a negative value to a positive one? Some singularities cross our contour and thus to maintain analyticity of $\cG^t(z,\zb)$, we need to deform the contour to make sure the singularities do not cross it. The possible singularities crossing the contour are at $a'=b'+a-b$ and $a'=b'-a+b$, where $x=\infty$, and at $a'=-b'+a-b$ and $a'=-b'-a+b$, where $x=1$. However, the latter two are in fact not singularities of the integrand, for the following reason. The integrand in \eqref{eq:Dispersion3App} is defined so that $f(x)=x^{3/2}{}_2F_1(\tfrac{1}{2},\tfrac{3}{2};2;1-x)$ inside $K_B$ is evaluated on the first sheet at the point $a'=b'=\epsilon$ (where $0<x<1$), and by analytic continuation along the rest of the contour. On the first sheet, $f(x)$ only has a branch cut at $x\in(-\infty,0]$, and is holomorphic at $x=1$. It is not hard to check that on our contour, $a'=-b'+a-b$ and $a'=-b'-a+b$ occurs at $x=1$ on the \emph{first} sheet, so there is no singularity there.

\smallskip

On the other hand, $a'=b'+a-b$ and $a'=b'-a+b$ are genuine singularities and must be avoided. We can take $a-b$ from $-|a-b|$ to $|a-b|$ along a complex semicircle around the origin, in which case the contour corresponding to the analytic continuation of $\cG^t(z,\zb)$ winds around the singularities. On the other hand, when $a>b$ the straight contour defines $-\cG^s(z,\zb)$, since $K_B(u,v;u',v') = -K_B(v,u;v',u')$. Therefore, $\cG^s(z,\zb) + \cG^t(z,\zb)$ is equal to the difference of the integral along the winding and straight contour. To evaluate this difference, note that $f(x)$ has a simple pole at $x=\infty$, together with a logarithmic branch cut
\be
x^{3/2}{}_2F_1(\tfrac{1}{2},\tfrac{3}{2};2;1-x) =\frac{4 x}{\pi} + \log(x)f_1(x)+f_2(x)\,,
\label{eq:fXExpansion}
\ee
where $f_{1,2}(x)$ are holomorphic at $x=\infty$. The contribution of the logarithmic cut to $\cG^s+\cG^t$ vanishes,\footnote{This can be shown by a similar contour argument as what follows for the simple poles, the only difference being that the final $b'$ integration gives zero since the integrand has no singularities in the region $\mathrm{Re}(b')>0$.} so we only need to take into account the simple poles at $a'=b'+a-b$ and $a'=b'-a+b$.
\be
\cG^s(z,\zb) + \cG^t(z,\zb) =\int\limits_{\epsilon-i\infty}^{\epsilon+i\infty}\frac{db'}{2\pi i}\left[\Res\limits_{a'=b'+a-b}-\Res\limits_{a'=b'-a+b}\right]4\pi^2a'b'K_B(b^2,a^2;b'^2,a'^2) \cG(w,\wb)\,.
\label{eq:gSgTApp}
\ee
Near the poles, we can keep only the first term on the RHS of \eqref{eq:fXExpansion}, which gives
\ba
4\pi^2a'b'&K_B(b^2,a^2;b'^2,a'^2) \sim\\
&\sim\frac{4(a' b')^{\frac{1}{2}}(a b)^{-\frac{1}{2}}\left(b^2-a^2+b'^2-a'^2\right)}{(a+b+a'+b')(a+b-a'-b')(a-b+a'-b')(a-b-a'+b')}\,,
\ea
so that
\ba
\Res\limits_{a'=b'+a-b}&\left[4\pi^2a'b'K_B(b^2,a^2;b'^2,a'^2) \cG(w,\wb)\right] =\\
&=-\frac{1}{b'-b}\sqrt{\frac{(a-b+b')b'}{a b}}\left.\cG(w,\wb)\right|_{a'=b'+a-b}\\
\Res\limits_{a'=b'-a+b}&\left[4\pi^2a'b'K_B(b^2,a^2;b'^2,a'^2) \cG(w,\wb)\right] =\\
&= -\frac{1}{b'+b}\sqrt{\frac{(b-a+b')b'}{a b}}\left.\cG(w,\wb)\right|_{a'=b'-a+b}\,.
\ea
The $b'$ integral in \eqref{eq:gSgTApp} can now be evaluated by closing the contour to the right. We can not close to the left since $\cG(w,\wb)$ has a branch cut there. Only the first residue, with pole at $b'=b$, contributes since $b>0$. It gives exactly $\cG(w,\wb)$ evaluated at $a'=a$, $b'=b$, in other words
\be
\cG^s(z,\zb) + \cG^t(z,\zb) = \cG(z,\zb)\,,
\ee
which completes the proof. What we have shown is that this equation gives the analytic continuation of either line of \eqref{eq:Dispersion3App} to the other side of the inequalities stated there, and demonstrates that both $\cG^s(z,\zb)$ and $\cG^t(z,\zb)$ are holomorphic in the entirety of $\mathcal{R}\times\mathcal{R}$, since $\cG(z,\zb)$ is.

\smallskip

\subsection{Recovering the original dispersion relation}\label{app:deformationToDDisc}
The purpose of this appendix is to explain in more detail that when $\cG(z,\zb)$ is single-valued in the Euclidean signature, then the contour definition \eqref{eq:Dispersion3App} of $\cG^s(z,\zb)$, $\cG^t(z,\zb)$ agrees with the original dispersion relation \eqref{eq:Dispersion1}. Let us again focus on $\cG^t(z,\zb)$, the proof for $\cG^s(z,\zb)$ being entirely equivalent. We will restrict to Lorentzian kinematics $z,\zb<0$, or equivalently $u,v>0$ with $\sqrt{v}\geq\sqrt{u}+1$.\footnote{The proof for more general kinematics follows by analytic continuation.} In particular, $\sqrt{v}>\sqrt{u}$, so \eqref{eq:Dispersion3App} applies.

\smallskip

We want to demonstrate using a contour deformation that the second line of \eqref{eq:Dispersion3App} is equivalent to
\be
\cG^t(z,\zb) = \iint\limits_{\mathrm{L}_{\mathrm{tu}}}\! du'dv'K(v,u;v',u')\dDisc_t[\cG(w,\wb)]\,,
\label{eq:GTApp}
\ee
where $\mathrm{L}_{\mathrm{tu}}=\{(u',v'):\,u'>1,\,0<v'<(\sqrt{u'}-1)^2\}$ is a Lorentzian diamond and
\be
K(v,u;v',u') = \frac{v-u+v'-u'}{64 \pi (u v u' v')^{\frac{3}{4}}} 
\left[x^{\frac{3}{2}}{}_2F_1\!\left(\tfrac{1}{2},\tfrac{3}{2};2;1-x\right) \theta(0<x<1) -4\delta(x-1)\right]\,,
\label{eq:kernelAppFull}
\ee
so that in fact the integration in \eqref{eq:GTApp} is restricted to the smaller region $\{(u',v'):\,u'>(\sqrt{u}+\sqrt{v})^2,\,0<v'\leq(\sqrt{u'}-\sqrt{u}-\sqrt{v})^2\}$. Recall that $\dDisc_t[\cG(w,\wb)]$ is defined as
\be
\dDisc_t[\cG(w,\wb)] = -\frac{1}{2}\left[\cG(w^{+},\wb^{+})+\cG(w^{-},\wb^{-})-\cG(w^{+},\wb^{-})-\cG(w^{-},\wb^{+})\right]\,,
\ee
where $w,\wb>1$ and we use notation $w^{\pm}=w\pm i\epsilon$, $\wb^{\pm}=\wb\pm i\epsilon$ for infinitesimal $\epsilon>0$.

\smallskip

Let us start by rewriting our end goal \eqref{eq:GTApp} as an integral in $w,\wb$ variables
\ba
\cG^t(z,\zb) = -\frac{1}{2}\int\limits_{1}^{\infty}\!\!dw\!\!\int\limits_{1}^{\infty}\!\!d\wb\,&(\wb-w)\theta(\wb>w)K(v,u;v',u')\times\\
\times&\left[\cG(w^{+},\wb^{+})+\cG(w^{-},\wb^{-})-\cG(w^{+},\wb^{-})-\cG(w^{-},\wb^{+})\right]\,.
\label{eq:GTApp2}
\ea
The indicator function $\theta(\wb>w)$ is present because $u',v'$ are invariant under $w\leftrightarrow\wb$ and thus the region $\{(w,\wb): w,\wb>1\}$ is a double cover of the region $\mathrm{L}_{\mathrm{tu}}$. In fact, without the factor $\theta(\wb>w)$, the integral in \eqref{eq:GTApp2} would vanish identically, due to being antisymmetric under $w\leftrightarrow\wb$. Indeed, for all $w,\wb>1$ we have
\ba
\cG(w^{+},\wb^{+}) &= \cG(\wb^{+},w^{+})\,,\\
\cG(w^{-},\wb^{-}) &= \cG(\wb^{-},w^{-})\,,\\
\cG(w^{+},\wb^{-}) &= \cG(\wb^{-},w^{+}) = \cG(\wb^{+},w^{-}) = \cG(w^{-},\wb^{+})\,,
\ea
where the first two lines follow from symmetry $\cG(w,\wb)=\cG(\wb,w)$ inside $\mathcal{R}\times\mathcal{R}$ and the third line from symmetry and single-valuedness in Euclidean signature.

\smallskip

The strategy of deriving \eqref{eq:GTApp2} from \eqref{eq:Dispersion3App} is to wrap the $\wb$ contour tightly on the branch cut $\wb\in[1,\infty)$, followed by wrapping the $w$ contour on the same branch cut $w\in[1,\infty)$. This naturally produces the four terms in the square bracket in \eqref{eq:GTApp2}, depending on whether $w,\wb$ are above or below the cut. The remaining task is to show that each term comes multiplied with the same factor $(w-\wb)\theta(\wb>w)K(v,u;v',u')/2$.

\smallskip

Consider first the term containing $\cG(w^+,\wb^+)$. After wrapping the $\wb$ contour in \eqref{eq:Dispersion3App}, the resulting $w$ integrand has various branch points located in $w>1$. The branch points can occur only at $x=0,1,\infty$, with all the a priori possibilities listed in \eqref{eq:xSingularities}. Recall that we are assuming $a,b>0$, $b>a+1$ and note that since both $w$ and $\wb$ are above the branch cut, we have $a'>0$, $b' = -\sqrt{(w-1)(\wb-1)}<0$ with $a'\geq-b'+1$. It follows that the singularities occur only at the following loci for $w,\wb>1$
\ba
x=0\,:\quad &w=1,\infty\,,\quad \wb=1,\infty\\
x=1\,:\quad &\sqrt{w}\sqrt{\wb}-\sqrt{w-1}\sqrt{\wb-1}=\sqrt{v}-\sqrt{u}\\
&\sqrt{w}\sqrt{\wb}+\sqrt{w-1}\sqrt{\wb-1}=\sqrt{v}+\sqrt{u}\\
x=\infty\,:\quad &\sqrt{w}\sqrt{\wb}-\sqrt{w-1}\sqrt{\wb-1}=\sqrt{v}+\sqrt{u}\\ 
&\sqrt{w}\sqrt{\wb}+\sqrt{w-1}\sqrt{\wb-1}=\sqrt{v}-\sqrt{u}\,.
\ea
These loci split the integration region $w,\wb>1$ into several subregions. We need to analytically continue $K_B(v,u;v',u')$ in the integrand of \eqref{eq:Dispersion3App} to each of these subregions. Antisymmetry of the integrand under $w\leftrightarrow\wb$ ensures that most of these contributions cancel. The only regions which make a non-vanishing contributions are
\ba
A &: (w>1) \wedge (\wb>1)\wedge(\wb>w)\wedge(\sqrt{w}\sqrt{\wb}-\sqrt{w-1}\sqrt{\wb-1}\geq\sqrt{v}+\sqrt{u})\\
B &: (w>1) \wedge (\wb>1)\wedge(\wb<w)\wedge(\sqrt{w}\sqrt{\wb}-\sqrt{w-1}\sqrt{\wb-1}\geq\sqrt{v}+\sqrt{u})\,.
\ea
$x(a,b;a',b')$ is negative in both $A$ and $B$, and is below the branch cut of
\be
f(x)=x^{\frac{3}{2}}{}_2F_1(\tfrac{1}{2},\tfrac{3}{2};2;1-x)\label{eq:fXApp}
\ee
in $A$ and above the cut in $B$. Thus in $A,B$, $f(x)$ becomes
\ba
A:\quad &f(x) \mapsto [-x(a,b;a',-|b'|)]^{\frac{3}{2}}{}_2F_1(\tfrac{1}{2},\tfrac{3}{2};2;1-x(a,b;a',-|b'|)+i\epsilon)\\
B:\quad &f(x) \mapsto -[-x(a,b;a',-|b'|)]^{\frac{3}{2}}{}_2F_1(\tfrac{1}{2},\tfrac{3}{2};2;1-x(a,b;a',-|b'|)-i\epsilon)\,.
\ea
We can combine the contribution from $A$ and are $B$ by changing the coordinates $w\leftrightarrow\wb$ in the latter. It follows that the total contribution of $f(x)$
becomes
\be
A+B:\quad f(x) \mapsto \left(\mbox{$\frac{x}{1-x}$}\right)^{\frac{3}{2}}
\left[{}_2F_1(\tfrac{1}{2},\tfrac{3}{2};2;\tfrac{1}{1-x}+i\epsilon)+{}_2F_1(\tfrac{1}{2},\tfrac{3}{2};2;\tfrac{1}{1-x}-i\epsilon)\right]\,,
\ee
where $x=x(a,b;a',|b'|)\in[0,1]$, and where we used the identity
\be
x(a,b;a',-b') = \frac{x(a,b;a',b')}{x(a,b;a',b')-1}\,.
\ee
Finally, we can simplify the result by using the following identity satisfied by the hypergeometric for $x\in[0,1]$ and $\epsilon\rightarrow 0^+$
\ba
(1-x)^{-\frac{3}{2}}
&\left[{}_2F_1(\tfrac{1}{2},\tfrac{3}{2};2;\tfrac{1}{1-x}+i\epsilon)+{}_2F_1(\tfrac{1}{2},\tfrac{3}{2};2;\tfrac{1}{1-x}-i\epsilon)\right] = \\
&2\,{}_2F_1(\tfrac{1}{2},\tfrac{3}{2};2;1-x) - 8 \delta(x-1)\,.
\ea
The delta function comes from the simple pole of $f(x)$ at infinity. In summary, we have shown that the part of integration where $w$ and $\wb$ are either both above or both below the branch cut contributes
\be
\cG^{t}(z,\zb)\supset -\frac{1}{2}\int\limits_{1}^{\infty}\!\!dw\!\!\int\limits_{1}^{\infty}\!\!d\wb\,(\wb-w)\theta(\wb>w)K(v,u;v',u')
\left[\cG(w^{+},\wb^{+})+\cG(w^{-},\wb^{-})\right]\,
\label{eq:GTHalfApp}
\ee
with $K(v,u;v',u')$ given by \eqref{eq:kernelAppFull}. This is in perfect agreement with \eqref{eq:GTApp2}.

\smallskip

What remains is to find the contribution to the integral when $w$ and $\wb$ are on opposite sides of the cut $w,\wb\in[1,\infty)$. In this situation $b'=|b'|=\sqrt{(w-1)(\wb-1)}$ is positive. After performing a similar analysis as in the previous case, we find that the only contribution can come from the regions
\ba
\widetilde{A} &: (w>1) \wedge (\wb>1)\wedge(\wb>w)\wedge(\sqrt{w}\sqrt{\wb}-\sqrt{w-1}\sqrt{\wb-1}\geq\sqrt{v}-\sqrt{u})\\
\widetilde{B} &: (w>1) \wedge (\wb>1)\wedge(\wb<w)\wedge(\sqrt{w}\sqrt{\wb}-\sqrt{w-1}\sqrt{\wb-1}\geq\sqrt{v}-\sqrt{u})\,,
\ea
where $x(a,b;a',b')\in(0,\infty)$. In region $\widetilde{A}$, $x$ is on the first sheet, and in $\widetilde{B}$, it is on the second sheet. Furthermore, the direction in which $x$ winds around the origin to get to the second sheet depends on whether we are looking at the contribution of $\cG(w^{+},\wb^{-})$ or $\cG(w^{-},\wb^{+})$. Since $\cG(w^{+},\wb^{-}) = \cG(w^{-},\wb^{+})$ for single-valued functions, we can simply combine these contributions. In summary, to get the combined contribution of $\cG(w^{+},\wb^{-})$ and $\cG(w^{-},\wb^{+})$ to the integral, we should make the following replacement for $f(x)$ (defined in \eqref{eq:fXApp})
\be
f(x) \mapsto f(x)-\frac{1}{2}f^{\circlearrowleft}(x-i \epsilon)-\frac{1}{2}f^{\circlearrowright}(x+i \epsilon)\,,
\label{eq:dDiscF}
\ee
where the circular arrows indicate the direction of analytic continuation around the origin. It is not hard to check that for $x\in(0,\infty)$
\be
f(x)-\frac{1}{2}f^{\circlearrowleft}(x-i \epsilon)-\frac{1}{2}f^{\circlearrowright}(x+i \epsilon) =
2\theta(x<1) f(x)- 8\delta(x-1)\,.
\ee
The indicator function $\theta(x<1)$ arises because the branch point $x=\infty$ is only logarithmic and thus the double discontinuity in \eqref{eq:dDiscF} annihilates $f(x)$ for $x>1$. In summary, the part of integration where $w$ and $\wb$ are on opposite sides of the branch cut contributes
\be
\cG^{t}(z,\zb)\supset \frac{1}{2}\int\limits_{1}^{\infty}\!\!dw\!\!\int\limits_{1}^{\infty}\!\!d\wb\,(\wb-w)\theta(\wb>w)K(v,u;v',u')
\left[\cG(w^{+},\wb^{-})+\cG(w^{-},\wb^{+})\right]\,,
\ee
again in perfect agreement with \eqref{eq:GTApp2}. The extra minus sign compared to \eqref{eq:GTHalfApp} arises because $w$ and $\wb$ run in opposite directions when on opposite sides of the cut. This completes the proof that \eqref{eq:Dispersion3App} and \eqref{eq:GTApp} are equivalent.

\subsection{Dispersion kernel from Mellin space}\label{app:Kmellin}
Below, we will consider the formula \eqref{eq:KMellin2} for the dispersion kernel arising from Mellin space, reproduced here
\ba
K_{\textrm{Mellin}}(u,v;u',v') = 
\frac{1}{2\pi^2u'v'}\iiint\!\frac{dp\,dq\,dp'}{(2\pi i)^3}
\frac{\Gamma(-p)^2\Gamma(-q)^2\Gamma(p'+1)^2}{(p'-p)\Gamma(p'-p-q)^2}\times\\
\times\left(\frac{u\phantom{'}}{v'}\right)^{p}\left(\frac{v\phantom{'}}{v'}\right)^{q}\left(\frac{v'}{u'}\right)^{p'}\,.
\ea
We will show that, as expected, $K_{\textrm{Mellin}}(u,v;u',v')$ vanishes for $\sqrt{v'}<\sqrt{u'}+\sqrt{u}+\sqrt{v}$, and that the contact term proportional to $\delta(\sqrt{v'}-\sqrt{u'}-\sqrt{u}-\sqrt{v})$ agrees with \eqref{eq:kernelsUV}. We assume $u,v,u',v'$ are real and positive.

\smallskip

The first step is to evaluate the integral over $q$. It vanishes for $v'<v$ since in that case we can deform the contour to the left, encountering no poles. When $v'> v$, we can use the formula
\be
\int\limits_{-\epsilon-i\infty}^{-\epsilon+i\infty}\!\!\frac{dq}{2\pi i}\frac{\Gamma(-q)^2}{\Gamma(r-q)^2}z^q = \frac{1}{\Gamma(2r)}(1-z)^{2r-1}{}_2F_1(r,r;2r;1-z)\,,
\ee
where $\epsilon>0$. We find
\ba
K_{\textrm{Mellin}}(u,v;u',v') = 
&\frac{1}{\pi^2u'v'}\iint\!\frac{dp\,dr}{(2\pi i)^2}
\frac{\Gamma(-p)^2\Gamma(p+r+1)^2}{\Gamma(2r+1)}\times\\
&\times\left(\tfrac{u\phantom{'}}{u'}\right)^{p}\left(\tfrac{v'}{u'}\right)^{r}\left(1-\tfrac{v}{v'}\right)^{2r-1}
\times{}_2F_1\!\left(r,r;2r;1-\tfrac{v}{v'}\right)\,,
\ea
where we changed variables from $p'$ to $r=p'-p$. The integral over $p$ can now be evaluated using the formula
\be
\int\limits_{-\epsilon-i\infty}^{-\epsilon+i\infty}\!\!\frac{dp}{2\pi i}\Gamma(-p)^2\Gamma(p+a)^2z^p = 
\frac{\Gamma(a)^4}{\Gamma(2a)}z^{-a}{}_2F_1\!\left(a,a;2a;\tfrac{z-1}{z}\right)\,,
\ee
where $0<\epsilon<\mathrm{Re}(a)$. The result is
\ba
K_{\textrm{Mellin}}(u,v;u',v') = 
&\frac{1}{\pi^2u(v'-v)}\int\!\frac{dr}{2\pi i}
\frac{\Gamma(r+1)^4}{\Gamma(2r+1)\Gamma(2r+2)}\left[\tfrac{(v'-v)^2}{u v'}\right]^{r}\times\\
&\times
{}_2F_1\!\left(r,r;2r;1-\tfrac{v\phantom{'}}{v'}\right){}_2F_1\!\left(r+1,r+1;2r+2;1-\tfrac{u'}{u\phantom{'}}\right)\,,
\ea
where the integral runs over any vertical line with $\mathrm{Re}(r)>0$. The only singularities of the integrand are at negative integers. Therefore, the integral vanishes provided the integrand decays as $r\rightarrow +\infty$. To diagnoze the asymptotics of the integrand, we need the asymptotics of the hypergeometric
\be
{}_2F_1(r,r;2r;1-z)\sim z^{-\frac{1}{4}} \left(\frac{\sqrt{z}+1}{2}\right)^{1-2 r}\quad\textrm{as}\quad r\rightarrow \infty\,.
\ee
In the end, we find the following $r\rightarrow\infty$ asymptotics of the integrand
\be
K_{\textrm{Mellin}}(u,v;u',v')|_{r\rightarrow\infty} =\int\!\!\frac{dr}{2\pi i}
\frac{\left(\frac{\sqrt{v'}-\sqrt{v}}{\sqrt{u}+\sqrt{u'}}\right)^{2 r}}{2 \pi (u u' v v')^{\frac{1}{4}}(\sqrt{u}+\sqrt{u'}) (\sqrt{v'}-\sqrt{v})}\,.
\label{eq:KMellinContact}
\ee
It follows $K_{\textrm{Mellin}}(u,v;u',v')$ vanishes for $\sqrt{v'}<\sqrt{u}+\sqrt{v}+\sqrt{u'}$ as expected. Furthermore, we can evaluate the integral on the RHS of \eqref{eq:KMellinContact} to find the contact term, since the latter only comes from the $r\rightarrow\infty$ asymptotics of the full integrand
\ba
K_{\textrm{Mellin}}(u,v;u',v')&\supset
\frac{\delta\left[\log\left(\frac{\sqrt{v'}-\sqrt{v}}{\sqrt{u}+\sqrt{u'}}\right)\right]}{4 \pi (u u' v v')^{\frac{1}{4}}(\sqrt{u}+\sqrt{u'}) (\sqrt{v'}-\sqrt{v})} =\\
&= \frac{\delta\left(\sqrt{v'}-\sqrt{u}-\sqrt{v}-\sqrt{u'}\right)}{4 \pi (u u' v v')^{\frac{1}{4}}(\sqrt{u}+\sqrt{u'})}\,,
\ea
in perfect agreement with \eqref{eq:kernelsUV}.

% !TEX root = ../main.tex

\section{Example decompositions}\label{app:decompositions}
In this appendix, we will compute the decomposition $\cG(u,v)=\cG^s(u,v) + \cG^t(u,v)$ for the connected four-point function of $\langle\phi^2\bar{\phi}^2\phi^2\bar{\phi}^2\rangle$, where $\phi$ is a free complex scalar in $d=3$ and $d=4$. In general $d$, the connected four-point function is
\be
G(x_1,x_2,x_3,x_4) = \frac{1}{(|x_{12}||x_{23}||x_{34}||x_{14}|)^{d-2}} \qquad\Rightarrow\qquad \cG(u,v) = (uv)^{-\frac{d-2}{2}}\,.
\ee
It is one of the simplest correlators for which $\dDisc_s[\cG]$, $\dDisc_t[\cG]$ and $\dDisc_s[\dDisc_t[\cG]]$ are all nonvanishing.

\subsection{$\cG(u,v) = 1/(uv)$}
In $d=4$, we have $\cG(u,v) = 1/(uv)$. $\cG(u,v)$ is u-channel superbounded and $\text{s}\leftrightarrow\text{t}$ symmetric. It follows that $\cG^t(u,v) = \cG^s(v,u)$ so it is enough to compute $\cG^s(u,v)$. To do that, we can use the dispersion relation in the form \eqref{eq:Dispersion1}. Note that
\be
\dDisc_s[G(x_1,x_2,x_3,x_4)] = -\frac{1}{2}\langle[\phi^2(x_1),\bar{\phi}^2(x_2)][\phi^2(x_3),\bar{\phi}^2(x_4)]\rangle\sim \frac{\delta(x^2_{12})\delta(x^2_{34})}{x^2_{23}x^2_{14}}\,,
\ee
so that $\dDisc_s[\cG(u',v')]$ is localized to $u'=0$. Performing the $v'$ integral in the dispersion relation leads to
\ba
\cG^s(u,v) &=  \frac{1}{u(v-u)} + \frac{\log(\frac{u}{v})}{(u-v)^2}\\
\cG^t(u,v) &=  \frac{1}{v(u-v)} + \frac{\log(\frac{v}{u})}{(u-v)^2}\,.
\label{eq:gST4D}
\ea
It is easy to check that indeed $\cG^s(u,v) + \cG^t(u,v) = 1/(uv)$. It may seem that $\cG^s(u,v)$ and $\cG^t(u,v)$ have a pole at $u=v$ but in fact the pole precisely cancels between the rational and logarithmic term, so that $\cG^{s,t}$ are both holomorphic for $z,\zb \in \mathcal{R}=\mathbb{C}\backslash((-\infty,0]\cup[1,\infty))$. Furthermore, $\cG^{s,t}$ are Euclidean single-valued around both s- and t-channel. Finally, it follows by expanding $\cG^s(u,v)$ around $v=0$ that $\dDisc_t[\cG^s] = \dDisc_s[\cG^t] = 0$ as required. $\cG^{s,t}$ are uniquely fixed by the properties listed in this paragraph.

\smallskip

Equivalently, $\cG^{s,t}$ can be found by computing the functional actions $\widehat{\alpha}^{s}_{i,j}[\cG]$, $\widehat{\beta}^{s}_{i,j}[\cG]$ since we have
\be
\cG^{t}(z,\zb) = \sum\limits_{i,j=0}^{\infty}\left[\widehat{\alpha}^{s}_{i,j}[\cG]+\widehat{\beta}^{s}_{i,j}[\cG]\frac{\log(z \zb)}{2}\right]z^{i}\zb^{j}\,.
\ee
By computing $\widehat{\alpha}^{s}_{i,j}[\cG]$, $\widehat{\beta}^{s}_{i,j}[\cG]$ with low values of $i,j$, we were able to guess the formulas
\ba
\widehat{\alpha}^{s}_{i,j}[\cG] &=\frac{\Gamma(i+j+2)}{\Gamma(i+1)\Gamma(j+1)}\left[H(i)+H(j)-2H(i+j+1)\right]+1\\
\widehat{\beta}^{s}_{i,j}[\cG] &=-\frac{2\,\Gamma(i+j+2)}{\Gamma(i+1)\Gamma(j+1)}\,,
\ea
where $H(i)$ is the harmonic number, which indeed resums to \eqref{eq:gST4D}.

\smallskip

Note that this example illustrates that there is no contradiction between $\cG = \cG^s + \cG^t$ and $\dDisc_t[\cG^s]=\dDisc_s[\cG^t] = 0$ even if the quadruple discontinuity of $\cG$ is nonvanishing
\be
\mathrm{qDisc}[\cG] = \dDisc_s[\dDisc_t[\cG]] \neq 0\,.
\ee
Indeed, in general we have
\be
\mathrm{qDisc}[\cG] = \dDisc_s[\dDisc_t[\cG]] = \dDisc_s[\dDisc_t[\cG^t]] \neq \dDisc_t[\dDisc_s[\cG^t]] = 0\,,
\ee
since $\dDisc_s$ and $\dDisc_t$ \emph{do not commute}. This may seem surprising since $\dDisc_s$ and $\dDisc_t$ can be computed by analytically continuing in different variables ($z$ and $\zb$ respectively). Noncommutativity arises because the analytic structure as a function of $z$ may depend on whether $\zb$ is on the first or second sheet. Indeed, $\cG^s(u,v)$ has a pole at $u=v$ on the second sheet, while there is no such pole on the first sheet.

\subsection{$\cG(u,v) = 1/\sqrt{uv}$}
For an example where $\dDisc_s[\cG(u,v)]$ does not localize to a delta function at $u=0$, we consider $d=3$ where
\be
\cG(z,\zb) = \frac{1}{\sqrt{u v}}\,.
\ee
The dispersion relation leads to
\ba
\cG^{s}(u,v) &=
\frac{2\,\mathrm{Li}_2\left(\sqrt{\frac{v}{u}}\right)-2\,\mathrm{Li}_2\left(-\sqrt{\frac{v}{u}}\right)+\log\left(\frac{v}{u}\right)\log\left(\frac{\sqrt{u}-\sqrt{v}}{\sqrt{u}+\sqrt{v}}\right)}{\pi ^2 \sqrt{uv}}\quad\text{for }v<u\\
\cG^{t}(u,v) &=
\frac{2\,\mathrm{Li}_2\left(\sqrt{\frac{u}{v}}\right)-2\,\mathrm{Li}_2\left(-\sqrt{\frac{u}{v}}\right)+\log\left(\frac{u}{v}\right)\log\left(\frac{\sqrt{v}-\sqrt{u}}{\sqrt{v}+\sqrt{u}}\right)}{\pi ^2 \sqrt{uv}}\quad\text{for }u<v\,.
\ea
$\cG^{s,t}(u,v)$ appear to have a logarithmic branch point at $u=v$ but in fact the branch cut precisely cancels between the first dilogarithm and $\log(\sqrt{u}-\sqrt{v})$. Thus $\cG^{s,t}(u,v)$ are holomorphic everywhere in $\mathcal{R}\times\mathcal{R}$. The above formulas also satisfy $\dDisc_s[\cG^t] = \dDisc_t[\cG^s] = 0$. Finally, the identity
\be
\cG^{s}(u,v) + \cG^{t}(u,v) = \frac{1}{\sqrt{uv}}
\ee
is a consequence of the dilogarithm identity
\be
\mathrm{Li}_2(x)+\mathrm{Li}_2\left(\frac{1}{x}\right)+\frac{1}{2} \log ^2(-x)+\frac{\pi ^2}{6} = 0\,.
\ee

% !TEX root = ../main.tex

\section{Details on Mellin-space sum rules}\label{app:mellinFunctionals}
The purpose of this appendix is to record the formula for the position-space kernel of functional $\widehat{B}_{2,\mT}$ introduced in Section \ref{sec:Subtractions} as the Mellin transform of $B_{2,v}$
\be
\widehat{B}_{2,\mT} = \Gamma(\Df-\tfrac{\mT}{2})^{-2}\Gamma(\tfrac{\mT}{2})^{-2}
\int\limits_{0}^{\infty}\frac{dv}{v}v^{\Df-\tfrac{\mT}{2}} B_{2,v}\,.
\ee
Recall the definition of $B_{2,v}$ as a double contour integral in eq. \eqref{eq:B2VDefinition}
\be
B_{2,v}[\mathcal{F}] =
\!\iint\limits_{C_-\times C_+}\!\!\frac{dwd\wb}{(2\pi i)^2}\,
\frac{(\wb-w)(v'-u')}{u'v'\sqrt{v^2-2(u'+v')v+(u'-v')^2}}
\mathcal{F}(w,\wb)\,.
\ee
Recall also its dDisc-manifesting form \eqref{eq:B2v dDisc}
\be \label{eq:B2v dDisc app}
B_{2,v} [\mathcal{F}] = 
 \int\limits_{v}^{\infty}dv'\!\!\!\!\!\!\!\!\int\limits_{0}^{(\sqrt{v'}-\sqrt{v})^2}\!\!\!\!\!\!\!\!\!du'
\frac{v'-u'}{\pi^2 u' v' \sqrt{v^2-2(u'+v')v+(u'-v')^2}}\,
\dDisc_{s}[\mathcal{F}(u',v')]
\,.
\ee
Since $\widehat{B}_{2,\mT}$ is the Mellin transform of $B_{2,v}$, it admits analogous representations
\ba
\widehat{B}_{2,\mT}[\mathcal{F}] &=
\!\iint\limits_{C_-\times C_+}\!\!\frac{dwd\wb}{(2\pi i)^2}\,
(\wb-w)H_{2,\mT}(u',v')
\mathcal{F}(w,\wb)\\
\widehat{B}_{2,\mT}[\mathcal{F}] &= 
 \int\limits_{0}^{\infty}\!\!dv'\!\!\int\limits_{0}^{v'}\!\!du'
\frac{1}{\pi^2}\widetilde{H}_{2,\mT}(u',v')\,
\dDisc_{s}[\mathcal{F}(u',v')].
\ea
It follows from these definitions that the position-space kernels are given by the following Mellin integrals
\be
H_{2,\mT}(u',v') = 
\Gamma(\Df-\tfrac{\mT}{2})^{-2}\Gamma(\tfrac{\mT}{2})^{-2}
\int\limits_{0}^{\infty}\frac{dv}{v}v^{\Df-\tfrac{\mT}{2}} 
\frac{v'-u'}{u'v'\sqrt{v^2-2(u'+v')v+(u'-v')^2}}
\ee
and
\be
\widetilde{H}_{2,\mT}(u',v') = 
\Gamma(\Df-\tfrac{\mT}{2})^{-2}\Gamma(\tfrac{\mT}{2})^{-2}
\!\!\!\!\!\!\!\!\int\limits_{0}^{(\sqrt{v'}-\sqrt{u'})^2}\!\!\!\!\!\!\!\!\frac{dv}{v}v^{\Df-\tfrac{\mT}{2}} 
\frac{v'-u'}{u'v'\sqrt{v^2-2(u'+v')v+(u'-v')^2}}\,.
\ee
These integrals can be done in a closed form. We find
\ba
H_{2,\mT}(u',v') =\frac{\Gamma (\tfrac{\mT}{2} -\Df +1)}{\Gamma( \tfrac{\mT}{2} )^2 \Gamma (\Df -\tfrac{\mT}{2} )}
\frac{v'-u'}{u'v'}|v'-u'|^{\Df -\tfrac{\mT}{2}-1 }\times\\
\times{}_2F_1\left(\frac{2\Df -\mT }{4},\frac{\mT -2\Df +2}{4};1;-\frac{4u'v'}{(v'-u')^2}\right)
\ea
and
\be
\widetilde{H}_{2,\mT}(u',v') =
\frac{1}{2\Gamma \left(\frac{\mT}{2}\right)^2 \Gamma (2 \Df -\mT)}
\frac{(v'-u')^{\Df -\frac{\mT}{2}}}{u' v'}
\,k_{\Df-\tfrac{\mT}{2}}\!\left(\tfrac{v'-u'}{v'}\right)\,,
\ee
where $k_{h}(z)=z^h{}_2F_1(h,h;2h;z)$.

% !TEX root = ../main.tex

\section{Details on evaluation of $B_{2,v}$ functionals}
\label{app:numericalbtwov}

\subsection{Direct numerical integration of dDisc}

The $B_{2,v}$ functionals can be computed as an integral over dDisc,
as just recorded in eq.~\eqref{eq:B2v dDisc app}.
For dimensions $\Delta>2\Df+J$, we find that the integral converges rapidly when employing,
for example, standard analytic expressions of the conformal blocks in $d=2$ and $d=4$.
We found particularly convenient to use the $W$ and $\Wb$ coordinates described below in eq.~\eqref{2d block}.
For smaller values of $\Delta$, it is necessary to perform an analytic continuation, which we now detail,
both for generic values of $v$ and for the series expansion around $v=1$.

\smallskip

Below the double-twist threshold, the divergences in eq.~\eqref{eq:B2v dDisc app} occur at $u'\to 0$.
This region can be regularized (without creating problems elsewhere)
by subtracting a finite number of powers of $u'$ in the series expansion of the block.
These subtractions then need to be computed analytically; this turns out to be possible, we find:
\ba
 \int\limits_{0}^{(\sqrt{v'}-\sqrt{v})^2}\!\!\!\!\!\!\!\!\!\frac{v'du'}{\pi^2 u' \sqrt{v^2-2(u'+v')v+(u'-v')^2}}
 \left(\frac{u'}{v'}\right)^p &=  \frac{\Gamma(p)\Gamma(p+1)}{\pi^2\Gamma(2p+1)}
(1-t)^{2p-1} {}_2F_1(p,p,2p,1-t) 
\\
&\equiv \mathcal{I}_p(t)
\ea
where $t=v/v'$. This formula is useful because the right-hand-side is analytic in the exponent $p$.
This enables one to play an ``add and subtract" game, where
power laws are subtracted from the integrand and added back in (analytically continued) integrated form:
\ba \label{B2v subtracted}
B_{2,v} [G^s_{\Delta,J}] &= 
 \int du' dv'
\frac{(v'-u')\dDisc_{s}[G^s_{\Delta,J}(u',v')] - v' \sum_p (u'/v')^p f_p(v')}
{\pi^2 u' v' \sqrt{v^2-2(u'+v')v+(u'-v')^2}}
\\ &\phantom{=}+\sum_p \int_0^1 \frac{dt}{t} \mathcal{I}_p(t) f_p(v/t) \,.
\ea
By including sufficiently many terms in the finite sum $\sum_p$ (where each exponent $p$ depends linearly on $\Delta$)
so that the first line converges, this formula gives the analytic continuation of $B_{2,v}$ to
any desired value of $\Delta$. A subtlety worth mentioning is that
the $t$ integral on the second line can itself diverge near $t=1$; this may be dealt with using
a second level of ``add and subtract" now using the simpler identity:
\be
 \int_0^1 dt (1-t)^{q} \equiv \frac{1}{q+1}\,.   \label{power integral}
\ee
We have tested our numerical implementation against the analytic results for identity
and near double-twist behavior recorded in eqs.~\eqref{eq:a2V}--\eqref{B2v identity}.

\smallskip

A comment is in order about the integration range.  The region in eq.~\eqref{B2v subtracted}
is on the left axis $w,\wb<0$, whereas standard expressions for conformal blocks
are typically given for $0<w<1$. In practice we deal with this by switching variables to
$W=\frac{w}{w-1}$ and $\Wb=\frac{\wb}{\wb-1}$ and
using the s $\leftrightarrow$ u transformation law of conformal blocks.
In our normalization \eqref{normalization} the blocks are thus computed as:
\be
 G^s_{\Delta,J}(w,\wb) = (-1)^J (w\wb)^{-\Df} \times G_{\Delta,J}(\tfrac{w}{w-1},\tfrac{\wb}{\wb-1})
\ee
where $G$ are standard (global) blocks, for example for identical external operators in $d=2$:
\be
 G_{\Delta,J}^{(d=2)}(W,\Wb) =
 \frac{k_{\Delta-J}(W)k_{\Delta+J}(\Wb)+(W\leftrightarrow \Wb)}{1+\delta_{J,0}}\,. \label{2d block}
\ee
We then transform the first line of eq.~(\ref{B2v subtracted}) to $W,\Wb$ coordinates.

\smallskip

For expanding around $v=1$ the procedure needs to be adapted to deal with 
singularities at $\wb=w$ as visible from eq.~\eqref{around 1}.
Consider an integral the generic form:
\be
 {\bf f} = \iint\limits_{C_-\times C_+}\!\!\frac{dwd\wb}{(2\pi i)^2}\,
\frac{f(w,\wb)}{(\wb-w)^{2q}} \label{generic singular}
\ee
where we will assume that $f(w,\wb)$ is symmetrical in $w\leftrightarrow \wb$ and analytic except
for possible poles at $w,\wb=0,1$.
On the $C_-\times C_+$ contour the denominator $1/(\wb-w)^2$ causes
no difficulty, however the integral does not manifest the double zeros.
We find that this contour is not very convenient for numerics since it suffers from large cancellations.
Instead, we deform the $\wb$ contour so as to integrate both variables over the left cut
$C_-\times C_-$.  The branches add up to a combination of principal values and discontinuities (see eq.~\eqref{Disc and P})
however all the discontinuities of the kernel give total derivatives thanks to symmetry of $f$,
by the argument given below eq.~\eqref{Disc and P}.
Recall that we should first aim to compute the integral for $\Delta$ and $\Df$
large enough that the integrand vanishes at $0$ and $\infty$ and integration-by-parts can be performed freely;
we can then analytically continue from there.
Thus only the principal value of the kernel contributes. Switching to the $W$ and $\Wb$ variables we
thus have an integral of the form
\be \label{singular funcs 1}
{\bf f}= \int_0^1 \frac{dW d\Wb}{2\pi^2}
\cP \frac{g(W,\Wb)}{(\Wb-W)^{2q}}
\ee
where $g$ includes a Jacobian and dDisc, explicitly,
\be
 g(W,\Wb) = \big[(1{-}W)(1{-}\Wb)\big]^{2q-2}\times \dDisc_s[f]\big(\tfrac{W}{W-1},\tfrac{\Wb}{\Wb-1}\big)\,.
\ee
The principal value prescription $\cP$
means to average over contours going and below the pole.
This is not yet convenient for numerics. 
The trick, we find, is to use symmetry in $W\leftrightarrow \Wb$
to restrict the range to $W<\Wb$, and then replace the $\cP$-distribution by the so-called $+$ distribution
and $\delta$-functions. Note that the $\cP$ prescription is equivalent to restricting the integration to $|W-\Wb|>\epsilon$
and throwing away singular powers and logarithms of $\epsilon$.
The $+$ prescription, on the other hand, is defined by subtracting from the integrand
negative powers of $(W-\Wb)$ in its Laurent series around $W=\Wb$; they are related as follows:
\ba \label{plus}
& \int_0^1 \frac{dW d\Wb}{2\pi^2}
\cP \left[\frac{g(W,\Wb)}{(\Wb-W)^{2q}}\right] 
-\int\limits_{0<W<\Wb<1} \frac{dWd\Wb g(W,\Wb)}{\pi^2(\Wb-W)^{2q}_+}
\\ &\hspace{30mm}=
\sum_{k=0}^{2q-1} \frac{1}{k(2q-1-k)!}\int_0^1 \frac{d\Wb}{\pi^2\Wb^k} \partial_W^{2q-1-k} g(W,\Wb)\Big|_{W=\Wb}\,,
\ea
where the second line is obtained simply by $\cP$-integrating the subtraction implied by the + prescription.
(Here when $k=0$, $\frac{1}{k\Wb^k}$ should be replaced by $\log \Wb$.)
Eq.~\eqref{plus} forms
the basis of our numerical implementation of sum rules of the form \eqref{generic singular}.
For $\a$-type and $\b$-type functionals, one can alternatively compute their actions using a combination of recursion relations and the Lorentzian inversion formula. We detail these methods in appendix~\ref{app:wsrecursion}.

\smallskip

The terms on the second line can be further simplified using integration by parts
when $g(W,\Wb)=g(\Wb,W)$, because for example $\partial_Wg(W,\Wb)_{W=\Wb} = \frac12\partial_{\Wb}g(\Wb,\Wb)$.
(In general, one can always eliminate the $\log \Wb$ term this way.)
We omit details and record only one concrete example relevant for $B_{2,1}'$:\footnote{
  The formula with the least number of derivatives appears to be:
  \be \mbox{eq.~\eqref{plus}}\simeq \sum_{k=1}^q \frac{c_k \partial_W^{2q-2k}g(W,\Wb)_{W=\Wb}}{k(2k-1)\Wb^{2k-1}(2q-2k)!}\ee
  where $c_k$ involves Bernoulli numbers: $c_k=-B_{2k}(2^{2k}-1)=\{-\tfrac12,\tfrac12,-\tfrac32,\tfrac{17}{2},\ldots\}$.}
\be \label{Pvalue for square}
\int_0^1 \frac{dW d\Wb}{2\pi^2}
\cP \left[\frac{g(W,\Wb)}{(\Wb-W)^2}\right] =\int\limits_{0<W<\Wb<1} \frac{dWd\Wb g(W,\Wb)}{\pi^2(\Wb-W)^{2}_+}
-\int_0^1 \frac{d\Wb}{2\pi^2\Wb} g(W,\Wb)\,.
\ee
For values of $\Delta$ where the integrals do not converge, we analytically continue each separately.
The + prescription is straightforward to analytically continue in $\Delta$, using
the following analytic integral of a power of $W$:
\be
 \int_0^{\Wb} \frac{dW\ W^p}{(\Wb-W)^q_+}  =  \frac{(-p)_{q-1}}{(q-1)!}\big(H(q-1)-H(p)\big)\times \Wb^{p+1-q}
\ee
where $H$ is  the harmonic number.  For one-dimensional integrals we simply use eq.~\eqref{power integral}.
Eqs.~(\ref{singular funcs 1}) and (\ref{Pvalue for square}) summarize our
method to compute integrals of the form (\ref{generic singular}), such as
$B_{2,1}$ and $B_{2,1}'$ in eqs.~\eqref{around 1}.
Finally, we evaluate $\widetilde{B}_{2,1}'$ using
\be
 \widetilde{B}_{2,1}' = \Df B_{2,1} +2 B_{2,1}'\,.
\ee
To give two examples, with $d=2$ and $\Df=\frac18$ we find:
\be
 \widetilde{B}_{2,1}'[G^s_{0,1}] = 0.807557\ldots
 \ ,\qquad
 \widetilde{B}_{2,1}'[G^s_{2,2}] =-10.585837\ldots
\ee

\subsection{Action of $\widetilde{B}_{2,v}$ on large-twist two-dimensional blocks}\label{app:heavy}
In this appendix we study analytically the limit  $\Delta-J\gg 1$ in $d=2$.
Because heavy blocks decay exponentially like $e^{-2\Delta/\sqrt{\wb}}$,
the $B_{2,v}$ functionals are then saturated by  the region $w,\wb\sim \Delta^2$: the u-channel Regge limit.

\smallskip

To study this limit we rescale $w\mapsto \wb t$ and take $\wb\to -\infty$.
The $B_{2,v}$ functional in eq.~\eqref{eq:B2v dDisc app} becomes simply
\be
 \lim\limits_{\Delta-J\gg 1} B_{2,v}[G^s_{\Delta,J}]
 = \int_0^{t_{\rm max}} 
 \frac{dt(1- t^2)}{\pi^2 t\sqrt{(1+t)^2-4t v}}
 \int_{-\infty}^0 \frac{d\wb}{\wb^2} \dDisc[G^s_{\Delta,J}(t\wb,\wb)]
\ee
where $t_{\rm max}$ is the smallest of the roots of the square root denominator
(or either of its complex roots when $v<1$).
The integral will be dominated by $\wb\sim -1/\Delta^2$, where the blocks simplify.
In $d=2$, the (global) blocks have the scaling limit \cite{Fitzpatrick:2012yx}:
\be
 G^s_{J,\Delta}(w,\wb) \to 4\frac{\Gamma(2h)}{\Gamma(h)^2}\frac{\Gamma(2\hb)}{\Gamma(\hb)^2}
 (w\wb)^{-\Df} 
 \big(
K_0(2\bar{h}/\sqrt{-w})K_0(2h/\sqrt{-\wb})+(h\leftrightarrow\bar{h})\big)
\label{heavy block}
\ee
where $h=\frac{\Delta-J}{2}$, $\bar{h}=\frac{\Delta+J}{2}$.
Note that in principle we could use the Stirling approximation to simplify the $\Gamma$-functions, but
this particular combination of $\Gamma$ factors occurs frequently and we prefer to keep it unexpanded.
In fact, we will find convenient to divide by the following overall shadow-symmetric combination of $\Gamma$-functions:
\ba \label{C factor}
 C^{\Df}_{\Delta,J} &\equiv
 \big(1 + \delta_{J,0}\big)
\frac{\Delta^2-J^2}{4}
 \frac{\Gamma\big(\tfrac{\Delta-J}{2}\big)^2}{\Gamma(\Delta-J)}
 \frac{\Gamma\big(\tfrac{\Delta+J}{2}\big)^2}{\Gamma(\Delta+J)}\times
 \\&\qquad\times
\Gamma\big(\tfrac{2\Df+\Delta-J}{2}\big)\Gamma\big(\tfrac{2\Df+2-\Delta-J}{2}\big)
\Gamma\big(\tfrac{2\Df+\Delta+J}{2}\big)\Gamma\big(\tfrac{2\Df+2-\Delta+J}{2}\big)\,,
\\
\big(C^{\Df}_{\Delta,J}\big)^{-1}&\sim 2\sin\left(\frac{\Delta-J-2\Df}{2}\right)^2 \frac{4^{\Delta-2}}{\pi^3(1+\delta_{0,J})}\left(\frac{\Delta^2-J^2}{4}\right)^{-\frac12-2\Df}\,.
\ea
The idea is that by respecting the shadow symmetry, we will ensure that subleading corrections
proceed in inverse powers of Casimir invariants rather than simply $1/\Delta$, hopefully
making the formula more accurate.
Indeed find empirically that dividing by $C^{\Df}_{\Delta,J}$ works better than dividing by the second line
at moderate values of $\Delta$.

\smallskip

The radial integral of the blocks in eq.~(\ref{heavy block}) gives the following Bessel moment:
\be
 \int_0^\infty \frac{d\wb}{\wb} \wb^{-p} K_0(2a/\sqrt{\wb})K_0(2b/\sqrt{\wb}) =
 \frac{\Gamma(p)^4}{4\Gamma(2p)}a^{-2p}\ {}_2F_1(p,p,2p,1-\tfrac{b^2}{a^2}),
\ee
and, dividing by the limit of eq.~\eqref{C factor}, we find:
\ba
\label{2d heavy}
\lim\limits_{\Delta-J\gg 1} B_{2,v}[G^{s(d=2)}_{\Delta,J}] &=
\frac{2}{C^{\Df}_{\Delta,J}}
\frac{\Gamma(p)^4}{\Gamma(2p)}
\int_0^{t_{\rm max}}
\frac{dt(1- t)}{t\sqrt{(1+t)^2-4t v}}
\frac{1+t}{\sqrt{t}}t^{\frac{p}{2}}\times
\\
&\qquad \times\left[\xi^p \ {}_2F_1(p,p,2p,1-t\xi^2) + (\xi\mapsto \xi^{-1})\right] 
\ea
where $p=2\Df+1$ and $\xi=\frac{\Delta-J}{\Delta+J}$.  In fact, all functionals derived from $B_{2,v}$
are given with similar formulas just with different $t$-dependent kernels, for example
\be
B_{2,1}:\ \frac{dt(1- t)}{t\sqrt{(1+t)^2-4t v}} \mapsto \frac{dt}{t}\,,\quad
B_{2,1}': 
\frac{dt(1- t)}{t\sqrt{(1+t)^2-4t v}} \mapsto \left(\frac{2dt}{(1-t)^2_+} -\delta(t-1)\right)\
\ee
with the $\delta$-function understood to have unit weight on the integration range $0\leq t\leq t_{\rm max}=1$.
We expect corrections to eq.~(\ref{2d heavy}) to be suppressed by inverse powers of Casimir invariants.
In particular, the formula should be reliable for any large-twist block, no matter its spin.

% !TEX root = ../main.tex

\newcommand\SO{\mathrm{SO}}
\newcommand\scD{\mathscr{D}}

\section{Recursion relations for Polyakov-Regge blocks from weight-shifting}\label{app:wsrecursion}

Many sum rules in this paper amount to decomposing Polyakov-Regge blocks into double-traces.
Here we describe recursion relations for Polyakov-Regge blocks using weight-shifting operators that are well-suited for computing their decompositions into double-traces. One virtue of this approach is that it can be straightforwardly extended to spinning operators. Below, we will explain the general method and discuss an example with external scalar primaries in detail.

Recall that a Polyakov-Regge block for a four-point function of scalars with dimensions $\De_\f$ satisfies
\begin{align}
\label{eq:polaykovreproducedone}
P_{\Delta, J}^{s|u} & = 
   G_{\Delta, J}^s -   \sum\limits_{n,\ell=0}^{\infty}\left\{\alpha^s_{n,\ell}[  G_{\Delta, J}^s  ]\,G^s_{\Delta_{n,\ell},\ell} +\beta^s_{n,\ell}[  G_{\Delta, J}^s]\,\partial_{\Delta}G^s_{\Delta_{n,\ell},\ell} \right\} \\
&= \sum\limits_{n,\ell=0}^{\infty}\left\{\alpha^t_{n,\ell}[ G_{\Delta, J}^s]\,G^t_{\Delta_{n,\ell},\ell} +\beta^t_{n,\ell}[ G_{\Delta, J}^s]\,\partial_{\Delta}G^t_{\Delta_{n,\ell},\ell} \right\},
\label{eq:polaykovreproducedtwo}
\end{align}
where $\De_{n,\ell}=2\De_\f+2n+\ell$. Thus, the expansion of $P_{\De,J}^{s|u}$ into double traces encodes the action of the functionals $\a_{n,\ell}^{s,t},\b_{n,\ell}^{s,t}$ on an individual conformal block $G_{\De,J}^s$. For a four-point function of more general operators, there are natural generalizations of (\ref{eq:polaykovreproducedone}) and (\ref{eq:polaykovreproducedtwo}), with functionals for each double-trace operator in the $s$- and $t$-channels. 

\smallskip

The Polyakov-Regge block $P_{\De,J}^{s|u}$ is uniquely determined by the properties listed in section~\ref{ssec:polyakovreggebootstrap}: (i) superboundedness in the $u$-channel, (ii) Euclidean single-valuedness, (iii) double-discontinuities satisfying
\ba 
{\rm dDisc}_s[P_{\Delta, J}^{s|u}]  & = {\rm dDisc}_s[G^s_{\Delta, J}]\,, \quad {\rm dDisc}_t[P_{\Delta, J}^{s|u}]  = 0\,. \ea
We can use these properties to derive relations between Polyakov-Regge blocks using weight-shifting operators.

\smallskip

Recall that a weight-shifting operator $\cD_A$ is a conformally-covariant differential operator transforming in a finite-dimensional representation $W$ of the conformal group $\SO(d,2)$ \cite{Karateev:2017jgd}. The index $A$ is an index for $W$: $A=1,\dots,\dim W$. Each weight-shifting operator has an associated weight $w_\cD$ for the Cartan subgroup of $\SO(d,2)$ acting on $W$. Given a primary $\cO$ with weight $w_\cO=(\De_\cO,J_\cO,\dots)$, the object $\cD_A \cO$ transforms like a primary with weight $w_\cO+w_\cD$ and an extra index $A$ under $\SO(d,2)$.

Let $\cD_{A,1}$ be a weight-shifting operator in the representation $W$ acting at the point $x_1$, and $\cD'^{A}_3$ be a weight-shifting operator in the dual representation $W^*$ acting at point $x_3$. For simplicity, we focus on the case where $\cD_{A,1}$ and $\cD'^{A}_3$ only shift the conformal dimension of the operator they act on
\begin{align}
w_\cD &= (\de_1,0,\dots) \nn\\
w_{\cD'} &= (\de_3,0,\dots).
\end{align}
However, our analysis easily generalizes to arbitrary weights.

\smallskip

Consider the {\it conformally invariant} bilocal operator $\scD=\cD_{A,1} \cD^{A}_3$. Acting on a four-point function $\cG$ of scalars with dimensions $(\De_1,\De_2,\De_3,\De_4)$, the object $\scD \cG$ transforms like a four-point function of scalars with dimensions
\be
(\De_1',\De_2',\De_3',\De_4') = (\De_1+\de_1,\De_2,\De_3+\de_3,\De_4).
\ee

Consider a Polyakov-Regge block $P_{\De,J}^{s|u;\De_i}$ for a four-point function of scalars with dimensions $\De_i$.
The key observation is that $P'=\scD P_{\De,J}^{s|u;\De_i}$ satisfies almost all the conditions required to be a Polyakov-Regge block for scalars with dimensions $\De_i'$. Firstly, $P'$ is conformally-invariant because $\scD$ is conformally-invariant. $P'$ is Euclidean single-valued because weight-shifting operators are differential operators whose coefficients are single-valued functions. Finally, $P'$ is $u$-channel superbounded for the following reason. Note that $\scD$ commutes with the conformal generators $L_1+L_3$ acting simultaneously on points $1$ and $3$. Consequently, it commutes with the conformal Casimirs in the $u$-channel. These Casimirs can be used to measure the effective spin in the $u$-channel Regge limit, and thus the Regge spin of $P'$ is the same as that of $P_{\De,J}^{s|u;\De_i}$.

\smallskip

However, $P'$ does not quite have the correct dDisc to be Polyakov-Regge block. Acting on a conformal block, $\scD$ gives a finite linear combination of new conformal blocks
\begin{align}
\label{eq:scdong}
\scD G^{s,\De_i}_{\De,J} &= \sum_{m,j} r^{\De_i}_{m,j}(\De,J) G^{s,\De_i'}_{\De+m,J+j}.
\end{align}
The quantum numbers $(\De+m,J+j)$ appearing on the right-hand side are the quantum numbers of traceless-symmetric tnesor primaries appearing in the tensor product $W\otimes V_{\De,J}$, where $V_{\De,J}$ is the generalized Verma module associated to a primary with dimension $\De$ and spin $J$. In particular, $m$ and $j$ range over a finite set of values. The coefficients in (\ref{eq:scdong}) can be straightforwardly computed by applying crossing transformations for weight-shifting operators as in \cite{Karateev:2017jgd,Karateev:2018oml}.

\smallskip

The dDisc of $\scD G^{s,\De_i}_{\De,J}$ is the same as that of the right-hand side of (\ref{eq:scdong}). (Note that $\scD$ takes double-trace blocks to double-trace blocks in the same channel.)  However, the dDisc of the right-hand side also agrees with the dDisc of a sum of Polyakov-Regge blocks. It follows that
\begin{align}
\label{eq:polyakovrecursion}
\scD P^{s|u;\De_i}_{\De,J} &= \sum_{m,j} r^{\De_i}_{m,j}(\De,J) P^{s|u;\De_i'}_{\De+m,J+j}.
\end{align}
This relation generalizes straightforwardly to Polyakov-Regge blocks of operators in arbitrary external and internal conformal representations. The relation (\ref{eq:polyakovrecursion}) is useful as a recursion relation in spin. We can choose $W$ such that the sum on the right-hand side comtains a single term with maximal spin $J+j$. Solving for this term determines the spin $J+j$ Polyakov-Regge block in terms of lower-spin Polyakov-Regge blocks.

\smallskip

Now consider the decomposition of $P^{s|u,\De_i}_{\De_0,J_0}$ into s-channel conformal blocks. It is convenient to write this in terms of a contour integral
\begin{align}
\label{eq:polyakovschandecomp}
P^{s|u;\De_i}_{\De_0,J_0} &= \sum_{J=0}^{J_0} \int_{\frac d 2-i\oo}^{\frac d 2+i\oo} \frac{d\De}{2\pi i} B^{s;\De_i}_{\De_0,J_0}(\De,J) G^{s,\De_i}_{\De,J}.
\end{align}
Note that the sum over $J$ truncates at $J_0$ because the Polyakov-Regge block is the sum of a Witten spin-$J_0$ exchange and lower-spin contact diagrams.
The values of $\a$-type and $\b$-type functionals are given by residues of $B^{s;\De_i}_{\De_0,J_0}(\De,J)$ at double-trace locations. Inserting this contour integral into (\ref{eq:polyakovrecursion}), and equating the coefficients of conformal blocks on both sides we find
\begin{align}
\sum_{m,j} B^{s;\De_i}_{\De_0,J_0}(\De-m,J-j)  r^{\De_i}_{m,j}(\De-m,J-j) &= \sum_{m,j} r^{\De_i}_{m,j}(\De_0,J_0) B^{s;\De_i'}_{\De_0+m,J_0+j}(\De,J).
\end{align}
Taking residues of both sides, we obtain recursion relations for $s$-channel double-trace functionals. 

\smallskip

Similarly, we can insert the decomposition of Polyakov-Regge blocks into $t$-channel blocks
\begin{align}
P^{s|u;\De_i}_{\De_0,J_0} &= \sum_{J=0}^\oo \int_{\frac d 2-i\oo}^{\frac d 2+i\oo} \frac{d\De}{2\pi i} B^{t;\De_i}_{\De_0,J_0}(\De,J) G^{t,\De_i}_{\De,J}.
\end{align}
The $t$-channel blocks satisfy their own analog of (\ref{eq:scdong})
\begin{align}
\label{eq:tcdong}
\scD G^{t,\De_i}_{\De,J} &= \sum_{m,j} r'^{\De_i}_{m,j}(\De,J) G^{t,\De_i'}_{\De+m,J+j}.
\end{align}
Plugging this into (\ref{eq:polyakovrecursion}) gives
\begin{align}
\sum_{m,j} B^{t,\De_i}_{\De_0,J_0}(\De-m,J-j)  r'^{\De_i}_{m,j}(\De-m,J-j) &= \sum_{m,j} r_{m,j}^{\De_i}(\De_0,J_0) B^{t;\De_i'}_{\De_0+m,J_0+j}(\De,J) 
\end{align}
Taking residues of both sides, we obtain recursion relations for $t$-channel double-trace functionals.

\subsection{An example}

Let us describe an example of the above methods in more detail. The simplest weight-shifting operator that only shifts dimensions is the embedding-space vector 
\begin{align}
X^A = (X^+,X^-,X^\mu)=(1,x^2,x^\mu)\in \mathbb{R}^{2,d}.
\end{align}
This can be interpreted as a 0-th order differential operator in the vector representation $W=\Box$, acting at $x$ and changing the dimension of the operator there by $-1$. We can define $\scD=-2X_1\.X_3=x_{13}^2$, which changes the dimensions by
\begin{align}
\scD : (\De_1,\De_2,\De_3,\De_4) &\to (\De_1-1,\De_2,\De_3-1,\De_4)=(\De_1',\De_2',\De_3',\De_4')
\end{align}
The tensor product $\Box\otimes V_{\De,J}$ contains traceless symmetric tensor representations $V_{\De+m,J+j}$ with $(m,j)\in\{(1,0),(-1,0),(0,1),(0,-1)\}$. The coefficients in (\ref{eq:scdong}) are \cite{Dolan:2011dv,Karateev:2017jgd}
\begin{align}
&r^{\De_i}_{-1,0}(\De,J) = 1, \nn\\
&r^{\De_i}_{1,0}(\De,J) = \nn\\
&\  \frac{(\Delta -1) (\Delta +2-d) (\Delta -\Delta _{12}+J) (\Delta -\Delta _{34}+J) (\Delta -\Delta _{12}-J+2-d) (\Delta -\Delta _{34}-J+2-d)}{16 (\frac{2-d}{2}+\Delta -1) (\frac{2-d}{2}+\Delta ) (\Delta +J-1) (\Delta +J) (\Delta -J+1-d) (\Delta -J+2-d)}, \nn\\
&r^{\De_i}_{0,1}(\De,J) =\frac{(\Delta -\Delta _{12}+J) (\Delta -\Delta _{34}+J)}{4 (\Delta +J-1) (\Delta +J)},
\nn\\
&r^{\De_i}_{0,-1}(\De,J) = \frac{J (d+J-3) (\Delta -\Delta _{12}-J+2-d) (\Delta -\Delta _{34}-J+2-d)}{4 (\frac{d-2}{2}+J-1) (\frac{d-2}{2}+J) (\Delta -J+1-d) (\Delta -J+2-d)}.
\end{align}
The $t$-channel decomposition coefficients $r'^{\De_i}_{m,j}(\De,J)$ can be obtained by swapping $1\leftrightarrow 3$ in (\ref{eq:scdong}), which gives
\begin{align}
r'^{\De_i}_{m,j}(\De,J) &= \left.r^{\De_i}_{m,j}(\De,J)\right|_{\De_1\leftrightarrow \De_3}.
\end{align}

Because there is a single term with maximal spin $J+1$ appearing on the right-hand side of (\ref{eq:polyakovrecursion}), we can replace $J\to J-1$ and obtain a recursion relation for $P^{s|u;\De_i}_{\De,J}$ in terms of Polyakov-Regge blocks with lower spins $J-1$ and $J-2$. The Polyakov-Regge block with spin $J=0$ is simply a scalar Witten exchange diagram. Furthermore, note that $r^{\De_i}_{0,-1}(\De,J)$ is proportional to $J$, which reflects the special case $\Box \otimes V_{\De,0} \ni V_{\De\pm 1,0}, V_{0,1}$. Consequently $J=0$ is a sufficient base case for the recursion (\ref{eq:polyakovrecursion}) to determine all Polyakov-Regge blocks with higher spin.\footnote{Similar techniques were used for computing Witten diagrams in \cite{Zhou:2020ptb}.}

\smallskip

One complication of using the operator $\scD=-2X_1\.X_3$ is that it changes the quantum numbers of external operators. Similarly, we can consider $\scD'=\cD^{+0}_{1,A} \cD^{+0,A}_3$, where $\cD^{+0}$ is the dimension-raising weight-shifting operator defined in \cite{Karateev:2017jgd}. This gives a complementary recursion relation that changes the external dimensions in the opposite way
\be
\scD':(\De_1,\De_2,\De_3,\De_4) \to (\De_1+1,\De_2,\De_3+1,\De_4).
\ee
Combining the two operators to form $\scD'\scD$, we obtain a recursion relation for the spin $J$ Polyakov-Regge block in terms of Polyakov-Regge blocks with spins $\{J-1,J-2,J-3,J-4\}$ with the same external dimensions.

\subsection{Relating to contact diagrams and the inversion formula}

The above procedure for computing Polyakov-Regge blocks leads to simple closed-form formulas for the action of $s$-channel functionals on $s$-channel blocks, following those of \cite{Mazac:2019shk}. The reason is that the base case of the recursion $P_{\De_0,0}^{s|u;\De_i}$ is an $s$-channel Witten scalar exchange, whose expansion in $s$-channel blocks is simple.

\smallskip

To compute the action of $t$-channel double-trace functionals on $s$-channel blocks, it is useful to relate them to the Lorentzian inversion formula. For convenience, we instead discuss the action of $s$-channel double-trace functionals on $t$-channel blocks, which is entirely equivalent. We would thus like to decompose $P^{t|u}_{\De,J}$ into $s$-channel double-traces.

\smallskip

Note that the $s$-channel Lorentzian inversion of a $t$-channel block computes the decomposition of $P^{t|s}_{\De,J}$ --- the Polyakov-Regge block that is superbounded in the $s$ channel --- into $s$-channel double-traces. Specifically let $\mathcal{I}_s^{(t)}(\De,J,\De',J)$ be the $s$-channel Lorentzian inversion of $G^t_{\De',J'}$. The decomposition of $P^{t|s}_{\De',J'}$ into double-traces is given by
\begin{align}
\label{eq:inversionresidues}
\mathtt{coefficient}(P^{t|s}_{\De',J'},\ptl_\De G_{\De_{n,\ell},\ell}^s) &= - \mathop{\mathrm{dRes}}_{\De=2\De_\f+2n+\ell} \mathcal{I}_s^{(t)}(\De,\ell,\De',J) 
\nn\\
\mathtt{coefficient}(P^{t|s}_{\De',J'},G_{\De_{n,\ell},\ell}^s) &= - \mathop{\mathrm{Res}}_{\De=2\De_\f+2n+\ell} \mathcal{I}_s^{(t)}(\De,\ell,\De',J) .
\end{align}
The Lorentzian inversion of 2d and 4d blocks is known exactly \cite{Liu:2018jhs}, and in 3-dimensions it can be computed efficiently numerically by decomposing 3d blocks into 2d blocks \cite{Hogervorst:2016hal,Albayrak:2019gnz,Liu:2020tpf,Caron-Huot:2020ouj}.

\smallskip

As discussed in section~\ref{ssec:polyakovreggebootstrap}, the Polyakov-Regge blocks $P_{\De,J}^{t|u}$ and $P_{\De,J}^{t|s}$ that are bounded in the $u$ and $s$ channels, respectively, differ by contact diagrams. This follows from the fact that
\begin{align}
P_{\De,J}^{t|u} &= W^t_{\De,J} +  \mathcal{C}^{t|u}_{\De,J},
\end{align}
where $\mathcal{C}^{t|u}_{\De,J}$ are contact interactions with spins $<J$ in all channels. Meanwhile, we have
\begin{align}
P_{\De,J}^{t|s} &= (-1)^J \left.P^{t|u}_{\De,J}\right|_{2\leftrightarrow 3} \nn\\
W_{\De,J}^{t} &= (-1)^J \left.W^{t}_{\De,J}\right|_{2\leftrightarrow 3},
\end{align}
where the first line follows from uniqueness of Polyakov-Regge blocks satisfying properties (i), (ii), and (iii), and the second line is a symmetry of Witten diagrams. Consequently,
\begin{align}
P_{\De,J}^{t|u} - P_{\De,J}^{t|s} &= \mathcal{C}^{t|u}_{\De,J} - (-1)^J \left.\mathcal{C}^{t|u}_{\De,J}\right|_{2\leftrightarrow 3}  = 2\Pi^t_{(-1)^{J+1}}\left[\mathcal{C}^{t|u}_{\De,J}\right],
\end{align}
where $\Pi^t_\pm[\cdot]$ denotes projection onto spin parity $\pm$ in the $t$-channel. To summarize, to compute the action of $s$-channel functionals on $t$-channel blocks, one can use the above recursion relations (or some other method) to obtain the contact diagram $\cC^{t|u}_{\De,J}$, which can then be decomposed into double-traces.  Because $\cC_{\De,J}^{t|u}$ has spin less than $J$ in all channels, it only contributes to $\a_{n,\ell}^s[G^t_{\De,J}]$ and $\b_{n,\ell}^s[G^t_{\De,J}]$ for $\ell<J$. The remaining contribution to $\a_{n,\ell}^s[G^t_{\De,J}]$ and $\b_{n,\ell}^s[G^t_{\De,J}]$ is given by (\ref{eq:inversionresidues}).

% !TEX root = ../main.tex

\section{Numerical implementation of the twist gap problem}
\label{app:extremalfunctionalnumerics}

In this appendix, we describe details of our numerical implementation of the twist-gap problem for section~\ref{sec:numericaltwistgap}. To test a spin-2 gap $\De_2$ for feasibility, we solve the following problem:
\be
\label{eq:ouroptimizationproblem}
\begin{array}{l}
\textrm{maximize $\w[G^s_{0,0}]$ such that}\\
\w \in \mathrm{Span}\{\w^\mathrm{deriv}_{m,n} \textrm{ with $m+n$ odd and $m+n\leq \Lambda$} \},\\
\w[G^s_{\De,J}] \geq 0 \quad \textrm{ for $J\neq 2$ even and $\De,J$ satisfying unitarity}, \\
\w[G^s_{\De,2}] \geq 0 \quad \textrm{ for $\De>\De_2$}, \\
\left.\w[\ptl_\De G^s_{\De,2}]\right|_{\De=\De_2} = 1.
\end{array}
\ee
The gap $\De_2$ is disallowed if $\max_\w \w[G^s_{0,0}]$ is positive, and allowed (at derivative order $\Lambda$) if $\max_\w \w[G^s_{0,0}]$ is negative. For $\Lambda=15,19,23,27,35$, we perform bisection in $\De_2$ to find the threshold between allowed and disallowed to an accuracy of $10^{-10}$, and determine the functional $\w$ at that threshold. (We use less accuracy at higher $\Lambda$, see below.) We use the program \texttt{scalar\_blocks}\footnote{\url{https://gitlab.com/bootstrapcollaboration/scalar\_blocks}} to compute conformal blocks and the semidefinite program solver \texttt{SDPB} \cite{Simmons-Duffin:2015qma,Landry:2019qug} to solve the optimization problem.

\smallskip

When testing feasibility, an alternative approach is to include $\w[G^s_{0,0}]\geq 0$ among the positivity constraints and use $0$ as the objective function. This approach can be faster when one simply wants to know the existence or nonexistence of a functional and isn't interested in the specific functional. By contrast, an optimization problem with nonzero objective function generically has a unique solution, and leads to more stable results. Furthermore, we found that the shape of the optimal functional $\w$ did not change much as $\De_2$ was varied close to the threshold between allowed and disallowed. Even for high derivative orders $\Lambda=43,51,59$, we checked that bisecting to within $10^{-4}$ of the threshold was sufficient to plot the extremal functional to higher resolution than the plots in this paper.

\newpage

%%%%%%%%%%%%%%%%%%%%%%%%%%%%%%%%
\bibliographystyle{./aux/ytphys}
\bibliography{./aux/refs}

\providecommand{\href}[2]{#2}\begingroup\raggedright\begin{thebibliography}{100}

\bibitem{Rattazzi:2008pe}
R.~Rattazzi, V.~S. Rychkov, E.~Tonni, and A.~Vichi, ``{Bounding scalar operator
  dimensions in 4D CFT},''
  \href{http://dx.doi.org/10.1088/1126-6708/2008/12/031}{{\em JHEP} {\bfseries
  12} (2008) 031},
\href{http://arxiv.org/abs/0807.0004}{{\ttfamily arXiv:0807.0004 [hep-th]}}.
%%CITATION = ARXIV:0807.0004;%%.

\bibitem{Poland:2016chs}
D.~Poland and D.~Simmons-Duffin, ``{The conformal bootstrap},''
  \href{http://dx.doi.org/10.1038/nphys3761}{{\em Nature Phys.} {\bfseries 12}
  no.~6, (2016) 535--539}.

\bibitem{Poland:2018epd}
D.~Poland, S.~Rychkov, and A.~Vichi, ``{The Conformal Bootstrap: Theory,
  Numerical Techniques, and Applications},''
  \href{http://dx.doi.org/10.1103/RevModPhys.91.015002}{{\em Rev. Mod. Phys.}
  {\bfseries 91} (2019) 015002},
\href{http://arxiv.org/abs/1805.04405}{{\ttfamily arXiv:1805.04405 [hep-th]}}.
%%CITATION = ARXIV:1805.04405;%%.

\bibitem{Chester:2019wfx}
S.~M. Chester, ``{Weizmann Lectures on the Numerical Conformal Bootstrap},''
  \href{http://arxiv.org/abs/1907.05147}{{\ttfamily arXiv:1907.05147
  [hep-th]}}.

\bibitem{Afkhami-Jeddi:2019zci}
N.~Afkhami-Jeddi, T.~Hartman, and A.~Tajdini, ``{Fast Conformal Bootstrap and
  Constraints on 3d Gravity},''
  \href{http://dx.doi.org/10.1007/JHEP05(2019)087}{{\em JHEP} {\bfseries 05}
  (2019) 087}, \href{http://arxiv.org/abs/1903.06272}{{\ttfamily
  arXiv:1903.06272 [hep-th]}}.

\bibitem{Homrich:2019cbt}
A.~Homrich, J.~Penedones, J.~Toledo, B.~C. van Rees, and P.~Vieira, ``{The
  S-matrix Bootstrap IV: Multiple Amplitudes},''
  \href{http://dx.doi.org/10.1007/JHEP11(2019)076}{{\em JHEP} {\bfseries 11}
  (2019) 076}, \href{http://arxiv.org/abs/1905.06905}{{\ttfamily
  arXiv:1905.06905 [hep-th]}}.

\bibitem{Agmon:2019imm}
N.~B. Agmon, S.~M. Chester, and S.~S. Pufu, ``{The M-theory Archipelago},''
  \href{http://dx.doi.org/10.1007/JHEP02(2020)010}{{\em JHEP} {\bfseries 02}
  (2020) 010}, \href{http://arxiv.org/abs/1907.13222}{{\ttfamily
  arXiv:1907.13222 [hep-th]}}.

\bibitem{Lin:2019vgi}
Y.-H. Lin, D.~Meltzer, S.-H. Shao, and A.~Stergiou, ``{Bounds on Triangle
  Anomalies in (3+1)d},''
  \href{http://dx.doi.org/10.1103/PhysRevD.101.125007}{{\em Phys. Rev. D}
  {\bfseries 101} no.~12, (2020) 125007},
  \href{http://arxiv.org/abs/1909.11676}{{\ttfamily arXiv:1909.11676
  [hep-th]}}.

\bibitem{Rong:2019qer}
J.~Rong and N.~Su, ``{Bootstrapping the $\mathcal{N}=1$ Wess-Zumino models in
  three dimensions},'' \href{http://arxiv.org/abs/1910.08578}{{\ttfamily
  arXiv:1910.08578 [hep-th]}}.

\bibitem{Reehorst:2019pzi}
M.~Reehorst, E.~Trevisani, and A.~Vichi, ``{Mixed Scalar-Current bootstrap in
  three dimensions},'' \href{http://arxiv.org/abs/1911.05747}{{\ttfamily
  arXiv:1911.05747 [hep-th]}}.

\bibitem{Chester:2019ifh}
S.~M. Chester, W.~Landry, J.~Liu, D.~Poland, D.~Simmons-Duffin, N.~Su, and
  A.~Vichi, ``{Carving out OPE space and precise $O(2)$ model critical
  exponents},'' \href{http://dx.doi.org/10.1007/JHEP06(2020)142}{{\em JHEP}
  {\bfseries 06} (2020) 142}, \href{http://arxiv.org/abs/1912.03324}{{\ttfamily
  arXiv:1912.03324 [hep-th]}}.

\bibitem{Henriksson:2020fqi}
J.~Henriksson, S.~R. Kousvos, and A.~Stergiou, ``{Analytic and Numerical
  Bootstrap of CFTs with $O(m)\times O(n)$ Global Symmetry in 3D},''
  \href{http://arxiv.org/abs/2004.14388}{{\ttfamily arXiv:2004.14388
  [hep-th]}}.

\bibitem{He:2020azu}
Y.-C. He, J.~Rong, and N.~Su, ``{Non-Wilson-Fisher kinks of $O(N)$ numerical
  bootstrap: from the deconfined phase transition to a putative new family of
  CFTs},'' \href{http://arxiv.org/abs/2005.04250}{{\ttfamily arXiv:2005.04250
  [hep-th]}}.

\bibitem{Li:2020bnb}
Z.~Li and D.~Poland, ``{Searching for gauge theories with the conformal
  bootstrap},'' \href{http://arxiv.org/abs/2005.01721}{{\ttfamily
  arXiv:2005.01721 [hep-th]}}.

\bibitem{Afkhami-Jeddi:2020hde}
N.~Afkhami-Jeddi, H.~Cohn, T.~Hartman, D.~de~Laat, and A.~Tajdini,
  ``{High-dimensional sphere packing and the modular bootstrap},''
  \href{http://arxiv.org/abs/2006.02560}{{\ttfamily arXiv:2006.02560
  [hep-th]}}.

\bibitem{Bonifacio:2020xoc}
J.~Bonifacio and K.~Hinterbichler, ``{Bootstrap Bounds on Closed Einstein
  Manifolds},'' \href{http://arxiv.org/abs/2007.10337}{{\ttfamily
  arXiv:2007.10337 [hep-th]}}.

\bibitem{Maldacena:2015waa}
J.~Maldacena, S.~H. Shenker, and D.~Stanford, ``{A bound on chaos},''
  \href{http://dx.doi.org/10.1007/JHEP08(2016)106}{{\em JHEP} {\bfseries 08}
  (2016) 106}, \href{http://arxiv.org/abs/1503.01409}{{\ttfamily
  arXiv:1503.01409 [hep-th]}}.

\bibitem{Hartman:2016lgu}
T.~Hartman, S.~Kundu, and A.~Tajdini, ``{Averaged Null Energy Condition from
  Causality},'' \href{http://dx.doi.org/10.1007/JHEP07(2017)066}{{\em JHEP}
  {\bfseries 07} (2017) 066}, \href{http://arxiv.org/abs/1610.05308}{{\ttfamily
  arXiv:1610.05308 [hep-th]}}.

\bibitem{Alday:2007mf}
L.~F. Alday and J.~M. Maldacena, ``{Comments on operators with large spin},''
  \href{http://dx.doi.org/10.1088/1126-6708/2007/11/019}{{\em JHEP} {\bfseries
  11} (2007) 019},
\href{http://arxiv.org/abs/0708.0672}{{\ttfamily arXiv:0708.0672 [hep-th]}}.
%%CITATION = ARXIV:0708.0672;%%.

\bibitem{Fitzpatrick:2012yx}
A.~L. Fitzpatrick, J.~Kaplan, D.~Poland, and D.~Simmons-Duffin, ``{The Analytic
  Bootstrap and AdS Superhorizon Locality},''
  \href{http://dx.doi.org/10.1007/JHEP12(2013)004}{{\em JHEP} {\bfseries 12}
  (2013) 004},
\href{http://arxiv.org/abs/1212.3616}{{\ttfamily arXiv:1212.3616 [hep-th]}}.
%%CITATION = ARXIV:1212.3616;%%.

\bibitem{Komargodski:2012ek}
Z.~Komargodski and A.~Zhiboedov, ``{Convexity and Liberation at Large Spin},''
  \href{http://dx.doi.org/10.1007/JHEP11(2013)140}{{\em JHEP} {\bfseries 11}
  (2013) 140},
\href{http://arxiv.org/abs/1212.4103}{{\ttfamily arXiv:1212.4103 [hep-th]}}.
%%CITATION = ARXIV:1212.4103;%%.

\bibitem{Alday:2013cwa}
L.~F. Alday and A.~Bissi, ``{Higher-spin correlators},''
  \href{http://dx.doi.org/10.1007/JHEP10(2013)202}{{\em JHEP} {\bfseries 10}
  (2013) 202}, \href{http://arxiv.org/abs/1305.4604}{{\ttfamily arXiv:1305.4604
  [hep-th]}}.

\bibitem{Alday:2015eya}
L.~F. Alday, A.~Bissi, and T.~Lukowski, ``{Large spin systematics in CFT},''
  \href{http://dx.doi.org/10.1007/JHEP11(2015)101}{{\em JHEP} {\bfseries 11}
  (2015) 101},
\href{http://arxiv.org/abs/1502.07707}{{\ttfamily arXiv:1502.07707 [hep-th]}}.
%%CITATION = ARXIV:1502.07707;%%.

\bibitem{Alday:2015ewa}
L.~F. Alday and A.~Zhiboedov, ``{An Algebraic Approach to the Analytic
  Bootstrap},'' \href{http://dx.doi.org/10.1007/JHEP04(2017)157}{{\em JHEP}
  {\bfseries 04} (2017) 157},
\href{http://arxiv.org/abs/1510.08091}{{\ttfamily arXiv:1510.08091 [hep-th]}}.
%%CITATION = ARXIV:1510.08091;%%.

\bibitem{Alday:2016njk}
L.~F. Alday, ``{Large Spin Perturbation Theory for Conformal Field Theories},''
  \href{http://dx.doi.org/10.1103/PhysRevLett.119.111601}{{\em Phys. Rev.
  Lett.} {\bfseries 119} no.~11, (2017) 111601},
\href{http://arxiv.org/abs/1611.01500}{{\ttfamily arXiv:1611.01500 [hep-th]}}.
%%CITATION = ARXIV:1611.01500;%%.

\bibitem{Alday:2016jfr}
L.~F. Alday, ``{Solving CFTs with Weakly Broken Higher Spin Symmetry},''
  \href{http://dx.doi.org/10.1007/JHEP10(2017)161}{{\em JHEP} {\bfseries 10}
  (2017) 161},
\href{http://arxiv.org/abs/1612.00696}{{\ttfamily arXiv:1612.00696 [hep-th]}}.
%%CITATION = ARXIV:1612.00696;%%.

\bibitem{Simmons-Duffin:2016wlq}
D.~Simmons-Duffin, ``{The Lightcone Bootstrap and the Spectrum of the 3d Ising
  CFT},'' \href{http://dx.doi.org/10.1007/JHEP03(2017)086}{{\em JHEP}
  {\bfseries 03} (2017) 086}, \href{http://arxiv.org/abs/1612.08471}{{\ttfamily
  arXiv:1612.08471 [hep-th]}}.

\bibitem{Carmi:2019cub}
D.~Carmi and S.~Caron-Huot, ``{A Conformal Dispersion Relation: Correlations
  from Absorption},''
\href{http://arxiv.org/abs/1910.12123}{{\ttfamily arXiv:1910.12123 [hep-th]}}.
%%CITATION = ARXIV:1910.12123;%%.

\bibitem{Caron-Huot:2017vep}
S.~Caron-Huot, ``{Analyticity in Spin in Conformal Theories},''
  \href{http://dx.doi.org/10.1007/JHEP09(2017)078}{{\em JHEP} {\bfseries 09}
  (2017) 078},
\href{http://arxiv.org/abs/1703.00278}{{\ttfamily arXiv:1703.00278 [hep-th]}}.
%%CITATION = ARXIV:1703.00278;%%.

\bibitem{Simmons-Duffin:2017nub}
D.~Simmons-Duffin, D.~Stanford, and E.~Witten, ``{A spacetime derivation of the
  Lorentzian OPE inversion formula},''
  \href{http://dx.doi.org/10.1007/JHEP07(2018)085}{{\em JHEP} {\bfseries 07}
  (2018) 085},
\href{http://arxiv.org/abs/1711.03816}{{\ttfamily arXiv:1711.03816 [hep-th]}}.
%%CITATION = ARXIV:1711.03816;%%.

\bibitem{Kravchuk:2018htv}
P.~Kravchuk and D.~Simmons-Duffin, ``{Light-ray operators in conformal field
  theory},'' \href{http://dx.doi.org/10.1007/JHEP11(2018)102}{{\em JHEP}
  {\bfseries 11} (2018) 102}, \href{http://arxiv.org/abs/1805.00098}{{\ttfamily
  arXiv:1805.00098 [hep-th]}}.
[,236(2018)].
%%CITATION = ARXIV:1805.00098;%%.

\bibitem{Albayrak:2019gnz}
S.~Albayrak, D.~Meltzer, and D.~Poland, ``{More Analytic Bootstrap:
  Nonperturbative Effects and Fermions},''
  \href{http://dx.doi.org/10.1007/JHEP08(2019)040}{{\em JHEP} {\bfseries 08}
  (2019) 040}, \href{http://arxiv.org/abs/1904.00032}{{\ttfamily
  arXiv:1904.00032 [hep-th]}}.

\bibitem{Liu:2020tpf}
J.~Liu, D.~Meltzer, D.~Poland, and D.~Simmons-Duffin, ``{The Lorentzian
  inversion formula and the spectrum of the 3d O(2) CFT},''
  \href{http://arxiv.org/abs/2007.07914}{{\ttfamily arXiv:2007.07914
  [hep-th]}}.

\bibitem{Caron-Huot:2020ouj}
S.~Caron-Huot, Y.~Gobeil, and Z.~Zahraee, ``{The leading trajectory in the 2+1D
  Ising CFT},'' \href{http://arxiv.org/abs/2007.11647}{{\ttfamily
  arXiv:2007.11647 [hep-th]}}.

\bibitem{Cordova:2017zej}
C.~Cordova, J.~Maldacena, and G.~J. Turiaci, ``{Bounds on OPE Coefficients from
  Interference Effects in the Conformal Collider},''
  \href{http://dx.doi.org/10.1007/JHEP11(2017)032}{{\em JHEP} {\bfseries 11}
  (2017) 032}, \href{http://arxiv.org/abs/1710.03199}{{\ttfamily
  arXiv:1710.03199 [hep-th]}}.

\bibitem{Gillioz:2018kwh}
M.~Gillioz, X.~Lu, and M.~A. Luty, ``{Graviton Scattering and a Sum Rule for
  the c Anomaly in 4D CFT},''
  \href{http://dx.doi.org/10.1007/JHEP09(2018)025}{{\em JHEP} {\bfseries 09}
  (2018) 025}, \href{http://arxiv.org/abs/1801.05807}{{\ttfamily
  arXiv:1801.05807 [hep-th]}}.

\bibitem{Mazac:2019shk}
D.~Maz\'{a}\v{c}, L.~Rastelli, and X.~Zhou, ``{A Basis of Analytic Functionals
  for CFTs in General Dimension},''
\href{http://arxiv.org/abs/1910.12855}{{\ttfamily arXiv:1910.12855 [hep-th]}}.
%%CITATION = ARXIV:1910.12855;%%.

\bibitem{Mazac:2016qev}
D.~Mazac, ``{Analytic bounds and emergence of AdS$_{2}$ physics from the
  conformal bootstrap},'' \href{http://dx.doi.org/10.1007/JHEP04(2017)146}{{\em
  JHEP} {\bfseries 04} (2017) 146},
\href{http://arxiv.org/abs/1611.10060}{{\ttfamily arXiv:1611.10060 [hep-th]}}.
%%CITATION = ARXIV:1611.10060;%%.

\bibitem{Mazac:2018mdx}
D.~Mazac and M.~F. Paulos, ``{The Analytic Functional Bootstrap I: 1D CFTs and
  2D S-Matrices},''
\href{http://arxiv.org/abs/1803.10233}{{\ttfamily arXiv:1803.10233 [hep-th]}}.
%%CITATION = ARXIV:1803.10233;%%.

\bibitem{Mazac:2018ycv}
D.~Mazac and M.~F. Paulos, ``{The Analytic Functional Bootstrap II: Natural
  Bases for the Crossing Equation},''
\href{http://arxiv.org/abs/1811.10646}{{\ttfamily arXiv:1811.10646 [hep-th]}}.
%%CITATION = ARXIV:1811.10646;%%.

\bibitem{Mazac:2018qmi}
D.~Mazac, ``{A Crossing-Symmetric OPE Inversion Formula},''
\href{http://arxiv.org/abs/1812.02254}{{\ttfamily arXiv:1812.02254 [hep-th]}}.
%%CITATION = ARXIV:1812.02254;%%.

\bibitem{Hartman:2019pcd}
T.~Hartman, D.~Maz{\'a}{\v c}, and L.~Rastelli, ``{Sphere Packing and Quantum
  Gravity},''
\href{http://arxiv.org/abs/1905.01319}{{\ttfamily arXiv:1905.01319 [hep-th]}}.
%%CITATION = ARXIV:1905.01319;%%.

\bibitem{Kaviraj:2018tfd}
A.~Kaviraj and M.~F. Paulos, ``{The Functional Bootstrap for Boundary CFT},''
  \href{http://dx.doi.org/10.1007/JHEP04(2020)135}{{\em JHEP} {\bfseries 04}
  (2020) 135}, \href{http://arxiv.org/abs/1812.04034}{{\ttfamily
  arXiv:1812.04034 [hep-th]}}.

\bibitem{Mazac:2018biw}
D.~Maz\'{a}\v{c}, L.~Rastelli, and X.~Zhou, ``{An analytic approach to
  BCFT$_{d}$},'' \href{http://dx.doi.org/10.1007/JHEP12(2019)004}{{\em JHEP}
  {\bfseries 12} (2019) 004}, \href{http://arxiv.org/abs/1812.09314}{{\ttfamily
  arXiv:1812.09314 [hep-th]}}.

\bibitem{Paulos:2019gtx}
M.~F. Paulos, ``{Analytic Functional Bootstrap for CFTs in $d>1$},''
\href{http://arxiv.org/abs/1910.08563}{{\ttfamily arXiv:1910.08563 [hep-th]}}.
%%CITATION = ARXIV:1910.08563;%%.

\bibitem{Mack:2009mi}
G.~Mack, ``{D-independent representation of Conformal Field Theories in D
  dimensions via transformation to auxiliary Dual Resonance Models. Scalar
  amplitudes},''
\href{http://arxiv.org/abs/0907.2407}{{\ttfamily arXiv:0907.2407 [hep-th]}}.
%%CITATION = ARXIV:0907.2407;%%.

\bibitem{Penedones:2010ue}
J.~Penedones, ``{Writing CFT correlation functions as AdS scattering
  amplitudes},'' \href{http://dx.doi.org/10.1007/JHEP03(2011)025}{{\em JHEP}
  {\bfseries 03} (2011) 025},
\href{http://arxiv.org/abs/1011.1485}{{\ttfamily arXiv:1011.1485 [hep-th]}}.
%%CITATION = ARXIV:1011.1485;%%.

\bibitem{Fitzpatrick:2011ia}
A.~Fitzpatrick, J.~Kaplan, J.~Penedones, S.~Raju, and B.~C. van Rees, ``{A
  Natural Language for AdS/CFT Correlators},''
  \href{http://dx.doi.org/10.1007/JHEP11(2011)095}{{\em JHEP} {\bfseries 11}
  (2011) 095}, \href{http://arxiv.org/abs/1107.1499}{{\ttfamily arXiv:1107.1499
  [hep-th]}}.

\bibitem{Penedones:2019tng}
J.~Penedones, J.~A. Silva, and A.~Zhiboedov, ``{Nonperturbative Mellin
  Amplitudes: Existence, Properties, Applications},''
\href{http://arxiv.org/abs/1912.11100}{{\ttfamily arXiv:1912.11100 [hep-th]}}.
%%CITATION = ARXIV:1912.11100;%%.

\bibitem{Sleight:2019ive}
C.~Sleight and M.~Taronna, ``{The Unique Polyakov Blocks},''
  \href{http://arxiv.org/abs/1912.07998}{{\ttfamily arXiv:1912.07998
  [hep-th]}}.

\bibitem{Gopakumar:2016wkt}
R.~Gopakumar, A.~Kaviraj, K.~Sen, and A.~Sinha, ``{Conformal Bootstrap in
  Mellin Space},'' \href{http://dx.doi.org/10.1103/PhysRevLett.118.081601}{{\em
  Phys. Rev. Lett.} {\bfseries 118} no.~8, (2017) 081601},
\href{http://arxiv.org/abs/1609.00572}{{\ttfamily arXiv:1609.00572 [hep-th]}}.
%%CITATION = ARXIV:1609.00572;%%.

\bibitem{Dey:2016mcs}
P.~Dey, A.~Kaviraj, and A.~Sinha, ``{Mellin space bootstrap for global
  symmetry},'' \href{http://dx.doi.org/10.1007/JHEP07(2017)019}{{\em JHEP}
  {\bfseries 07} (2017) 019}, \href{http://arxiv.org/abs/1612.05032}{{\ttfamily
  arXiv:1612.05032 [hep-th]}}.

\bibitem{Dey:2017fab}
P.~Dey, K.~Ghosh, and A.~Sinha, ``{Simplifying large spin bootstrap in Mellin
  space},'' \href{http://dx.doi.org/10.1007/JHEP01(2018)152}{{\em JHEP}
  {\bfseries 01} (2018) 152},
\href{http://arxiv.org/abs/1709.06110}{{\ttfamily arXiv:1709.06110 [hep-th]}}.
%%CITATION = ARXIV:1709.06110;%%.

\bibitem{Dey:2017oim}
P.~Dey and A.~Kaviraj, ``{Towards a Bootstrap approach to higher orders of
  epsilon expansion},'' \href{http://dx.doi.org/10.1007/JHEP02(2018)153}{{\em
  JHEP} {\bfseries 02} (2018) 153},
  \href{http://arxiv.org/abs/1711.01173}{{\ttfamily arXiv:1711.01173
  [hep-th]}}.

\bibitem{Gopakumar:2018xqi}
R.~Gopakumar and A.~Sinha, ``{On the Polyakov-Mellin bootstrap},''
\href{http://arxiv.org/abs/1809.10975}{{\ttfamily arXiv:1809.10975 [hep-th]}}.
%%CITATION = ARXIV:1809.10975;%%.

\bibitem{Polyakov:1974gs}
A.~M. Polyakov, ``{Nonhamiltonian approach to conformal quantum field
  theory},'' {\em Zh. Eksp. Teor. Fiz.} {\bfseries 66} (1974) 23--42.
[Sov. Phys. JETP39,9(1974)].
%%CITATION = ZETFA,66,23;%%.

\bibitem{CHMRSD}
S.~Caron-Huot, D.~Maz\'{a}\v{c}, L.~Rastelli, and D.~Simmons-Duffin, ``{work in
  progress}.''.

\bibitem{Kologlu:2019bco}
M.~Kologlu, P.~Kravchuk, D.~Simmons-Duffin, and A.~Zhiboedov, ``{Shocks,
  Superconvergence, and a Stringy Equivalence Principle},''
\href{http://arxiv.org/abs/1904.05905}{{\ttfamily arXiv:1904.05905 [hep-th]}}.
%%CITATION = ARXIV:1904.05905;%%.

\bibitem{Pappadopulo:2012jk}
D.~Pappadopulo, S.~Rychkov, J.~Espin, and R.~Rattazzi, ``{OPE Convergence in
  Conformal Field Theory},''
  \href{http://dx.doi.org/10.1103/PhysRevD.86.105043}{{\em Phys. Rev. D}
  {\bfseries 86} (2012) 105043},
  \href{http://arxiv.org/abs/1208.6449}{{\ttfamily arXiv:1208.6449 [hep-th]}}.

\bibitem{Costa:2012cb}
M.~S. Costa, V.~Goncalves, and J.~Penedones, ``{Conformal Regge theory},''
  \href{http://dx.doi.org/10.1007/JHEP12(2012)091}{{\em JHEP} {\bfseries 12}
  (2012) 091},
\href{http://arxiv.org/abs/1209.4355}{{\ttfamily arXiv:1209.4355 [hep-th]}}.
%%CITATION = ARXIV:1209.4355;%%.

\bibitem{Heemskerk:2009pn}
I.~Heemskerk, J.~Penedones, J.~Polchinski, and J.~Sully, ``{Holography from
  Conformal Field Theory},''
  \href{http://dx.doi.org/10.1088/1126-6708/2009/10/079}{{\em JHEP} {\bfseries
  10} (2009) 079}, \href{http://arxiv.org/abs/0907.0151}{{\ttfamily
  arXiv:0907.0151 [hep-th]}}.

\bibitem{Dolan:2003hv}
F.~Dolan and H.~Osborn, ``{Conformal partial waves and the operator product
  expansion},'' \href{http://dx.doi.org/10.1016/j.nuclphysb.2003.11.016}{{\em
  Nucl. Phys. B} {\bfseries 678} (2004) 491--507},
  \href{http://arxiv.org/abs/hep-th/0309180}{{\ttfamily arXiv:hep-th/0309180}}.

\bibitem{Beem:2013qxa}
C.~Beem, L.~Rastelli, and B.~C. van Rees, ``{The $\mathcal N=4$ Superconformal
  Bootstrap},'' \href{http://dx.doi.org/10.1103/PhysRevLett.111.071601}{{\em
  Phys. Rev. Lett.} {\bfseries 111} (2013) 071601},
  \href{http://arxiv.org/abs/1304.1803}{{\ttfamily arXiv:1304.1803 [hep-th]}}.

\bibitem{Caron-Huot:2018kta}
S.~Caron-Huot and A.-K. Trinh, ``{All tree-level correlators in
  AdS$_{5}$$\times$S$_{5}$ supergravity: hidden ten-dimensional conformal
  symmetry},'' \href{http://dx.doi.org/10.1007/JHEP01(2019)196}{{\em JHEP}
  {\bfseries 01} (2019) 196},
\href{http://arxiv.org/abs/1809.09173}{{\ttfamily arXiv:1809.09173 [hep-th]}}.
%%CITATION = ARXIV:1809.09173;%%.

\bibitem{Fitzpatrick:2011dm}
A.~Fitzpatrick and J.~Kaplan, ``{Unitarity and the Holographic S-Matrix},''
  \href{http://dx.doi.org/10.1007/JHEP10(2012)032}{{\em JHEP} {\bfseries 10}
  (2012) 032}, \href{http://arxiv.org/abs/1112.4845}{{\ttfamily arXiv:1112.4845
  [hep-th]}}.

\bibitem{Gopakumar:2016cpb}
R.~Gopakumar, A.~Kaviraj, K.~Sen, and A.~Sinha, ``{A Mellin space approach to
  the conformal bootstrap},''
  \href{http://dx.doi.org/10.1007/JHEP05(2017)027}{{\em JHEP} {\bfseries 05}
  (2017) 027},
\href{http://arxiv.org/abs/1611.08407}{{\ttfamily arXiv:1611.08407 [hep-th]}}.
%%CITATION = ARXIV:1611.08407;%%.

\bibitem{Sleight:2018ryu}
C.~Sleight and M.~Taronna, ``{Anomalous Dimensions from Crossing Kernels},''
  \href{http://dx.doi.org/10.1007/JHEP11(2018)089}{{\em JHEP} {\bfseries 11}
  (2018) 089},
\href{http://arxiv.org/abs/1807.05941}{{\ttfamily arXiv:1807.05941 [hep-th]}}.
%%CITATION = ARXIV:1807.05941;%%.

\bibitem{Kravchuk:2020scc}
P.~Kravchuk, J.~Qiao, and S.~Rychkov, ``{Distributions in CFT. Part I.
  Cross-ratio space},'' \href{http://dx.doi.org/10.1007/JHEP05(2020)137}{{\em
  JHEP} {\bfseries 05} (2020) 137},
  \href{http://arxiv.org/abs/2001.08778}{{\ttfamily arXiv:2001.08778
  [hep-th]}}.

\bibitem{Hartman:2015lfa}
T.~Hartman, S.~Jain, and S.~Kundu, ``{Causality Constraints in Conformal Field
  Theory},'' \href{http://dx.doi.org/10.1007/JHEP05(2016)099}{{\em JHEP}
  {\bfseries 05} (2016) 099},
\href{http://arxiv.org/abs/1509.00014}{{\ttfamily arXiv:1509.00014 [hep-th]}}.
%%CITATION = ARXIV:1509.00014;%%.

\bibitem{Hofman:2008ar}
D.~M. Hofman and J.~Maldacena, ``{Conformal collider physics: Energy and charge
  correlations},'' \href{http://dx.doi.org/10.1088/1126-6708/2008/05/012}{{\em
  JHEP} {\bfseries 05} (2008) 012},
  \href{http://arxiv.org/abs/0803.1467}{{\ttfamily arXiv:0803.1467 [hep-th]}}.

\bibitem{Kologlu:2019mfz}
M.~Kologlu, P.~Kravchuk, D.~Simmons-Duffin, and A.~Zhiboedov, ``{The light-ray
  OPE and conformal colliders},''
  \href{http://arxiv.org/abs/1905.01311}{{\ttfamily arXiv:1905.01311
  [hep-th]}}.

\bibitem{Henn:2019gkr}
J.~Henn, E.~Sokatchev, K.~Yan, and A.~Zhiboedov, ``{Energy-energy correlation
  in $N$=4 super Yang-Mills theory at next-to-next-to-leading order},''
  \href{http://dx.doi.org/10.1103/PhysRevD.100.036010}{{\em Phys. Rev. D}
  {\bfseries 100} no.~3, (2019) 036010},
  \href{http://arxiv.org/abs/1903.05314}{{\ttfamily arXiv:1903.05314
  [hep-th]}}.

\bibitem{Rychkov:2016mrc}
S.~Rychkov, D.~Simmons-Duffin, and B.~Zan, ``{Non-gaussianity of the critical
  3d Ising model},'' \href{http://dx.doi.org/10.21468/SciPostPhys.2.1.001}{{\em
  SciPost Phys.} {\bfseries 2} no.~1, (2017) 001},
  \href{http://arxiv.org/abs/1612.02436}{{\ttfamily arXiv:1612.02436
  [hep-th]}}.

\bibitem{Adams:2006sv}
A.~Adams, N.~Arkani-Hamed, S.~Dubovsky, A.~Nicolis, and R.~Rattazzi,
  ``{Causality, analyticity and an IR obstruction to UV completion},''
  \href{http://dx.doi.org/10.1088/1126-6708/2006/10/014}{{\em JHEP} {\bfseries
  10} (2006) 014}, \href{http://arxiv.org/abs/hep-th/0602178}{{\ttfamily
  arXiv:hep-th/0602178}}.

\bibitem{Hoffmann:2000mx}
L.~Hoffmann, A.~C. Petkou, and W.~Ruhl, ``{Aspects of the conformal operator
  product expansion in AdS / CFT correspondence},''
  \href{http://dx.doi.org/10.4310/ATMP.2000.v4.n3.a3}{{\em Adv. Theor. Math.
  Phys.} {\bfseries 4} (2002) 571--615},
  \href{http://arxiv.org/abs/hep-th/0002154}{{\ttfamily arXiv:hep-th/0002154}}.

\bibitem{Dolan:2000ut}
F.~Dolan and H.~Osborn, ``{Conformal four point functions and the operator
  product expansion},''
  \href{http://dx.doi.org/10.1016/S0550-3213(01)00013-X}{{\em Nucl. Phys. B}
  {\bfseries 599} (2001) 459--496},
  \href{http://arxiv.org/abs/hep-th/0011040}{{\ttfamily arXiv:hep-th/0011040}}.

\bibitem{Sleight:2017fpc}
C.~Sleight and M.~Taronna, ``{Spinning Witten Diagrams},''
  \href{http://dx.doi.org/10.1007/JHEP06(2017)100}{{\em JHEP} {\bfseries 06}
  (2017) 100}, \href{http://arxiv.org/abs/1702.08619}{{\ttfamily
  arXiv:1702.08619 [hep-th]}}.

\bibitem{Sleight:2018epi}
C.~Sleight and M.~Taronna, ``{Spinning Mellin Bootstrap: Conformal Partial
  Waves, Crossing Kernels and Applications},''
  \href{http://dx.doi.org/10.1002/prop.201800038}{{\em Fortsch. Phys.}
  {\bfseries 66} no.~8-9, (2018) 1800038},
\href{http://arxiv.org/abs/1804.09334}{{\ttfamily arXiv:1804.09334 [hep-th]}}.
%%CITATION = ARXIV:1804.09334;%%.

\bibitem{Liu:2018jhs}
J.~Liu, E.~Perlmutter, V.~Rosenhaus, and D.~Simmons-Duffin, ``{$d$-dimensional
  SYK, AdS Loops, and $6j$ Symbols},''
  \href{http://dx.doi.org/10.1007/JHEP03(2019)052}{{\em JHEP} {\bfseries 03}
  (2019) 052}, \href{http://arxiv.org/abs/1808.00612}{{\ttfamily
  arXiv:1808.00612 [hep-th]}}.

\bibitem{Zhou:2018sfz}
X.~Zhou, ``{Recursion Relations in Witten Diagrams and Conformal Partial
  Waves},'' \href{http://dx.doi.org/10.1007/JHEP05(2019)006}{{\em JHEP}
  {\bfseries 05} (2019) 006},
\href{http://arxiv.org/abs/1812.01006}{{\ttfamily arXiv:1812.01006 [hep-th]}}.
%%CITATION = ARXIV:1812.01006;%%.

\bibitem{Sleight:2019hfp}
C.~Sleight and M.~Taronna, ``{Bootstrapping Inflationary Correlators in Mellin
  Space},'' \href{http://dx.doi.org/10.1007/JHEP02(2020)098}{{\em JHEP}
  {\bfseries 02} (2020) 098}, \href{http://arxiv.org/abs/1907.01143}{{\ttfamily
  arXiv:1907.01143 [hep-th]}}.

\bibitem{Sleight:2020obc}
C.~Sleight and M.~Taronna, ``{From AdS to dS Exchanges: Spectral
  Representation, Mellin Amplitudes and Crossing},''
  \href{http://arxiv.org/abs/2007.09993}{{\ttfamily arXiv:2007.09993
  [hep-th]}}.

\bibitem{Sinha:2020win}
A.~Sinha and A.~Zahed, ``{Crossing Symmetric Dispersion Relations in QFTs},''
  \href{http://dx.doi.org/10.1103/PhysRevLett.126.181601}{{\em Phys. Rev.
  Lett.} {\bfseries 126} no.~18, (2021) 181601},
  \href{http://arxiv.org/abs/2012.04877}{{\ttfamily arXiv:2012.04877
  [hep-th]}}.

\bibitem{Gopakumar:2021dvg}
R.~Gopakumar, A.~Sinha, and A.~Zahed, ``{Crossing Symmetric Dispersion
  Relations for Mellin Amplitudes},''
  \href{http://arxiv.org/abs/2101.09017}{{\ttfamily arXiv:2101.09017
  [hep-th]}}.

\bibitem{Auberson:1973pf}
G.~Auberson and N.~N. Khuri, ``{Rigorous parametric dispersion representation
  with three-channel symmetry},''
  \href{http://dx.doi.org/10.1103/PhysRevD.6.2953}{{\em Phys. Rev. D}
  {\bfseries 6} (1972) 2953--2966}.

\bibitem{Alday:2016mxe}
L.~F. Alday and A.~Bissi, ``{Crossing symmetry and Higher spin towers},''
  \href{http://dx.doi.org/10.1007/JHEP12(2017)118}{{\em JHEP} {\bfseries 12}
  (2017) 118}, \href{http://arxiv.org/abs/1603.05150}{{\ttfamily
  arXiv:1603.05150 [hep-th]}}.

\bibitem{ElShowk:2012ht}
S.~El-Showk, M.~F. Paulos, D.~Poland, S.~Rychkov, D.~Simmons-Duffin, and
  A.~Vichi, ``{Solving the 3D Ising Model with the Conformal Bootstrap},''
  \href{http://dx.doi.org/10.1103/PhysRevD.86.025022}{{\em Phys. Rev. D}
  {\bfseries 86} (2012) 025022},
  \href{http://arxiv.org/abs/1203.6064}{{\ttfamily arXiv:1203.6064 [hep-th]}}.

\bibitem{Cornalba:2006xm}
L.~Cornalba, M.~S. Costa, J.~Penedones, and R.~Schiappa, ``{Eikonal
  Approximation in AdS/CFT: Conformal Partial Waves and Finite N Four-Point
  Functions},'' \href{http://dx.doi.org/10.1016/j.nuclphysb.2007.01.007}{{\em
  Nucl. Phys. B} {\bfseries 767} (2007) 327--351},
  \href{http://arxiv.org/abs/hep-th/0611123}{{\ttfamily arXiv:hep-th/0611123}}.

\bibitem{Poland:2010wg}
D.~Poland and D.~Simmons-Duffin, ``{Bounds on 4D Conformal and Superconformal
  Field Theories},'' \href{http://dx.doi.org/10.1007/JHEP05(2011)017}{{\em
  JHEP} {\bfseries 05} (2011) 017},
  \href{http://arxiv.org/abs/1009.2087}{{\ttfamily arXiv:1009.2087 [hep-th]}}.

\bibitem{ElShowk:2012hu}
S.~El-Showk and M.~F. Paulos, ``{Bootstrapping Conformal Field Theories with
  the Extremal Functional Method},''
  \href{http://dx.doi.org/10.1103/PhysRevLett.111.241601}{{\em Phys. Rev.
  Lett.} {\bfseries 111} no.~24, (2013) 241601},
  \href{http://arxiv.org/abs/1211.2810}{{\ttfamily arXiv:1211.2810 [hep-th]}}.

\bibitem{El-Showk:2014dwa}
S.~El-Showk, M.~F. Paulos, D.~Poland, S.~Rychkov, D.~Simmons-Duffin, and
  A.~Vichi, ``{Solving the 3d Ising Model with the Conformal Bootstrap II.
  c-Minimization and Precise Critical Exponents},''
  \href{http://dx.doi.org/10.1007/s10955-014-1042-7}{{\em J. Stat. Phys.}
  {\bfseries 157} (2014) 869}, \href{http://arxiv.org/abs/1403.4545}{{\ttfamily
  arXiv:1403.4545 [hep-th]}}.

\bibitem{Rychkov:2017tpc}
J.~Qiao and S.~Rychkov, ``{Cut-touching linear functionals in the conformal
  bootstrap},'' \href{http://dx.doi.org/10.1007/JHEP06(2017)076}{{\em JHEP}
  {\bfseries 06} (2017) 076}, \href{http://arxiv.org/abs/1705.01357}{{\ttfamily
  arXiv:1705.01357 [hep-th]}}.

\bibitem{Simmons-Duffin:2015qma}
D.~Simmons-Duffin, ``{A Semidefinite Program Solver for the Conformal
  Bootstrap},'' \href{http://dx.doi.org/10.1007/JHEP06(2015)174}{{\em JHEP}
  {\bfseries 06} (2015) 174}, \href{http://arxiv.org/abs/1502.02033}{{\ttfamily
  arXiv:1502.02033 [hep-th]}}.

\bibitem{Landry:2019qug}
W.~Landry and D.~Simmons-Duffin, ``{Scaling the semidefinite program solver
  SDPB},'' \href{http://arxiv.org/abs/1909.09745}{{\ttfamily arXiv:1909.09745
  [hep-th]}}.

\bibitem{Paulos:2019fkw}
M.~F. Paulos and B.~Zan, ``{A functional approach to the numerical conformal
  bootstrap},'' \href{http://arxiv.org/abs/1904.03193}{{\ttfamily
  arXiv:1904.03193 [hep-th]}}.

\bibitem{Gillioz:2019iye}
M.~Gillioz, X.~Lu, M.~A. Luty, and G.~Mikaberidze, ``{Convergent Momentum-Space
  OPE and Bootstrap Equations in Conformal Field Theory},''
  \href{http://dx.doi.org/10.1007/JHEP03(2020)102}{{\em JHEP} {\bfseries 03}
  (2020) 102}, \href{http://arxiv.org/abs/1912.05550}{{\ttfamily
  arXiv:1912.05550 [hep-th]}}.

\bibitem{Gillioz:2020mdd}
M.~Gillioz, M.~Meineri, and J.~Penedones, ``{A Scattering Amplitude in
  Conformal Field Theory},'' \href{http://arxiv.org/abs/2003.07361}{{\ttfamily
  arXiv:2003.07361 [hep-th]}}.

\bibitem{CutkoskyFuture}
D.~Meltzer and A.~Sivaramakrishnan, ``{Cutkosky Rules and Unitarity in
  AdS/CFT},'' {\em {to appear}} .

\bibitem{Iliesiu:2018zlz}
L.~Iliesiu, M.~Kolo\u{g}lu, and D.~Simmons-Duffin, ``{Bootstrapping the 3d
  Ising model at finite temperature},''
  \href{http://dx.doi.org/10.1007/JHEP12(2019)072}{{\em JHEP} {\bfseries 12}
  (2019) 072}, \href{http://arxiv.org/abs/1811.05451}{{\ttfamily
  arXiv:1811.05451 [hep-th]}}.

\bibitem{Qiao:2017xif}
J.~Qiao and S.~Rychkov, ``{A tauberian theorem for the conformal bootstrap},''
  \href{http://dx.doi.org/10.1007/JHEP12(2017)119}{{\em JHEP} {\bfseries 12}
  (2017) 119},
\href{http://arxiv.org/abs/1709.00008}{{\ttfamily arXiv:1709.00008 [hep-th]}}.
%%CITATION = ARXIV:1709.00008;%%.

\bibitem{Mukhametzhanov:2018zja}
B.~Mukhametzhanov and A.~Zhiboedov, ``{Analytic Euclidean Bootstrap},''
\href{http://arxiv.org/abs/1808.03212}{{\ttfamily arXiv:1808.03212 [hep-th]}}.
%%CITATION = ARXIV:1808.03212;%%.

\bibitem{Mukhametzhanov:2019pzy}
B.~Mukhametzhanov and A.~Zhiboedov, ``{Modular Invariance, Tauberian Theorems,
  and Microcanonical Entropy},''
\href{http://arxiv.org/abs/1904.06359}{{\ttfamily arXiv:1904.06359 [hep-th]}}.
%%CITATION = ARXIV:1904.06359;%%.

\bibitem{Pal:2019zzr}
S.~Pal and Z.~Sun, ``{Tauberian-Cardy formula with spin},''
  \href{http://dx.doi.org/10.1007/JHEP01(2020)135}{{\em JHEP} {\bfseries 01}
  (2020) 135}, \href{http://arxiv.org/abs/1910.07727}{{\ttfamily
  arXiv:1910.07727 [hep-th]}}.

\bibitem{Mukhametzhanov:2020swe}
B.~Mukhametzhanov and S.~Pal, ``{Beurling-Selberg Extremization and Modular
  Bootstrap at High Energies},''
  \href{http://dx.doi.org/10.21468/SciPostPhys.8.6.088}{{\em SciPost Phys.}
  {\bfseries 8} no.~6, (2020) 088},
  \href{http://arxiv.org/abs/2003.14316}{{\ttfamily arXiv:2003.14316
  [hep-th]}}.

\bibitem{Karateev:2017jgd}
D.~Karateev, P.~Kravchuk, and D.~Simmons-Duffin, ``{Weight Shifting Operators
  and Conformal Blocks},''
  \href{http://dx.doi.org/10.1007/JHEP02(2018)081}{{\em JHEP} {\bfseries 02}
  (2018) 081}, \href{http://arxiv.org/abs/1706.07813}{{\ttfamily
  arXiv:1706.07813 [hep-th]}}.

\bibitem{Karateev:2018oml}
D.~Karateev, P.~Kravchuk, and D.~Simmons-Duffin, ``{Harmonic Analysis and Mean
  Field Theory},'' \href{http://dx.doi.org/10.1007/JHEP10(2019)217}{{\em JHEP}
  {\bfseries 10} (2019) 217}, \href{http://arxiv.org/abs/1809.05111}{{\ttfamily
  arXiv:1809.05111 [hep-th]}}.

\bibitem{Dolan:2011dv}
F.~Dolan and H.~Osborn, ``{Conformal Partial Waves: Further Mathematical
  Results},'' \href{http://arxiv.org/abs/1108.6194}{{\ttfamily arXiv:1108.6194
  [hep-th]}}.

\bibitem{Zhou:2020ptb}
X.~Zhou, ``{How to Succeed at Witten Diagram Recursions without Really
  Trying},'' \href{http://arxiv.org/abs/2005.03031}{{\ttfamily arXiv:2005.03031
  [hep-th]}}.

\bibitem{Hogervorst:2016hal}
M.~Hogervorst, ``{Dimensional Reduction for Conformal Blocks},''
  \href{http://dx.doi.org/10.1007/JHEP09(2016)017}{{\em JHEP} {\bfseries 09}
  (2016) 017}, \href{http://arxiv.org/abs/1604.08913}{{\ttfamily
  arXiv:1604.08913 [hep-th]}}.

\end{thebibliography}\endgroup
%%%%%%%%%%%%%%%%%%%%%%%%%%%%%%%%

\end{document}